\documentclass[titlepage,11pt]{article}
\usepackage[USenglish]{babel}
\usepackage{multicol}
\usepackage{a4wide}
\usepackage[all]{xy}
\usepackage{amssymb}
\usepackage{amsmath}
\usepackage{latexsym}
\usepackage{pifont}
\usepackage{pdfsync}
\usepackage{euler}
\usepackage{color}
\usepackage{enumerate}
\usepackage{float,fancyvrb,verbatim,multicol}
\usepackage{ifthen}
\usepackage[parfill]{parskip}    
\usepackage{xr}
\usepackage{makeidx}
\usepackage[utf8]{inputenc} 
\title{Categories and all that --- A Tutorial}
\author{Ernst-Erich Doberkat\\Chair for Software
  Technology\\Technische Universit\"at Dortmund\\\texttt{doberkat@acm.org}}
\date{\today}
\newcommand{\labelImpl}[2]{\ensuremath{\ref{#1}~\Rightarrow~\ref{#2}}}

\newcommand{\Klasse}[2]{\left[#1\right]_{#2}}
\newcommand{\Faktor}[2]{{#1}/{#2}}
\newcommand{\fMap}[1]{\eta_{#1}}
\newcommand{\Bild}[2]{{#1}\left[#2\right]}
\newcommand{\InvBild}[2]{\Bild{#1^{-1}}{#2}}
\newcommand{\Kern}[1]{\mathsf{ker}\left(#1\right)}
\newcommand{\Folge}[1]{(#1_n)_{n \in \Nat}}

\newcommand{\theTheory}[2]{Th_{#1}({#2})}
\newcommand{\supp}{\mathsf{supp}}

\newcommand{\spaceFont}[1]{\mathfrak{#1}}

\newcommand{\Prob}[1]{\spaceFont{P}\left(#1\right)}
\newcommand{\SubProb}[1]{\spaceFont{S}\left(#1\right)}
\newcommand{\Functor}[2]{\ensuremath{\spaceFont{#1}\left(#2\right)}}
\newcommand{\Category}[1]{\ensuremath{\mathbf{#1}}}
\newcommand{\SubProbSenza}{\spaceFont{S}}
\newcommand{\ProbSenza}{\spaceFont{P}}
\newcommand{\PowerSet}[1]{\ensuremath{\mathcal{P}\left(#1\right)}}
\newcommand{\PowerSenza}{\ensuremath{\mathcal{P}}}
\newcommand{\FunctorSenza}[1]{\ensuremath{\spaceFont{#1}}}

\newcommand{\Borel}[1]{\ensuremath{{\mathcal B}(#1)}}

\edef\LinkeKlammer{\lbrack\!\lbrack}
\edef\RechteKlammer{\rbrack\!\rbrack}
\newcommand{\Gilt}[1][\phi]{\ensuremath{\LinkeKlammer#1\RechteKlammer}}

\newcommand{\Trans}{\rightsquigarrow}

\newcommand{\Eins}[1]{\mathtt{1}\!\!{\mathtt{l}}_{#1}}

\newtheorem{definition}{Definition}[section]
\newcommand{\BeginDefinition}[1]{%
  \begin{definition}\label{#1}
}
\newcommand{\EndDefinition}{\end{definition}}

\newtheorem{example}[definition]{Example}
\newcommand{\BeginExample}[1]{%
  \begin{example}\label{#1}\rm
}
\newcommand{\EndExample}{--- \end{example}}

\newtheorem{observation}[definition]{Observation}
\newcommand{\BeginObservation}[1]{
  \begin{observation}\label{#1}\rm
}
\newcommand{\EndObservation}{--- \end{observation}}

\newtheorem{theorem}[definition]{Theorem}
\newcommand{\BeginTheorem}[1]{%
  \begin{theorem}\label{#1}
}
\newcommand{\EndTheorem}{\end{theorem}}

\newtheorem{corollary}[definition]{Corollary}
\newcommand{\BeginCorollary}[1]{
  \begin{corollary}\label{#1}
}

\newtheorem{proposition}[definition]{Proposition}
\newcommand{\BeginProposition}[1]{%
  \begin{proposition}\label{#1}
}
\newcommand{\EndProposition}{\end{proposition}}

\newcommand{\EndCorollary}{\end{corollary}}
\newtheorem{lemma}[definition]{Lemma}
\newcommand{\BeginLemma}[1]{%
  \begin{lemma}\label{#1}
}
\newcommand{\EndLemma}{\end{lemma}}

\newtheorem{claim}{Claim}
\newcommand{\BeginClaim}[1]{%
  \begin{claim}\label{#1}
}
\newcommand{\EndClaim}{\end{claim}}

\newenvironment{proof}{\textbf{Proof\ }}{\ensuremath{\QED}}
\newcommand{\BeginProof}{\begin{proof}}
\newcommand{\EndProof}{\end{proof}}

\newenvironment{remark}{\textbf{Remark:\ }}{}
\newcommand{\BeginRemark}{\begin{remark}}
\newcommand{\EndRemark}{\QED\end{remark}}
\newcommand{\QED}{%
\ensuremath{\dashv}
}

\newcommand{\pReal}{\mathbb{R}_{+}}
\newcommand{\Nat}{\mathbb{N}}

\newcommand{\unit}{\ensuremath{\eta}}
\newcommand{\multi}{\ensuremath{\mu}}

\def\Ganz{\mathbb{Z}}
\makeatletter
\def\@axx#1{\ensuremath{\mathbb{(#1)}}}
\def\AC{\@axx{AC}}
\def\WO{\@axx{WO}}
\def\ZL{\@axx{ZL}}
\def\MP{\@axx{MP}}
\def\MI{\@axx{MI}}
\def\AD{\@axx{AD}}
\makeatother

\newcommand{\isEquiv}[3]{\ensuremath{{#1}\ {#3}\ {#2}}}
\newenvironment{theExercises}
{\begin{exercise}\rm}
{\end{exercise}}
\newtheorem{exercise}{Exercise}
\newcommand{\BeginExercise}[1]{%
  \begin{theExercises}\label{#1}
}
\newcommand{\EndExercise}{\end{theExercises}}
\newcommand{\Exercise}[2]{\BeginExercise{#1}{#2}\EndExercise}
\newcommand{\Solution}[2]{\paragraph{Solution for
    Exercise~\ref{#1}}{#2}}
\def\endEx{{\Large\ding{44}}}
\renewcommand{\EndExample}{\endEx \end{example}}

\def\smallBox#1{\parbox[t]{.49\linewidth}{\vspace{1pt}#1}}

\def\CatFont{\mathbf}
\def\Category#1{\ensuremath{\CatFont{#1}}}
\renewcommand{\spaceFont}[1]{\CatFont{#1}}
\def\SubProbSenza{\ensuremath{\mathbb{S}}}
\def\catK{\Category{K}}
\def\catL{\Category{L}}
\def\catM{\Category{M}}

\def\catC{\Category{C}}
\def\catD{\Category{D}}
\def\catSET{\Category{Set}}
\def\catMeas{\Category{Meas}}
\def\obj#1{\ensuremath{|#1|}}
\def\objK{\obj{\catK}}
\def\objL{\obj{\catL}}
\def\hom#1{\ensuremath{\mathrm{hom_{#1}}}}
\newcommand\homK[1]{\hom{\catK}(#1)}
\newcommand\homL[1]{\hom{\catL}(#1)}
\def\funD{\FunctorSenza{D}}
\def\funF{\FunctorSenza{F}}
\def\funG{\FunctorSenza{G}}
\def\funH{\FunctorSenza{H}}
\def\funS{\FunctorSenza{S}}
\def\funR{\FunctorSenza{R}}
\def\funT{\FunctorSenza{T}}
\def\funV{\FunctorSenza{V}}
\newcommand\homSET[1]{\hom{\catSET}(#1)}
\renewcommand{\Prob}[1]{\SubProbSenza\left(#1\right)}

\def\frameFont{\mathfrak}
\def\fmF{\frameFont{F}}
\def\fmG{\frameFont{G}}
\def\fmM{\frameFont{M}}
\def\fmN{\frameFont{N}}
\def\calN{{\cal N}}
\def\calM{{\cal M}}
\def\DefSect{
\ifthenelse{\boolean{isBook}}{
\def\Section{\chapter}
\def\Subsection{\section}
\def\Subsubsection{\subsection}
}{
\def\Section{\section}
\def\Subsection{\subsection}
\def\Subsubsection{\subsubsection}
}
}
\def\lift{\ensuremath{\mathbb{L}}}
\newboolean{isBook}
\setboolean{isBook}{false}
\DefSect

\def\phi{\varphi}
\usepackage{fancyhdr}
\makeindex
\begin{document}
\maketitle
\tableofcontents\newpage
\def\Folder{Categs}
\Section{Categories}
\label{sec:categories}

Many areas of Mathematics show surprising structural similarities,
which suggests that it might be interesting and helpful to focus on an
abstract view, unifying concepts. This abstract view looks at the
mathematical objects from the outside and studies the relationship
between them, for example groups (as objects) and homomorphisms (as an
indicator of their relationship), or topological spaces together with
continuous maps, or ordered sets with monotone maps, the list could be
extended. It leads to the general notion of a category. A category is
based on a class of objects together with morphisms for each two
objects, which can be composed; composition follows some laws which
are considered evident and natural. 

This is an approach which has considerable appeal to a software
engineer as well. In software engineering, the implementation details
of a software system are usually not particularly important from an
architectural point of view, they are encapsulated in a component. In
contrast, the relationship of components to each other is of interest because
this knowledge is necessary for composing a system from its
components.Roughly speaking,  the architecture of a software system is 
characterized both through its components and their interaction,
the static part of which can be described through what we may perceive as
morphisms. 

This has been recognized fairly early in the software
architecture community, witnessed by the April 1995 issue of the
\emph{IEEE Transactions on Software Engineering}, which was devoted to
software architecture and introduced some categorical language in
discussing architectures. So the language of categories offers some
attractions to software engineers, as can be seen from,
e.g.,~\cite{Fiadeiro,Barbosa,EED-PipesAndFilters}. We will also see
that the tool set of modal logics, another area which is important to
software construction, profits substantially from constructions which
are firmly grounded in categories.

We will discuss categories here and introduce the reader to the basic
constructions. The world of categories is too rich to be captured in
these few pages, so we have made an attempt to provide the reader with
some basic proficiency in categories, helping in learning her or him
to get a grasp on the recent literature. This modest goal is attained
by blending the abstract mathematical development with a plethora of
examples. Exercisesprovide an opportunity to practice understanding and
to give hints at developments not spelled out in the text.

We give a brief overview over the contents.

\paragraph{Overview.}
\label{sec:overview}

The definition of a category and a discussion of their most elementary
properties is done in Section~\ref{sec:cat-basic-defs}, examples show
that categories are indeed a very versatile and general instrument for
mathematical modelling. Section~\ref{sec:elem-constr} discusses
constructions like product and coproduct, which are familiar from
other contexts, in this new language, and we look at pushouts and
pullbacks, first in the familiar context of sets, then in a more
general setting. Functors are introduced in Section~\ref{sec:functors}
for relating categories to each other, and natural transformations
permit functors to enter a relationship. We show also that set valued
functors play a special r\^ole, which gives occasion to investigate
more deeply the hom-sets of a category. Products and coproducts have
an appearance again, but this time as instances of the more general
concept of limits resp. colimits. 

Monads and Kleisli tripel as very
special functors are introduced and discussed in
Section~\ref{sec:monads}, their relationship is investigated, and some
examples are given, which provide an idea about the usefulness of
this concept; a small section on monads in the programming language
\texttt{Haskell} provides a pointer to the practical use of
monads. Next we show that monads are generated from adjunctions. This
important concept is introduced and discussed in~\ref{sec:algebras};
we define adjunctions, show by examples that adjunctions are a
colorfully blooming and nourished flower in the garden of Mathematics,
and give an alternative formulation in terms of units and counits; we
then show that each adjunction gives us a monad, and that each monad
also generates an adjunction. The latter part is interesting since it
gives the occasion of introducing the algebras for a monad; we discuss
two examples fairly extensively, indicating what such algebras might
look like. 

While an algebra provides a morphism $\funF a \to a$, a
coalgebra provides a morphism $a\to \funF a$. This is introduced an
discussed in Section~\ref{sec:coalgs}, many examples show that this
concept models a broad variety of applications in the area of
systems. Coalgebras and their properties are studied, among them
bisimulations, a concept which originates from the theory of concurrent
systems and which is captured now coalgebraically. The Kripke models
for modal logics provide an excellent play ground for coalgebras, so
they are introduced in Section~\ref{sec:modal-logics}, examples show
the broad applicability of this concept (but neighborhood models as a
generalization are introduced as well). We go a bit beyond a mere
application of coalgebras and give also the construction of the
canonical model through Lindenbaum's construction of maximally
consistent sets, which by the way provide an application of transfinite
induction as well. We finally show that coalgebras may be put to use
when constructing coalgebraic logics, a very fruitful and general
approach to modal logics and their generalizations.

\Subsection{Basic Definitions}
\label{sec:cat-basic-defs}

We will define what a category is, and give some examples for
categories. It shows that this is a very general notion, covering also many
formal structures that are studied in theoretical computer science. A
very rough description would be to say that a category is a bunch of
objects which are related to each other, the relationships being
called morphisms. This gives already the gist of the
definition --- objects which are related to each other. But the
relationship has to be made a bit more precise for being amenable to
further investigation. So this is the definition of a category.

\BeginDefinition{def-category-basic}
A category $\catK$ consists of a class $\objK$ of \emph{\index{category!object}objects} and
for any objects $a, b$ in  $\obj{\catK}$ of a set $\homK{a, b}$ of
\emph{\index{category!morphism}morphisms} with a \emph{\index{category!composition}composition operation}
$\circ$, mapping $\homK{b, c}\times\homK{a, b}$ to $\homK{a, c}$ with
the following properties
\begin{description}
\item[Identity] For every object $a$ in $\objK$ there exists a morphism
  $id_{a}\in \homK{a, a}$ with $f\circ id_{a} = f = id_{b}\circ f$,
  whenever $f\in\homK{a, b}$,
\item[Associativity] If $f\in\homK{a, b}, g\in\homK{b, c}$ and $h\in\homK{c, d}$,
  then $h\circ(g\circ f) = (h\circ g)\circ f$. 
\end{description}
\EndDefinition
Note that we do not think that a category is based on a set of objects
(which would yield difficulties) but rather than on a class.  In fact,
if we would insist on having a set of objects, we could not talk about
the category of sets, which is an important species for a category. We
insist, however, on having \emph{sets} of morphisms, because we want
morphisms to be somewhat clearly represented. Usually we write for
$f\in\homK{a, b}$ also $f: a\to b$, if the context is clear. Thus if
$f: a\to b$ and $g: b \to c$, then $g \circ f: a \to c$; one may think
that first $f$ is applied (or executed), and then $g$ is applied to
the result of $f$. Note the order in which the application is written
down: $g\circ f$ means that $g$ is applied to the result of $f$. The
first postulate says that there is an \emph{identity morphism}
$id_{a}: a\to a$ for each object $a$ of $\catK$ which does not have an
effect on the other morphisms upon composition, so no matter if you do
$id_{a}$ first and then morphism $f: a\to b$, or if you do $f$ first
and then $id_{b}$, you end up with the same result as if doing only
$f$. Associativity is depicted through this diagram
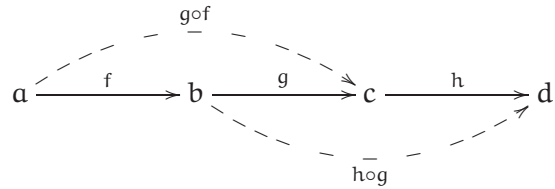
\begin{equation*}
  \xymatrix{a\ar[rr]^{f}\ar@/^2pc/@{-->}[rrrr]^{g\circ f} && b\ar[rr]^{g}\ar@/_2pc/@{-->}[rrrr]_{h\circ g} && c\ar[rr]^{h} && d
}
\end{equation*}
Hence is you take the fast train $g\circ f$ from $a$ to $c$ first (no
stop at $b$) and
then switch to train $h$ or is you travel first with $f$ from $a$ to
$b$ and then change to the fast train $h\circ g$ (no stop at $c$), you will end up with
the same result.

Given $f\in\homK{a, b}$, we call object $a$ the \emph{\index{morphism!domain}domain}, object $b$ the
\emph{\index{morphism!codomain}codomain} of morphism $f$.

Let us have a look at some examples.
\BeginExample{ex-cat-sets-maps}
The category $\index{$\catSET$}\catSET$ is the most important of them all. It has sets
as its class of objects, and the morphisms $\homSET{a, b}$ are just
the maps from set $a$ to set $b$. The identity map $id_{a}: a \to a$ maps each
element to itself, and composition is just composition of maps, which
is  associative:
\begin{align*}
  \bigl(f\circ (g\circ h)\bigr)(x) & = f\bigl(g\circ h(x)\bigr)\\
& = f\bigl(g(h(x)) \bigr)) \\
& = \bigl(f\circ g \bigr))(h(x))\\
& = \bigl((f\circ g)\circ h\bigr)(x)
\end{align*}
\EndExample
The next example shows that one class of objects can carry more than
one definition of morphisms.
\BeginExample{ex-cat-sets-rels}
The category $\Category{Rel}$ has sets as its class of objects. Given
sets $a$ and $b$, 
 $f\in \hom{\Category{Rel}}(a, b)$ is a morphism from $a$ to $b$ iff
 $f\subseteq a\times b$ is a relation. Given set $a$, define
 \begin{equation*}
   id_{a} := \{\langle x, x\rangle\mid x \in a\}
 \end{equation*}
as the identity relation, and define for $f\in\hom{\Category{Rel}}(a,
b), g\in\hom{\Category{Rel}}(b, c)$ the composition as 
\begin{equation*}
  g\circ f := \{\langle x, z\rangle \mid \text{ there exists } y \in b
  \text{ with } \langle x, y\rangle \in f\text{ and }\langle y, z\rangle\in g\}
\end{equation*}
Because existential quantifiers can be interchanged, composition is
associative, and $id_{a}$ serves in fact as the identity element for composition.
\EndExample
But morphisms do not need to be maps or relations.

\BeginExample{ex-cat-ord}
Let $(P, \leq)$ be a partially ordered set. Define $\index{$\Category{P}$}\Category{P}$ by
taking the class $\obj{\Category{P}}$ of objects as $P$, and put
\begin{equation*}
\hom{\Category{P}}(p, q) :=
  \begin{cases}
    \{\langle p, q\rangle\}, & \text{if }p \leq q\\
\emptyset, &\text{otherwise}.
  \end{cases}
\end{equation*}
Then $id_{p}$ is $\langle p, p\rangle$, the only element of
$\hom{\Category{P}}(p, p)$, and if $f: p\to q, g: q\to r$, thus $p\leq
q$ and $q\leq r$, hence by transitivity $p\leq r$, so that we put
$g\circ f := \langle p, r\rangle$. Let $h: r\to s$, then 
\begin{align*}
  h\circ (g\circ f) & = h\circ \langle p, r\rangle\\
& = \langle p, s\rangle \\
& = \langle q, s\rangle\circ f\\
& = (h\circ g)\circ f
\end{align*}
It is clear that $id_{p} = \langle p, p\rangle$ serves as a neutral
element. 
\EndExample

A directed graph generates a category through all its finite
paths. Composition of two paths is then just their combination,
indicating movement from one node to another, possibly via
intermediate nodes. But we also have to cater for the situation that
we want to stay in a node.

\BeginExample{ex-graph-paths}
Let ${\cal G} = (V, E)$ be a directed graph. Recall that a \emph{\index{graph!path}path}
$\langle p_{0}, \dots, p_{n}\rangle$ is a finite sequence of nodes
such that adjacent nodes form an edge, i.e., such that $\langle p_{i},
p_{i+1}\rangle\in E$ for $0\leq i < n$; each node $a$ has an empty
path $\langle a, a\rangle$ attached to it, which may or may not be an
edge in the graph. The objects of the category
$\index{$F({\cal G})$}F({\cal G})$ are the nodes $V$ of ${\cal G}$, and a morphism $a\to
b$ in $F({\cal G})$ is a path connecting $a$ with  $b$ in ${\cal G}$,
hence a path $\langle p_{0}, \dots, p_{n}\rangle$ with
$p_{0} = a$ and $p_{n} = b$. The empty path serves as the identity
morphism, the composition of morphism is just their concatenation;
this is plainly associative. This category is  called the
\emph{\index{graph!free category}\index{category!free}free category generated by graph ${\cal G}$}.
\EndExample

These two examples base categories on a set of objects; they are
instances of small categories. A category is called \emph{\index{category!small}small} iff the
objects form a set (rather than a class). 

The discrete category is a trivial but helpful example.

\BeginExample{ex-discr-categ}
Let $X\not= \emptyset$ be a set, and define a category $\catK$ through
$\objK := X$ with 
\begin{equation*}
  \homK{x, y} :=
  \begin{cases}
    id_{x}, & x = y\\
\emptyset, & \text{otherwise}
  \end{cases}
\end{equation*}
This is the \emph{\index{category!discrete}discrete category} on $X$.
\EndExample

Algebraic structures furnish a rich and plentiful source for examples. Let us have a
look at groups, and at Boolean algebras.

\BeginExample{ex-cat-group}
The category of groups has as objects all groups $(G,\cdot)$, and as
morphisms $f: (G, \cdot) \to (H, *)$ all maps $f: G\to H$ which are
group homomorphisms, i.e., for which $f(1_{G}) = 1_{H}$ (with $1_{G},
1_{H}$ as the respective neutral elements), for which $f(a^{-1}) =
(f(a))^{-1}$ and $f(a\cdot b) = f(a)*f(b)$ always holds. The identity
morphism $id_{(G, \cdot)}$ is the identity map, and composition of
homomorphisms is composition of maps. Because composition is inherited
from category $\catSET$, we do not have to check for associativity or
for identity. 

Note that we did not give the category a particular name, it is simple
referred to as to the \emph{category of  groups}. 
\EndExample

\BeginExample{ex-cat-boolalg}
Similarly, the category of Boolean algebras has Boolean algebras as
objects, and a morphism $f: G\to H$ for the Boolean algebras $G$ and
$H$ is a map $f$ between the carrier sets with these properties:
\begin{align*}
  f(-a) & = -f(a)\\
f(a\wedge b) & = f(a)\wedge f(b)\\
f(\top) & = \top
\end{align*}
(hence also $f(\bot) = \bot$, and $f(a\vee b) = f(a)\vee f(b)$). Again, composition of morphisms is
composition of maps, and the identity morphism is just the identity
map.  
\EndExample
The next example deals with transition systems. Formally, a transition
system is a directed graph. But whereas discussing a graph 
puts the emphasis usually on its paths, a transition system is
concerned more with the study of, well, the transition from one state
to another one, hence the focus is usually stronger localized. This is
reflected also when defining morphisms, which, as we will see, come in
two flavors.

\BeginExample{ex-cat-transition-syst}
A \emph{\index{transition system}transition system} $(S, \rightsquigarrow_{S})$ is
a set $S$ of states together with a transition relation
$\rightsquigarrow_{S}\ \subseteq S\times S$. Intuitively, $s\rightsquigarrow_{S} s'$ iff there is
a transition from $s$ to $s'$. Transition systems form a category: the
objects are transition systems, and a morphism $f: (S, \rightsquigarrow_{S})\to(T,
\rightsquigarrow_{T})$ is a map $f: S\to T$ such that $s\rightsquigarrow_{S} s'$ implies
$f(s)\rightsquigarrow_{T} f(s')$. This means that a transition from $s$ to $s'$ in
$(S, \rightsquigarrow_{S})$ entails a transition from $f(s)$ to $f(s')$ in the
transition system $(T, \rightsquigarrow_{T})$. Note that the defining condition for
$f$ can be written as $\rightsquigarrow_{S}\ \subseteq\ \InvBild{(f\times
  f)}{\rightsquigarrow_{T}}$ with $f\times f: \langle s, s'\rangle \mapsto \langle
f(s), f(s')\rangle$.
\EndExample

The morphisms in Example~\ref{ex-cat-transition-syst} are interesting
from a relational point of view. We will require an additional property
which, roughly speaking, makes sure that we not only transport
transitions through morphisms, but that we are also able to capture
transitions which emanate from the image of a state. So we want to be
sure that, if $f(s)\Trans_{T} t$, we obtain this transition from a
transition arising from $s$ in the original system. This idea is
formulated in the next example, it will arise again in a very natural
manner in Example~\ref{ex-transition-syst-act-coalg-morph} in the
context of coalgebras. 

\BeginExample{ex-cat-transition-syst-bound}
We continue with transition systems, so we define a category which has
transition systems as objects. A \emph{\index{morphism!bounded}morphism} $f: (S,
\rightsquigarrow_{S})\to(T,\rightsquigarrow_{T})$ in the present category is a map $f: S \to T$ such that for all $s,
s'\in S, t\in T$
\begin{description}
\item[\textit{Forward}:] $s\rightsquigarrow_{S}s'$ implies $f(s)\rightsquigarrow_{T}f(s')$ ,
\item[\textit{Backward}:] if $f(s)\rightsquigarrow_{T}t'$, then there exists $s'\in S$ with $f(s')=t'$
  and $s\rightsquigarrow_{S}s'$. 
\end{description}
The forward condition is already known from
Example~\ref{ex-cat-transition-syst}, the backward condition is
new. It states that if we start a transition from some $f(s)$ in $T$, then this
transition originates from some transition starting from $s$ in
$S$; to distinguish these morphisms from the ones considered in
Example~\ref{ex-cat-transition-syst}, they are called \emph{bounded}
morphisms. The identity map $S\to S$ yields a bounded morphism, and
the composition of bounded morphisms is a bounded morphism again. In
fact, let $f: (S, \rightsquigarrow_{S})\to(T, \rightsquigarrow_{T}), g: (T, \rightsquigarrow_{T})\to(U,
\rightsquigarrow_{U})$ be bounded morphisms, and assume that 
$
g(f(s))\rightsquigarrow_{U}u'.
$
Then we can find $t'\in T$ with $g(t')=u'$ and $f(s)\rightsquigarrow_{T}t'$, hence we
find $s'\in S$ with $f(s') = t'$ and $s\rightsquigarrow_{S}s'$. 

Bounded morphisms are of interest in the study of models for modal
logics~\cite{Blackburn-Rijke-Venema}, see Lemma~\ref{char-frame-morph}. 
\EndExample

The next examples reverse arrows when it comes to define
morphisms. The examples so far observed the effects of maps in the
direction in which the maps were defined. We will, however, also have
an opportunity to look back, and to see what properties the inverse
image of a map is supposed to have. We study this in the context of
measurable, and of topological spaces.

\BeginExample{ex-cat-meas-space}
Let $S$ be a set, and assume that ${\cal A}$ is a $\sigma$-algebra on
$S$. Then the pair $(S, {\cal A})$ is called a
\emph{\index{space!measurable}measurable space}, the elements of the
$\sigma$-algebra are sometimes called ${\cal A}$-measurable sets. The
category $\index{$\Category{Meas}$}\Category{Meas}$ has as objects all measurable spaces. 

Given two measurable spaces $(S, {\cal A})$ and $(T, {\cal B})$, a map
$f: S\to T$ is called a \emph{morphism of measurable spaces} iff $f$
is \emph{${\cal A}$-${\cal B}$-\index{map!measurable}measurable}. This
means that $\InvBild{f}{B}\in{\cal A}$ for all $B\in{\cal B}$, hence
the set $ \{s\in S \mid f(s) \in B\} $ is an ${\cal A}$-measurable set
for each ${\cal B}$-measurable set $B$. Each $\sigma$-algebra is a
Boolean algebra, but the definition of a morphism of measurable spaces
does not entail that such a morphism induces a morphism of Boolean
algebras (see Example~\ref{ex-cat-boolalg}). Consequently the behavior of
$f^{-1}$ rather than the one of $f$ determines whether $f$ belongs to
the distinguished set of morphisms.

Thus the ${\cal A}$-${\cal B}$-measurable maps $f: S \to T$ are the
morphisms $f: (S, {\cal A})\to(T, {\cal B})$ in category
$\Category{Meas}$.  The identity morphism on $(S, {\cal A})$ is the
identity map (this map is measurable because $\InvBild{id}{A} =
A\in{\cal A}$ for each $A\in{\cal A}$). Composition of measurable maps
yields a measurable map again: let $f: (S, {\cal A})\to(T, {\cal B})$
and $g: (T, {\cal B})\to(U, {\cal C})$, then $ \InvBild{(g\circ f)}{D}
= \InvBild{f}{\InvBild{g}{D}} \in {\cal A}, $ for $D\in{\cal C}$,
because $\InvBild{g}{D}\in{\cal B}$. It is clear that composition is
associative, since it is based on composition of ordinary maps.
\EndExample

The next example deals with topologies, which are of course also sets
of subsets. Continuity is formulated similar to measurably in terms
of the inverse rather than the direct image.

\BeginExample{ex-cat-top-space}
Let $S$ be a set and ${\cal G}$ be a topology on $S$; hence ${\cal
  G}\subseteq\PowerSet{S}$ such that $\emptyset, S\in{\cal G}$, ${\cal
  G}$ is closed under finite intersections and arbitrary unions. Then
$(S, {\cal G})$ is called a \emph{\index{space!topological}topological
  space}. Given another topological space $(T, {\cal H})$, a map $f:
S\to T$ is called ${\cal G}$-${\cal H}$-\emph{\index{map!continuous}continuous} iff the inverse
image of an open set is open again, i.e., iff 
$\InvBild{f}{G}\in {\cal H}$ for all $G\in{\cal G}$. Category
$\index{$\Category{Top}$}\Category{Top}$ of topological spaces has all topological
spaces as objects, and continuous maps as morphisms. The identity $(S, {\cal
  G})\to (S, {\cal G})$ is certainly continuous. Again, it follows that the
composition of morphisms yields a morphism, and that their composition
is associative.  
\EndExample

Now that we know what a category is, we start constructing new categories from given ones.  We begin by building on category $\Category{Meas}$ another interesting category, indicating that a category can be used as a building block for another one.

\BeginExample{ex-cat-meas-space-stochrel}
A measurable space $(S, {\cal A})$ together with a probability measure
$\mu$ on ${\cal A}$ is called a
\emph{\index{space!probability}probability space} and written as $(S,
{\cal A}, \mu)$. The category $\index{$\Category{Prob}$}\Category{Prob}$ of all probability
spaces has ---~ you guessed it~--- as objects all probability spaces;
a morphism $f: (S, {\cal A}, \mu)\to(T, {\cal B}, \nu)$ is a morphism
$f: (S, {\cal A})\to(T, {\cal B})$ in $\Category{Meas}$ for the underlying measurable
spaces such that $ \nu(B) = \mu(\InvBild{f}{B}) $ holds for all
$B\in{\cal B}$. Thus the $\nu$-probability for event $B\in{\cal B}$ is
the same as the $\mu$-probability for all those $s\in S$ the image of
which is in $B$. Note that $\InvBild{f}{B}\in{\cal A}$ due to $f$
being a morphism in $\Category{Meas}$, so that $\mu(\InvBild{f}{B})$
is in fact defined.
\EndExample

We go a bit further and combine two measurable spaces into a third one; this requires adjusting the notion of a morphism, which are in this new category basically pairs of morphisms from the underlying category. This shows the flexibility with which we may --- and do --- manipulate morphisms. 

\BeginExample{ex-cat-stoch-rel}
Denote for the measurable space $(S, {\cal A})$ by $\Prob{S, {\cal
    A}}$ the set of all subprobability measures. Define
\begin{align*}
 \beta_{S}(A, r) & := \{\mu\in\Prob{S, {\cal
    A}} \mid \mu(A) \geq r\},\\
w({\cal A}) & := \sigma(\{\beta_{S}(A, r) \mid A\in {\cal A}, 0 \leq r \leq 1\})
\end{align*}
Thus $\beta_{S}(A, r)$ denotes all probability measures which evaluate the set $A$ not
smaller than $r$, and $w({\cal A})$ collects all these sets into a
$\sigma$-algebra; $w$ alludes to
``weak'', $w({\cal A})$ is sometimes called the \emph{\index{$\sigma$-algebra!weak-*}weak-*-$\sigma$-algebra} or
the \emph{weak $\sigma$-algebra} associated with ${\cal A}$
as the $\sigma$-algebra generated by the family of sets.
This makes $(\Prob{S, {\cal A}}, w({\cal A}))$ a measurable space, based
on the probabilities over $(S, {\cal A})$.

Let $(T, {\cal B})$ be another measurable space. A map $K:
S\to\Prob{T, {\cal B}}$ is ${\cal A}$-$w({\cal B})$-measurable iff $\{s\in S \mid K(s)(B) \geq r\}\in {\cal A}$ for all $B\in{\cal
  B}$; this follows from Exercise~\ref{ex-meas-generator}. We take as objects for
our category the triplets $\bigl((S, {\cal A}), (T, {\cal B}), K\bigr)$,
where $(S, {\cal A})$ and $(T, {\cal B})$ are measurable spaces and
$K: S\to\Prob{T, {\cal B}}$ is ${\cal A}$-$w({\cal B})$-measurable. A
morphism $(f, g): \bigl((S, {\cal A}), (T, {\cal B}), K\bigr)\to
\bigl((S', {\cal A}'), (T', {\cal B}'), K'\bigr)$ is a pair of
morphisms $f: (S, {\cal A})\to (S', {\cal A}')$ and $g: (T, {\cal
  B})\to(T', {\cal B}')$ such that 
\begin{equation*}
  K(s)(\InvBild{g}{B'}) = K'(f(s))(B')
\end{equation*}
holds for all $s\in S$ and for all $B'\in{\cal B}'$. 

The composition of morphisms is defined component wise:
\begin{equation*}
  (f', g')\circ(f, g) := (f'\circ f, g'\circ g).
\end{equation*}
Note that $f'\circ f$ and $g'\circ g$ refer to the composition of
maps, while $(f', g')\circ(f, g)$ refers to the newly defined
composition in our new spic-and-span category (we should probably use
another symbol, but no confusion can arise, since the new composition
operates on pairs). The identity morphism for $\bigl((S, {\cal A}),
(T, {\cal B}), K\bigr)$ is just the pair $(id_{S}, id_{T})$. Because
the composition of maps is associative, composition in our new
category is associative as well, and because $(id_{S}, id_{T})$
is composed from identities, it is also an identity.

This category is sometimes called the \emph{category of stochastic
relation}s. 
\EndExample

Before continuing, we introduce commutative diagrams. Suppose that we
have in a category $\catK$ morphisms $f: a \to b$ and $g: b \to
c$. The combined morphism $g\circ f$ is represented graphically as
\begin{equation*}
  \xymatrix{
a\ar[rr]^{f}\ar[dr]_{g\circ f} && b\ar[dl]^{g}\\
&c
}
\end{equation*}
If the morphisms $h: a\to d$ and $\ell: d\to c$ satisfy $g\circ f =
\ell\circ h$, we have a \emph{\index{diagram!commutative}commutative
  diagram};  in this case we  do not draw out the morphism in the
diagonal. 
\begin{equation*}
  \xymatrix{
a\ar[rr]^{f}\ar[d]_{h}\ar@{..>}[drr]^{g\circ f}_{\ell\circ h} && b\ar[d]^{g}\\
d\ar[rr]_{\ell} && c
}
\end{equation*}

We consider automata next, to get some feeling for the handling of
commutative diagrams, and as an illustration for an important
formalism looked at through the glasses of categories.

\BeginExample{ex-det-automaton}
Given sets $X$ and $S$ of inputs and states, respectively, an
\emph{\index{automaton}automaton} $(X, S, \delta)$ is defined by a map
$\delta: X\times S \to S$. The interpretation is that $\delta(x, s)$
is the new state after input $x\in X$ in state $s\in
S$. Reformulating, $\delta(x): s \mapsto \delta(x, s)$ is perceived as
a map $S\to S$ for each $x\in X$, so that the new state now is written
as $\delta(x)(s)$; manipulating a map with two arguments in this way is called
\emph{currying} and will be considered in greater detail in
Example~\ref{curry-is-adjoint}. The objects of our category of
automata are the automata, and an
\emph{\index{morphism!automaton}automaton morphism} $f: (X, S,
\delta)\to (X, S', \delta')$ is a map $f: S \to S'$ such that this
diagram commutes for all $x\in X$:
\begin{equation*}
\xymatrix{
S\ar[rr]^{f}\ar[d]_{\delta(x)} &&S'\ar[d]^{\delta'(x)}\\
S\ar[rr]_{f} && S'
}
\end{equation*}
Hence we have $f(\delta(x)(s)) = \delta'(x)(f(s))$ for each $x\in X$
and $s\in S$; this means that computing the new state and mapping it through $f$
yields the same result as computing the new state for the mapped
one. The identity map $S\to S$ yields a morphism, hence automata
form a category. 

Note that morphisms are defined only for automata with the same input
alphabet. This reflects the observation that the input alphabet is
usually given by the environment, while the set of states represents a
model about the automata's behavior, hence is at our disposal for manipulation. 
\EndExample

Whereas we constructed above new categories from given one in an ad
hoc manner, categories also yield new categories systematically. This
is a simple example.

\BeginExample{ex-down-under}
Let $\catK$ be a category; fix an object $x$ on $\catK$. The objects
of our new category are the morphisms $f\in\homK{a, x}$ for an object
$a$. Given objects $f\in\homK{a, x}$ and $g\in\homK{b, x}$ in the new
category, a morphism $\phi:f \to g$ is a morphism $\phi\in\homK{a, b}$
with $f = g\circ \phi$, so that this diagram commutes
\begin{equation*}
\xymatrix{
a\ar[rr]^{\phi}\ar[dr]_{f} && b\ar[dl]^{g}\\
&x&
}
\end{equation*}
Composition is inherited from $\catK$. The identity $id_{f}: f\to f$
is $id_{a}\in\homK{a, a}$, provided $f\in\homK{a, x}$. Since the
composition in $\catK$ is associative, we have only to make sure that
the composition of two morphisms is a morphism again. This can be read
off the following diagram: $(\phi\circ\psi)\circ h =
\phi\circ(\psi\circ h) = \phi\circ g = f$.
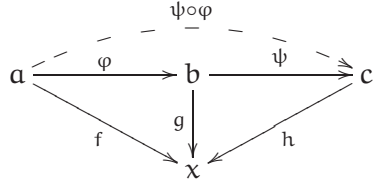
\begin{equation*}
\xymatrix{
a\ar[rr]^{\phi}\ar[drr]_{f}\ar@/^1.5pc/@{-->}[rrrr]^{\psi\circ\phi} && b\ar[rr]^{\psi}\ar[d]_{g} && c\ar[dll]^{h}\\
&&x
}
\end{equation*}
This category is sometimes called the
\emph{\index{category!slice}slice category} $\Faktor{\catK}{x}$; the object $x$
is interpreted as an index, so that a morphism $f: a\to  x$ serves as
an indexing function. A morphism $\phi: a\to  b$ in
$\Faktor{\catK}{x}$ is then compatible with the index operation. 
\EndExample

The next example reverses arrows while at the same time maintaining the same class of objects. 
\BeginExample{ex-opposite-category}
Let $\catK$ be a category. We define $\index{$\catK^{op}$}\catK^{op}$, the category
\emph{\index{category!dual}dual} to $\catK$, in the following way: the
objects are the same as for the original category, hence
$\obj{\catK^{op}} = \objK$, and the arrows are reversed, hence we put
$
\hom{\catK^{op}}(a, b) := \hom{\catK}(b, a)$ 
for the objects $a, b$; the identity remains the same. We have to define composition in this new
category. Let $f\in\hom{\catK^{op}}(a, b)$ and
$g\in\hom{\catK^{op}}(b, c)$, then $g\ast f := f\circ g
\in\hom{\catK^{op}}(a, c)$. It is readily verified that $\ast$
satisfies all the laws for composition from
Definition~\ref{def-category-basic}. 

The dual category is sometimes helpful because it permits to cast
notions into a uniform framework.  
\EndExample

\BeginExample{ex-opposite-category-rel}
Let us look at $\index{$\Category{Rel}$}\Category{Rel}$ again. The morphisms
$\hom{\Category{Rel}^{op}}(S, T)$ from $S$ to $T$ in
$\Category{Rel}^{op}$ are just the morphisms $\hom{\Category{Rel}}(T,
S)$ in $\Category{Rel}$. Take $f\in\hom{\Category{Rel}^{op}}(S, T)$,
then $f\subseteq T\times S$, hence $f^{t}\subseteq S\times T$, where
relation 
\begin{equation*}
f^{t} := \{\langle s, t\rangle \mid \langle t, s\rangle \in f\}
\end{equation*}
is the transposed of relation $f$. The map $f\mapsto
f^{t}$ is injective and compatible with composition, moreover it maps
$\hom{\Category{Rel}^{op}}(S, T)$ onto $\hom{\Category{Rel}}(T,
S)$. But this means that $\Category{Rel}^{op}$ is essentially the same
as $\Category{Rel}$.
\EndExample

It is sometimes helpful to combine two categories into a product:

\BeginLemma{product-category}
Given categories $\catK$ and $\catL$, define the objects of
$\catK\times\catL$ as pairs $\langle a, b\rangle$, where $a$ is an
object in $\catK$, and $b$ is an object in $\catL$. A morphism
$\langle a, b\rangle\to \langle a', b'\rangle$ in $\catK\times\catL$
is comprised of morphisms $a\to a'$ in $\catK$ and $b\to b'$ in
$\catL$. Then $\catK\times\catL$ is a category. \QED
\EndLemma

We have a closer look at morphisms now. Experience tells us that
injective and surjective maps are fairly helpful, so a
characterization in a category might be desirable. There is a small
but not insignificant catch, however. We have seen that morphisms are
not always maps, so that we are forced to find a characterization
purely in terms of composition and equality, because this is all we
have in a category. The following characterization of injective maps
provides a clue for a more general definition.

\BeginProposition{char-injective-cancel}
Let $f: X\to Y$ be a map, then these statements are equivalent.
\begin{enumerate}
\item\label{char-injective-cancel-item:1} $f$ is injective.
\item\label{char-injective-cancel-item:2} If $A$ is an arbitrary set,  $g_1, g_2: A \to X$ are maps with $f\circ g_1 = f\circ g_2$, then $g_1 = g_2$
\end{enumerate}
\EndProposition

\BeginProof
\labelImpl{char-injective-cancel-item:1}{char-injective-cancel-item:2}: Assume $f$ is injective and $f\circ g_1 = f\circ g_2$, but $g_1 \not= g_2$. Thus there exists $x\in A$ with $g_1(x) \not=g_2(x)$. But $f(g_1(x)) = f(g_2(x))$, and since $f$ is injective, $g_1(x) = g_2(x)$. This is a contradiction.

\labelImpl{char-injective-cancel-item:2}{char-injective-cancel-item:1}: Assume the condition holds, but $f$ is not injective. Then there exists $x_1 \not= x_2$ with $f(x_1) = f(x_2)$.  Let $A := \{\star\}$ and put $g_1(\star) := x_1$, $g_2(\star) := x_2$,  thus 
$
f(x_1) = (f\circ g_1)(\star) =   (f\circ g_2)(\star) =   f(x_2).
$
By the condition $g_1 = g_2$, thus $x_1 = x_2$. Another contradiction.
\EndProof

This leads to a definition of the category version of injectivity as a
morphism which is  cancellable on the left.

\BeginDefinition{cat-def-mono}
Let $\catK$ be a category, $a, b$ objects in $\catK$. Then $f: a \to
b$ is called a
\emph{\index{category!monomorphism}\index{morphism!monomorphism}monomorphism}
(or a \emph{\index{monic}monic}) iff whenever $g_1, g_2: x \to a$ are morphisms with $f\circ g_1 = f\circ g_2$, then $g_1 = g_2$.
\EndDefinition

These are some simple properties of monomorphisms, which are also sometimes
called monos.

\BeginLemma{cat-prop-monos}
In a category $\catK$,
\begin{enumerate}
\item The identity is a monomorphism.
\item The composition of two monomorphisms is a monomorphism again.
\item If $k \circ f$ is a monomorphism for some morphism $k$, then $f$ is a monomorphism.
\end{enumerate}
\EndLemma

\BeginProof
The first part is trivial. Let $f: a \to b$ and $g: b \to c$ both monos. Assume $h_1, h_2: x \to a$ with 
$
h_1 \circ (g\circ f) = h_2 \circ (g\circ f).
$
We want to show $h_1 = h_2$. By associativity
$
(h_1 \circ g) \circ f = (h_2 \circ g)\circ f.
$
Because $f$ is a mono, we conclude $h_1\circ g = h_2\circ g$, because $g$ is a mono, we see
$
h_1 = h_2.
$

Finally, let $f: a \to b$ and $k: b \to c$. Assume $h_1, h_2: x \to a$ with 
$
f \circ h_1 = f \circ h_2.
$
We claim $h_1 = h_2$. Now 
$
f\circ h_1 = f\circ h_2
$
implies
$
k\circ f \circ h_1  = k\circ f \circ h_2.
$
Thus $h_1 = h_2$. 
\EndProof

In the same way we characterize surjectivity purely in terms of
composition (exhibiting a nice symmetry between the two notions). 

\BeginProposition{char-surjective-cancel}
Let $f: X\to Y$ be a map, then these statements are equivalent.
\begin{enumerate}
\item\label{char-surjective-cancel-item:1} $f$ is surjective.
\item\label{char-surjective-cancel-item:2} If $B$ is an arbitrary set,
  $g_1, g_2: Y \to B$ are maps with $g_{1}\circ f = g_{2}\circ f$, then $g_1 = g_2$
\end{enumerate}
\EndProposition

\BeginProof
\labelImpl{char-surjective-cancel-item:1}{char-surjective-cancel-item:2}:
Assume $f$ is surjective, $g_1\circ f = g_2\circ f$, but $g_1(y) \not=
g_2(y)$ for some $y$. If we can find $x\in X$ with $f(x) = y$, then 
$
g_1(y) = (g_1\circ f)(x) = (g_2\circ f)(x) = g_2(y),
$
which would be a contradiction. Thus $y\notin\Bild{f}{X}$, hence $f$ is not onto.

\labelImpl{char-surjective-cancel-item:2}{char-surjective-cancel-item:1}:
Assume that there exists $y\in Y$ with $y\notin\Bild{f}{X}$. Define $g_1, g_2: Y \to \{0, 1, 2\}$ through

\smallBox{\begin{equation*}
g_1(y) := 
\begin{cases}
 0,     & \text{if } y \in \Bild{f}{X}, \\
 1,    & \text{otherwise}.
\end{cases}
\end{equation*}
}\smallBox{
\begin{equation*}
g_2(y) := 
\begin{cases}
 0,     & \text{if } y \in \Bild{f}{X}, \\
 2,    & \text{otherwise}.
\end{cases}
\end{equation*}
}

Then $g_1\circ f = g_2\circ f$, but $g_1 \not= g_2$. This is a contradiction.
\EndProof

This suggests a definition of surjectivity through
a morphism which is right cancellable.

\BeginDefinition{cat-def-epi}
Let $\catK$ be a category, $a, b$ objects in $\catK$. Then $f: a \to
b$ is called a
\emph{\index{category!epimorphism}\index{morphism!epimorphism}epimorphism}
(or an \emph{\index{epic}epic}) iff whenever $g_1, g_2: b \to c$ are morphisms with $g_1\circ f = g_2\circ f$, then $g_1 = g_2$.
\EndDefinition

These are some important properties of epimorphisms, which are
sometimes called epis:
\BeginLemma{}
In a category $\catK$,
\begin{enumerate}
\item The identity is an epimorphism.
\item The composition of two epimorphisms is an epimorphism again.
\item If $f \circ k$ is an epimorphism for some morphism $k$, then $f$ is an epimorphism.
\end{enumerate}
\EndLemma

\BeginProof
We sketch the proof only for the the third part:
\begin{equation*}
g_1\circ f = g_2\circ f 
\Rightarrow
g_1\circ f \circ k = g_2\circ f \circ k 
\Rightarrow
g_1 = g_2.
\end{equation*}
\EndProof

This is a small application of the decomposition of a map into an
epimorphism and a monomorphism.

\BeginProposition{cat-epi-mono-decomp}
Let $f: X \to Y$ be a map. Then there exists a  factorization of $f$ into $m \circ e$ with $e$ an epimorphism and $m$ a monomorphism. 
\begin{equation*}
\xymatrix{
X\ar[rr]^f\ar[dr]_e && Y\\
&M\ar[ur]_m
}
\end{equation*}
\EndProposition

The idea of the proof may best be described in terms of $X$ as inputs,
$Y$ as outputs of system $f$. We collect all inputs with the same
functionality, and assign each collection the functionality through
which it is defined.

\BeginProof
Define 
\begin{equation*}\label{page:def-kernel}
\Kern{f} := \{\langle x_1, x_2\rangle \mid f(x_1) = f(x_2)\}
\end{equation*}
(the \emph{\index{map!kernel}kernel} of $f$). This is an equivalence
relation on $X$ (\emph{reflexivity}: $\langle x, x\rangle\in \Kern{f}$
for all $x$, \emph{symmetry}: if $\langle x_{1}, x_{2}\rangle\in\Kern{f}$ then $\langle x_{2}, x_{1}\rangle\in\Kern{f}$; \emph{transitivity}: $\langle x_{1}, x_{2}\rangle\in\Kern{f}$ and $\langle x_{2}, x_{3}\rangle\in\Kern{f}$ together imply $\langle x_{1}, x_{3}\rangle\in\Kern{f}$). 

Define 
\begin{equation*}
e:  
  \begin{cases}
    X &\to \Faktor{X}{\Kern{f}},\\
    x &\mapsto \Klasse{x}{\Kern{f}}
  \end{cases}
\end{equation*}
then $e$ is an epimorphism. In fact, if $g_{1}\circ e = g_{2}\circ e$
for $g_{1}, g_{2}: \Faktor{X}{\Kern{f}}\to B$ for some set $B$, then
$g_{1}(t) = g_{2}(t)$ for all $t\in\Faktor{X}{\Kern{f}}$, hence
$g_{1}= g_{2}$. 

Moreover  
\begin{equation*}
m: 
  \begin{cases}
\Faktor{X}{\Kern{f}} &\to Y \\
\Klasse{x}{\Kern{f}} &\mapsto f(x)   
  \end{cases}
\end{equation*}
is well defined, since if $\Klasse{x}{\Kern{f}} = \Klasse{x'}{\Kern{f}}$,
then $f(x) = f(x')$, and a monomorphism. In fact, if $m\circ g_{1} =
m\circ g_{2}$ for arbitrary $g_{1}, g_{2}: A\to \Faktor{X}{\Kern{f}}$ for
some set $A$, then $f(g_{1}(a)) = f(g_{2}(a))$ for all $a$, hence 
$\langle g_{1}(a), g_{2}(a)\rangle \in\Kern{f}$. But this means
$
\Klasse{g_{1}(a)}{\Kern{f}} = \Klasse{g_{2}(a)}{\Kern{f}}
$
for all $a\in A$, so $g_{1}= g_{2}$. 
Evidently $f = m \circ e$. 
\EndProof

Looking a bit harder at the diagram, we find that we can say even
more, viz., that the decomposition is unique up to isomorphism.

\BeginCorollary{cat-epi-mono-decomp-cor}
If the map $f: X \to Y$ can be written as $f = e\circ m = e'\circ m'$
with epimorphisms $e, e'$ and monomorphisms $m, m'$, then there is a
bijection $b$ with $e' = b\circ e$ and $m = m'\circ b$. 
\EndCorollary

\BeginProof
Since the composition of bijections is a bijection again, we may and
do assume without loss of generality that $e: X\to
\Faktor{X}{\Kern{f}}$ maps $x$ to its class $\Klasse{x}{\Kern{f}}$,
and that $m: \Faktor{X}{\Kern{f}}\to Y$ maps $\Klasse{x}{\Kern{f}}$ to
$f(x)$. Then we have this diagram for the primed factorization $e':
X\to Z$ and $m': Z\to Y$:
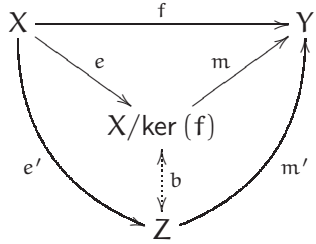
\begin{equation*}
\xymatrix{
X\ar[rr]^{f}\ar[dr]^{e}\ar@/^-1.5pc/[ddr]_{e'} && Y\\
&\Faktor{X}{\Kern{f}}\ar[ur]^{m}\ar@{<..>}[d]^{b}\\
&Z\ar@/^-1.5pc/[uur]_{m'}
}
\end{equation*}
Note that 
\begin{align*}
 \Klasse{x}{\Kern{f}} \not=\Klasse{x'}{\Kern{f}} 
&\Leftrightarrow
f(x) \not= f(x')\\
&\Leftrightarrow 
m'(e'(x)) \not= m'(e'(x'))\\
&\Leftrightarrow
e(x)\not= e(x')
\end{align*}
Thus defining 
$
b(\Klasse{x}{\Kern{f}}) := e'(x)
$
gives an injective map $\Faktor{X}{\Kern{f}}\to Z$. Given $z\in Z$,
there exists $x\in X$ with $e'(x) = z$, hence $b(\Klasse{x}{\Kern{f}})
= z$, thus $b$ is onto. Finally, 
$
m'(b(\Klasse{x}{\Kern{f}})) = m'(e'(x)) = f(x) =
m(\Klasse{x}{\Kern{f}}).
$
\EndProof

This factorization of a morphism is called an \emph{\index{morphism!epi/mono factorization}epi/mono factorization},
and we just have shown that such a factorization is unique up to
isomorphisms (a.k.a. bijections in $\catSET$). 

The following example shows that epimorphisms are not necessarily
surjective, even if they are maps.

\BeginExample{ex-monoid-epi}
Recall that $(M, *)$ is a \emph{\index{monoid}monoid} iff $*: M\times M\to M$ is
associative with a neutral element $0_{M}$. For example, $(\Ganz, +)$
and $(\Nat, \cdot)$ are monoids, so is the set $X^{*}$ of all strings over
alphabet $X$ with concatenation as composition (with the empty string
as neutral element). A \emph{morphism} $f: (M, *)\to (N, \ddag)$ is a map
$f: M\to N$ such that $f(a*b) = f(a)\ddag f(b)$, and $f(0_{M}) = 0_{N}.$

Now let $f: (\Ganz, +)\to (N, \ddag)$ be a morphism, then $f$ is uniquely
determined by the value $f(1)$. This is so since $m = 1 + \dots + 1$ ($m$ times) for $m > 0$, thus 
$
f(m) = f(1 +\dots + 1) = f(1)\ddag\dots\ddag f(1).
$
Also $f(-1)\ddag f(1) = f(-1 + 1) = f(0)$, so $f(-1)$ is
inverse to $f(1)$, hence $f(-m)$ is inverse to $f(m)$. Consequently,
if two morphisms map $1$ to the same value, then the morphisms are identical.

Note that the inclusion $i: x \mapsto x$ is a morphism $i: (\Nat_{0},
+)\to (\Ganz, +)$. We claim that $i$ is an epimorphism. Let
$g_{1}\circ i = g_{2}\circ i$ for some morphisms $g_{1}, g_{2}:
(\Ganz, +)\to (M, *)$. Then 
$
g_{1}(1) = (g_{1}\circ i)(1) = (g_{2}\circ i)(1) = g_{2}(1).
$
Hence $g_{1}= g_{2}$. Thus epimorphisms are not necessarily
surjective. 
\EndExample

Composition induces maps between the hom sets\label{ref-hom-sets} of a category, which we
are going to study now. Specifically, let $\catK$ be a fixed category,
take objects $a$ and $b$ and fix for the moment a morphism $f: a \to
b$. Then $g \mapsto f\circ g$ maps $\homK{x, a}$ to $\homK{x, b}$, and
$h\mapsto h\circ f$ maps $\homK{b, x}$ to $\homK{a, x}$ for each
object $x$. We investigate $g\mapsto f\circ g$ first. Define for an
object $x$ of $\catK$ the map
\begin{equation*}
  \homK{x, f}:
  \begin{cases}
    \homK{x, a} & \to \homK{x, b}\\
g &\mapsto f\circ g
  \end{cases}
\end{equation*}
Then $\homK{x, f}$ defines a map between morphisms, and we can
determine through this map whether or not $f$ is a monomorphism.

\BeginLemma{hom-yields-injective}
$f: a \to  b$ is a monomorphism iff $\homK{x, f}$ is injective for all
objects $x$.
\EndLemma

\BeginProof
This follows immediately from the observation 
\begin{equation*}
f\circ g_{1} = f\circ g_{2} \Leftrightarrow \homK{x, f}(g_{1}) =
\homK{x, f}(g_{2}).
\end{equation*}
\EndProof

Dually, define for an object $x$ of $\catK$ the map
\begin{equation*}
  \homK{f, x}:
  \begin{cases}
    \homK{b, x} & \to \homK{a, x}\\
g &\mapsto g\circ f
  \end{cases}
\end{equation*}
Note that we change directions here: $f: a \to  b$ corresponds to
$\homK{b, x}\to \homK{a, x}$. Note also that we did reuse the name
$\homK{\cdot}$; but no confusion should arise, because the signature
tells us which  map we specifically have in
mind. Lemma~\ref{hom-yields-injective} seems to suggest that
surjectivity of $\homK{f, x}$ and $f$ being an epimorphism are
related. This, however, is not the case. But try this:

\BeginLemma{hom-yields-surjective}
$f: a\to b$ is an epimorphism iff $\homK{f, x}$ is injective for each
object $x$. 
\EndLemma

\BeginProof
$\homK{f, x}(g_{1}) = \homK{f, x}(g_{2})$ is equivalent to $g_{1}\circ
f = g_{2}\circ f$.   
\EndProof

Not surprisingly, an isomorphism is an invertible morphism; this is
described in our scenario as follows.

\BeginDefinition{morph-isomorphism}
$f: a\to b$ is called an \emph{\index{morphism!isomorphism}isomorphism} iff there exists a morphism $g:
b\to a$ such that $g\circ f = id_{a}$ and $f\circ g = id_{b}$. 
\EndDefinition

It is clear that morphism $g$ is in this case uniquely determined: let $g$ and
$g'$ be morphisms with the property above, then we obtain
$
g = g\circ id_{b} = g\circ (f\circ g') = (g\circ f)\circ g' =
id_{a}\circ g' = g'.
$  

When we are in the category $\catSET$ of sets with maps, an
isomorphism $f$ is bijective. In fact, let $g$ be chosen to $f$
according to  Definition~\ref{morph-isomorphism}, then 
\begin{align*}
  h_{1}\circ f = h_{2}\circ f & \Rightarrow  h_{1}\circ f \circ g =
  h_{2}\circ f\circ g \Rightarrow h_{1} = h_{2},\\
f\circ g_{1} = f\circ g_{2} & \Rightarrow g\circ f\circ g_{1}= g\circ
f\circ g_{2} \Rightarrow  g_{1} = g_{2},
\end{align*}
so that the first line makes $f$ an epimorphism, and the second one a
monomorphism. 

The following lemma is often helpful (and serves as an example of the
popular art of \emph{\index{diagram!chasing}diagram chasing}).

\BeginLemma{diagram-chasing}
Assume that in this diagram
\begin{equation*}
\xymatrix{
a\ar[rr]^{f}\ar[d]_{k} && b\ar[rr]^{g}\ar[d]_{\ell} && c\ar[d]^{m}\\
x\ar[rr]_{r} && y\ar[rr]_{s} && z
}
\end{equation*}
the outer diagram commutes, that the leftmost diagram commutes, and
that $f$ is an epimorphism. Then the rightmost diagram commutes as well.
\EndLemma

\BeginProof
In order to show that $m\circ g = s\circ \ell$ it is enough to show
that $m\circ g\circ f = s\circ \ell\circ f$, because we then can
cancel $f$, since $f$ is an epi. But now
\begin{align*}
  (m\circ g)\circ f & = m\circ (g\circ f)\\
& = (s\circ r)\circ k &&\text{(commutativity of the outer diagram)}\\
& = s\circ (r\circ k)\\
& = s\circ (\ell\circ f) &&\text{(commutativity of the leftmost
  diagram)}\\
& = (s\circ \ell)\circ f
\end{align*}
Now cancel $f$.
\EndProof


\Subsection{Elementary Constructions}
\label{sec:elem-constr}

In this section we deal with some elementary constructions, showing
mainly how some important constructions for sets can be carried over
to categories, hence are available in more general
structures. Specifically, we will study products and sums (coproducts)
as well as pullbacks and pushouts. We will not study more general
constructs at present, in particular we will not have a look at limits and
colimits. Once products and pullbacks are understood, the step to
limits should not be too complicated, similarly for colimits, as the reader can see in the brief discussion in Section~\ref{sec:limit-and-colimits}.

We fix a category $\catK$.

\Subsubsection{Products and Coproducts}
\label{sec:prod-coprod}
The Cartesian product of two sets is just the set of pairs. In a
general category we do not have a characterization through sets and
their elements at our disposal, so we have to fill this gap by going
back to morphisms. Thus we require a characterization of product
through morphisms. The first thought is using the projections $\langle
x, y\rangle \mapsto x$ and $\langle x, y\rangle \mapsto y$, since a
pair can be reconstructed through its projections. But this is not
specific enough. An additional characterization of the projections is
obtained through factoring: if there is another pair of maps
pretending to be projections, they better be related to the
``genuine'' projections. This is what the next definition expresses.

\BeginDefinition{categ-product}
Given objects $a$ and $b$ in $\catK$. An object $c$  is called the
\emph{\index{category!product}product of $a$ and $b$ }iff 
\begin{enumerate}
\item there exist morphisms $\pi_{a}: c\to a$ and $\pi_{b}: c \to  b$,
\item for each object $d$ and morphisms $\tau_{a}: d\to a$ and
  $\tau_{b}: d\to b$ there exists a unique morphism $\sigma: d\to c$
  such that $\tau_{a} = \pi_{a}\circ \sigma$ and $\tau_{b} =
  \pi_{b}\circ \sigma$. 
\end{enumerate}
Morphisms $\pi_{a}$ and $\pi_{b}$ are called  \emph{\index{category!product!projection}projections} to $a$
resp. $b$.
\EndDefinition
Thus $\tau_{a}$ and $\tau_{b}$ factor uniquely through $\pi_{a}$ and
$\pi_{b}$. Note that we insist on having a unique factor, and that the
factor should be the same for both pretenders. We will see in a minute
why this is a sensible assumption. If it exists, the product of objects $a$ and $b$
is denoted by $a\times b$; the projections $\pi_{a}$ and $\pi_{b}$ are
usually understood and not  mentioned explicitly. 

This diagram\label{product-diagram} depicts the situation:
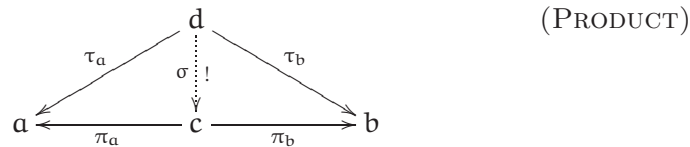
\begin{equation*}
\xymatrix{
&&d\ar[dll]_{\tau_{a}}\ar@{..>}[d]_{\sigma}^{!}\ar[drr]^{\tau_{b}}&&&&\text{\textsc{(Product)}}\\
a && c\ar[ll]^{\pi_{a}}\ar[rr]_{\pi_{b}} && b
}
\end{equation*}

\BeginLemma{cat-prod-unique}
If the product of two objects exists, it is unique up to isomorphism.
\EndLemma

\BeginProof
Let $a$ and $b$ be the objects in question, also assume that $c_{1}$
and $c_{2}$ are products with morphisms $\pi_{i, a}\to a$ and $\pi_{i,
b}\to b$ as the corresponding morphisms, $i = 1, 2$. 

Because $c_{1}$ together with $\pi_{1, a}$ and $\pi_{1, b}$ is a
product, we find a unique morphism $\xi: c_{2}\to c_{1}$ with $\pi_{2,
a} = \pi_{1, a}\circ \xi$ and $\pi_{2, b} = \pi_{1, b}\circ \xi$;
similarly, we find a unique morphism $\zeta: c_{1}\to c_{2}$ with $\pi_{1,
a} = \pi_{2, a}\circ \zeta$ and $\pi_{1, b} = \pi_{2, b}\circ \zeta$. 

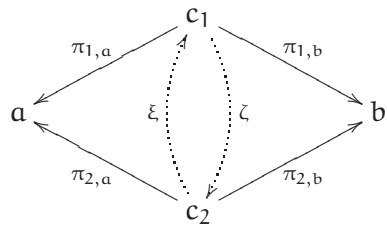
\begin{equation*}
\xymatrix{
&& c_{1}\ar[dll]_{\pi_{1, a}}\ar[drr]^{\pi_{1, b}}\ar@/^1pc/@{..>}[dd]^{\zeta}\\
a && && b\\
&&c_{2}\ar[ull]^{\pi_{2, a}}\ar[urr]_{\pi_{2, b}}\ar@/^1pc/@{..>}[uu]^{\xi}
}
\end{equation*}

Now look at $\xi\circ \zeta$: We obtain
\begin{align*}
  \pi_{1, a} \circ \xi\circ \zeta & = \pi_{2, a} \circ \zeta = \pi_{1,
    a}\\
\pi_{1, b} \circ \xi\circ \zeta & = \pi_{2, b} \circ \zeta = \pi_{1, b}
\end{align*}
Then uniqueness of the factorization implies that $\xi\circ \zeta =
id_{c_{1}}$, similarly, $\zeta\circ \xi = id_{c_{2}}$. Thus $\xi$ and
$\zeta$ are isomorphisms. 
\EndProof

Let us have a look at some examples, first and foremost sets.

\BeginExample{ex-prod-in-sets}
Consider the category $\catSET$ with maps as morphisms. Given sets $A$
and $B$, we claim that $A\times B$ together with the projections
$\pi_{A}: \langle a, b\rangle \mapsto a$ and $\pi_{B}: \langle a,
b\rangle \mapsto b$ constitute the product of $A$ and $B$ in
$\catSET$. In fact, if $\tau_{A}: D\to A$ and $\tau_{B}: D\to B$ are
maps for some set $D$, then $\sigma: d \mapsto \langle \tau_{A}(d),
\tau_{B}(d)\rangle$ satisfies the equations $\tau_{A} = \pi_{A}\circ
\sigma$, $\tau_{B}= \pi_{B}\circ \sigma$, and it is clear that this is
the only way to factor, so $\sigma$ is uniquely determined. 
\EndExample

If sets carry an additional structure, this demands additional
attention.

\BeginExample{ex-prod-in-meas}
Let $(S, {\cal A})$ and $(T, {\cal B})$ be measurable spaces, so we
are now in the category $\Category{Meas}$ of measurable spaces with
measurable maps as morphisms, see Example~\ref{ex-cat-meas-space}. For
constructing a product one is tempted to take the product $S\times T$
is $\catSET$ and to find a suitable $\sigma$-algebra ${\cal C}$ on
$S\times T$ such that the projections $\pi_{S}$ and $\pi_{T}$ become
measurable. Thus ${\cal C}$ would have to contain
$\InvBild{\pi_{S}}{A} = A\times T$ and $\InvBild{\pi_{T}}{B} = S\times
B$ for each $A\in{\cal A}$ and each $B\in{\cal B}$. Because a
$\sigma$-algebra is closed under intersections, ${\cal C}$ would have
to contain all measurable rectangles $A\times B$ with sides in ${\cal
  A}$ and ${\cal B}$. So let's try this:
\begin{equation*}
  {\cal C} := \sigma(\{A\times B \mid A\in{\cal A}, B\in{\cal B}\})
\end{equation*}
Then clearly $\pi_{S}: (S\times T, {\cal C}) \to (S, {\cal A})$ and
$\pi_{T}: (S\times T, {\cal C}) \to (T, {\cal B})$ are morphisms in
$\Category{Meas}$. Now let $(D, {\cal D})$ be a measurable space with
morphisms $\tau_{S}: D\to S$ and $\tau_{T}: D\to T$, and define 
$\sigma$ as above through $\sigma(d) := \langle \tau_{S}(d),
\tau_{T}(d)\rangle$. We claim that $\sigma$ is a morphism in
$\Category{Meas}$. It has to be shown that 
$
\InvBild{\sigma}{C}\in{\cal D}
$ 
for all $C\in{\cal C}$. We have a look at all elements of ${\cal C}$
for which this is true, and we define
\begin{equation*}
{\cal G} := \{C\in{\cal C} \mid \InvBild{\sigma}{C}\in{\cal D}\}.
\end{equation*}
If we can show that ${\cal G} = {\cal C}$, we are done. 
It is evident that ${\cal G}$ is a $\sigma$-algebra, because the
inverse image of a map respects countable Boolean operations. Moreover, if
$A\in{\cal A}$ and $B\in{\cal B}$, then 
$
\InvBild{\sigma}{A\times B} =
\InvBild{\tau_{S}}{A}\cap\InvBild{\tau_{T}}{B}\in{\cal D},
$
so that $A\times B\in{\cal G}$, provided $A\in{\cal A}, B\in{\cal
  B}$. But now we have
\begin{equation*}
  {\cal C} = \sigma(\{A\times B \mid A\in{\cal A}, B\in{\cal
    B}\})\subseteq {\cal G}\subseteq{\cal C}.
\end{equation*}
Hence each element of ${\cal C}$ is a member of ${\cal G}$, thus
$\sigma$ is ${\cal D}$-${\cal C}$-measurable. Again, the construction
shows that there is no other possibility for defining $\sigma$. Hence
we have shown that two objects in the category  $\Category{Meas}$ of measurable
spaces with measurable maps have a product.

The $\sigma$-algebra ${\cal C}$ which is constructed above is usually
denoted by ${\cal A}\otimes{\cal B}$ and called the
\emph{product\index{$\sigma$-algebra!product} $\sigma$-algebra} of ${\cal A}$ and ${\cal B}$.  
\EndExample

\BeginExample{prob-has-no-prods}
While the category $\Category{Meas}$ has products, the situation
changes when taking probability measures into account, hence when
changing to the category $\Category{Prob}$ of probability spaces, see
Example~\ref{ex-cat-meas-space-stochrel}. Recall that the product
measure $\mu\otimes\nu$ of two probability measures $\mu$ on 
$\sigma$-algebra ${\cal A}$ resp. $\nu$ on ${\cal B}$ is the unique
probability measure on the product $\sigma$-algebra ${\cal
  A}\otimes{\cal B}$ with $(\mu\otimes\nu)(A\times B) =
\mu(A)\cdot\nu(B)$ for $A\in{\cal A}$ and $B\in{\cal B}$, in
particular, $\pi_{S}: (S\times T, {\cal A}\otimes {\cal B},
\mu\otimes\nu) \to (S, {\cal A}, \mu)$ and $\pi_{T}: (S\times T, {\cal A}\otimes {\cal B},
\mu\otimes\nu) \to (T, {\cal B}, \nu)$ are morphisms in
$\Category{Prob}$. 

Now define $S := T := [0, 1]$ and take in each case
the smallest $\sigma$-algebra which is generated by the open intervals
as a $\sigma$-algebra, hence put ${\cal A} :=
{\cal B} := \Borel{[0, 1]}$; $\lambda$ is Lebesgue measure on
$\Borel{[0, 1]}$. Define 
\begin{equation*}
  \kappa(E) := \lambda(\{x\in [0, 1] \mid \langle x, x\rangle \in E\})
\end{equation*}
for $E\in{\cal A}\otimes{\cal B}$ (well, we have to show that $\{x\in
[0, 1] \mid \langle x, x\rangle \in E\}\in \Borel{[0, 1]}$, whenever
$E\in {\cal A}\otimes{\cal B}$. This is not difficult and relegated to
Exercise~\ref{ex-aus-measurability}). Then $\pi_{S}: (S\times T, {\cal
  A}\otimes{\cal B}, \kappa)\to (S, {\cal A}, \lambda)$ and $\pi_{T}: (S\times T, {\cal
  A}\otimes{\cal B}, \kappa)\to (T, {\cal B}, \lambda)$ are morphisms
in $\Category{Prob}$, because
$
\kappa(\InvBild{\pi_{S}}{G}) = \kappa(G\times T)
= \lambda(\{x\in [0, 1] \mid \langle x, x\rangle \in G\times T\}) =
  \lambda(G)
$
for $G\in\Borel{S}$. If we could find a morphism $f: (S\times T, {\cal
  A}\otimes{\cal B}, \kappa)\to (S\times T, {\cal
  A}\otimes{\cal B}, \lambda\otimes\lambda)$ factoring through the
projections, $f$ would have to be the identity; thus would imply that
$\kappa = \lambda\otimes\lambda$, but this is not the case: take 
$
E := [1/2, 1]\times[0, 1/3],
$
then $\kappa(E) = 0$, but $(\lambda\otimes\lambda)(E) = 1/6$. 

Thus we conclude that the category $\Category{Prob}$ of probability spaces does not have
products. 
\EndExample

The product topology on the Cartesian product of the carrier sets of topological spaces is familiar, open sets in the product just contain open rectangles. The categorical view is that of a product in the category of topological spaces.

\BeginExample{ex-prod-in-top}
Let $(T, {\cal G})$ and $(T, {\cal H})$ be topological spaces, and
equip the Cartesian product $S\times T$ with the product topology
${\cal G}\times{\cal H}$. This is the smallest topology on $S\times T$
which contains all the open rectangles $G\times H$ with $G\in {\cal
  G}$ and $H\in {\cal H}$. We claim that this is a product in the
category $\Category{Top}$ of topological spaces. In fact, the
projections $\pi_{S}:
S\times T\to S$ and $\pi_{T}: S\times T\to T$ are continuous, because,
e.g, $\InvBild{\pi_{S}}{G} = G\times T\in {\cal G}\times{\cal H}$. Now
let $(D, {\cal D})$ be a topological space with continuous maps
$\tau_{S}: D\to S$ and $\tau_{T}: D\to T$, and define $\sigma: D\to
S\times T$ through  $\sigma: d \mapsto
\langle\tau_{S}(d), \tau_{T}(d)\rangle$. Then
$\InvBild{\sigma}{G\times H} =
\InvBild{\tau_{S}}{G}\cap\InvBild{\tau_{T}}{H}\in {\cal D}$, and since
the inverse image of a topology under a map is a topology again,
$\sigma: (D, {\cal D})\to (S\times T, {\cal G}\times {\cal H})$ is
continuous. Again, this is the only way to define a morphism $\sigma$ so that
$\tau_{S} = \pi_{S}\circ \sigma$ and $\tau_{T}= \pi_{T}\circ
\sigma$. 
\EndExample

The category coming from a partially ordered set from
Example~\ref{ex-cat-ord} is investigated next.

\BeginExample{ex-cat-ord-prod}
Let $(P, \leq)$ be a partially ordered set, considered as a category
$\Category{P}$. Let $a, b\in P$, and assume that $a$ and $b$ have a
product $x$ in $\Category{P}$. Thus there exist morphisms $\pi_{a}:
x\to a$ and $\pi_{b}: x\to b$, which means by the definition of this
category that $x\leq a$ and $x\leq b$ hold, hence that $x$ is a lower
bound to $\{a, b\}$. Moreover, if $y$ is such that there are
morphisms $\tau_{a}: y\to a$ and $\tau_{b}: y\to b$, then there exists
a unique $\sigma: y\to x$ with $\tau_{a} = \pi_{a}\circ \sigma$ and
$\tau_{b}: \pi_{b}\circ \sigma$. Translated into $(P, \leq)$, this
means that if $y\leq a$ and $y \leq b$, then $y\leq x$ (morphisms in
$\Category{Meas}$ are unique, if they exist). Hence the product $x$ is
just the greatest lower bound of $\{a, b\}$.

So the product corresponds to the infimum. This example demonstrates again
that products do not necessarily exist in a category.
\EndExample

Given morphisms $f: x\to a$ and $g: x\to b$, and assuming that the
product $a\times b$ exists, we want to ``lift'' $f$ and $g$ to the
product, i. e., we want to find a morphism $h: x\to a\times b$ with $f
= \pi_{a}\circ h$ and $g = \pi_{b}\circ h$. Let us see how this is
done in $\catSET$: Here $f: X\to A$ and $g: X\to B$ are maps, and one
defines the lifted map $h: X\to A\times B$ through $h: x \mapsto
\langle f(x), g(x)\rangle$, so that the conditions on the projections
is satisfied. The next lemma states that this is always possible in a
unique way.

\BeginLemma{lift-to-prod}
Assume that the product $a\times b$ exists for the objects $a$ and
$b$. Let $f: x\to a$ and $g: x \to b$ be morphisms. Then there exists a
unique morphism $q: x\to a\times b$ such that $f = \pi_{a}\circ q$ and
$g = \pi_{b}\circ q$. Morphism $q$ is denoted by $f\times g$. 
\EndLemma

\BeginProof
The diagram looks like this:
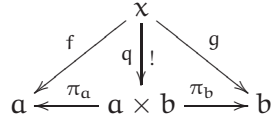
\begin{equation*}
\xymatrix{
&x\ar[dl]_{f}\ar[dr]^{g}\ar[d]_{q}^{!}\\
a&a\times b\ar[l]_{\pi_{a}}\ar[r]^{\pi_{b}}&b}
\end{equation*}
Because $f: x\to a$ and $g: x\to b$, there exists a unique $q: x\to
a\times b$ with  $f = \pi_{a}\circ q$ and
$g = \pi_{b}\circ q$. This follows from the definition of the
product.
\EndProof

Let us look at the product through our $\hom{\catK}$-glasses. If
$a\times b$ exists, and if $\tau_{a}: d\to a$ and $\tau_{b}: d\to b$
are morphisms, we know that there is a \emph{unique} $\sigma: d\to a\times b$
rendering this diagram commutative
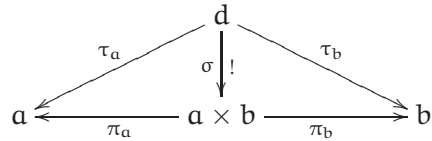
\begin{equation*}
\xymatrix{
&&d\ar[dll]_{\tau_{a}}\ar[d]_{\sigma}^{!}\ar[drr]^{\tau_{b}}\\
a && a\times b\ar[ll]^{\pi_{a}}\ar[rr]_{\pi_{b}} && b
}
\end{equation*}
Thus the map 
\begin{equation*}
  p_{d}:
  \begin{cases}
    \homK{d, a}\times\homK{d, b}&\to \homK{d, a\times b}\\
\langle \tau_{a}, \tau_{b}\rangle & \mapsto \sigma
  \end{cases}
\end{equation*}
is well defined. In fact, we can say more

\BeginProposition{pd-is-bijective}
$p_{d}$ is a bijection.
\EndProposition

\BeginProof
Assume $\sigma = p_{d}(f, g) = p_{d}(f', g')$. Then $f = \pi_{a}\circ
\sigma = f'$ and $g = \pi_{b}\circ \sigma = g'$. Thus $\langle f,
g\rangle = \langle f', g'\rangle$. Hence $p_{d}$ is
injective. Similarly, one shows that $p_{d}$ is surjective: Let
$h\in\homK{d, a\times b}$, then $\pi_{a}\circ h: d\to a$ and
$\pi_{b}\circ h: d\to b$ are morphisms, so there exists a unique $h':
d\to a\times b$ with $\pi_{a}\circ h' = \pi_{a}\circ h$ and
$\pi_{b}\circ h'=\pi_{b}\circ h$. Uniqueness implies that $h = h'$, so
$h$ occurs in the image of $p_{d}$.
\EndProof

Let us consider the dual construction.

\BeginDefinition{cat-sum}
Given objects $a$ and $b$ in category $\catK$, the object $s$ together
with morphisms $i_{a}: a\to s$ and $i_{b}: b\to s$ is called the
\emph{\index{category!coproduct}coproduct}(or the \emph{\index{category!sum}sum})
of $a$ and $b$ iff for each object $t$ with morphisms $j_{a}: a\to t$
and $j_{b}: b\to t$ there exists a unique morphism $r: s\to t$ such that
$j_{a} = r\circ i_{a}$ and $j_{b}= r\circ i_{b}$. Morphisms $i_{a}$
and $i_{b}$ are called \emph{\index{category!coproduct!injection}injections}, the
coproduct of $a$ and $b$ is denoted by $a+b$. 
\EndDefinition
This is the corresponding\label{coproduct-diagram} diagram:

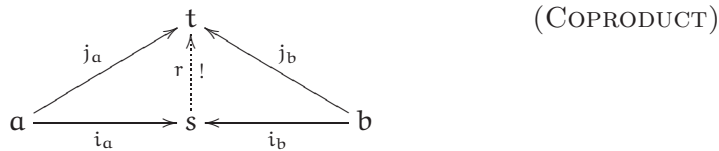
\begin{equation*}
\xymatrix{
&&t&&&&\text{\textsc{(Coproduct)}}\\
a\ar[urr]^{j_{a}}\ar[rr]_{i_{a}} && s\ar@{..>}[u]^{r}_{!} && b\ar[ull]_{j_{b}}\ar[ll]^{i_{b}}
}
\end{equation*}

Let us have a look at some examples. 

\BeginExample{ex-cat-ord-coprod}
Let $(P, \leq)$ be a partially ordered set, and consider category
$\Category{P}$, as in Example~\ref{ex-cat-ord-prod}. The coproduct of
the elements $a$ and $b$ is just the supremum $\sup\{a, b\}$. This is
shown with exactly the same arguments which have been used in
Example~\ref{ex-cat-ord-prod} for showing the the product of two
elements corresponds to their infimum.
\EndExample

And then there is of course category $\catSET$. 

\BeginExample{ex-coprod-set}
Let $A$ and $B$ be disjoint sets. Then $S := A\cup B$ together with

\smallBox{
  \begin{equation*}
    i_{A}:
    \begin{cases}
      A & \to  S\\
a &\mapsto a
    \end{cases}
  \end{equation*}
}\smallBox{
  \begin{equation*}
    i_{B}:
    \begin{cases}
      B & \to  S\\
b &\mapsto b
    \end{cases}
  \end{equation*}
}

form the coproduct of $A$ and $B$. In fact, if $T$ is a set with maps
$j_{A}: A\to T$ and $j_{B}: B\to T$, then define
\begin{equation*}
  r: 
  \begin{cases}
    S & \to T\\
s & \mapsto j_{A}(a), \text{ if } s = i_{A}(a),\\
s & \mapsto j_{B}(b), \text{ if } s = i_{B}(b)
  \end{cases}
\end{equation*}
Then $j_{A} = r\circ i_{A}$ and $j_{B}= r\circ i_{B}$, and these
definitions are the only possible ones. 

Note that we needed for this construction to work disjointness of the
participating sets. Consider for example $A := \{-1, 0\}$, $B := \{0,
1\}$ and let $T := \{-1, 0, 1\}$ with $j_{A}(x) := -1$, $j_{B}(x) :=
+1$. No matter where we embed $A$ and $B$, we cannot factor $j_{A}$
and $j_{B}$ uniquely.

If the sets are not disjoint, we first do a preprocessing step and
embed them, so that the embedded sets are disjoint. The injections
have to be adjusted accordingly. So this construction would work:
Given sets $A$ and $B$, define $S := \{\langle a, 1\rangle \mid a\in
A\}\cup\{\langle b, 2\rangle \mid b\in B\}$ with $i_{A}: a\mapsto
\langle a, 1\rangle$ and $b \mapsto \langle b, 2\rangle$. Note that we
do not take a product like $S\times\{1\}$, but rather use a very
specific construction; this is so since the product is determined
uniquely only by isomorphism, so we might not have gained anything by
using it. Of course, one has to be sure that the sum is not dependent
in an essential way on this embedding.
\EndExample

The question of uniqueness is answered through this observation. It
relates the coproduct in $\catK$ to the product in the dual category
$\catK^{op}$ (see Example~\ref{ex-opposite-category}). 

\BeginProposition{coprod-dual-to-prod}
The coproduct $s$ of objects $a$ and $b$ with injections $i_{a}: a\to
s$ and $i_{b}: b\to s$ in category $\catK$ is the product in category
$\catK^{op}$ with projections $i_{a}: s\to^{op} a$ and $i_{s}: s\to^{op} b$.
\EndProposition


\BeginProof
Revert in diagram \textsc{(Coproduct)} on page~\pageref{coproduct-diagram} to obtain diagram
\textsc{(Product)} on page~\pageref{product-diagram}.
\EndProof

\BeginCorollary{uniq-coprod}
If the coproduct of two objects in a category exists, it is unique up
to isomorphisms.
\EndCorollary

\BeginProof
Proposition~\ref{coprod-dual-to-prod} together with
Lemma~\ref{cat-prod-unique}. 
\EndProof

Let us have a look at the coproduct for topological spaces.

\BeginExample{coprod-top-spaces}
Given topological spaces $(S, {\cal G})$ and $(T,{\cal H})$, we may
and do assume that $S$ and $T$ are disjoint. Otherwise wrap the
elements of the sets accordingly; put
\begin{align*}
  A^{\dag} & := \{\langle a, 1\rangle \mid a \in A\},\\
B^{\ddag} & := \{\langle b, 2\rangle \mid b \in B\},
\end{align*}
and consider the
topological spaces $(S^{\dag},\{G^{\dag} \mid G\in{\cal G}\})$
and $(T^{\ddag},\{H^{\ddag} \mid H\in{\cal H}\})$ instead of $(S,
{\cal G})$ and $(T,{\cal H})$. Define on
the coproduct $S+T$ of $S$ and $T$ in $\catSET$ with injections
$i_{S}$ and $i_{T}$ the topology 
\begin{equation*}
{\cal G}+{\cal H} := \{W\subseteq S+T \mid \InvBild{i_{S}}{W}\in{\cal
G}\text{ and } \InvBild{i_{T}}{W}\in{\cal H}\}.
\end{equation*}
This is  a topology: Both $\emptyset$ and $S+T$ are members of
${\cal G}+{\cal H}$, and since ${\cal G}$ and ${\cal H}$ are
topologies, ${\cal G}+{\cal H}$ is closed under finite intersections
and arbitrary unions. Moreover, both $i_{S}: (S, {\cal G})\to (S+T,
{\cal G}+{\cal H})$ and $i_{T}: (T, {\cal H})\to (S+T, {\cal G}+{\cal
  H})$ are continuous; in fact, ${\cal G}+{\cal H}$ is the smallest
topology on $S+T$ with this property. 

Now assume that $j_{S}: (S,
{\cal G})\to (R, {\cal R})$ and $j_{T}: (T,
{\cal H})\to (R, {\cal R})$ are continuous maps, and let $r: S+T\to R$
be the unique map determined by the coproduct in $\catSET$. Wouldn't
is be nice if $r$ is continuous? Actually, it is. Let $W\in{\cal R}$
be open in $R$, then $\InvBild{i_{S}}{\InvBild{r}{W}} = \InvBild{(r\circ
  i_{S})}{W} = \InvBild{j_{S}}{W}\in{\cal G}$, similarly,
$\InvBild{i_{S}}{\InvBild{r}{W}}\in{\cal H}$, thus by definition,
$\InvBild{r}{W}\in {\cal G}+{\cal H}$. Hence we have found the
factorization $j_{S}= r\circ i_{S}$ and $j_{T}= r\circ i_{T}$ in the
category $\Category{Top}$. This
factorization is unique, because it is inherited from the unique
factorization in $\catSET$. Hence we have shown that $\Category{Top}$
has finite coproducts. 
\EndExample

A similar construction applies to the category of measurable spaces.

\BeginExample{coprod-meas-space}
Let $(S, {\cal A})$ and $(T, {\cal B})$ be measurable spaces; we may
assume again that the carrier sets $S$ and $T$ are disjoint. Take the
injections $i_{S}: S\to S+T$ and $i_{T}: T\to S+T$ from $\catSET$.  Then 
\begin{equation*}
  {\cal A}+{\cal B} := \{W\subseteq S+T \mid
  \InvBild{i_{S}}{W}\in{\cal A}\text{ and } \InvBild{i_{T}}{W}\in{\cal B}\}
\end{equation*}
is a $\sigma$-algebra, $i_{S}: (S, {\cal A})\to (S+T, {\cal A}+{\cal
  B})$ and  $i_{T}: (T, {\cal B})\to (S+T, {\cal A}+{\cal
  B})$ are measurable. The unique factorization property is
established in exactly the same way as for $\Category{Top}$. 
\EndExample

\BeginExample{coprod-relations}
Let us consider the category $\Category{Rel}$ of relations, which is
based on sets as objects. If $S$ and $T$ are sets, we again may and do
assume that they are disjoint. Then $S+T = S\cup T$ together with the
injections 
\begin{align*}
  I_{S} & := \{\langle s, i_{S}(s)\rangle \mid s\in S\},\\
I_{T} & := \{\langle t, i_{T}(t)\rangle \mid t \in T\}
\end{align*}
form the coproduct, where $i_{S}$ and $i_{T}$ are the injections into
$S+T$ from $\catSET$. In fact, we have to show that we can find for
given relations $q_{S}\subseteq S\times D$ and $q_{T}\subseteq T\times
D$ a unique relation $Q\subseteq (S+T)\times D$ with $q_{S} =
I_{S}\circ Q$ and $q_{T} = I_{T}\circ Q$. The choice is fairly
straightforward: Define
\begin{equation*}
  Q := \{\langle i_{S}(s), q\rangle \mid \langle s, q\rangle \in
  q_{S}\}
\cup 
\{\langle i_{T}(t), q\rangle \mid \langle t, q\rangle \in q_{T}\}.
\end{equation*}
Thus
\begin{equation*}
\langle s, q\rangle \in I_{S}\circ Q
\Leftrightarrow
\text{there exists } x \text{ with }\langle s, x\rangle \in I_{S}\text{ and } \langle x,
q\rangle \in Q
\Leftrightarrow \langle s, q\rangle\in q_{S}.
\end{equation*}
Hence $q_{S} = I_{S}\circ Q$, similarly, $q_{T} = I_{T}\circ Q$. It is
clear that no other choice is possible. 

Consequently, the coproduct is the same as in $\catSET$. 
\EndExample

We have just seen in a simple example that dualizing, i.e., going to
the dual category, is very helpful. Instead of proving directly that
the coproduct is uniquely determined up to isomorphism, if it exists,
we turned to the dual category and reused the already established
result that the product is uniquely determined, casting it into a new
context. The duality, however, is a purely structural property, it usually does
not help us with specific constructions. This could be seen when we
wanted to construct the coproduct of two sets; it did not help here
that we knew how to construct the product of two sets, even though
product and coproduct are intimately related through dualization. We
will make the same observation when we deal with pullbacks and
pushouts.

\Subsubsection{Pullbacks and Pushouts}
\label{sec:pullback-pushout}
Sometimes one wants to complete the square as in the diagram below on
the left hand side:

\smallBox{
\begin{equation*}
\xymatrix{
&&b\ar[d]^{g}\\
a\ar[rr]_{f} && c
}
\end{equation*}
}\smallBox{
\begin{equation*}
\xymatrix{
d\ar@{-->}[rr]^{i_{2}}\ar@{-->}[d]_{i_{1}}&&b\ar[d]^{g}\\
a\ar[rr]_{f} && c
}
\end{equation*}
}

Hence one wants to find an object $d$ together with morphisms $i_{1}:
d\to a$ and $i_{2}: d\to b$ rendering the diagram on the right hand
side commutative. This completion should be as coarse as possible in
this sense. If we have another objects, say, $e$ with morphisms
$j_{1}: e\to a$ and $j_{2}: e\to b$ such that $f\circ j_{1} = g\circ
j_{2}$, then we want to be able to uniquely factor through $i_{1}$ and
$i_{2}$.


This is captured in the following definition.

\BeginDefinition{def-pullback}
Let $f: a\to c$ and $g: b\to c$ be morphisms in $\catK$ with the same
codomain. An object $d$ together with morphisms $i_{1}: d\to a$ and
$i_{2}: d\to b$ is called a \emph{\index{category!pullback}\index{pullback}pullback} of $f$ and $g$ iff
\begin{enumerate}
\item $f\circ i_{1} = g\circ i_{2}$,
\item If $e$ is an object with morphisms
$j_{1}: e\to a$ and $j_{2}: e\to b$ such that $f\circ j_{1} = g\circ
j_{2}$, then there exists a unique morphism $h: e\to d$ such that 
$j_{1} = i_{1}\circ h$ and $j_{2} = i_{2} \circ h.$
\end{enumerate}
If we postulate the existence of the morphism $h: e\to
d$, but do not insist on its uniqueness, then $d$ with $i_{1}$ and $i_{2}$
is called a \emph{\index{pullback!weak}\index{category!weak pullback}weak pullback}. 
\EndDefinition

A diagram for a pullback looks like this

\begin{equation*}
\xymatrix{
e\ar@/^1pc/@{-->}[drrr]^{j_{2}}\ar@/^-1pc/@{-->}[ddr]_{j_{1}}\ar@{..>}[dr]^{h}_{!}\\
& d\ar@{-->}[rr]^{i_{2}}\ar@{-->}[d]_{i_{1}} && b\ar[d]^{g}\\
& a\ar[rr]_{f} && c
}
\end{equation*}
It is clear that a pullback is unique up to isomorphism; this is shown
in exactly the same way as in Lemma~\ref{cat-prod-unique}. Let us have
a look at $\catSET$ as an important example to get a first impression
on the inner workings of a pullback.

\BeginExample{pullback-in-set}
Let $f: X\to Z$ and $g: Y\to Z$ be maps. We claim that 
\begin{equation*}
  P  := \{\langle x, y\rangle \in X\times Y\mid f(x) = g(y)\}
\end{equation*}
together with the projections $\pi_{X}: \langle x, y\rangle \mapsto x$
and  $\pi_{Y}:  \langle x, y\rangle \mapsto y$
is a pullback for $f$ and $g$. 

Let $\langle x, y\rangle\in P$, then
\begin{equation*}
  (f\circ \pi_{X})(x, y) = f(x) = g(y) = (g\circ \pi_{Y})(x, y),
\end{equation*}
so that the first condition is satisfied. Now assume that $j_{X}: T\to
X$ and $j_{Y}: T\to Y$ satisfies 
$
f(j_{X}(t)) = g(j_{Y}(t))
$
for all $t\in T$. Thus $\langle j_{X}(t), j_{Y}(t)\rangle\in P$ for
all $t$, and defining 
$
r(t) := \langle j_{X}(t), j_{Y}(t)\rangle,
$
we obtain $j_{X} = \pi_{X}\circ r$ and $j_{Y} = \pi_{Y}\circ r$. Moreover,
this is the only possibility to define a factor map with the desired
property. 

An interesting special case occurs for $X = Y$ and $f = g$. Then $P =
\Kern{f}$, so that the kernel of a map occurs as a pullback in category $\catSET$. 
\EndExample

As an illustration for the use of a pullback construction, look at
this simple statement.

\BeginLemma{pullback-yields-mono}
Assume that $d$ with morphisms $i_{a}: d\to a$ and $i_{b}: d\to b$ is
a pullback for $f: a\to c$ and $g: b\to c$. If $g$ is a mono, so is
$i_{a}$. 
\EndLemma

\BeginProof
Let $g_{1}, g_{2}: e\to d$ be morphisms with $i_{a}\circ g_{1} =
i_{a}\circ g_{2}$. We have to show that $g_{1} = g_{2}$ holds. If we
know that $i_{b}\circ g_{1} = i_{b}\circ g_{2},$ we may use the
definition of a pullback and capitalize on the uniqueness of the
factorization. But let's see. 

From $i_{a}\circ g_{1} = i_{b}\circ g_{2}$ we conclude $f\circ
i_{a}\circ g_{1} = f\circ i_{b}\circ g_{2}$, and because $f\circ i_{a}
= g\circ i_{b}$, we obtain $g\circ i_{b}\circ g_{1} = g\circ i_{b}
\circ g_{2}$. Since $g$ is a mono, we may cancel on the left of this
equation, and we obtain, as desired,  $i_{b}\circ g_{1} = i_{b}\circ g_{2}.$

But since we have a pullback, there exists a unique $h: e\to d$ with
$i_{a}\circ g_{1} = i_{a}\circ h\ \bigl(= i_{a}\circ g_{2} \bigr)$ and
$i_{b}\circ g_{1} = i_{b}\circ h\ \bigl(= i_{b}\circ g_{2}\bigr)$. We
see that the morphisms $g_{1}$, $g_{2}$ and $h$ have the same properties with
respect to factoring, so they must be identical by uniqueness. Hence $g_{1} = h =
g_{2}$, and we are done.
\EndProof

This is another simple example for the use of a pullback in $\catSET$.

\BeginExample{factor-as-a-pullback}
Let $R$ be an equivalence relation on a set $X$ with projections
$\pi_{1}: \langle x_{1}, x_{2}\rangle \mapsto x_{1}$; the second projection $\pi_{2}: R \to X$ is defined similarly. Then 
\begin{equation*}
\xymatrix{
R\ar[d]_{\pi_{1}}\ar[rr]^{\pi_{2}}&& X\ar[d]^{\fMap{R}}\\
X\ar[rr]_{\fMap{R}}&&\Faktor{X}{R}
}
\end{equation*}
(with $\fMap{R}: x \mapsto \Klasse{x}{R}$) is a pullback diagram. In
fact, the diagram commutes. Let $\alpha, \beta: M\to X$ be
maps with $\alpha\circ \fMap{R} = \beta\circ \fMap{R}$, thus
$\Klasse{\alpha(m)}{R} = \Klasse{\beta(m)}{R}$ for all $m\in M$; hence
$\langle \alpha(m), \beta(m)\rangle \in R$ for all $m$. The
only map $\tau: M\to R$ with $\alpha = \pi_{1}\circ \tau$ and $\beta =
\pi_{2}\circ \tau$ is $\tau(m) := \langle \alpha(m),
\beta(m)\rangle$. 
\EndExample

Pullbacks are compatible with products in a sense which we will make
precise in a moment. Before we do that, however, we need an auxiliary
statement:

\BeginLemma{double-product-constr}
Assume that the products $a\times a'$ and $b\times b'$ exist in
category $\catK$. Given morphisms $f: a\to b$ and $f': a'\to b'$,
there exists a unique morphism $f\times f': a\times a'\to
b\times b'$ such that 
\begin{align*}
  \pi_{b}\circ f\times f' & = f\circ \pi_{a}\\
\pi_{b'}\circ f\times f' & = f'\circ \pi_{a'}
\end{align*}
\EndLemma

\BeginProof
Apply the definition of a product to the morphisms $f\circ \pi_{a}:
a\times a'\to b$ and $f'\circ \pi_{a'}: a\times a'\to b'$.
\EndProof

Thus the morphism $f\times f'$ constructed in the lemma renders both parts of
this diagram commutative. 
\begin{equation*}
\xymatrix{
a\ar[d]_{f} && a\times a'\ar[d]_{f\times f'}\ar[ll]_{\pi_{a}}\ar[rr]^{\pi_{a'}} && a'\ar[d]^{f'}\\
b && b\times b'\ar[ll]^{\pi_{b}}\ar[rr]_{\pi_{b'}}  && b'
}
\end{equation*}
Denoting this morphism as $f\times f'$, we note that $\times$ is overloaded
for morphisms, a look at domains and codomains indicates without ambiguity,
however, which version is intended. 

Quite apart from its general interest, this is what we need
Lemma~\ref{double-product-constr} for.

\BeginLemma{pullback-vs-products}
Assume that we have these pullbacks
\begin{equation*}
\xymatrix{
a\ar[rr]^{f}\ar[d]_{g} && b\ar[d]^{h} &&& a'\ar[rr]^{f'}\ar[d]_{g'} && b'\ar[d]^{h'}\\
c\ar[rr]_{k} && d &&& c'\ar[rr]_{k'} && d'
}
\end{equation*}
Then this is a pullback diagram as well
\begin{equation*}
\xymatrix{
a\times a'\ar[rr]^{f\times f'}\ar[d]_{g\times g'} && b\times b'\ar[d]^{h\times h'}\\
c\times c'\ar[rr]_{k\times k'} && d\times d'
}
\end{equation*}
\EndLemma

\BeginProof
1.
We show first that the diagram commutes. It is sufficient to compute
the projections, from uniqueness then equality will follow. Allora:
\begin{align*}
  \pi_{d}\circ (h\times h')\circ (f\times f') & = (h\circ \pi_{a})\circ
  (f\times f') = h\circ f\circ \pi_{a}\\
\pi_{d}\circ (k\times k')\circ (g\times g') & = k\circ \pi_{c}\circ
  (g\times g') = k\circ g\circ \pi_{a} = h\circ f\circ \pi_{a}.
\end{align*}
A similar computation is carried out for $\pi_{a'}$. 

2.
Let $j: t\to c\times c'$ and $\ell: t\to b\times b'$ be morphisms such
that $(k\times k')\circ j = (h\times h')\circ  \ell$, then we claim
that there exists a unique morphism $r: t\to a\times a'$ such that $j
= (g\times g')\circ r$ and $\ell = (f\times f')\circ r.$ The plan is
to obtain $r$ from the projections and then show that this morphism is
unique. 

3.
We show that this diagram commutes
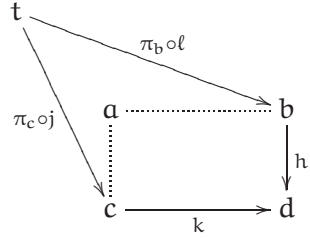
\begin{equation*}
\xymatrix{
t\ar[drrr]^{\pi_{b}\circ  \ell}\ar[ddr]_{\pi_{c}\circ j}\\
& a\ar@{..}[rr]\ar@{..}[d] && b\ar[d]^{h}\\
& c\ar[rr]_{k} && d
}
\end{equation*}
We have
\begin{align*}
  k\circ (\pi_{c}\circ j) & = (k\circ \pi_{c})\circ j\\
& = (\pi_{d}\circ k\times k')\circ j\\
& =\pi_{d}\circ (k\times k'\circ j)\\
& \stackrel{(\ddag)}{=}  \pi_{d}\circ (h\times h'\circ \ell)\\
& = (\pi_{d}\circ h\times h')\circ \ell\\
& = (h\circ \pi_{b})\circ \ell\\
& = h\circ (\pi_{b}\circ \ell)
\end{align*}
In $(\ddag)$ we use Lemma~\ref{double-product-constr}. Using the
primed part of Lemma~\ref{double-product-constr} we obtain $k'\circ
(\pi_{c'}\circ j) = h'\circ (\pi_{b'}\circ \ell).$

Because the left hand side in the assumption is a pullback diagram,
there exists a unique morphism $\rho: t\to a$ with 
$
\pi_{c}\circ j  = g\circ \rho,
\pi_{b}\circ \ell  = f\circ \rho.
$
Similarly, there exists a unique morphism $\rho': t\to a'$ with 
$
\pi_{c'}\circ j  = g'\circ \rho',
\pi_{b'}\circ \ell  = f'\circ \rho'.
$

4.
Put $r := \rho\times\rho'$, then $r: t\to a\times a'$, and we have
this diagram

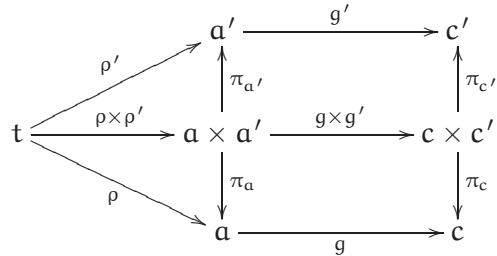
\begin{equation*}
\xymatrix{
&&a'\ar[rr]^{g'}&&c'\\
t\ar[urr]^{\rho'}\ar[drr]_{\rho}\ar[rr]^{\rho\times\rho'}
&&
a\times a'\ar[u]_{\pi_{a'}}\ar[d]^{\pi_{a}}\ar[rr]^{g\times g'} 
&&
c\times c'\ar[u]_{\pi_{c'}}\ar[d]^{\pi_{c}}\\
&& a\ar[rr]_{g} && c
}
\end{equation*}
Hence
\begin{align*}
  \pi_{c}\circ (g\times g')\circ (\rho\times\rho') & = g\circ
  \pi_{a}\circ (\rho\times\rho') = g\circ \rho = \pi_{c}\circ j\\
\pi_{c'}\circ (g\times g')\circ (\rho\times\rho') & = g'\circ
\pi_{a'}\circ (\rho\times\rho') = g'\circ \rho' = \pi_{c'}\circ j.
\end{align*}
Because a morphism into a product is uniquely determined by its
projections, we conclude that $(g\times g')\circ (\rho\times\rho') =
j$. Similarly, we obtain $(f\times f')\circ (\rho\times\rho') = \ell$.

5.
Thus $r = \rho\times\rho'$ can be used for factoring; it remains to
show that this is the only possible choice. In fact, let $\sigma: t\to
a\times a'$ be a morphism with $(g\times g')\circ \sigma = j$ and
$(f\times f')\circ \sigma = \ell$, then it is enough to show that
$\pi_{a}\circ \sigma$ has the same properties as $\rho$, and that
$\pi_{a'}\circ \sigma$ has the same properties as $\rho'$. Calculating
the composition with $g$ resp. $f$, we obtain
\begin{align*}
  g\circ \pi_{a}\circ \sigma & = \pi_{c}\circ (g\times g')\circ \sigma
  = \pi_{c}\circ j\\
f\circ \pi_{a}\circ \sigma & = \pi_{d}\circ (f\times f')\circ \sigma = \pi_{b}\circ \ell
\end{align*}
This implies $\pi_{a}\circ \sigma = \rho$ by uniqueness of $\rho$, the
same argument implies $\pi_{a'}\circ \sigma = \rho'$. But this means
$\sigma = \rho\times\rho'$, and uniqueness is established.
\EndProof

Let's dualize. The pullback was defined so that the upper left
corner of a diagram is filled in an essentially unique way, the dual construction will have to fill
the lower right corner of a diagram in the same way. But by reversing
arrows, we convert a diagram in which the lower right corner is
missing into a diagram without an upper left corner:

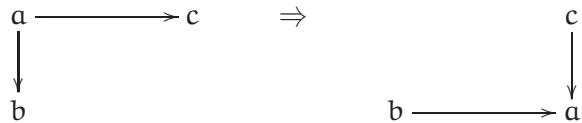
\begin{equation*}
\xymatrix{
a\ar[d]\ar[rr]&& c &\Rightarrow&   && c\ar[d]\\
b &&  & & b\ar[rr] && a
}
\end{equation*}
The corresponding construction is called a pushout.

\BeginDefinition{def-pushout}
Let $f: a\to b$ and $g: a\to c$ be morphisms in category $\catK$ with
the same domain. An object $d$ together with morphisms $p_{b}: b\to d$
and $p_{c}: c\to d$ is called the
\emph{\index{category!pushout}pushout} of $f$ and $g$ iff these
conditions are satisfied:
\begin{enumerate}
\item $p_{b}\circ f = p_{c}\circ g$
\item if $q_{b}: b\to e$ and $q_{c}: c\to e$ are morphisms such that
  $q_{b}\circ f = q_{c}\circ g$, then there exists a unique morphism
  $h: d\to e$ such that $q_{b} = h\circ p_{b}$ and $q_{c} = h\circ
  q_{c}$. 
\end{enumerate}
\EndDefinition

This diagram obviously looks like this:

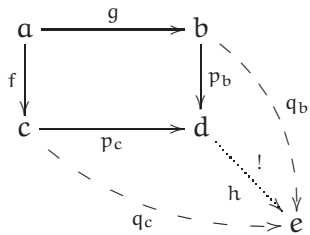
\begin{equation*}
\xymatrix{
a\ar[rr]^{g}\ar[d]_{f} && b\ar[d]^{p_{b}}\ar@/^1pc/@{-->}[ddr]^{q_{b}}\\
c\ar[rr]_{p_{c}}\ar@/^-1pc/@{-->}[drrr]_{q_{c}} && d\ar@{..>}[dr]^{!}_{h}\\
&&&e
}
\end{equation*}

It is clear that the pushout of $f\in\homK{a, b}$  and $g\in\homK{a,
  c}$ is the pullback of $f\in\hom{\catK^{op}}(b, a)$ and of
$g\in\hom{\catK^{op}}(c, a)$ in the dual category. This, however, does
not really provide assistance when constructing a pushout. Let us consider
specifically the category $\catSET$ of sets with maps as morphisms. We know that dualizing a
product yields a sum, but it is not quite clear how to proceed
further. The next example tells us what to do.

\BeginExample{pushout-in-set}
We are in the category $\catSET$ of sets with maps as morphisms
now. Consider maps $f: A\to B$ and $g: A\to C$. Construct on the sum
$B+C$ the smallest equivalence relation $R$ which contains 
$
R_{0} := \{\langle (i_{B}\circ f)(a), (i_{C}\circ g)(a)\rangle \mid a\in A\}.
$
Here $i_{B}$ and $i_{C}$ are the injections of $B$ resp. $C$ into the
sum. Let $D$ the factor space $\Faktor{(A+B)}{R}$ with $p_{B}: b
\mapsto \Klasse{i_{B}(b)}{R}$ and $p_{C}: c
\mapsto \Klasse{i_{C}(C)}{R}$. The construction yields $p_{B}\circ f =
p_{C}\circ g$, because $R$ identifies the embedded elements $f(a)$ and
$g(a)$ for any $a\in A$. 

Now assume that $q_{B}: B\to E$ and $q_{C}: C\to E$ are maps with $q_{B}\circ f
= q_{C}\circ g$. Let $q: D\to E$ be the unique map with $q\circ i_{B}
= q_{B}$ and $q\circ i_{C} = q_{C}$ (Lemma~\ref{lift-to-prod} together
with Proposition~\ref{coprod-dual-to-prod}). Then $R_{0}\subseteq
\Kern{q}$: Let $a\in A$, then 
\begin{equation*}
  q(i_{B}(f(a))) = q_{B}(f(a)) = q_{C}(g(a)) = q(i_{C}(f(a)),
\end{equation*}
so that $\langle i_{B}(f(a)), i_{C}(f(a))\rangle \in \Kern{q}$. 
Because $\Kern{q}$ is an equivalence relation on $D$, we conclude
$R\subseteq\Kern{q}$. Thus $h(\Klasse{x}{R}) := q(x)$ defines a map
$\Faktor{D}{R}\to E$ with 
\begin{align*}
  h(p_{B}(b)) & = h(\Klasse{i_{B}(b)}{R}) = q(i_{B}(b)) = q_{B}(b),\\
h(p_{C}(c)) & = h(\Klasse{i_{C}(c)}{R}) = q(i_{C}(c)) = q_{C}(c)
\end{align*}
for $b\in B$ and $c\in C$. It is clear that there is no other way to
define a map $h$ with the desired properties. 
\EndExample

So we have shown that the pushout in the category $\catSET$ of sets
with maps exists. To illustrate the construction, consider the pushout
of two factor maps. In this example $\rho\vee\tau$ denotes the
smallest equivalence relation which contains the equivalence relations
$\rho$ and $\tau$.

\BeginExample{factor-as-pushout}
Let $\rho$ and $\tau$ be equivalence relations on
a set $X$ with factor maps $\fMap{\rho}: X\to \Faktor{X}{\rho}$ and
$\fMap{\tau}: X\to \Faktor{X}{\tau}$. Then the pushout of these maps
is $\Faktor{X}{(\rho\vee\tau)}$ with $\zeta_{\rho}: \Klasse{x}{\rho}\mapsto
\Klasse{x}{\rho\vee\tau}$ and $\zeta_{\tau}:\Klasse{x}{\tau}\mapsto
\Klasse{x}{\rho\vee\tau}$ as the associated maps. In fact, we have
$\fMap{\rho}\circ \zeta_{\rho} = \fMap{\tau}\circ \zeta_{\tau}$, so
the first property is satisfied. Now let  $t_{\rho}:
\Faktor{X}{\rho}\to E$ and $t_{\tau}:
\Faktor{X}{\tau}\to E$ be maps with $t_{\rho}\circ \fMap{\rho} =
t_{\tau}\circ \fMap{\tau}$ for a set $E$, then $h: \Klasse{x}{\rho\vee\tau}\mapsto
t_{\rho}(\Klasse{x}{\rho})$ maps $\Faktor{X}{(\rho\vee\tau)}$ to $E$
with plainly $t_{\rho} = h\circ \zeta_{\rho}$ and $t_{\tau} = h\circ
\zeta_{\tau}$; moreover, $h$ is uniquely determined by this
property. Because the pushout is up to isomorphism uniquely determined
by Lemma~\ref{cat-prod-unique} and Proposition~\ref{coprod-dual-to-prod},
we have shown that the supremum of two equivalence relations in the
lattice of equivalence relations can be
computed through the pushout of its components.   
\EndExample

\Subsection{Functors and Natural Transformations}
\label{sec:functors}

We introduce functors which help in transporting information between
categories in a way similar to morphisms, which are thought to
transport information between objects. Of course, we will have to
observe some properties in order to capture in a formal way the intuitive understanding of a
functor as a structure preserving element. Functors
themselves can be related, leading to the notion of a natural
transformation. Given a category, there is a plethora of functors and
natural transformations provided by the hom sets; this is studied in
some detail, first, because it is a built-in in every category, second
because the Yoneda Lemma relates this rich structure to set based
functors, which in turn will be used when studying adjunctions.
 
\Subsubsection{Functors}
\label{sec:functors-nat-trans}

Loosely speaking, a functor is a pair of structure preserving maps between categories: it maps
one category to another one in a compatible way. A bit more precise, a functor $\funF$ between categories
$\catK$ and $\catL$ assigns to each object $a$ in category $\catK$ an
object $\funF(a)$ in $\catL$, and it assigns each morphism $f: a\to b$
in $\catK$ a morphism $\funF(f): \funF(a)\to \funF(b)$ in $\catL$;
some obvious properties have to be observed. In this way it is
possible to compare categories, and to carry properties from one
category to another one. To be more specific:

\BeginDefinition{def-functor}
A \emph{\index{functor}functor} $\funF: \catK\to \catL$ assigns to each object $a$ in
category $\catK$ an object $\funF(a)$ in category $\catL$ and maps
each hom set $\homK{a, b}$ of $\catK$ to the hom set $\homL{\funF(a),
  \funF(b)}$ of $\catL$ subject to these conditions
\begin{itemize}
\item $\funF(id_{a}) = id_{\funF(a)}$ for each object $a$ of $\catK$,
\item if $f: a\to b$ and $g:b\to c$ are morphisms in $\catK$, then
  $\funF(g\circ f) = \funF(g)\circ \funF(f)$. 
\end{itemize}
A functor $\funF: \catK\to \catK$ is called an \emph{\index{functor!endofunctor}endofunctor} on
$\catK$. 
\EndDefinition

The first condition says that the identity morphisms in $\catK$ are
mapped to the identity morphisms in $\catL$, and the second condition
tell us that $\funF$ has to be compatible with composition in the
respective categories. Note that for specifying a functor, we have to
say what the functor does with objects, and how the functor transforms
morphisms. By the way, we often write $\funF(a)$ as $\funF a$, and
$\funF(f)$ as $\funF f$.

Let us have a look at some examples. Trivial examples for functors include the
\emph{\index{functor!identity}identity functor} $Id_{\catK}$, which maps objects resp. morphisms to
itself, and the \emph{\index{functor!constant}constant functor} $\Delta_{x}$ for an object $x$, which
maps every object to $x$, and every morphism to $id_{x}$. 

\BeginExample{powerset-functor}
Consider the category $\catSET$ of sets with maps as morphisms. Given
set $X$, $\PowerSenza X$ is a set again; define $\PowerSenza(f)(A) :=
\Bild{f}{A}$ for the map $f: X\to Y$ and for $A\subseteq X$, then
$\PowerSenza f: \PowerSenza X\to \PowerSenza Y$. We check the laws for
a functor:
\begin{itemize}
\item $\PowerSenza (id_{X})(A) = \Bild{id_{X}}{A} = A =
  id_{\PowerSenza X}(A)$, so
    that $\PowerSenza id_{X} = id_{\PowerSenza X}$.
  \item let $f: X\to Y$ and $g: Y\to Z$, then $\PowerSenza f:
    \PowerSenza X\to \PowerSenza Y$ and $\PowerSenza g:
    \PowerSenza Y\to \PowerSenza Z$ with
    \begin{align*}
      (\PowerSenza(g)\circ \PowerSenza(f))(A)
& = \PowerSenza(g)(\PowerSenza(f)(A))\\
& = \Bild{g}{\Bild{f}{A}}\\
& = \{g(f(a))\mid a \in A\}\\
& = \Bild{(g\circ f)}{A}\\
& = \PowerSenza(g\circ f)(A)
    \end{align*}
for $A\subseteq X$. Thus the \emph{\index{functor!power set}power set functor} $\PowerSenza$ is compatible with composition
of maps. 
\end{itemize}
\EndExample

\BeginExample{hom-sets-functors}
Given a category $\catK$ and an object $a$ of $\catK$, associate 
\begin{align*}
  a_{+}: x & \mapsto \homK{a, x}\\
a^{+}: x & \mapsto \homK{x, a}.
\end{align*}
with $a$ together with the maps on hom-sets $\homK{a, \cdot}$
resp. $\homK{\cdot, a}$. Then $a^{+}$ is a
functor $\catK\to \catSET$. 

In fact, given morphism $f: x\to y$, we have $a_{+} f: \homK{a, x}\to \homK{a,
  y}$, taking $g$ into $f\circ g$. Plainly, $a_{+} (id_{x}) =
id_{\homK{a, x}} = id_{a_{+}(x)}$, and 
\begin{equation*}
  a_{+}(g\circ f)(h) = (g\circ f)\circ h = g\circ (f\circ h) = a_{+}(g)(a_{+}(f)(h)),
\end{equation*}
if $f: x\to y, g: y\to z$ and $h: a\to x$. 
\EndExample

Functors come in handy when we want to forget part of the structure.

\BeginExample{forgetful-measurable}
Let $\Category{Meas}$ be the category of measurable spaces. Assign
to each measurable space $(X, {\cal C})$ its carrier set $X$, and to
each morphism $f: (X, {\cal C})\to (Y, {\cal D})$ the corresponding
map $f: X\to Y$. It is immediately checked that this constitutes a
functor $\Category{Meas}\to \catSET$. Similarly, we might forget the
topological structure by assigning each topological space its carrier
set, and assign each continuous map to itself. These functors are sometimes
called \emph{\index{functor!forgetful}forgetful functors}. 
\EndExample

The following example twists Example~\ref{forgetful-measurable}
a little bit.

\BeginExample{forgetful-measurable-contra} 
{
  \def\fnV{\FunctorSenza{B}} \def\catM{\Category{Meas}} Assign to each
  measurable space $(X, {\cal C})$ its $\sigma$-algebra $\fnV(X, {\cal
    C}) := {\cal C}$. Let $f: (X, {\cal C})\to (Y, {\cal D})$ be a
  morphism in $\catM$; put $\fnV(f) := f^{-1}$, then $\fnV(f): \fnV(Y,
  {\cal D})\to \fnV(X, {\cal C}),$ because $f$ is ${\cal C}$-${\cal
    D}$-measurable. We plainly have $ \fnV(id_{X, {\cal C}}) =
  id_{\fnV(X, {\cal C})} $ and $ \fnV(g\circ f) = (g\circ f)^{-1} =
  f^{-1}\circ g^{-1} = \fnV(f)\circ \fnV(g), $ so $\fnV: \catM\to
  \catSET$ is no functor, although it behaves like one. \textsc{Don't
    panic!} If we reverse arrows, things work out properly: $\fnV:
  \catM\to \catSET^{op}$ is, as we have just shown, a functor (the dual
  $\catK^{op}$ of a category $\catK$ has been introduced in
  Example~\ref{ex-opposite-category}).

This functor could be called the \emph{Borel functor} (the measurable sets
are sometimes also called the Borel sets).
}\EndExample

\BeginDefinition{contravariant-functor}
A functor $\funF: \catK\to \catL^{op}$ is called a \emph{\index{functor!contravariant}contravariant
functor} between $\catK$ and $\catL$; in contrast, a functor according
to Definition~\ref{def-functor} is called
\emph{\index{functor!covariant}covariant}.  
\EndDefinition

If we talk about functors, we always mean the covariant flavor,
contravariance is mentioned explicitly. 

Let us complete the discussion from Example~\ref{hom-sets-functors} by
considering $a^{+}$, which takes $f: x\to y$ to $a^{+} f: \homK{y,
  a}\to \homK{x, a}$ through $g \mapsto g\circ f$. $a^{+}$ maps the
identity on $x$ to the identity on $\homK{x, a}$. If $g: y\to z$, we
have 
\begin{equation*}
  a^{+}(f)(a^{+}(g)(h)) = a^{+}(f)(h\circ g) = h\circ g\circ f = a^{+}(g\circ f)(h)
\end{equation*}
for 
$h: z\to a$. Thus $a^{+}$ is a contravariant functor $\catK\to \catSET$, while its cousin $a_{+}$ is covariant.
 
Functors may also be used to model structures.

\BeginExample{functor-sequences}
{\def\fnS{\FunctorSenza{S}}
Consider this functor $\fnS: \catSET\to \catSET$ which assigns each
set $X$ the set $X^{\Nat}$ of all sequences over $X$; the map $f: X\to
Y$ is assigned the map $\fnS: \Folge{x}\mapsto \bigl(f(x_{n})\bigr)_{n\in\Nat}$. 
Evidently, $id_{X}$ is mapped to $id_{X^{\Nat}}$, and it is easily
checked that $\fnS(g\circ f) = \fnS(g)\circ \fnS(f)$. Hence $\fnS$
constitutes an endofunctor on $\catSET$. 
}\EndExample

\BeginExample{sequences-term-o-nonterm}
Similarly, define the endofunctor $\funF$ on $\catSET$ by assigning
$X$ to $X^{\Nat}\cup X^{*}$ with $X^{*}$ as the set of all finite
sequences over $X$. Then $\funF X$ has all finite or infinite
sequences over the set $X$. Let $f: X\to Y$ be a map, and let
$(x_{i})_{i\in I}\in\funF X$ be a finite or infinite sequence, then
put $(\funF f)(x_{i})_{i\in I} := \bigl(f(x_{i})\bigl)_{i\in
  I}\in\funF Y$. It  is not difficult to see that $\funF$
satisfies the laws for a functor. 
\EndExample

The next example deals with automata which produce an output (in
contrast to Example~\ref{ex-det-automaton} where we mainly had state
transitions in view).

\BeginExample{automata-with-output}
An \emph{\index{automaton!with output}automaton with output} $(A, B, X, \delta)$ has an input alphabet $A$, an output
alphabet $B$ and a set $X$ of states with a map $\delta: X\times A\to
X\times B$; $\delta(x, a) = \langle x', b\rangle$ yields the next
state $x'$ and the output $b$ if the input is $a$ in state $x$. A morphism $f: (X, A, B, \delta)\to (Y, A, B, \tau)$ of
automata is a map $f: X\to Y$ such that $\tau(f(x), a) = (f\times
id_{B})(\delta(x, a))$ for all $x\in X, a\in A$, thus $(f\times
id_{B})\circ \delta = \tau\circ (f\times id_{A})$. This yields
apparently a category $\Category{AutO}$, the category of automata with
output. 

We want to expose the state space $X$ in order to make it a parameter
to an automata, because input and output alphabets are given from the
outside, so for modeling purposes only states are at our disposal. 
Hence we reformulate $\delta$ and take it as a map $\delta_{*}: X\to (X\times
B)^{A}$ with $\delta_{*}(x)(a) := \delta(x, a)$.  Now $f: (X, A, B, \delta)\to (Y, A, B, \tau)$ is a morphism
iff this diagram commutes
\begin{equation*}
\xymatrix{
X\ar[rr]^{f}\ar[d]_{\delta_{*}} && Y\ar[d]^{\tau_{*}}\\
(X\times B)^{A}\ar[rr]_{f^{\bullet}}&&(Y\times B)^{A}
}
\end{equation*}
with
$
f^{\bullet}(t)(a) := (f\times id_{B})(t(a)).
$
Let's see why this is the case. Given $x\in X, a \in A$, we have
\begin{equation*}
  f^{\bullet}(\delta_{*}(x))(a) = (f\times id_{B})(\delta_{*}(x)(a)) =
  (f\times id_{B})(\delta(x, a)) = \tau(f(x), a) = \tau_{*}(f(x))(a),
\end{equation*}
thus $f^{\bullet}(\delta_{*}(x)) = \tau_{*}(f(x))$ for all $x\in X$,
hence $f^{\bullet}\circ \delta_{*} = \tau_{*}\circ f$, so the diagram
is commutative indeed. Define 
$
\funF(X) := (X\times A)^{B},
$
for an object $(A, B, X, \delta)$ in
category $\Category{AutO}$  and and put 
$
\funF(f) := f^{\bullet}
$
for the automaton morphism $f: (X, A, B, \delta)\to (Y, A, B,
\tau)$, thus $\funF(f)$  renders this diagram commutative:

\begin{equation*}
\xymatrix{
&A\ar[dl]_{t}\ar[dr]^{\funF(f)(t)}\\
X\times B\ar[rr]_{f\times id_{B}}&&Y\times B
}
\end{equation*}

We claim that $\funF: \Category{AutO}\to \catSET$ is
functor. Let $g: (Y, A, B, \tau)\to (Z, A, B, \theta)$ be a morphism,
then $\funF(g)$ makes this diagram
commutative  for all $s\in (Y\times B)^{A}$

\begin{equation*}
\xymatrix{
&A\ar[dl]_{s}\ar[dr]^{\funF(g)(s)}\\
X\times B\ar[rr]_{g\times id_{B}}&&Y\times B
}
\end{equation*}
In particular, we have for $s:= \funF(f)(t)$ with an arbitrary
$t\in(X\times B)^{A}$ this commutative diagram
\begin{equation*}
\xymatrix{
&A\ar[dl]_{\funF(f)(t)}\ar[dr]^{\funF(g)( \funF(f)(t))}\\
Y\times B\ar[rr]_{g\times id_{B}}&&Z\times B
}
\end{equation*}
Thus the outer diagram commutes

\begin{equation*}
\xymatrix{
&&A\ar[dll]_{t}\ar[d]_{\funF(f)(t)}\ar[drr]^{\funF(g)( \funF(f)(t))}\\
X\times B\ar[rr]_{f\times id_{B}} && Y\times B\ar[rr]_{g\times id_{B}} && Z\times B
}
\end{equation*}
Consequently, we have
\begin{align*}
  \funF(g)(\funF(f)(t)) & = (g\times id_{B})\circ (f\times
  id_{B})\circ t\\
& = ((g\circ f)\times id_{B})\circ t\\
& = \funF(g\circ f)(t).
\end{align*}
Now $\funF(id_{X}) = id_{(X\times B)^{A}}$ is trivial, so that we have
established indeed that $\funF: \Category{AutO}\to \catSET$ is a
functor, assigning states to possible state transitions. 
\EndExample

\BeginExample{ex-transition-syst-act}
A \emph{labeled \index{transition system!labeled}transition system}
is a collection of transitions indexed by a set of actions. Formally, given a set $A$ of actions, $\bigl(S,
(\Trans_{a})_{a\in A}\bigr)$ is a labeled transition system iff $\Trans_{a}\
\subseteq S\times S$ for all $a\in A$. Thus
state $s$ may go into state $s'$ after action $a\in A$; this is
written as  $\isEquiv{s}{s'}{\Trans_{a}}$. A morphism $f: \bigl(S,
(\Trans_{S, a})_{a\in A}\bigr)\to \bigl(T,
(\Trans_{T, a})_{a\in A}\bigr)$ of transition systems is a map $f:
S\to T$ such that $\isEquiv{s}{s'}{\Trans_{S, a}}$ implies
$\isEquiv{f(s)}{f(s')}{\Trans_{T, a}}$ for all actions $a$,
cp. Example~\ref{ex-cat-transition-syst}. 

We model a transition system $\bigl(S, (\Trans_{a})_{a\in A}\bigr)$ as
a map $F: S\to \PowerSet{A\times S}$ with $F(s)
:= \{\langle a, s'\rangle \mid \isEquiv{s}{s'}{\Trans_{a}}\}$  (or, conversely, $\Trans_{a} = \{\langle s,
s'\rangle \mid \langle a, s'\rangle\in F(s)\}$), thus
$F(s)\subseteq A\times S$ collects actions and new states. This
suggests defining a map $\funF(S) := \PowerSet{A\times S}$ which can
be made a functor once we have decided what to do with morphisms $f:
\bigl(S, (\Trans_{S, a})_{a\in A}\bigr)\to \bigl(T, (\Trans_{T,
  a})_{a\in A}\bigr)$. Take $V\subseteq A\times S$ and define 
$
\funF(f)(V) := \{\langle a, f(s)\rangle \mid \langle a, s\rangle\in
V\},
$
(clearly we want to leave the actions alone). Then we have 
\begin{align*}
 \funF(g\circ f)(V) & = \{\langle a, g(f(s))\rangle \mid \langle a,
 s\rangle\in V\}\\
& = \{\langle a, g(y)\rangle \mid \langle a, y\rangle \in
\funF(f)(V)\}\\
& = \funF(g)(\funF(f)(V))
\end{align*}
for a morphism 
$
g: \bigl(T, (\Trans_{T, a})_{a\in A}\bigr)\to \bigl(U, (\Trans_{U,
  a})_{a\in A}\bigr)
$. Thus we have shown that $\funF(g\circ f) = \funF(g)\circ \funF(f)$
holds. Because $\funF$ maps the identity to the identity, $\funF$ is a
functor from the category of labeled transition systems to
$\catSET$. 
\EndExample

The next examples deal with functors induced by probabilities.

\BeginExample{functor-discrete-probs}
{\def\funD{\FunctorSenza{D}}
Given a set $X$, define the support $\supp(p)$ for a map $p: X\to [0, 1]$
as $\supp(p) := \{x\in X \mid p(x)\not=0\}$. A \emph{\index{probability!discrete}discrete probability}
$p$ on $X$ is a map $p: X\to [0, 1]$ with finite support such that 
\begin{equation*}
  \sum_{x\in X}p(x) := \sup\{\sum_{x\in F}p(x) \mid
  F\subseteq\supp(p)\} = 1.
\end{equation*}
Denote by 
\begin{equation*}
\funD(X) := \{p: X\to [0, 1] \mid p\text{ is a discrete
  probability}\}
\end{equation*}
the set of all discrete probabilities. Let $f: X\to
Y$ be a map, and define 
\begin{equation*}
  \funD(f)(p)(y) := \sum_{\{x\in X\mid f(x) = y\}} p(x).
\end{equation*}
Because $\funD(f)(p)(y) > 0$ iff $y\in \Bild{f}{supp(p)}$, $\funD(f)(p): Y\to [0, 1]$ has finite support, and 
\begin{equation*}
  \sum_{y\in Y}\funD(f)(p)(y) = \sum_{y\in Y} \sum_{\{x\in X\mid f(x) = y\}} p(x) = \sum_{x\in X}p(x) = 1.
\end{equation*}
It is clear that $\funD(id_{X})(p) = p$, so we have to check whether
$\funD(g\circ f) = \funD(g)\circ \funD(f)$ holds. 

We use a little trick for this, which will turn out to be helpful
later as well. Define
\begin{equation*}
  p(A) := \sup \{\sum_{x\in F}p(x) \mid F\subseteq A\cap\supp(p)\}
\end{equation*}
for $p\in \funD(X)$ and  $A\subseteq X$, then $p$ is a probability
measure on $\PowerSenza X$. Then 
$
\funD(f)(p)(y) = p(\InvBild{f}{\{y\}}),
$
and
$
\funD(f)(B) = p(\InvBild{f}{B})
$
for $B\subseteq Y$. 
Thus we obtain for the maps $f: X\to Y$ and $g: Y\to Z$

\begin{align*}
  \funD(g\circ f)(p)(z) 
&=  p\bigl(\InvBild{(g\circ f)}{\{z\}}\bigr)\\
& = p\bigr(\InvBild{f}{\InvBild{g}{\{z\}}}\bigl)\\
& = \funD(f)(p)(\InvBild{g}{\{z\}})\\
& = \funD(f)(\funD(g)(p))(z)
\end{align*}
Thus $\funD(g\circ f) = \funD(g)\circ \funD(f)$, as claimed. 

Hence $\funD$ is an endofunctor on $\catSET$, the
\emph{\index{probability functor!discrete}discrete probability
  functor}. It is immediate that all the arguments above hold also for
probabilities the support of which is countable; but since we will
discuss an interesting example on page~\pageref{sec:alger-discr-prob} which deal with the
finite case, we stick to that here. 
}\EndExample

There is a continuous version of this functor as well. We generalize
things a bit and formulate the example for subprobabilities. 

\BeginExample{functor-cont-probs}
{\def\catM{\Category{Meas}}
\def\funM{\SubProbSenza}
\def\mbX{X, {\cal A}}
\def\mbY{Y, {\cal B}}
\def\mbZ{Z, {\cal C}}
We are now working in the category $\catM$ of measurable spaces with
measurable maps as morphisms. 
Given a measurable space $(\mbX)$,
the set $\funM(\mbX)$
of all subprobability measures
is a measurable space with the weak $\sigma$-algebra $w({\cal A})$
associated with ${\cal A}$, see Example~\ref{ex-cat-stoch-rel}. Hence
$\funM$ maps measurable spaces to measurable spaces. Define for a morphism
$f: (\mbX)\to (\mbY)$
\begin{equation*}
  \funM(f)(\mu)(B) := \mu(\InvBild{f}{B})
\end{equation*}
for $B\in{\cal B}$. Then $\funM(f): \funM(\mbX)\to \funM(\mbY)$ is $w({\cal A})$-$w({\cal B})$-measurable. 
Now let $g: (\mbY)\to (\mbZ)$ be
a morphism in $\catM$, then we show as in
Example~\ref{functor-discrete-probs} that
\begin{equation*}
  \funM(g\circ f)(\mu)(C) = \mu(\InvBild{f}{\InvBild{g}{C}}) = \funM(f)(\funM(g)(\mu))(C),
\end{equation*}
for $C\in{\cal C}$, thus 
$
\funM(g\circ f) = \funM(g)\circ \funM(f).
$
Since $\funM$ preserves the identity, $\funM: \catM\to \catM$ is an
endofunctor, the (continuous space) \emph{\index{probability
    functor!continuous space}probability functor}. 
}\EndExample

The next two examples deal with upper closed sets, the first one with
these sets proper, the second one with a more refined version, viz.,
with ultrafilters. Upper closed sets are used, e.g., for the
interpretation of game logic, a variant of modal logics, see Example~\ref{ex-spec-rel-6}.  

\BeginExample{upper-closed-functor}
{\def\funV{\FunctorSenza{V}}
\def\funP{\PowerSenza}
Call a subset $V\subseteq\funP S$ \emph{\index{upper closed}upper closed} iff $A\in V$ and
$A\subseteq B$ together imply $B\in V$; for example, each filter is
upper closed. Denote by 
\begin{equation*}
  \funV S := \{V\subseteq \funP S \mid V\text{ is upper closed}\}
\end{equation*}
the set of all upper closed subsets of $\funP S$. Given $f: S\to T$,
define
\begin{equation*}
  (\funV f)(V) := \{W\subseteq \funP T \mid \InvBild{f}{W}\in V\}
\end{equation*}
for $V\in\funV S$. Let $W\in \funV(V)$ and $W_{0}\supseteq W$, then
$\InvBild{f}{W}\subseteq\InvBild{f}{W_{0}}$, so that
$\InvBild{f}{W_{0}}\in V$, hence $\funV f: \funV S\to \funV W$. It is
easy to see that $\funV (g\circ f) = \funV(g)\circ \funV (f)$,
provided $f: S\to T$, and $g: T\to V$. Moreover, $\funV(id_{S}) =
id_{\funV(S)}$. Hence $\funV$ is an endofunctor on the category
$\catSET$ of sets with maps as morphisms. 
}\EndExample

Ultrafilters are upper closed, but are much more complex than plain
upper closed sets, since they are filters, and they are maximal. Thus
we have to look a bit closer at the properties which the functor is to
represent.

\BeginExample{ex-ultrafilter-functor}
{
\def\funU{\FunctorSenza{U}}
\def\funP{\PowerSenza}
Let 
\begin{equation*}
\funU S := \{q \mid q\text{ is an ultrafilter over } S\}
\end{equation*}
assign
to each set $S$ its ultrafilters, to be more precise, all ultrafilters
of the power set of $S$. This is the object part of an endofunctor
over the category $\catSET$ with maps as morphisms. Given a map $f:
S\to T$, we have to define $\funU f: \funU S \to \funU T$. Before
doing so, a preliminary consideration will help.

One first notes that, given two Boolean algebras $B$ and $B'$ and a
Boolean algebra morphism $\gamma: B\to B'$, $\gamma^{-1}$ maps
ultrafilters over $B'$ to ultrafilters over $B$. In fact, let $w$ be an
ultrafilter over $B'$, put $v := \InvBild{\gamma}{w}$; we go quickly over
the properties of an ultrafilter should have. First, $v$ does not
contain the bottom element $\bot_{B}$ of $B$, for otherwise $\bot_{B'}
=\gamma(\bot_{B})\in w$. If $a\in v$ and $b\geq a$, then
$\gamma(b)\geq \gamma(a)\in w$, hence $\gamma(b)\in w$, thus $b\in v$;
plainly, $v$ is closed under $\wedge$. Now assume $a\not\in v$, then
$\gamma(a)\not\in w$, hence $\gamma(-a)=-\gamma(a)\in w$, since $w$ is
an ultrafilter. Consequently, $-a\in v$. This establishes the claim.

Given a map $f: S\to T$, define $F_{f}: \funP T \to  \funP S$ through
$F_{f} := f^{-1}$. This is a homomorphism of the Boolean algebras
$\funP T$ and $\funP S$, thus $F_{f}^{-1}$ maps $\funU S$ to $\funU
T$. Put $\funU(f) := F_{f}^{-1}$; note that we reverse the arrows'
directions twice. It is clear that $\funU(id_{S})
= id_{\funU(S)}$, and if $g:T\to Z$, then
\begin{equation*}
  \funU(g\circ f) = F_{g\circ f}^{-1} = (F_{f}\circ F_{g})^{-1} =
  F_{g}^{-1}\circ F_{f}^{-1} = \funU(g)\circ \funU(f).
\end{equation*}
This shows that $\funU$ is an endofunctor on the category $\catSET$ of
sets with maps as morphisms ($\funU$ is sometimes denoted by $\beta$).
}\EndExample

We can use functors for constructing new categories from given
ones. As an example we define the comma category associated with two
functors. 

\BeginDefinition{comma-category}
Let $\funF: \catK\to \catL$ and $\funG: \catM\to \catL$ be
functors. The \emph{\index{category!comma}comma category} $(\funF, \funG)$ associated with
$\funF$ and $\funG$ has as objects the triplets $\langle a, f,
b\rangle$ with objects $a$ from $\catK$, $b$ from $\catM$, and
morphisms $f: \funF a\to \funG b$. A morphism $(\phi, \psi): \langle
a, f, b\rangle\to \langle a', f', b'\rangle$ is a pair of morphisms $\phi:
a\to a'$ of $\catK$ and $\psi: b\to b'$ of $\catM$ such that this
diagram commutes
\begin{equation*}
\xymatrix{
\funF a\ar[d]_{f}\ar[rr]^{\funF \phi}&& \funF a'\ar[d]^{f'}\\
\funG b\ar[rr]_{\funG \psi} && \funG b'
}
\end{equation*}
Composition of morphism is component wise. 
\EndDefinition
The slice category $\Faktor{\catK}{x}$ defined in
Example~\ref{ex-down-under} is apparently the comma category
$(Id_{\catK}, \Delta_{x})$.

Functors can be composed, yielding a new functor. The proof for this
statement is straightforward.

\BeginProposition{functor-compos}
Let $\FunctorSenza{F}: \Category{C}\to \Category{D}$ and
$\FunctorSenza{G}: \Category{D}\to \Category{E}$ be functors. Define
$(\FunctorSenza{G}\circ \FunctorSenza{F}) a :=
\FunctorSenza{G}(\FunctorSenza{F} a)$ for an object $a$ of
$\Category{C}$, and $(\FunctorSenza{G}\circ \FunctorSenza{F}) f :=
\FunctorSenza{G}(\FunctorSenza{F} f)$ for a morphism $f: a\to b$ in
$\Category{C}$, then $\FunctorSenza{G}\circ \FunctorSenza{F}: \Category{C}\to \Category{E}$ is a
functor. 
\QED
\EndProposition

\Subsubsection{Natural Transformations}
\label{sec:nat-transf}
We see that we can compose functors in an obvious way. This raises the
question whether or not functors themselves form a category. But we do
not yet have morphisms between functors at out disposal. Natural
transformations will assume this r\^ole. Nevertheless, the question
remains, but it will not be answered in the positive; this is so
because morphisms between objects should form a set, and it will be
clear that his is not the case.  Pumplün~\cite{Pumpluen-Kategorien}
points at some difficulties that might arise and arrives at the
pragmatic view that for practical problems this question is not
particularly relevant.  

But let us introduce natural transformations between functors
$\FunctorSenza{F}, \FunctorSenza{G}$ now. The basic idea is that for each object $a$, 
$\FunctorSenza{F}a$ is transformed into $\FunctorSenza{G}a$ in a way
which is compatible with the structure of the participating
categories.

\BeginDefinition{nat-transf}
Let $\funF, \funG:\catK\to \catL$
be covariant functors. A family $\eta = (\eta_{a})_{a\in \objK}$ is called a
\emph{\index{category!natural transformation}\index{natural transformation}natural transformation} $\eta: \funF\to \funG$ iff $\eta_{a}: \funF
a\to \funG a$ is a morphism in $\catL$ for all objects $a$ in $\catK$
such that this diagram commutes for any morphism $f: a\to b$ in $\catK$
\begin{equation*}
\xymatrix{
a\ar[d]^{f}&&\funF a\ar[rr]^{\eta_{a}}\ar[d]_{\funF f} && \funG a\ar[d]^{\funG f}\\
b&&\funF b\ar[rr]_{\eta_{b}} && \funG b
}
\end{equation*}
\EndDefinition

Thus a natural transformation $\eta: \funF\to \funG$ is a family of
morphisms, indexed by the objects of the common domain of $\funF$ and
$\funG$; $\eta_{a}$ is called the \emph{\index{category!natural transformation!component}\index{natural transformation!component}component of $\eta$ at $a$}. 

If $\funF$ and $\funG$ are both contravariant functors $\catK\to
\catL$, we may perceive them as covariant functors $\catK\to
\catL^{op}$, so that we get for the contravariant case this diagram: 
\begin{equation*}
\xymatrix{
a\ar[d]^{f}&&\funF a\ar[rr]^{\eta_{a}} && \funG a\\
b&&\funF b\ar[rr]_{\eta_{b}}\ar[u]^{\funF f} && \funG b\ar[u]_{\funG f}
}
\end{equation*}

Let us have a look at some examples.

\BeginExample{nat-transf-hom}
$a_{+}: x \mapsto \homK{a, x}$ yields a (covariant) functor $\catK\to
\catSET$ for each object $a$ in $\catK$, see
Example~\ref{hom-sets-functors} (just for simplifying notation, we use again
$a_{+}$ rather than $\homK{a, -}$, see
page~\pageref{ref-hom-sets}). Let $\tau: b\to a$ be a morphism in
$\catK$, then this induces a natural transformation $\eta_{\tau}:
a_{+}\to b_{+}$ with
\begin{equation*}
\eta_{\tau, x}:
  \begin{cases}
   a_{+}(x)&\to b_{+}(x)\\
g & \mapsto g\circ \tau 
  \end{cases}
\end{equation*}
In fact, look at this diagram with a $\catK$-morphism
$f: x\to y$:
\begin{equation*}
\xymatrix{
x\ar[d]_{f} && a_{+}(x)\ar[rr]^{\eta_{\tau, x}}\ar[d]_{a_{+}(f)} && b_{+}(x)\ar[d]^{b_{+}(f)}\\
y && a_{+}(y)\ar[rr]_{\eta_{\tau, y}} && b_{+}(y)
}
\end{equation*}
Then we have for $h\in a_{+}(x) = \homK{a, x}$ 
\begin{align*}
(\eta_{\tau, y} \circ a_{+}(f))(h) & = \eta_{\tau, y}(f\circ h)\\
& = (f\circ h)\circ \tau\\
& = f\circ (h\circ \tau)\\
& = b_{+}(f)(\eta_{\tau, x}(h))\\
& = (b_{+}(f)\circ \eta_{\tau, x})(h) 
\end{align*}
Hence $\eta_{\tau}$ is in fact a natural transformation.
\EndExample

This is an example in the category of groups:

\BeginExample{nat-transf-groups}
Let $\catK$ be the category of groups (see
Example~\ref{ex-cat-group}). It is not difficult to see that $\catK$
has products. Define for a group $H$ the map $\funF_{H}(G) :=
H\times G$ on objects, and if $f: G\to G'$ is a morphism in $\catK$, define
$\funF_{H}(f): H\times G\to H\times G'$ through $\funF_{H}(f): \langle h,
g\rangle \mapsto \langle h, f(g)\rangle$. Then $\funF_{H}$ is an
endofunctor on $\catK$. Now let $\phi: H\to K$ be a morphism. Then
$\phi$ induces a natural transformation $\eta_{\phi}$ upon setting
\begin{equation*}
\eta_{\phi, G}
  \begin{cases}
    \funF_{H} &\to \funF_{K}\\
\langle h, g\rangle & \mapsto \langle \phi(h), g\rangle.
  \end{cases}
\end{equation*}
In fact, let $\psi: L\to L'$ be a group homomorphism, then this
diagram commutes

\begin{equation*}
\xymatrix{
L\ar[d]_{\psi} && \funF_{H}\ L\ar[rr]^{\eta_{\phi,
    L}}\ar[d]_{\funF_{H}\ \psi} && \funF_{K}\ L\ar[d]^{\funF_{K}\ \psi}\\
L' && \funF_{H}\ L'\ar[rr]_{\eta_{\phi, L'}} && \funF_{K}\ L'
}
\end{equation*}
To see this, take $\langle h, \ell\rangle\in\funF_{H}\ L =  H\times L$,
and chase it through the diagram:
\begin{equation*}
  (\eta_{\phi, L'}\circ \funF_{H}\ \psi)(h, \ell) = \langle \phi(h),
  \phi(\ell)\rangle = (\funF_{K}(\psi)\circ \eta_{\phi, L})(h, \ell).
\end{equation*}
\EndExample

Consider as a example a comma category $(\funF, \funG)$
(Definition~\ref{comma-category}). There are functors akin to a
projection which permit to recover the original functors, and which
are connected through a natural transformation. To be specific:

\BeginProposition{comma-cat-functors}
Let $\funF: \catK\to \catL$ and $\funG: \catM\to \catL$ be
functors. Then there are functors $\funS: (\funF, \funG)\to \catK$ and
$\funR: (\funF, \funG)\to \catL$ rendering this diagram commutative:
\begin{equation*}
\xymatrix{
(\funF, \funG)\ar[d]_{\funS}\ar[rr]^{\funR}&&\catM\ar[d]^{\funG}\\
\catK\ar[rr]_{\funF}&&\catL
}
\end{equation*}
There exists a natural transformation $\eta: \funF\circ \funS\to
\funG\circ \funR$.  
\EndProposition

\BeginProof
Put for the object $\langle a, f, b\rangle$ of $(\funF, \funG)$ and
the morphism $(\phi, \psi)$
\begin{align*}
  \funS \langle a, f, b\rangle & := a, &
\funS (\phi, \psi) & := \phi,\\
\funR \langle a, f, b\rangle & := b, &
\funR (\phi, \psi) & := \psi.
\end{align*}
Then it is clear that the desired equality holds. Moreover,
$\eta_{\langle a, f, b\rangle} := f$ is the desired natural
transformation. The crucial diagram commutes by the definition of
morphisms in the comma category. 
\EndProof

\BeginExample{prod-homsets}
Assume that the product $a\times b$ for the objects $a$ and $b$ in
category $\catK$ exists, then Proposition~\ref{pd-is-bijective} tells
us that we have for each object $d$ a bijection $p_{d}: \homK{d,
  a}\times\homK{d, b}\to \homK{d, a\times b}$. 
Thus $(\pi_{a}\circ p_{d})(f, g) = f$ and $(\pi_{b}\circ p_{d})(f,
g) = g$ for every morphism $f: d\to a$ and $g: d\to b$. Actually,
$p_{d}$ is the component of a natural transformation $p: \funF\to
\funG$ with $\funF := \homK{-, a}\times\homK{-, b}$ and $\funG :=
\homK{-, a\times b}$ (note that this is short hand for the obvious
assignments to objects and functors). Both $\funF$ and $\funG$ are
\emph{contra}variant functors from $\catK$ to $\catSET$. So in order
to establish naturalness, we have to establish that the following
diagram commutes

\begin{equation*}
\xymatrix{
c\ar[d]_{f} && \homK{c, a}\times\homK{c, b}\ar[rrr]^{p_{c}}&&&\homK{c, a\times
  b}\\
d && \homK{d, a}\times\homK{d, b}\ar[rrr]_{p_{d}}\ar[u]^{\funF\ f}&&&\homK{d, a\times
  b}\ar[u]_{\funG\ f}
}
\end{equation*}
Now take $\langle g, h\rangle\in\homK{d, a}\times\homK{d, b}$, then
\begin{align*}
  \pi_{a}\bigl((p_{c}\circ \funF\ f)(g, h)\bigr) & = g\circ f =
  \pi_{a}\bigl((\funG\ f)\circ p_{d}\bigr)(g, h),\\
\pi_{b}\bigl((p_{c}\circ \funF\ f)(g, h)\bigr) & = h\circ f = \pi_{b}\bigl((\funG\ f)\circ p_{d}\bigr)(g, h).
\end{align*}
From this, commutativity follows. 
\EndExample

We will ---~for the sake of illustration~--- define two ways of composing natural transformations. One
is somewhat canonical, since it is based on the composition of
morphisms, the other one is a bit tricky, since it involves the functors
directly. Let us have a look at the direct one first.\index{natural transformation!horizontal composition}

\BeginLemma{compos-nat-trans}
Let $\eta: \funF\to \funG$ and $\zeta: \funG\to \funH$ be natural
transformations. Then $(\tau\circ \zeta)(a) := \tau(a)\circ \zeta(a)$
defines a natural transformation $\tau\circ\zeta: \funF\to \funH$. 
\EndLemma

\BeginProof
Let $\catK$ be the domain of functor $\funF$, and assume that $f: a\to
b$ is a morphism in $\catK$. Then we have this diagram
\begin{equation*}
\xymatrix{
a\ar[d]_{f} &&\funF a\ar[d]_{\funF f}\ar[rr]^{\eta_{a}} && \funG a\ar[d]_{\funG g}\ar[rr]^{\zeta_{a}} &&
\funH a\ar[d]^{\funH f}\\
b && \funF b\ar[rr]_{\eta_{b}} && \funG b\ar[rr]_{\zeta_{b}} && \funH b
}
\end{equation*}
Then
\begin{equation*}
  \funH(f)\circ (\tau\circ \zeta)_{a} 
= \funH(f)\circ \tau_{a}\circ \zeta_{a}
= \tau_{b}\circ \funG(f)\circ \zeta_{a}
= \tau_{b}\circ \zeta_{b}\circ \funF(f)
= (\tau\circ \zeta)_{b}\circ \funF(f).
\end{equation*}
Hence the outer diagram commutes. 
\EndProof

The next composition is slightly more involved.

\BeginProposition{godement-product}
Given natural transformations $\eta: \funF\to \funG$ and $\tau:
\funS\to \funR$ for functors $\funF, \funG: \catK\to \catL$ and
$\funS, \funR: \catL\to \catM$. Then $\tau_{\funG a}\circ \funS(\eta_{a})
= \funR(\eta_{a})\circ \tau_{\funF a}$ always holds. Put
$(\tau\ast\eta)_{a} := \tau_{\funG a}\circ \funS(\eta_{a})$. Then
$\tau\ast\eta$ defines a natural transformation $\funS\circ \funF\to
\funR\circ \funG$. $\tau\ast\eta$ is called the \emph{\index{Godement
    product}\index{natural transformation!Godement
    product}Godement product} of $\eta$ and $\tau$.
\index{natural transformation!vertical composition}
\EndProposition

\BeginProof
1.
Because $\eta_{a}: \funF a \to  \funG a$, this diagram commutes by
naturality of $\tau$:

\begin{equation*}
\xymatrix{
\funS(\funF a)\ar[d]_{\funS(\eta_{a})}\ar[rr]^{\tau_{\funF a}} && \funR(\funF a)\ar[d]^{\funR(\eta_{a})}\\
\funS(\funG a)\ar[rr]_{\tau_{\funG a}} && \funR(\funG a)
}
\end{equation*}
This establishes the first claim. 

2.
Now let $f: a\to b$ be a morphism in
$\catK$, then the outer diagram commutes, since $\funS$ is a functor,
and since $\tau$ is a natural transformation.

\begin{equation*}
\xymatrix{
\funS(\funF a)\ar[d]_{\funS(\funF f)}\ar[rr]^{\funF(\eta_{a})}\ar@/^2pc/@{-->}[rrrr]^{(\tau\ast\eta)_{a}} 
&& \funS(\funG a) \ar[d]_{\funS(\funG f)}\ar[rr]^{\tau_{\funG a}}
&& \funR(\funG a)\ar[d]^{\funR(\funG f)}\\
\funS(\funF b)\ar[rr]_{\funF(\eta_{b})}\ar@/^-2pc/@{-->}[rrrr]_{(\tau\ast\eta)_{b}}  
&& \funS(\funG b) \ar[rr]_{\tau_{\funG b}}
&& \funR(\funG b)
}
\end{equation*}
Hence $\tau\ast\eta: \funS\circ \funF\to \funR\circ \funG$ is natural indeed. 
\EndProof

In~\cite{MacLane}, $\eta\circ \tau$ is called the \emph{vertical}, and
$\eta\ast\tau$ the \emph{horizontal} composition of the natural
transformations $\eta$ and $\tau$. If $\eta:
\funF\to \funG$ is a natural transformation, then the morphisms $(\funF\eta)(a) :=
\funF \eta_{a}: (\funF\circ \funF)(a)\to (\funF\circ \funG)(a)$ and
$(\eta\funF)(a) := \eta_{\funF a}: (\funF\circ \funF)(a)\to
(\funG\circ \funF)(a)$ are available.

We know from Example~\ref{hom-sets-functors} that $\homK{a, -}$
defines a covariant set valued functor; suppose we
haven another set valued functor $\funF: \catK\to \catSET$. Can we
somehow compare these functors? This question looks on first sight
quite strange, because we do not have any yardstick to compare these
functors against. On second thought, we might use natural
transformations for such an endeavor. It turns out that for any object
$a$ of $\catK$ the set $\funF a$ is essentially given by the natural
transformations $\eta: \homK{a, -} \to \funF$. We will show now that
there exists a bijective assignment between $\funF a$ and these
natural transformations. The reader might wonder about this somewhat
intricate formulation; it is due to the observation that these natural
transformation in general do not form a set but rather a class, so
that we cannot set up a proper bijection (which would require sets as
the basic scenario).

\BeginLemma{yoneda-lemma}
Let $\funF: \catK\to \catSET$ be a functor; given the object $a$ of
$\catK$ and a natural transformation $\eta: \homK{a, -}\to F$, define
the \emph{\index{Yoneda isomorphism}Yoneda isomorphism}
\begin{equation*}
  y_{a, \funF}(\eta) := \eta(a)(id_{a}) \in \funF\ a
\end{equation*}
Then $y_{a, \funF}$ is bijective (i.e., onto, and one-to-one).
\EndLemma

\BeginProof
0.
The assertion is established by defining for each $t\in \funF a$ a natural transformation
$\sigma_{a, \funF}(t): \homK{a, -}\to F$ which is inverse to $y_{a, \funF}$. 

1.
Given an object $b$ of $\catK$ and $t\in\funF\ a$, put
\begin{equation*}
  \bigl(\sigma_{a, \funF}(t)\bigr)_{b} := \sigma_{a, \funF}(t)(b):
  \begin{cases}
    \homK{a, b} & \to \funF\ b\\
f & \mapsto (\funF\ f)(t)
  \end{cases}
\end{equation*}
(note that $\funF\ f: \funF\ a \to \funF\ b$ for $f: a\to b$, hence
$(\funF\ f)(t)\in\funF\ b$). This defines a natural transformation
$\sigma_{a, \funF}(t): \homK{a, -}\to \funF$. In fact, if $f: b\to b'$,
then 
\begin{align*}
  \sigma_{a, \funF}(t)(b')(\homK{a, f} g) & = \sigma_{a, \funF}(t)(b')(f\circ g)\\
& = \funF(f\circ g)(t)\\
& = (\funF\ f)(\funF(g)(t))\\
& = (\funF\ f)(\sigma_{a, \funF}(t)(b)(g)).
\end{align*}
Hence $\sigma_{a, \funF}(t)(b')\circ \homK{a, f} = (\funF\ f)\circ
\sigma_{a, \funF}(t)(b)$. 

2.
We obtain 
\begin{align*}
  (y_{a, \funF}\circ \sigma_{a, \funF})(t) & = y_{a, \funF}(\sigma_{a, \funF}(t))\\
& = \sigma_{a, \funF}(t)(a)(id_{a})\\
& = (\funF\ id_{a})(t)\\
& = id_{\funF a}(t)\\
& = t
\end{align*}
That's not too bad, so let us try to establish that $\sigma_{a,
  \funF}\circ y_{a, \funF}$ is the identity as well. Given a natural
transformation $\eta: \homK{a, -}\to F$, we obtain
\begin{equation*}
  (\sigma_{a, \funF}\circ y_{a, \funF})(\eta) = \sigma_{a, \funF}(y_{a, \funF}(\eta))
  = \sigma_{a, \funF}(\eta_{a}(id_{a})).
\end{equation*}
Thus we have to evaluate $\sigma_{a, \funF}(\eta_{a}(id_{a}))$. Take an
object $b$ and a morphism $f: a\to b$, then
\begin{align*}
  \sigma_{a, \funF}(\eta_{a}(id_{a})(b)(f)) & = (\funF
  f)(\eta_{a}(id_{a}))\\
& = (\funF(f)\circ \eta_{a})(id_{a})\\
& = (\eta_{b}\circ \homK{a, f})(id_{a})&&\text{ ($\eta$ is natural)}\\
& = \eta_{b}(f)&&\text{ (since $\homK{a, f}\circ id_{a} = f\circ id_{a}$
  = f)}
\end{align*}
Thus $\sigma_{a, \funF}(\eta_{a}(id_{a})) = \eta$. Consequently we have
shown that $y_{a, \funF}$ is left and right invertible, hence is a
bijection. 
\EndProof

Now consider the set valued functor $\homK{b, -}$, then the Yoneda
embedding says that $\homK{b, a}$ can be mapped bijectively to the
natural transformations from $\homK{a, -}$ to $\homK{b, -}$. This
entails these natural transformations being essentially the
morphisms $b\to a$, and, conversely, each morphism $b\to a$ yields a
natural transformation $\homK{a, -}\to \homK{b, -}$. The following
statement makes this observation precise.

\BeginProposition{yoneda-repr}
Given a natural transformation $\eta: \homK{a, -}\to \homK{b, -}$,
there exists a unique morphism $g: b\to a$ such that $\eta_{c}(h) =
h\circ g$ for every object $c$ and every morphism $h:a \to c$ (thus $\eta
= \homK{g, -}$).  
\EndProposition

\BeginProof
0.
Let $y := y_{a, \homK{a, -}}$ and $\sigma := \sigma_{a, \homK{a,
    -}}$. Then $y$ is a bijection with $(y\circ \sigma)(\eta) = \eta$
and $(\sigma\circ y)(h) = h$. 

1.
Put 
$
g := \eta_{a}(id_{a}),
$
then $g\in\homK{b, a}$, since $\eta_{a}: \homK{a, a}\to \homK{b, a}$
and $id_{a}\in\homK{a, a}$. Now let $h\in\homK{a, c}$, then 
\begin{align*}
  \eta_{c}(h) & = \sigma\bigl(\eta_{a}(id_{a})\bigr)(c)(h) && \text{ (since $\eta = y\circ \sigma$)}\\
& = \sigma(g)(c)(h)&& \text{ (Definition of $\sigma$)} \\
& = \homK{b, g}(h)&&\text{ ($\homK{b,-}$ is the target functor)}  \\
& = h\circ g
\end{align*}

2.
If $\eta = \homK{g, -}$, then $\eta_{a}(id_{a}) = \homK{g, id_{a}} =
  id_{a}\circ g = g$, so $g: b\to a$ is uniquely determined. 
\EndProof

A final example comes from measurable spaces, dealing with the
weak-$\sigma$-algebra. We have defined in
Example~\ref{forgetful-measurable-contra} the contravariant functor
which assigns to each measurable space its $\sigma$-algebra, and we
have defined in Example~\ref{ex-cat-stoch-rel} the weak
$\sigma$-algebra on its set of probability measures together with a
set of generators. We show that this set of generators yields a family
of natural transformations between the two contravariant functors
involved.

\BeginExample{beta-is-natural}
{
  \def\fnV{\FunctorSenza{B}} 
\def\fnW{\FunctorSenza{W}}
\def\catM{\Category{Meas}}
\def\eS{(S, {\cal A})}
\def\Te{(T, {\cal B})}
The contravariant functor $\fnV: \catM\to \catSET$ assigns to each measurable
space its $\sigma$-algebra, and to each measurable map its
inverse. Denote by $\fnW := \ProbSenza\circ \fnV$ the functor that assigns
to each measurable space the weak $\sigma$-algebra on its probability
measures; $\fnW: \catM\to \catSET$ is contravariant as well. Recall from Example~\ref{ex-cat-stoch-rel} that 
the set
\begin{equation*}
 \beta_{S}(A, r) := \{\mu\in\Prob{S, {\cal
    A}} \mid \mu(A) \geq r\}
\end{equation*}
denotes the set of all probability measures which evaluate the
measurable set $A$ not smaller than a given $r$, and that the weak
$\sigma$-algebra on $\Prob{S, {\cal A}}$ is generated by all these
sets. We claim that $\beta(\cdot, r)$ is a natural transformation
$\fnV\to \fnW$. Thus we have to show that this diagram commutes 
\begin{equation*}
\xymatrix{
\eS\ar[d]_{f}&&\fnV\eS\ar[rr]^{\beta_{S}(\cdot, r)}&&\fnW\eS\\
\Te&&\fnV\Te\ar[rr]_{\beta_{T}(\cdot, r)}\ar[u]^{\fnV\ f}&&\fnW\Te\ar[u]_{\fnW\ f}
}
\end{equation*}

Recall that we have $\fnV(f)(C) = \InvBild{f}{C}$ for $C\in{\cal B}$,
and that $\fnW(f)(D) = \InvBild{\Prob{f}}{D}$, if $D\subseteq
\Prob{T, {\cal B}}$ is measurable. 
Now, given $C\in{\cal B}$,  by expanding definitions we obtain
\begin{align*}
  \mu\in\fnW(f)(\beta_{T}(C, r)) 
& \Leftrightarrow
\mu\in\InvBild{\Prob{f}}{\beta_{T}(C, r)}\\
& \Leftrightarrow
\Prob{f}(\mu) \in\beta_{T}(C, r)\\
& \Leftrightarrow
\Prob{f}(\mu)(C) \geq r\\
& \Leftrightarrow
\mu(\InvBild{f}{C}) \geq r\\
& \Leftrightarrow
\mu\in\beta_{S}(\fnV(f)(C), r)
\end{align*}
Thus the diagram commutes in fact, and we have established that the
generators for the weak $\sigma$-algebra come from a natural
transformation. 
}
\EndExample

\Subsubsection{Limits and Colimits}
\label{sec:limit-and-colimits}

We have defined some constructions which permit to build new objects
in a category from given ones, e.g., the product from two objects or
the pushout. Each time we had some universal condition which had to be
satisfied. 

We will discuss the general construction very briefly and refer
the reader to~\cite{MacLane, Barr+Wells, Pumpluen-Kategorien}, where
they are studied in great detail.

\BeginDefinition{def-cone}
Given a functor $\funF: \catK\to \catL$, a \emph{\index{cone}cone on
  $\funF$} consists of an object $c$ in $\catL$ and of a family of
morphisms $p_{d}: c\to \funF d$ in $\catL$ for each object $d$ in
$\catK$ such that $p_{d'} = (\funF g)\circ p_{d}$ for each morphism
$g: d\to d'$ in $\catK$. 
\EndDefinition

So a cone $(c, (p_{d})_{d\in \objK})$ on $\funF$ looks like, well, a
cone:
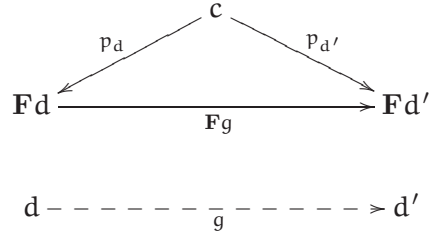
\begin{equation*}
\xymatrix{
&&c\ar[dll]_{p_{d}}\ar[drr]^{p_{d'}}\\
\funF d\ar[rrrr]_{\funF g}&&&&\funF d'\\
d\ar@{-->}[rrrr]_{g}&&&&d'
}
\end{equation*}
 
A limiting cone provides a factorization for each other cone, to be
specific

\BeginDefinition{limiting-cone}
Let $\funF: \catK\to \catL$ be a functor. The cone $(c, (p_{d})_{d\in
  \objK})$ is a \emph{\index{limit}\index{cone!limit}limit of $\funF$} iff for every cone $(e,
(q_{d})_{d\in \objK})$ on $\funF$ there exists a unique morphism $f:
e\to c$ such that $q_{d} = p_{d}\circ f$ for each object $d$ in
$\catK$.  
\EndDefinition

Thus we have locally this situation for each morphism $g: d\to d'$ in
$\catK$:
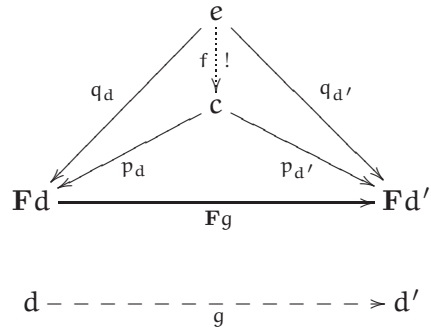
\begin{equation*}
\xymatrix{
&&e\ar@{..>}[d]^{!}_{f}\ar[ddll]_{q_{d}}\ar[ddrr]^{q_{d'}}\\
&&c\ar[dll]^{p_{d}}\ar[drr]_{p_{d'}}\\
\funF d\ar[rrrr]_{\funF g}&&&&\funF d'\\
d\ar@{-->}[rrrr]_{g}&&&&d'
}
\end{equation*}
The unique factorization probably gives already a clue at the
application of this concept. Let us look at some examples.

\BeginExample{prod-as-limit}
Let $X := \{1, 2\}$ and $\catK$ be the discrete category on $X$ (see
Example~\ref{ex-discr-categ}). Put $\funF 1 := a$ and $\funF 2 := b$
for the objects $a, b\in\objL$. Assume that the product $a\times b$
with projections $\pi_{a}$ and $\pi_{b}$ exists in $\catL$, and put
$p_{1} := \pi_{a}$, $p_{2} := \pi_{b}$. Then $(a\times b, p_{1},
p_{2})$ is a limit of $\funF$. Clearly, this is a cone on $\funF$, and
if $q_{1}: e\to a$ and $q_{2}: e\to b$ are morphisms, there exists a
unique morphism $f: e\to a\times b$ with $q_{1} = p_{1}\circ f$ and
$q_{2} = p_{2}\circ f$ by the definition of a product. 
\EndExample

The next example shows that a pullback can be interpreted as a limit.

\BeginExample{pullback-as-limit}
Let $a, b, c$ objects in category $\catL$ with morphisms $f: a\to c$
and $g: b\to c$. Define category $\catK$ by $\objK := \{a, b, c\}$,
the hom-sets are defined as follows
\begin{equation*}
  \homK{x, y} :=
  \begin{cases}
    \{id_{x}\}, & x = y\\
    \{f\}, & x = a, y = c\\
    \{g\}, & x = b, y = c\\
    \emptyset, & \text{ otherwise}
  \end{cases}
\end{equation*}
Let $\funF$ be the identity on $\objK$ with $\funF f := f, \funF g :=
g$, and $\funF id_{x} := id_{x}$ for $x\in \objK$. If object $p$
together with morphisms $t_{a}: p\to a$ and $t_{b}: p\to b$ is a
pullback for $f$ and $g$, then it is immediate that $(p, t_{a}, t_{b},
t_{c})$ is a limit cone for $\funF$, where $t_{c} := f\circ t_{a} =
g\circ t_{b}$. 
\EndExample

Dualizing the concept of a cone, we obtain cocones.

\BeginDefinition{def-cocone}
Given a functor $\funF: \catK\to \catL$, an object $c\in\objL$ together with 
morphisms $s_{d}: \funF D\to c$ for each object $d$ of $\catK$ such that
$s_{d} = s_{d'}\circ \funF g$ for each morphism $g: d\to d'$ is called
a \emph{\index{cocone}cocone on $\funF$}.
\EndDefinition

Thus we have this situation
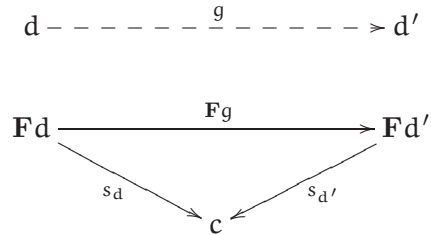
\begin{equation*}
\xymatrix{
d\ar@{-->}[rrrr]^{g} &&&& d'\\
\funF d\ar[rrrr]^{\funF g}\ar[drr]_{s_{d}} &&&& \funF d'\ar[dll]^{s_{d'}}\\
&&c
}
\end{equation*}

A colimit is then defined for a cocone.

\BeginDefinition{def-colimit}
A cocone $(c, (s_{d})_{d\in\objK})$ is called a \emph{\index{cocone!colimit}\index{colimit}colimit} for the
functor $\funF: \catK\to \catL$ iff for every cocone $(e,
(t_{d})_{d\in\objK})$ for $\funF$ there exists a unique morphism $f:
c\to e$ such that $t_{d} = f\circ  s_{d}$ for every object $d\in\objK$. 
\EndDefinition

So this yields
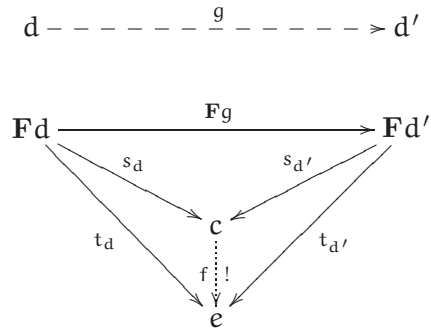
\begin{equation*}
\xymatrix{
d\ar@{-->}[rrrr]^{g} &&&& d'\\
\funF d\ar[rrrr]^{\funF g}\ar[drr]^{s_{d}}\ar[ddrr]_{t_{d}} 
&&&& \funF d'\ar[dll]_{s_{d'}}\ar[ddll]^{t_{d'}}\\
&& c\ar@{..>}[d]_{f}^{!}\\
&& e
}
\end{equation*}

Let us have a look at coproducts as an example.

\BeginExample{coprod-as-colimit}
Let $a$ and $b$ be objects in category $\catL$ and assume that their
coproduct $a+b$ with injections $j_{a}$ and $j_{b}$ exists in
$\catL$. Take again $I := \{1, 2\}$ and let $\catK$ be the discrete
category over $I$. Put $\funF 1 := a$ and $\funF 2 := b$, then it
follows from the definition of the coproduct that the cocone $(a+b,
j_{a}, j_{b})$ is a colimit for $\funF$.
\EndExample

One shows that the pushout can be represented as a colimit in the same
way as in Example~\ref{pullback-as-limit} for the representation of
the pullback as a limit.

Both limits and colimits are powerful general concepts for
representing important constructions with and on categories. We will
encounter them later on, albeit mostly indirectly.


\Subsection{Monads and Kleisli Tripels}
\label{sec:monads}

We have now functors and natural transformations at our disposal, and
we will put then to work. The first application we will tackle
concerns monads. Moggi's work~\cite{Moggi-Inf+Control, Moggi-LectNotes}
shows a connection between monads and computation which we will
discuss now. Kleisli tripels as a practical disguise for monads are
introduced first, and it will be shown through Manes' Theorem that they are
equivalent in the sense that each Kleisli tripel generates a monad,
and vice versa in a reversible construction. Some examples for monads
follow, and we will finally have a brief look at the monadic
construction in the programming language \texttt{Haskell}.

\Subsubsection{Kleisli Tripels}
\label{sec:kleisli-trip}

Assume that we work in a category $\catK$ and interpret values and
computations of a programming language in $\catK$. We need to
distinguish between the values of a type $a$ and the computations of
type $a$, which are of type $\funT a$. For example
\begin{description}
\item[\emph{Non-deterministic computations}] Taking the values from a set $A$
  yields computations of type $\funT A = \PowerSenza_{f}(A)$, where the
  latter denotes all finite subsets of $A$.
\item[\emph{Probabilistic computations}] Taking values from a set $A$ will
  give computations in the set $\funT A = \FunctorSenza{D} A$ of all discrete probabilities on
  $A$, see Example~\ref{functor-discrete-probs}.
\item[\emph{Exceptions}] Here values of type $A$ will result in values taken
  from $\funT A = A+E$ with $E$ as the set of
  \emph{\index{exception}exceptions}.
\item[\emph{Side effects}] Let $L$ is the set of addresses in the store and
  $U$ the set of all storage cells, a computation of type $A$ will
  assign each element of $U^{L}$ an element of $A$ or another element of
  $U^{L}$, thus we have $\funT A = (A+U^{L})^{U^{L}}$. 
\item[\emph{Interactive input}] Let $U$ be the set of characters, then $\funT
  A$ is the set of all trees with finite fan out, so that the internal
  nodes have labels from $U$, and the leaves have labels taken from $A$.
\end{description}

In order to model this, we require an embedding of the values taken
from $a$ into the computations of type $\funT a$, which is represented
as a morphism $\eta_{a}: a\to \funT a$. Moreover, we want to be able
to ``lift'' values to computations in this sense: if $f: a\to \funT b$
is a map from values to computations, we want to extend $f$ to a map
$f^{*}: \funT a\to \funT b$ from computations to computations (thus we
will be able to combine computations in a modular
fashion). Understanding a morphism $a\to \funT b$ as a program
performing computations of type $b$ on values of type $a$, this
lifting will then permit performing computations of type $b$ depending
on computations of type $a$.

This leads to the definition of a Kleisli tripel.

\BeginDefinition{kleisli-tripel}
Let $\catK$ be a category. A \emph{\index{Kleisli!tripel}Kleisli tripel} $(\funT, \eta, -^{*})$
over $\catK$ consists of a map $\funT: \objK\to \objK$ on objects, a
morphism $\eta_{a}: a\to \funT a$ for each object $a$, an operation ${}^{*}$ such that
$f^{*}: \funT a\to \funT b$, if $f: a\to \funT b$ with the following
properties:
\begin{dingautolist}{192}
\item\label{kleisli-tripel-1} $\eta_{a}^{*} = id_{\funT a}$.
\item\label{kleisli-tripel-2} $f^{*}\circ \eta_{a} = f$, provided $f: a\to \funT b$.
\item\label{kleisli-tripel-3} $g^{*}\circ f^{*} = (g^{*}\circ f)^{*}$ for $f: a\to \funT b$
  and $g: b\to \funT c$. 
\end{dingautolist}
\EndDefinition

Let us discuss briefly these properties of a Kleisli tripel. The first
property says that lifting the embedding $\eta_{a}: a\to \funT a$ will
give the identity on $\funT a$. The second condition says that
applying the lifted morphism $f^{*}$ to an embedded value $\eta_{a}$
will yield the same value as the given $f$. The third condition says
that combining lifted morphisms is the same as lifting the
lifted second morphism applied to the value of the first morphism. 

The category associated with a Kleisli tripel has the same objects as
the originally given category (which is not too much of a surprise),
but morphisms will correspond to programs: a program which performs a
computation of type $b$ on values of type $a$. Hence a morphism in
this new category is of type $a\to \funT b$ in the given one.  

\BeginDefinition{kleisli-category}
Given a Kleisli tripel $(\funT, \eta, -^{*})$ over category $\catK$, the
\emph{\index{Kleisli!category}Kleisli category} $\catK_{\funT}$ is defined as follows
\begin{itemize}
\item $\obj{\catK_{\funT}} = \objK$, thus $\catK_{\funT}$ has the same objects
  as $\catK$.
\item $\hom{\catK_{\funT}}(a, b) = \homK{a, \funT b}$, hence $f$ is a
  morphism $a\to b$ in $\catK_{\funT}$ iff $f:a \to \funT b$ is a
  morphism in $\catK$.
\item The identity for $a$ in $\catK_{\funT}$ is $\eta_{a}: a \to
  \funT a$.
\item The composition $g \ast f$ of $f\in\hom{\catK_{\funT}}(a, b)$
  and $g\in\hom{\catK_{\funT}}(b, c)$ is defined through $g\ast f :=
  g^{*}\circ f$. 
\end{itemize}
\EndDefinition

We have to show that Kleisli composition is associative: in fact, we
have 
\begin{align*}
  (h\ast g)\ast f & = (h\ast g)^{*}\circ f \\
& = (h^{*}\circ g)^{*}\circ f &&\text{(definition of $h\ast g$)}\\
& = h^{*}\circ g^{*}\circ f && \text{(property \ref{kleisli-tripel-3})}\\
& = h^{*}\circ (g\ast f)&&\text{(definition of $g\ast f$)}\\
& = h\ast (f\ast g)
\end{align*}
Thus $\catK_{\funT}$ is in fact a category. The map on objects in a
Kleisli category extends to a functor (note that we did not postulate
for a Kleisli triple that $\funT f$ is defined for morphisms). This functor is
associated with two natural transformations which together form a
monad. We will first define what a monad formally is, and then discuss the construction
in some detail. 

\Subsubsection{Monads}
\label{sec:monad-defn}
\BeginDefinition{monad}
A \emph{\index{monad}monad} over a category $\catK$ is a triple $(\funT, \eta, \mu)$ with
these properties:
\begin{dingautolist}{202}
\item\label{monad-cond-1} $\funT$ is an endofunctor on $\catK$.
\item\label{monad-cond-2} $\eta: Id_{\catK}\to \funT$ and $\mu: \funT^{2}\to \funT$ are
  natural transformations. $\eta$ is called the \emph{\index{monad!unit}unit}, $\mu$ the
  \emph{\index{monad!multiplication}multiplication} of the monad. 
\item\label{monad-cond-3} These diagrams commute
\begin{equation*}
\xymatrix{
\funT^{3}\ a\ar[d]_{\funT\mu_{a}}\ar[rr]^{\mu_{\funT a}}&&
\funT^{2}\ a\ar[d]^{\mu_{a}}
&&\funT\ a\ar[rr]^{\eta_{\funT a}}\ar[drr]_{id_{\funT a}} 
&& \funT^{2}\ a\ar[d]_{\mu_{a}} 
&& \funT\ a\ar[ll]_{\funT\eta_{a}}\ar[dll]^{id_{\funT a}}\\
\funT^{2}\ a\ar[rr]_{\mu_{a}} && \funT\ a &&&&\funT\ a
}
\end{equation*}
\end{dingautolist}
\EndDefinition

Each Kleisli triple generates a monad, and vice versa. This is what
\index{Theorem!Manes}Manes' Theorem says:

\BeginTheorem{manes}
Given a category $\catK$, there is a one-one correspondence between Kleisli triples and monads.
\EndTheorem

\BeginProof
1.
Let $(\funT, \eta, -^{*})$ be a Kleisli triple. We will extend $\funT$
to a functor $\catK\to \catK$, and define the multiplication; the
monad's unit will be $\eta$. Define
\begin{align*}
  \funT f &:= (\eta_{b}\circ f)^{*}, \text{ if } f: a\to b,\\
\mu_{a} & := (id_{\funT a})^{*}.
\end{align*}
Then $\mu$ is a natural transformation $\funT^{2}\to \funT$. Clearly,
$\mu_{a}: \funT^{2} a \to \funT a$ is a morphism. Let $f: a\to b$ be a
morphism in $\catK$, then we have 
\begin{align*}
  \mu_{b}\circ \funT^{2} f & = id_{\funT b}^{*}\circ (\eta_{\funT
    b}\circ (\eta_{b}\circ f)^{*})^{*}\\
& = \bigl(id_{\funT b}^{*}\circ (\eta_{\funT b}\circ (\eta_{b}\circ
f)^{*})\bigr)^{*}&& \text{(by~\ref{kleisli-tripel-3})}\\
& = \bigl(id_{\funT b}\circ (\eta_{b}\circ f)^{*}\bigr)^{*}&&\text{(since $id_{\funT b}^{*}\circ \eta_{\funT b} = id_{\funT b}$)}\\
& = {(\eta_{b}\circ f)^{*}}^{*}
\end{align*}
Similarly, we obtain
\begin{equation*}
  (\funT f)\circ \mu_{a} = (\eta_{b}\circ f)^{*}\circ id_{\funT a}^{*}
  = \bigl((\eta_{b}\circ f)^{*}\circ id_{\funT a}\bigr)^{*} = {(\eta_{b}\circ f)^{*}}^{*}.
\end{equation*}
Hence $\mu: \funT^{2}\to \funT$ is natural. Because we obtain for the
morphisms $f: a\to b$ and $g: b\to c$ the identity
\begin{equation*}
  (\funT g)\circ (\funT f) = (\eta_{c}\circ g)^{*}\circ (\eta_{b}\circ
  f)^{*} = \bigl((\eta_{c}\circ g)^{*}\circ \eta_{b}\circ f\bigr)^{*} =
  (\eta_{c}\circ g\circ f)^{*} = \funT (g\circ f)
\end{equation*}
and since by~\ref{kleisli-tripel-1}
\begin{equation*}
  \funT\ id_{a} = (\eta_{a}\circ id_{\funT a})^{*} = \eta_{a}^{*} =
  id_{\funT a},
\end{equation*}
we conclude that $\funT$ is an endofunctor on $\catK$. 

We check the laws for unit and multiplication according
to~\ref{monad-cond-3}. One notes first that
\begin{equation*}
  \mu_{a}\circ \eta_{\funT a} = id_{\funT a}^{*}\circ \eta_{\funT a}
  \stackrel{(\ddag)}{=} id_{\funT a}
\end{equation*}
(in equation $(\ddag)$ we use \ref{kleisli-tripel-2}), and that
\begin{equation*}
  \mu_{a}\circ \funT a = id_{\funT a}^{*}(\eta_{\funT a}\circ
  \eta_{a})^{*} = (id_{\funT a}^{*}\circ \eta_{\funT a}\circ
  \eta_{a})^{*} \stackrel{(\dag)}{=} \eta_{a}^{*} = (\eta_{a}\circ id_{a})^{*} = T(id_{a})
\end{equation*}
(in equation $(\dag)$ we use \ref{kleisli-tripel-2} again). Hence the
rightmost diagram in~\ref{monad-cond-3} commutes. Turning to the
leftmost diagram, we note that
\begin{equation*}
  \mu_{a}\circ \mu_{\funT a} = id_{\funT a}^{*}\circ id_{\funT^{2}
    a}^{*} = (id_{\funT a}^{*}\circ id_{\funT^{2}a})^{*} \stackrel{(\%)}{=} \mu_{a}^{*},
\end{equation*}
using~\ref{kleisli-tripel-3} in equation $(\%)$. On the other hand, 
\begin{equation*}
 \mu_{a}\circ (\funT\ \mu_{a})  = id_{\funT a}^{*}\circ (\funT\ id_{\funT a}^{*}) = id_{\funT a}^{*}\circ
  (\eta_{\funT a}\circ id_{\funT a}^{*})^{*} = {id_{\funT a}^{*}}^{*}
  = \mu_{a}^{*},
\end{equation*}
because $id_{\funT a}^{*}\circ \eta_{\funT a} = id_{\funT a}$
by~\ref{kleisli-tripel-2}. Hence the leftmost diagram commutes as
well, and we have defined a monad indeed. 

2.  
To establish the converse, define $f^{*} := \mu_{b}\circ (\funT
f)$ for the morphism $f: a\to \funT b$. We obtain from the right
hand triangle $\eta_{a}^{*} = \mu_{a}\circ (\funT\eta_{a}) = id_{\funT
  a}$, thus~\ref{kleisli-tripel-1} holds. Since $\eta: Id_{\catK}\to
\funT$ is natural, we have $(\funT f)\circ \eta_{a} = \eta_{\funT
  b}\circ f$ for $f: a\to \funT b$. Hence
\begin{equation*}
  f^{*}\circ \eta_{a} = \mu_{b}\circ(\funT f)\circ \eta_{a} =
  \mu_{b}\circ \eta_{\funT b}\circ f = f
\end{equation*}
by the left hand side of the right triangle, giving~\ref{kleisli-tripel-2}. Finally, note that due to
$\mu: \funT^{*}\to \funT$ being natural, we have for $g: b\to \funT c$
the commutative diagram 
\begin{equation*}
\xymatrix{
\funT^{2}b\ar[d]_{\funT^{2}g }\ar[rr]^{\mu_{b}}&&\funT b\ar[d]^{\funT g}\\
\funT^{3}c\ar[rr]_{\mu_{\funT c}}&&\funT^{2}c}
\end{equation*}
Then
\begin{align*}
  g^{*}\circ f^{*}
& = \mu_{c}\circ (\funT g)\circ \mu_{b}\circ (\funT f) \\
& = \mu_{c}\circ \mu_{\funT c}\circ (\funT^{2}g)\circ (\funT f)
&& \text{(since $(\funT g)\circ \mu_{b} = \mu_{\funT c}\circ \funT^{2} g$)}\\
& = \mu_{c}\circ (\funT \mu_{c})\circ \funT(\funT(g) \circ f)
&&\text{(since $\mu_{c}\circ \mu_{\funT c} = \mu_{c}\circ
  (\funT\mu_{c})$)}\\
& = \mu_{c}\circ \funT(\mu_{c}\circ \funT(g)\circ f)\\
& = \mu_{c}\circ \funT(g^{*}\circ f)\\
& = (g^{*}\circ f)^{*} 
\end{align*}
This establishes~\ref{kleisli-tripel-3} and shows that this defines a Kleisli tripel.
\EndProof

Taking a Kleisli tripel and producing a monad from it, one suspects
that one might end up with a different Kleisli triple for the
generated monad. But this is not the case; just for the record:

\BeginCorollary{cor-manes}
If the monad is given by a Kleisli tripel, then the Kleisli triple
defined by the monad coincides with the given one. Similarly, if the
Kleisli tripel is given by the monad, then the monad defined by the
Kleisli tripel coincides with the given one.
\EndCorollary

\BeginProof
We use the notation from above. 
Given the monad, put $f^{+} := id_{\funT b}\circ (\eta_{b}\circ f)^{*}$, then 
\begin{equation*}
f^{+} = \mu_{\funT b}\circ (\eta_{b}\circ f)^{*} = (id_{\funT a}\circ
f)^{*} = f^{*}.
\end{equation*}
On the other hand, given the Kleisli tripel, put $\funT_{0} f := (\eta_{b}\circ f)^{*}$, then
\begin{equation*}
  \funT_{0} f = \mu_{b}\circ \funT(\eta_{b}\circ f) = \mu_{b}\circ
  \funT(\eta_{b}) = \funT f.
\end{equation*}
\EndProof

Let us have a look at some examples. Theorem~\ref{manes} tells us that
the specification of a Kleisli tripel will give us the monad, and vice
versa. Thus we are free to specify one or the other; usually the
specification of the Kleisli tripel is shorter and more concise.

\BeginExample{nondet-comput}
Nondeterministic computations may be modelled through a map $f: S\to
\PowerSet{T}$: given a state (or an input, or whatever) from set $S$,
the set $f(s)$ describes the set of all possible outcomes. Thus we
work in category $\catSET$ with maps as morphisms and take the power set functor
$\PowerSenza$ as the functor. Define 
\begin{align*}
  \eta_{S}(x) & := \{x\},\\
f^{*}(B) & := \bigcup_{x\in B} f(x)
\end{align*}
for the set $S$, for $B\subseteq S$ and the map $f: S\to \PowerSet{T}$. Then clearly
$\eta_{S}: S\to \PowerSet{S}$, and $f^{*}: \PowerSet{S}\to
\PowerSet{T}$. We check the laws for a Kleisli tripel:
\begin{dingautolist}{192}
 \item Since $\eta_{S}^{*}(B) = \bigcup_{x\in B} \eta_{S}(x) = B$, we
   see that $\eta_{S}^{*} = id_{\PowerSet{S}}$
 \item It is clear that $f^{*}\circ \eta_{a} = f$ holds for $f: S\to
   \PowerSet{S}$. 
 \item Let $f: S\to \PowerSet{T}$ and $g: T\to \PowerSet{U}$, then 
   \begin{align*}
     u \in (g^{*}\circ f^{*})(B)
& \Leftrightarrow 
u\in g(y)\text{ for some $x\in B$ and some $y\in f(x)$}\\
&\Leftrightarrow 
u \in g^{*}(f(x))\text{ for some $x\in B$}
   \end{align*}
Thus $(g^{*}\circ f^{*})(B) = (g^{*}\circ f)^{*}(B)$. 
\end{dingautolist}
Hence the laws for a Kleisli tripel are satisfied. Let us just compute
$\mu_{S} = id_{\PowerSet{S}}^{*}$: Given $\beta\in\PowerSet{\PowerSet{S}}$,
we obtain
\begin{equation*}
  \mu_{S}(\beta) = id_{\PowerSet{S}}^{*} =
  \bigcup_{B\in\beta} B
= \bigcup \beta.
\end{equation*}

The same argumentation can be carried out when the power set functor is
replaced by the finite power set functor $\PowerSenza_{f}: S \mapsto
\{A\subseteq S \mid A \text{ is finite}\}$ with the obvious definition
of $\PowerSenza_{f}$ on maps.
\EndExample

In contrast to nondeterministic computations, probabilistic ones argue
with probability distributions. We consider the discrete case first, and
here we focus on probabilities with finite support.

\BeginExample{discr-prob-monad}
{\def\funD{\FunctorSenza{D}}
We work in the category $\catSET$ of sets with maps as morphisms and
consider the \index{probability functor!discrete}discrete probability functor $\funD S := \{p: S\to [0, 1] \mid p\text{ is a discrete
  probability}\}$, see
Example~\ref{functor-discrete-probs}.  Let $f: S\to \funD S$ be a map
and $p\in\funD S$, put
\begin{equation*}
  f^{*}(p)(s) := \sum_{t\in S} f(t)(s)p(t).
\end{equation*}
Then 
\begin{equation*}
\sum_{s\in S}  f^{*}(p)(s) = \sum_{s}\sum_{t}f(t)(s)p(t) =
\sum_{t}\sum_{s} f(t)(s)p(t) = \sum_{t} p(t) = 1,
\end{equation*}
hence $f^{*}: \funD S\to \funD S$. Note that the set 
$\{\langle s, t\rangle \in S\times T\mid f(s)(t)p(s) > 0\}$ is
finite, because $p$ has finite support, and because each $f(s)$
has finite support as well. Since each of the summands is
non-negative, we may reorder the summations at our convenience. Define moreover 
\begin{equation*}
  \eta_{S}(s)(s') := d_{S}(s)(s') := 
  \begin{cases}
    1, & s = s'\\
0, & \text{otherwise},
  \end{cases}
\end{equation*}
so that $\eta_{S}(s)$ is the discrete \emph{\index{Dirac measure}Dirac measure} on $s$. Then
\begin{dingautolist}{192}
\item $\eta_{S}^{*}(p)(s) = \sum_{s'}d_{S}(s)(s')p(s') = p(s)$, hence we may conclude
  that $\eta_{S}^{*}\circ p = p$.
\item $f^{*}(\eta_{S})(s) = f(s)$ is immediate.
\item Let $f: S\to \funD T$ and $g: T\to \funD U$, then we have for
  $p\in\funD S$ and $u\in U$
  \begin{align*}
    (g^{*}\circ f^{*})(p)(u) 
& = \sum_{t\in T}g(t)(u)f^{*}(p)(u)\\
& = \sum_{t\in T}\sum_{s\in S}g(t)(u)f(s)(t)p(s)\\
& = \sum_{\langle s, t\rangle\in S\times T}g(t)(u)f(s)(t)p(s)\\
& = \sum_{s\in S}\bigl[\sum_{t\in T} g(t)(u)f(s)(t)\bigr]p(s)\\
& = \sum_{s\in S}g^{*}(f(s))(u)p(s)\\
& = (g^{*}\circ f)^{*}(p)(u)
  \end{align*}
\end{dingautolist}
Again, we are not bound to any particular order of summation. 

We obtain for $M\in(\funD\circ \funD)S$\begin{equation*}
  \mu_{S}(M)(s) = id_{\funD S}^{*}(M)(s) =
  \sum_{q\in\funD(S)}M(q)\cdot q(s).
\end{equation*}
The last sum extends over a finite set, because the support of $M$ is
finite.  }
\EndExample

Since programs may fail to halt, one works sometimes in models which
are formulated in terms of subprobabilities rather than
probabilities. This is what we consider next, extending the previous
example to the case of general measurable spaces.

\BeginExample{prob-comput} { \def\catM{\Category{Meas}}
  \def\funM{\SubProbSenza} \def\mbX{X, {\cal A}} \def\mbY{Y, {\cal B}}
  \def\mbZ{Z, {\cal C}} We work in the category of measurable spaces
  with measurable maps as morphisms, see
  Example~\ref{ex-cat-meas-space}. In Example~\ref{functor-cont-probs}
  the subprobability functor\index{probability functor!continuous} was
  introduced, and it was shown that for a measurable space $S$ the set
  $\funM S$ of all subprobabilities is a measurable space again (we
  omit in this example the $\sigma$-algebra from notation, a
  measurable space is for the time being a pair consisting of a
  carrier set and a $\sigma$-algebra on it). A probabilistic
  computation $f$ on the measurable spaces $S$ and $T$ produces from
  an input of an element of $S$ a subprobability distribution $f(s)$
  on $T$, hence an element of $\funM T$. We want $f$ to be a morphism
  in $\catM$, so $f: S\to \funM T$ is assumed to be measurable.

We know from Example~\ref{ex-cat-stoch-rel} and
Exercise~\ref{ex-meas-generator} that $f: S\to \funM T$ is measurable
iff these conditions are satisfied:
\begin{enumerate}
\item $f(s)\in\funM (T)$ for all $s\in S$, thus $f(s)$ is a
  subprobability on (the measurable sets of) $T$.
\item For each measurable set $D$ in $T$, the map $s \mapsto f(s)(D)$
  is measurable.
\end{enumerate}


Returning to the definition of a Kleisli tripel, we define for the
measurable space $S$, $f: S\to \funM T$, 
\begin{align*}
  e_{S} & := \delta_{S},\\
f^{*}(\mu)(B) & := \int_{S} f(s)(B)\ \mu(ds)&& (\mu\in\funM S,
B\subseteq T\text{ measurable}).
\end{align*}
Thus $e_{S}(x) = \delta_{S}(x)$, the \index{Dirac measure}Dirac measure associated with
$x$, and $f^{*}: \funM S\to \funM S$ is a morphism (in this example,
we write $e$ for the unit, and $m$ for the multiplication). Note that $f^{*}(\mu)\in \SubProb{T}$ in the scenario above; in order to see whether the properties of a Kleisli tripel are satisfied, we need to know how to integrate with this measure. Standard arguments show that
\begin{equation}
\label{int-arg-Kleisli}
  \int_{T} h~df^{*}(\mu) = \int_{S}\int_{T}h(t)~f(s)(dt)~\mu(ds),
\end{equation} 
whenever $h: T\to \pReal$ is measurable and bounded. 

Let us again check the properties a Kleisli tripel. Fix  $B$ as a measurable subset of $S$, $f: S\to \funM
S$ and $g: T\to \funM U$ as morphisms in $\catM$. 
\begin{dingautolist}{192}
\item Let $\mu\in\funM S$, then
  \begin{equation*}
e_{S}^{*}(\mu)(B) = \int_{S} \delta_{S}(x)(B)\ \mu(dx) = \mu(B),
\end{equation*}
hence $e_{S}^{*} = id_{\funM S}$.
\item If $x\in S$, then 
  \begin{equation*}
    f^{*}(e_{S}(x))(B) = \int_{S} f(s)(B)\ \delta_{S}(x)(ds) = f(x)(B),
  \end{equation*}
since $\int h\ d\delta_{S}(x) = h(x)$ for every measurable map $h$. Thus
$f^{*}\circ e_{S} = f$.
\item Given $\mu\in\funM S$, we have
  \begin{align*}
    (g^{*}\circ f^{*})(\mu)(B) 
& = g^{*}(f^{*}(\mu))(B)\\
& = \int_{T} g(t)(B)\ f^{*}(\mu)(dt)\\
& \stackrel{(\ref{int-arg-Kleisli})}{=} \int_{S}\int_{T} g(t)(B)\ f(s)(dt)\ \mu(ds)\\
& = \int_{S} g^{*}(f(s))(B)\ \mu(ds)\\
& = (g^{*}\circ f)^{*}(\mu)(B)
  \end{align*}
Thus $g^{*}\circ f^{*} = (g^{*}\circ f)^{*}$.
\end{dingautolist}
Hence $(\funM, e, -^{*})$ forms a Kleisli triple over the category
$\catM$ of measurable spaces.

Let us finally determine the monad's multiplication. We have for
$M\in(\funM\circ \funM) S$ and the measurable set $B\subseteq S$
\begin{equation*}
 m_{S}(M)(B) = id_{\funM(S)}^{*}(M)(B) = \int_{\funM(S)} \tau(B)\ M(d\tau)
\end{equation*}
}\EndExample

The underlying monad has been investigated by M. Giry, so it is called
in her honor the \index{Giry monad}\emph{Giry monad}. It is used
extensively as the machinery on which Markov transition systems are
based. 

The next example shows that ultrafilter define a monad as well. 

\BeginExample{kleisli-ultrafilter}
{
\def\funU{\FunctorSenza{U}}
\def\funP{\PowerSenza}
Let $\funU$ be the ultrafilter functor on $\catSET$, see
Example~\ref{ex-ultrafilter-functor}. Define for the set $S$ and the
map $f: S\to \funU T$
\begin{align*}
  \eta_{S}(s) & := \{A\subseteq S \mid s \in A\},\\
f^{*}(U) & := \{B\subseteq T \mid \{s\in S \mid B \in f(s)\}\in U\},
\end{align*}
provided $U\in\funU S$ is an ultrafilter. Then $\emptyset\not\in
f^{*}(U)$, since $\emptyset\not\in U$. $\eta_{S}(s)$ is the principal ultrafilter associated with $s\in
S$, hence $\eta_{S}: S\to \funU S$. Because the intersection of two sets is a member of an ultrafilter iff both sets are elements of it, 
\begin{equation*}
\{s\in S \mid B_{1}\cap
B_{2}\in f(s)\} = \{s\in S \mid B_{1}\in f(s)\}\cap\{s\in S \mid B_{2}\in
f(s)\},
\end{equation*}
$f^{*}(U)$
is closed under intersections, moreover,
$B\subseteq C$ and $B\in f^{*}(U)$ imply $C\in f^{*}(U)$. If $B\not\in
f^{*}(U)$, then $\{s\in S \mid f(s)\in B\}\not\in U$, hence $\{s\in S
\mid B\not\in f(s)\}\in U$, thus $S\setminus B\in f^{*}(U)$, and vice
versa. Hence $f^{*}(U)$ is an ultrafilter, thus $f^{*}: \funU S\to \funU
T$. 

We check whether $(\funU, \eta, -^{*})$ is a Kleisli tripel. 
\begin{dingautolist}{192}
\item Since $B\in \eta_{S}^{*}$ iff $\{s\in S \mid s \in B\}\in U$, we
  conclude that $\eta_{S}^{*} = id_{\funU S}$.
\item Similarly, if $f: S\to \funU T$ and $s\in S$, then $B \in
  (f^{*}\circ \eta_{S})(s)$ iff $B \in f(s)$, hence $f^{*}\circ
  \eta_{S} = f.$
\item Let $f: S\to \funU T$ and $g: T\to \funU W$. Then 
  \begin{align*}
    B \in (g^{*}\circ f^{*})(U) 
& \Leftrightarrow
\{s\in S \mid \{t\in T \mid B\in g(t)\}\in f(s)\}\in U\\
& \Leftrightarrow
B \in (g^{*}\circ f)^{*}(U)
  \end{align*}
for $U\in \funU S$. Consequently, $g^{*}\circ f^{*} = (g^{*}\circ f)^{*}.$
\end{dingautolist}
Let us compute the monad's multiplication. Define for $B\subseteq S$
the set 
$
[B] := \{C\in \funU S \mid B\in C\}
$
as the set of all ultrafilters on $S$ which contain $B$ as an element,
then an easy computation shows
\begin{equation*}
  \mu_{S}(V) = id_{\funU S}^{*}(V) = \{B\subseteq S \mid [B]\in V\}
\end{equation*}
for $V\in(\funU\circ\funU) S$.  
}\EndExample

\BeginExample{upper-closed-kleisli}
{\def\funV{\FunctorSenza{V}}
\def\funP{\PowerSenza}
This example deals with upper closed subsets of the power set of a
set, see Example~\ref{upper-closed-functor}. Let again
\begin{equation*}
  \funV S := \{V\subseteq\funP S \mid V\text{ is upper closed}\}
\end{equation*}
be the endofunctor on $\catSET$ which assigns to set $S$ all upper
closed subsets of $\funP S$. We define the components of a Kleisli
tripel as follows: $\eta_{S}(s)$ is the principal ultrafilter
generated by $s\in S$, which is upper closed, and if $f: S\to \funV T$
is a map, we put 
\begin{equation*}
  f^{*}(V) := \{B\subseteq T \mid \{s\in S \mid B\in f(s)\}\in V\}
\end{equation*}
for $V\in \funV T$, see in Example~\ref{kleisli-ultrafilter}. 

The argumentation in Example~\ref{kleisli-ultrafilter} carries over
and shows that this defines a Kleisli tripel. 
}
\EndExample

These examples show that monads and Kleisli tripels are constructions
which model many computationally interesting subjects. After looking
at the practical side of this, we return to the discussion of the
relationship of monads to adjunctions, another important construction.

{
\floatstyle{boxed}
\newfloat{Programm}{hbtp}{lop}[section]
\floatname{Programm}{Implementierung}
\dimen255=\the\leftmargin
\CustomVerbatimEnvironment%
{MeinVerbatim}{Verbatim}{xleftmargin=.7\dimen255,baselinestretch=.9}
\Subsubsection{Monads in \texttt{Haskell}}
\label{sec:monads-haskell}

The functional programming language \texttt{Haskell} thrives on the construction
of \index{monad!\texttt{Haskell}}monads. We have a brief look. 

\texttt{Haskell} permits the definition of type classes; the definition of a type class requires
the specification of the types on which the class is based, and the
signature of the functions defined by this class. The definition of
class \texttt{Monad} is given below (actually, it is rather a specification of Kleisli
tripels). 
\begin{MeinVerbatim}
class Monad m where
  (>>=)  :: m a -> (a -> m b) -> m b
  return :: a -> m a
  (>>)   :: m a -> m b -> m b
  fail   :: String -> m a
\end{MeinVerbatim}
Thus class \texttt{Monad} is based on type constructor \texttt{m}, it specifies four functions of
which \texttt{>>=} and \texttt{return} are the most interesting. The
first one is called \emph{bind} and used as an infix operator: given \texttt{x}
of type \texttt{m a} and a function \texttt{f} of type \texttt{a -> m
  b}, the evaluation of \texttt{x >>= f}
will yield a result of type \texttt{m b}. This corresponds to
$f^{*}$. The function \texttt{return} takes a value of type \texttt{a} and
evaluates to a value of type \texttt{m a}; hence it corresponds to
$\eta_{a}$ (the name \texttt{return} is probably not a fortunate
choice). The function \texttt{>>}, usually used as an infix operator
as well, is defined by default in terms of \texttt{>>=}, and function \texttt{fail} serves to handling
exceptions; both functions will not concern us here.  

Not every conceivable definition of the functions \texttt{return} and the bind
function \texttt{>>=} are suitable for the definition of a
monad. These are the laws the \texttt{Haskell} programmer has to enforce, and
it becomes evident that these are just the laws for a Kleisli tripel
from Definition~\ref{kleisli-tripel}:
\begin{MeinVerbatim}
return x >>= f            == f x
p >>= return              == p  
p >>= (\x -> (f x >>= g)) == (p >>= (\x -> f x)) >>= g
\end{MeinVerbatim}
(here \texttt{x} is not free in \texttt{g}; note that 
\verb+\x -> f x+
is \texttt{Haskell}'s way of expressing the anonymous function $\lambda x.f x$). The compiler for Haskell
cannot check these laws, so the programmer has to make sure that they
hold.

We demonstrate the concept with a simple example. Lists\label{the-list-monad} are a popular
data structure. They are declared as a monad in this way:
\begin{MeinVerbatim}
instance Monad [] where
	return t = [t]
	x >>= f  = concat (map f x)
\end{MeinVerbatim} 
This makes the polymorphic type constructor \texttt{[]} for lists a monad; it
specifies essentially that \texttt{f} is mapped over the list
\texttt{x} (note that \texttt{x} has to be a list, and \texttt{f} a function of the
list's base type to a list); this yields a
list of lists which then will be flattened through an application of
function \texttt{concat}. This example should clarify things:
\begin{MeinVerbatim}
>>> q = (\w -> [0 .. w])
>>> [0 .. 2] >>= q
[0,0,1,0,1,2]
\end{MeinVerbatim}
The effect is explained in the following way: The definition of the
bind operation \texttt{>>=} requires the computation of 
\begin{MeinVerbatim}
concat (map q [0 .. 2]) = concat [(q 0), (q 1), (q 2)] 
= concat [[0], [0, 1], [0, 1, 2]]
= [0,0,1,0,1,2].
\end{MeinVerbatim}
We check the laws of a monad.
\begin{itemize}
\item We have \texttt{return x == [x]}, hence 
\begin{MeinVerbatim}
return x >>= f == [x] >>= f 
               == concat (map f [x]) 
               == concat [f x] 
               == f x
\end{MeinVerbatim}
\item Similarly, if  \texttt{p} is a given list, then
\begin{MeinVerbatim}
p >>= return == concat (map return p) 
             == concat [[x] | x <- p] 
             == [x | x <- p] 
             == p
\end{MeinVerbatim}
\item For the third law, if \texttt{p} is the empty list, then the left and the
  right hand side are empty as well. Hence let us assume that 
\texttt{p = [x1, .., xn]}. We obtain for the left hand side
\begin{MeinVerbatim}
p >>= (\x -> (f x >>= g)) 
== concat (map (\x -> (f x >>= g)) p) 
== concat (concat [map g (f x) | x <- p]), 
\end{MeinVerbatim}
and for the right hand side
\begin{MeinVerbatim}
(concat [f x | x <- p]) >>= g 
== ((f x1) ++ (f x2) ++ .. ++ (f xn)) >>= g
== concat (map g ((f x1) ++ (f x2) ++ .. ++ (f xn)))
== concat (concat [map g (f x) | x <- p])
\end{MeinVerbatim}
(this argumentation could of course be made more precise through a
proof by induction on the length of list \texttt{p}, but this would lead us too far from the present discussion).  
\end{itemize}
Kleisli composition \texttt{>=>} can be defined in a monad as follows:
\begin{MeinVerbatim}
(>=>) :: Monad m => (a -> m b) -> (b -> mc) -> (a -> m c)
f >=> g = \x -> (f x) >>= g
\end{MeinVerbatim}
This gives in the first line a type declaration for operation
\texttt{>=>} by indicating that the infix operator \texttt{>=>} takes two
arguments, viz., a function the signature of which is \texttt{a -> m b}, a
second one with signature \texttt{b -> m c}, and that the result will
be a function of type \texttt{a -> m c}, as expected. The precondition
to this type declaration is that \texttt{m} is a monad. The body of
the function will use the bind operator for binding \texttt{f x} to \texttt{g}; this
results in a function depending on \texttt{x}.  
It  can be shown that this composition is associative. 
}

\Subsection{Adjunctions and Algebras}
\label{sec:algebras}
\def\unit{\eta}
\def\multi{\mu}

\paragraph*{Adjunctions.}
{
\def\homLa#1#2{\homL{#1,#2}}
\def\homKa#1#2{\homK{#1,#2}}
\def\xxF#1{\funF #1}
\def\xxG#1{\funG #1}
\def\xxT#1{\funT #1}
\def\Eins#1{Id_{#1}}
\def\epsilon{\varepsilon}
We define the basic notion of an adjunction and show that
an adjunction defines a pair of natural transformations
through universal arrows (which is sometimes taken as the
basis for adjunctions).

\BeginDefinition{adjunction}
Let $\catK$ and $\catL$ be categories. Then
$
(\funF, \funG, \varphi)
$
is called an \emph{adjunction}\index{adjunction} iff
\begin{enumerate}
  \item $\funF: \catK \rightarrow \catL$ and
  $\funG: \catL \rightarrow \catK$ are functors,
  \item for each object $a$ in $\catL$ and $x$ in $\catK$ there
is a bijection
$$
\varphi_{x, a}:
\homLa{\xxF{x}}{a} \rightarrow \homKa{x}{\xxG{a}}
$$
which is natural in $x$ and $a$.
\end{enumerate}
$\funF$ is called the \emph{left adjoint to $\funG$}, $\funG$ is called the
\emph{right adjoint to $\funF$}.
\index{left adjoint}\index{adjoint!left}\index{right adjoint}\index{adjoint!right}
\EndDefinition

That $\varphi_{x, a}$ is natural for each $x, a$ means that for all
morphisms $f: a \rightarrow b$ in $\catL$ and $g: y \rightarrow x$ in
$\catK$ both diagrams commute:
\begin{equation*}
\xymatrix{
\homLa{\xxF{x}}{a}\ar[rr]^{\phi_{x, a}}\ar[d]_{f_{*}}&&
\homKa{x}{\xxG{a}}\ar[d]^{(\funG f)_{*}} 
&&\homLa{\xxF{x}}{a}\ar[rr]^{\phi_{x, a}}\ar[d]_{(\funF g)^{*}}
&&\homKa{x}{\xxG{a}}\ar[d]^{g^{*}}
\\
\homLa{\xxF{x}}{b}\ar[rr]_{\phi_{x, b}} && \homKa{x}{\xxG{b}}
&&\homLa{\xxF{y}}{a}\ar[rr]_{\phi_{y, a}}&&\homKa{y}{\xxG{a}}
}
\end{equation*}

Here $f_* := \homLa{\xxF{x}}{f}$ and
$g^*:= \homKa{g}{\xxG{a}}$ are the hom-set functors associated with $f$ resp.
$g$, similar for $(\xxG{f})_*$ and for $(\xxF{g})^*$; for the hom-set functors
see Example~\ref{hom-sets-functors}.

Let us have a look at a simple example: Currying.

\BeginExample{curry-is-adjoint}
A map $f: X\times Y\to Z$ is sometimes considered as a map $f: X\to (Y\to Z)$, so that $f(x, y)$ is considered as the value at $y$ for the map $f(x) := \lambda b.f(x, b)$. This helpful technique is called \emph{\index{currying}currying} and will be discussed now.

Fix a set $E$ and define the endofunctors $\funF, \funG: \catSET\to \catSET$ by $\funF := -\times E$ resp. $\funG := -^{E}$. Thus we have in particular $(\funF f)(x, e) := \langle f(x), e\rangle$ and $(\funG f)(g)(e) := f(g(e))$, whenever $f: X\to Y$ is a map. 

Define the map $\phi_{X, A}: \hom{\catSET}(\funF X, A)\to \hom{\catSET}(X,\funG A)$ by $\phi_{X, A}(k)(x)(e) := k(x, e)$. Then $\phi_{X, A}$ is a bijection. In fact, let $k_{1}, k_{2}: \funF X\to A$ be different maps, then $k_{1}(x, e)\not= k_{2}(x, e)$ for some $\langle x, e\rangle \in X\times E$, hence $\phi_{X, A}(k_{1})(x)(e) \not= \phi_{X, A}(k_{2})(x)(e)$, so that $\phi_{X, A}$ is one-to-one. Let $\ell: X\to \funG A$ be a map, then $\ell = \phi_{X, A}(k)$ with $ k(x, e) := \ell(x)(e)$. Thus $\phi_{X, A}$ is onto.

In order to show that $\phi$ is natural both in $X$ and in $A$, take maps $f: A\to B$ and $g: Y\to X$ and trace $k\in\hom{\catSET}(\funF X, A)$ through the diagrams in Definition~\ref{adjunction}. We have
\begin{equation*}
\phi_{X, B}(f_{*}(k))(x)(e) = f_{*}(k)(x, e) = f(k(x, e)) = f(\phi_{X, A}(k)(x, e)) = (\funG f)_{*}(\phi_{X, A}(k)(x)(e).
\end{equation*}
Similarly, 
\begin{equation*}
g^{*}(\phi_{X, A}(k))(y)(e) = k(g(y), e) = (\funF g)^{*}(k)(y, e) = \phi_{Y, A}((\funF g)^{*}(k))(y)(e).
\end{equation*}
This shows that $(\funF, \funG, \phi)$ with $\phi$ as the currying function is an adjunction. 
\EndExample

Another popular example is furnished through the diagonal functor.

\BeginExample{diag-has-adjunct}
{
\def\catp{\catK\times\catK}
Let $\catK$ be a category such that for any two objects $a$ and $b$ their product $a\times b$ exists. Recall the definition of the Cartesian product of categories from Lemma~\ref{product-category}. Define the diagonal functor $\Delta: \catK\to \catp$ through $\Delta a := \langle a, a\rangle$ for objects and $\Delta f := \langle f, f\rangle$ for morphism $f$. Conversely, define $\funT: \catp\to \catK$ by putting $\funT (a, b) := a\times b$ for objects, and $\funT \langle f, g\rangle := f\times g$ for morphism $\langle f, g\rangle$. 

Let $\langle k_{1}, k_{2}\rangle \in \hom{\catp}(\Delta a, \langle
b_{1}, b_{2}\rangle)$, hence we have morphisms $k_{1}: a\to b_{1}$ and
$k_{2}: a\to b_{2}$. By the definition of the product, there exists a
unique morphism $k: a\to b_{1}\times b_{2}$ with $k_{1} = \pi_{1}\circ
k$ and $k_{2} = \pi_{2}\circ k$, where $\pi_{i}: b_{1}\times b_{2}\to
b_{i}$ are the projections, $i = 1, 2$. Define $\phi_{a, b_{1}\times
  b_{2}}(k_{1}, k_{2}) := k$, then it is immediate that 
$
\phi_{a, b_{1}\times  b_{2}}: \hom{\catp}(\Delta a, \langle
b_{1}, b_{2}\rangle)\to \hom{\catK}(a, \funT (b_{1}, b_{2})
$
is a bijection. 

Let $\langle f_{1}, f_{2}\rangle :\langle a_{1}, a_{2}\rangle\to \langle b_{1}, b_{2}\rangle$ be a morphism, then the diagram
\begin{equation*}
\xymatrix{
\hom{\catp}(\Delta x, \langle a_{1}, a_{2}\rangle) 
\ar[d]_{\langle f_{1}, f_{2}\rangle_{*}}
\ar[rr]^{\phi_{x, \langle a_{1}, a_{2}\rangle}}
&&
\hom{\catK}(x, a_{1}\times a_{2})
\ar[d]^{(\funT (f_{1}, f_{2}))_{*}}\\ 
\hom{\catp}(\Delta x, \langle b_{1}, b_{2}\rangle)
\ar[rr]_{\phi_{x, \langle b_{1}, b_{2}\rangle}}
&&
\hom{\catK}(x, b_{1}\times b_{2})
}
\end{equation*}
splits into the two commutative diagrams
\begin{equation*}
\xymatrix{
\hom{\catK}(x, a_{i})
\ar[d]_{f_{i, *}}
\ar[rr]^{\pi_{i}\circ \phi_{x, \langle a_{1}, a_{2}\rangle}}
&&
\hom{\catK}(x, a_{i})
\ar[d]^{(\pi_{i}\circ (\funT (f_{1}, f_{2})))_{*}}\\
\hom{\catK}(x, b_{i})
\ar[rr]_{\pi_{i}\circ \phi_{x, \langle b_{1}, b_{2}\rangle}}
&&
\hom{\catK}(x, b_{i})
}
\end{equation*}
for $i = 1, 2$, hence is commutative itself. One argue similarly for a morphism $g: b\to a$. Thus the bijection $\phi$ is natural. 

Hence we have found out that $(\Delta, \funT, \phi)$ is an adjunction, so that the diagonal functor has the product functor as an adjoint. 
}
\EndExample

A map $f:  X\to Y$ between sets provides us with another example,
which is the special case of a Galois connection (recall that a pair
$f: P\to Q$ and $g: Q\to P$ of monotone maps between the partially
ordered ets $P$ and $Q$ form a \emph{\index{Galois connection}Galois connection} iff $f(p)\geq q
\Leftrightarrow p \leq g(q)$ for all $p\in P, q\in Q$.).

\BeginExample{adj-through-map}
Let $X$ and $Y$ be sets, then the inclusion on $\PowerSenza{X}$ resp. $\PowerSenza{Y}$
makes these sets cate\-gories, see Example~\ref{ex-cat-ord}. Given a map$f:X\to  Y$, define $f_{?}: \PowerSenza{X}\to \PowerSenza{Y}$ as the
direct image $f_{?}(A) := \Bild{f}{A}$ and $f_{!}: \PowerSenza{Y}\to
\PowerSenza{X}$ as the inverse image $f_{!}(B) := \InvBild{f}{B}$. Now
we have for $A\subseteq X$ and $B\subseteq Y$
\begin{equation*}
  B \subseteq f_{?}(A) \Leftrightarrow B\subseteq \Bild{f}{A}
\Leftrightarrow
\InvBild{f}{B}\subseteq A \Leftrightarrow f_{!}(B)\subseteq A.
\end{equation*}
This means in terms of the hom-sets that $\hom{\PowerSenza{Y}}(B,
f_{?}(A)) \not= \emptyset$ iff $\hom{\PowerSenza{X}}(f_{!}(B), A)\not=
  \emptyset$. Hence this gives an adjunction $(f_{!}, f_{?}, \phi)$.
\EndExample

Back to the general development. This auxiliary statement will help in
some computations.

\BeginLemma{aux-repr-adjunction}
Let $(\funF, \funG, \phi)$ be an adjunction, $f: a\to b$ and $g: y\to x$ be morphisms in $\catL$ resp. $\catK$. Then we have 
\begin{align*}
  (\funG f)\circ \phi_{x, a}(t) & = \phi_{x, b}(f\circ t),\\
\phi_{x, a}(t)\circ g &= \phi_{y, a}(t\circ \funF g)
\end{align*}
for each morphism $t: \funF x\to a$ in $\catL$.
\EndLemma

\BeginProof
Chase $t$ through the left hand diagram of Definition~\ref{adjunction} to obtain
\begin{equation*}
  ((\funG f)_{*}\circ \phi_{x, a})(t) = (\funG f)\circ \phi_{x, a}(t) = \phi_{x, b}(f_{*}(t)) = \phi_{x, b}(f\circ t).
\end{equation*}
This yields the first equation, the second is obtained from tracing $t$ through the diagram on the right hand side.
\EndProof

An adjunction induces natural transformations which make this important
construction easier to handle, and which helps indicating connections
of adjunctions to monads and Eilenberg-Moore algebras in the sequel. Before
entering the discussion, universal arrows are introduced.

\BeginDefinition{universal-arrows}
Let $\funS: \catC \rightarrow \catD$ be a functor, and
$c$ an object in $\catC$.
\begin{enumerate}
  \item the pair $\langle r, u\rangle$ is called a \emph{universal arrow from $c$ to
$\funS$} iff $r$ is an object in $\catC$ and $u: c \rightarrow \funS r$ is a morphism in $\catD$ such that for any arrow $f: c \rightarrow \funS d$ there exists a
unique arrow $f': r \rightarrow d$ in $\catC$ such that $f =(\funS f') \circ u$.
  \item the pair $\langle r, v\rangle$ is called a \emph{universal arrow from $\FunctorSenza{S}$ to
$c$} iff $r$ is an object in $\catC$ and $v:\funS r \rightarrow c$ is a morphism in $\catD$ such that for any arrow $f:\funS d \rightarrow c$ there exists a
unique arrow $f': d \rightarrow r$ in $\catC$ such that $f = v
\circ (\funS f')$.
\end{enumerate}
\EndDefinition

Thus, if the pair $\langle r, u\rangle$ is universal from $c$ to
$\funS$, then each arrow $c \rightarrow \funS d$
in $\catC$ factors uniquely through the $\funS$-image
of an arrow $r \rightarrow d$ in $\catC$. Similarly,
if the pair $\langle r, v\rangle$  is universal from $\funS$
to $c$, then each $\catD$-arrow $\funS d \rightarrow c$ factors
uniquely through the $\funS$-image of an $\catC$-arrow $d \rightarrow r$. These diagrams depict the situation for a universal arrow $u: c\to \funS r$ resp. a universal arrow $v: \funS r\to c$.
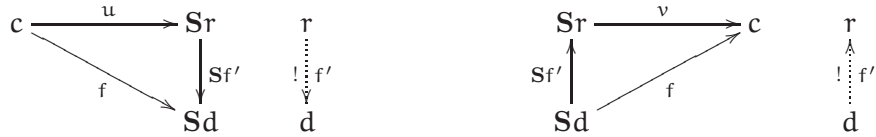
\begin{equation*}
\xymatrix{
c\ar[rr]^{u}\ar[drr]_{f} && \funS r\ar[d]^{\funS f'}&r\ar@{..>}[d]_{!}^{f'}
&&&
\funS r\ar[rr]^{v} && c&r
\\
&&\funS d&d
&&&
\funS d\ar[u]^{\funS f'}\ar[urr]_{f}&&&d\ar@{..>}[u]_{f'}^{!}
}
\end{equation*}

This is a characterization of a universal arrow from $c$ to $\funS$.

\BeginLemma{char-universal-arrow}
Let $\funS: \catC\to \catD$ be a functor. Then $\langle r, u\rangle$ is an universal arrow from $c$ to $\funS$ iff the function $\psi_{d}$ which maps each morphism $f': r\to d$ to the morphism $(\funS f')\circ u$ is a natural bijection $\hom{\catC}(r, d) \to \hom{\catD}(c, \funS d)$.
\EndLemma

\BeginProof
1.
If $\langle r, u\rangle$ is an universal arrow, then bijectivity of $\psi_{d}$ is just a reformulation of the definition. It is also clear that $\psi_{d}$ is natural in $d$, because if $g: d\to d'$ is a morphism, then $\funS(g'\circ f')\circ u = (\funS g')\circ (\funS g)\circ u$.

2.
Now assume that $\psi_{d}: \hom{\catC}(r, d) \to \hom{\catD}(c, \funS d)$ is a bijection for each $d$, and choose in particular $r = d$. Define $u := \psi_{r}(id_{r})$, then $u: c\to \funS r$ is a morphism in $\catD$. Consider this diagram for an arbitrary $f': r\to d$
\begin{equation*}
\xymatrix{
\hom{\catC}(r, r)\ar[d]_{\hom{\catC}(r, f')}\ar[rr]^{\psi_{r}}
&&\hom{\catD}(c, \funS r)\ar[d]^{\hom{\catD}(s, \funS f')}\\
\hom{\catC}(r, d)\ar[rr]_{\psi_{d}}&&\hom{\catD}(c, \funS d)
}
\end{equation*}

Given a morphism $f: c\to \funS d$ in $\catD$, there exists a unique morphism $f': r\to d$ such that $f =\psi_{d}(f')$, because $\psi_{d}$ is a bijection. Then we have
 \begin{align*}
 f & = \psi_{d}(f')\\
& = \bigl(\psi_{d}\circ \hom{\catC}(r, f')\bigr)(id_{r})\\
 & = \bigl(\hom{\catD}(c, \funS f')\circ \psi_{r}\bigr)(id_{r})&&\text{(commutativity)}\\
 & = \hom{\catD}(c, \funS f')\circ u&& (u = \psi_{r}(id_{r}))\\
 & = (\funS f')\circ u.
 \end{align*}
\EndProof

Universal arrows will be used now for a characterization of adjunctions
in terms of natural transformations (we will sometimes omit the indices for the
natural transformation $\varphi$ that comes with an adjunction).

\BeginTheorem{char-adjunction-general}
Let $(\funF, \funG, \varphi)$ be an adjunction for the functors
$
\funF: \catK \rightarrow \catL
$
and
$
\funG: \catL \rightarrow \catK.
$
Then there exist natural transformations
$
\eta: Id_{\catK} \to  \funG \circ \funF
$
and
$
\varepsilon: \funF \circ \funG \rightarrow Id_{\catL}
$
with these properties:
\begin{enumerate}
  \item \label{char-adjunction-general-a}
the pair $\langle \funF x, \eta_x\rangle$ is a universal arrow from $x$ to $\funG$ for each $x$ in $\catK$,
and
$
\varphi(f) = \xxG{f} \circ \eta_x
$
holds for each $f: \xxF{x} \rightarrow a$,
\item\label{char-adjunction-general-b}
  the pair $\langle \funG a, \varepsilon_a\rangle$ is universal from $\funF$ to $a$ for each $a$ in $\catL$,
and
$
\varphi^{-1}(g) = \varepsilon_a\circ\xxF{g}
$
holds for each $g: x \rightarrow \xxG{a}$,
\item\label{char-adjunction-general-c}
the composites
\begin{equation*}
\xymatrix{
\funG\ar[rr]^{\eta\funG}\ar[drr]_{id_{\funG}} 
&& \funG \circ \funF \circ \funG \ar[d]^{\funG\varepsilon}
&&& 
\funF\ar[rr]^{\funF\eta}\ar[drr]_{id_{\funF}} 
&& \funF \circ \funG \circ \funF\ar[d]^{\varepsilon\funF}\\
&&\funG
&&&
&& \funF
}
\end{equation*}
are the identities for $\funG$ resp. $\funF$.
\end{enumerate}
\EndTheorem

\BeginProof
1.
Put $\eta_{x} := \phi_{x, \funF x}(id_{\funF x})$, then $\eta_{x}: x\to \funG\funF x$. In order to show that $\langle \funF x, \eta_{x}\rangle$ is a universal arrow from $x$ to $\funG$, we take a morphism $f: x\to \funG a$ for some object $a$ in $\catL$. Since $(\funF, \funG, \phi)$ is an adjunction, we know that there exists a unique morphism $f': \funF x\to a$ such that $\phi_{x, a}(f') = f$. We have also this commutative diagram
\begin{equation*}
\xymatrix{
\hom{\catK}(\funF x, \funF x)\ar[d]_{\hom{}(\funF x, f')}\ar[rr]^{\phi_{x, \funF x}}
&&\hom{\catL}(x, \funG\funF x)\ar[d]^{\hom{\catL}(x, \funG f')}\\
\hom{\catK}(\funF x,a)\ar[rr]_{\phi_{x, a}}&&\hom{\catL}(x, \funG a)
}
\end{equation*}
Thus
\begin{align*}
  (\funG f')\circ \eta_{x} & = 
\bigl(\hom{\catL}(x, \funG f')\circ \phi_{x, \funF x}\bigr)(id_{\funF x}) \\
& = \bigl(\phi_{x, a}\circ \hom{\catK}(\funF x, f')\bigr)(id_{\funF x})\\
& = \phi_{x, a}(f')\\
& = f
\end{align*}

2.
$\eta: Id_{\catK}\to \funG\circ \funF$ is a natural transformation. Let $h: x\to y$ be a morphism in $\catK$, then we have by Lemma~\ref{aux-repr-adjunction}
\begin{align*}
  \funG(\funF h)\circ \eta_{x} & = \funG(\funF h)\circ \phi_{x,\funF x}(id_{\funF x})\\
& = \phi_{x, \funF y}(\funF h\circ id_{\funF x})\\
& = \phi_{x, \funF y}(id_{\funF y}\circ \funF h)\\
& = \phi_{y, \funF y}(id_{\funF y})\circ h\\
& = \eta_{y}\circ h.
\end{align*}

3.
Put $\epsilon_{a} := \phi_{\funG a, a}^{-1}(id_{\funG a})$ for the object $a$ in $\catL$, then the properties for $\epsilon$ are proved in exactly the same way as for those of $\eta$.

4.
From
$
\varphi_{x, a}(f) = \xxG{f} \circ \eta_x
$
we obtain
\begin{equation*}
id_{\xxG{a}} = \varphi(\varepsilon_a) = \funG{\varepsilon_a} \circ \eta_{\xxG{a}}
= (\funG\epsilon\circ \eta\funG)(a),
\end{equation*}
so that
$
\funG\varepsilon\circ \eta\funG
$
is the identity transformation on $\funG$. Similarly,
$
\eta\funF\circ \funF\varepsilon
$
is the identity for $\funF$.
\EndProof

The transformation $\eta$ is sometimes called the \emph{\index{adjunction!unit}unit} of the adjunction,
whereas $\varepsilon$ is called its \emph{\index{adjunction!counit}counit}.
The converse to Theorem~\ref{char-adjunction-general} holds as well: from
two transformations $\eta$ and $\varepsilon$  with the signatures as above
one can construct an adjunction. The proof is a fairly straightforward verification.

\BeginProposition{conv-char-adjunction-general}
Let
$
\funF: \catK \rightarrow \catL
$
and
$
\funG: \catL \rightarrow \catK
$
be functors, and assume that natural transformations
$
\eta: \Eins{\catK}\to  \funG \circ \funF
$
and
$
\varepsilon: \funF \circ \funG \rightarrow\Eins{\catL}
$
are given so that
$
(\funG\varepsilon) \circ (\eta\funG)
$
is the identity of $\funG$, and
$
(\varepsilon\funF)\circ(\funF\eta)
$
is the identity of $\funF$. Define $\varphi_{x, a}(k) := (\xxG{k}) \circ \eta_x$,
whenever $k: \xxF{x} \rightarrow a$ is a morphism in $\catL$.
Then
$
(\funF, \funG, \varphi)
$
defines an adjunction.
\EndProposition

\BeginProof
1.
Define $\theta_{x, a}(\ell) := \epsilon_{a}\circ \funF g$ for $\ell: x\to \funG a$, then we have 
\begin{align*}
  \phi_{x, a}(\theta_{x, a}(g)) 
& = \funG(\epsilon_{a}\circ \funF g)\circ \eta_{x}\\
& = (\funG\epsilon_{a})\circ (\funG \funF g)\circ \eta_{x}\\
& = (\funG\epsilon_{a})\circ \eta_{\funG a}\circ g && \text{ ($\eta$ is natural)}\\
& = \bigl((\funG\epsilon\circ \eta\funG) a\bigr)\circ g\\
& = id_{\funG a} g\\
& = g\\
\end{align*}
Thus $\phi_{x, a}\circ \theta_{x, a} = id_{\hom{\catL}(x, \funG a)}$. Similarly, one shows that $\theta_{x, a}\circ \phi_{x, a} = id_{\hom{\catK}(\funF x, a)}$, so that $\phi_{x, a}$ is a bijection.

2.
We have to show that $\varphi_{x, a}$ is natural for each $x, a$, so take a 
morphism $f: a \rightarrow b$ in $\catL$ and chase $k: \funF x\to a$ through this diagram.
\begin{equation*}
\xymatrix{
\homLa{\xxF{x}}{a}\ar[rr]^{\phi_{x, a}}\ar[d]_{f_{*}}
&&
\homKa{x}{\xxG{a}}\ar[d]^{(\funG f)_{*}} 
\\
\homLa{\xxF{x}}{b}\ar[rr]_{\phi_{x, b}} 
&& \homKa{x}{\xxG{b}}
}
\end{equation*}

Then $\bigl((\funG f)_{*}\circ \phi_{x, a}\bigr)(k) = (\funG f\circ \funG k)\circ \eta_{x}= \funG (f\circ k)\circ \eta_{x} = \phi_{x, b}(f_{*}\circ k).$

\EndProof

Thus for identifying an adjunction it is sufficient to
identify its unit and its counit. This includes verifying the identity
laws of the functors for the corresponding compositions. The following
example has another look at currying (Example~\ref{curry-is-adjoint}), demonstrating the approach and
suggesting that identifying unit and counit is sometimes easier than
working with the originally given definition.

\BeginExample{currying-again}
Continuing Example~\ref{curry-is-adjoint}, we take the definitions of the endofunctors $\funF$ and $\funG$ from there. Define for the set $X$ the natural transformations $\eta: Id_{\catSET}\to \funG\circ \funF$ and $\epsilon: \funF\circ \funG\to Id_{\catSET}$ through
\begin{equation*}
  \eta_{X}:
  \begin{cases}
    X &\to  (X\times E)^{E}\\
x & \mapsto \lambda e.\langle x, e\rangle
  \end{cases}
\end{equation*}
and
\begin{equation*}
  \epsilon_{X}:
  \begin{cases}
    (X\times E)^{E}\times E &\to  X\\
\langle g, e\rangle & \mapsto g(e)
  \end{cases}
\end{equation*}
Note that we have $(\funG f)(h) = f\circ h$ for $f: X^{E}\to Y^{E}$ and $h\in X^{E}$, so that we obtain 
\begin{align*}
  \funG \epsilon_{X}(\eta_{\funG X}(g))(e)
& = (\epsilon_{X}\circ \eta_{\funG X}(g))(e)\\
& = \epsilon_{X}(\eta_{\funG X}(g))(e) \\
& = \epsilon_{X}(\eta_{\funG X}(g)(e))\\
& = \epsilon_{X}(g, e)\\
& = g(e),
\end{align*}
whenever $e\in E$ and $g\in\funG X = X^{E}$, hence $(\funG \epsilon)\circ (\eta\funG) = id_{\funG}$. One shows similarly that $(\epsilon\funF)\circ (\funF\eta) = id_{\funF}$ through
\begin{equation*}
  \epsilon_{\funF X}(\funF \eta_{X}(x, e)) = \eta_{X}(x)(e) = \langle x, e\rangle.
\end{equation*}
\EndExample


Now let $(\funF, \funG, \varphi)$ be an adjunction with
functors $\funF: \catK \to \catL$ and
$\funG: \catL \to \catK$, the unit
$\eta$ and the counit $\varepsilon$. Define the functor $\funT$ through
$
\funT := \funG \circ \funF.
$
Then
$
\funT: \catK \to \catK
$
defines an endofunctor on category $\catK$ with
$
\multi_a := \left(\funG\varepsilon\funF\right)(a) =  \xxG{\varepsilon_{\xxF{a}}}
$
as a morphism
$
\multi_a: \Functor{T^2}{a} \to \xxT{a}.
$
Because
$
\varepsilon_a: \xxF{\xxG{a}} \to a
$
is a morphism in $\catL$, and because
$
\varepsilon: \funF \circ \funG \to \Eins{\catL}
$
is natural, the diagram
\begin{equation*}
\xymatrix{
(\funF\circ \funG\circ \funF\circ \funG)a\ar[rr]^{\varepsilon_{(\funF\circ \funG)a}}\ar[d]_{(\funF\circ \funG)\varepsilon_{a}}
&&(\funF\circ \funG)a\ar[d]^{\varepsilon_{a}}\\
(\funF\circ \funG)a\ar[rr]_{\varepsilon_{a}}&&a
}
\end{equation*}

is commutative. This means that this diagram
\begin{equation*}
\xymatrix{
\funF\circ \funG\circ \funF\circ \funG\ar[rr]^{\varepsilon(\funF\circ \funG)}\ar[d]_{(\funF\circ \funG)\varepsilon}
&&\funF\circ \funG\ar[d]^{\varepsilon}\\
\funF\circ \funG\ar[rr]_{\varepsilon}&&\Eins{\catK}
}
\end{equation*}
of functors and natural transformations commutes. Multiplying from the left with $\funG$ and from the right with $\funF$
gives this diagram.
\begin{equation*}
\xymatrix{
\funG\circ \funF\circ \funG\circ \funF\circ \funG\circ \funF
\ar[rrr]^{\funG\varepsilon(\funF\circ \funG\circ \funF)}\ar[d]_{(\funG\circ \funF\circ \funG)\varepsilon\funF}
&&&\funG\circ \funF\circ \funG\circ \funF\ar[d]^{\funG\varepsilon\funF}\\
\funG\circ \funF\circ \funG\circ \funF\ar[rrr]_{\funG\varepsilon\funF}&&&\funG\circ \funF
}
\end{equation*}
Because $\funT \mu = (\funG\circ \funF\circ \funG)\varepsilon \funF$, and $\funG \varepsilon(\funF\circ \funG\circ \funF) = \mu\funT$, this diagram can be written as 
\begin{equation*}
\xymatrix{
\funT^{3}\ar[rr]^{\funT\mu}\ar[d]_{\mu\funT}&&\funT^{2}\ar[d]^{\mu}\\
\funT^{2}\ar[rr]_{\mu}&&\funT
}
\end{equation*}
This gives the commutativity of the left hand diagram in Definition~\ref{monad}. Because $\funG\varepsilon\circ \eta\funG$ is the identity on $\funG$, we obtain
\begin{equation*}
  \funG\varepsilon_{\funF a}\circ \eta_{\funG\funF a} = (\funG\varepsilon\circ \eta\funG)(\funF a) = \funG\funF a,
\end{equation*}
which implies that the diagram 
\begin{equation*}
\xymatrix{
\funT a\ar[rr]^{\mu_{\funT a}}\ar[drr]_{id_{\funT a}} && \funT^{2}a\ar[d]^{\mu_{a}}\\
&&\funT a
}
\end{equation*}
commutes. On the other hand, we know that $\varepsilon\funF\circ \funF\eta$ is the identity on $\funF$; this yields 
\begin{equation*}
  \funG\varepsilon_{\funF a}\circ \funG\funF\eta_{a} = \funG(\varepsilon\funF\circ \funF\eta)a = \funG\funF a.
\end{equation*}
Hence we may complement the last diagram:
\begin{equation*}
\xymatrix{
\funT a\ar[rr]^{\mu_{\funT a}}\ar[drr]_{id_{\funT a}} 
&& \funT^{2}a\ar[d]^{\mu_{a}}
&& \funT a\ar[ll]_{\funT\eta_{a}}\ar[dll]^{id_{\funT a}}\\
&&\funT a
}
\end{equation*}
This gives the right hand side diagram  in Definition~\ref{monad}. We have shown
\BeginProposition{adjunction-def-monad}
Each adjunction defines a monad.
\QED
\EndProposition

}

It turns out that we not only may proceed from an adjunction to a monad,
but that it is also possible to traverse this path in the other
direction. We will show that a monad defines an adjunction. In order
to do that, we have to represent the functorial part of a monad as the
composition of two other functors, so we need a second category for
this. The algebras which are defined for a monad provide us with this
category. So we will define algebras (and in a later chapter, their
counterparts, coalgebras), and we will study them. This will help us
in showing that each monad defines an adjunction. Finally, we will
have a look at two examples for algebras, in order to illuminate this
concept.

Given a monad $ (\funT, \unit, \multi) $ in
a category $\catK$, a pair $ \langle x, h\rangle $ consisting
of an object $x$ and a morphism $h: \funT x \to x$
in $\catK$ is called an \emph{Eilenberg-Moore
algebra}\index{Eilenberg-Moore algebra}\index{algebra!Eilenberg-Moore} for the monad
iff the following diagrams commute
\begin{equation*}
\xymatrix{
\funT^{2}x\ar[rr]^{\funT h}\ar[d]_{\multi_{x}} && \funT x\ar[d]^{h}
&&&x\ar[rr]^{\unit_{x}}\ar[drr]_{id_{x}}&&\funT x\ar[d]^{h}
\\
\funT x\ar[rr]_{h} && x
&&&&&x
}
\end{equation*}
The morphism $h$ is called the \emph{\index{algebra!structure
    morphism}structure morphism} of the algebra, $x$ its carrier. 

An \emph{algebra morphism}\index{Eilenberg-Moore
algebra!morphism}\index{algebra!morphism}
 $ f: \langle x,
h\rangle \rightarrow \langle x', h'\rangle $ between the algebras
$\langle x, h\rangle$ and $\langle x', h'\rangle$ is a morphism
$f: x \rightarrow x'$ in $\catK$ which renders the diagram
\begin{equation*}
\xymatrix{
\funT x\ar[d]_{\funT f}\ar[rr]^{h} && x\ar[d]^{f}\\
\funT x'\ar[rr]_{h'}&& x'
}
\end{equation*}
commutative. Eilenberg-Moore algebras together with their morphisms form a category
$\Category{Alg}_{(\funT, \unit, \multi)}$.
We will usually omit the reference to the monad. Fix for the moment $(\funT, \unit, \multi)$ as a monad in
category $\catK$, and let
$
\Category{Alg}:= \Category{Alg}_{(\funT, \unit, \multi)}
$
be the associated category of Eilenberg-Moore algebras\label{alg-eilenberg-moore}.

We give some simple examples.

\BeginLemma{e-m-morphs-1}
The pair $\langle \funT x, \multi_x\rangle$ is a
$\funT$-algebra for each $x$ in $\catK$.
\EndLemma

\BeginProof
This is immediate from the laws for $\unit$ and $\multi$ in a monad.
\EndProof

These algebras are usually called the \emph{free algebras}\index{algebra!free} for
the monad. Morphisms in the base category $\catK$ translate into morphisms in $\Category{Alg}$
through functor $\funT$.

\BeginLemma{e-m-morphs-2}
If $f: x \rightarrow y$ is a morphism in $\catK$, then
$
\funT f: \langle\funT x, \multi_x\rangle\rightarrow\langle\funT{y}, \multi_y\rangle
$
is a morphism in $\Category{Alg}$. If $\langle x, h\rangle$ is an algebra, then $h: \langle \funT x, \mu_{x}\rangle\to \langle x, h\rangle$ is a morphism in $\Category{Alg}$. 
\EndLemma

\BeginProof
Because $\multi: \funT^{2}\to  \funT$ is a natural transformation, we see
$
\multi_y \circ \funT^{2}{f} =(\funT f) \circ \multi_x.
$
This is just the defining equation for a morphism in $\Category{Alg}$. The second assertion follows also from the defining equation of an algebra morphism.
\EndProof


We will identify the algebras for the power set monad now, which are closely connected to semi-lattices. Recall that an
ordered set $(X, \leq)$ is a $\sup$-\emph{\index{semi lattice}semi lattice} iff each subset has
its supremum in $X$. 

\BeginExample{manes-monad}
The algebras for the monad $(\PowerSenza, \unit,
\multi) $ in the category $\catSET$ of sets with maps (sometimes
called the \emph{\index{monad!Manes}Manes monad}) may be identified with the
complete $\sup$-semi lattices. We will show this now. 

Assume first that $\leq$ is a partial order on a
set $X$ that is $\sup$-complete, so that $\sup A$ exists for each $A
\subseteq X.$ Define
$
h(A) := \sup A,
$
then we have for each ${\cal A} \in \PowerSet{\PowerSet{X}}$ from the
familiar properties of the supremum
\begin{equation*}
\sup (\bigcup {\cal A}) = \sup\ \{\sup a \mid  a \in A\}.
\end{equation*}
This translates into
$
\bigl(h \circ \multi_X\bigr)({\cal A}) = \bigl(h \circ
  (\PowerSenza h)\bigr)({\cal A}).
$
Because
$
x = \sup\{x\}
$
holds for each $x \in X$, we see that $\langle X, h\rangle$ defines an
algebra.

Assume on the other hand that $\langle X, h\rangle$ is an
algebra, and put 
\begin{equation*}
x \leq x'
\Leftrightarrow
h(\{x, x'\}) = x'
\end{equation*}
for $x, x' \in X$.
This defines a partial order: reflexivity and antisymmetry are
obvious. Transitivity is seen as follows: assume $x \leq x'$ and
$x'\leq x''$, then
\begin{multline*}
  h(\{x, x''\})
 =
  h(\{h(\{x\}), h(\{x', x''\}))
 = 
  \bigl(h \circ (\PowerSenza h)\bigr)(\{\{x\}, \{x', x''\}\})\\
  = 
  \bigl(h \circ \multi_X\bigr)(\{\{x\}, \{x', x''\}\})
  = 
  h(\{x, x', x''\})
 = 
  \bigl(h \circ \multi_X\bigr)(\{\{x, x'\}, \{x', x''\}\})\\
  = 
  \bigl(h \circ (\PowerSenza h)\bigr)(\{\{x, x'\}, \{x', x''\}\})
  = 
  h(\{x', x''\})
  = 
  x''.
\end{multline*}
It is clear from
$
\{x\} \cup \emptyset = \{x\}
$
for every $x \in X$ that $h(\emptyset)$ is the smallest
element. Finally, it has to be shown that $h(A)$ is the smallest upper
bound for $A \subseteq X$ in the order $\leq$. We may assume that $A
\not= \emptyset$. Suppose that $x \leq t$ holds for all $x \in A$,
then
\begin{multline*}
  h(A \cup \{t\})
 =
  h\bigl(\bigcup_{x \in A} \{x, t\}\bigr)
 =
  \bigl(h \circ \multi_X\bigr)(\{\{x, t\}\mid  x \in A\})\\
  =
  \bigl(h \circ (\PowerSenza h)\bigr)(\{\{x, t\}\mid  x \in A\})
 =
  h\bigl(\{h(\{x, t\})\ x \in A\}\bigr)
 =
  h(\{t\}) 
 =
  t.
\end{multline*}
Thus, if $x \leq t$ for all $x \in A$, hence $h(A) \leq t$, thus $h(A)$
is an upper bound to $A$, and similarly, $h(A)$ is the smallest upper
bound.
\EndExample

We have shown that each adjunction defines a monad, and now turn to
the converse. In fact, we will show that each monad defines an
adjunction the monad of which is the given monad. Fix the monad $(\funT, \unit, \multi)$ over category $\catK$, and define as above 
$
\Category{Alg}:= \Category{Alg}_{(\funT, \unit, \multi)}
$
as the category of Eilenberg-Moore algebras. We want to define an adjunction, so by Proposition~\ref{conv-char-adjunction-general} it will be most convenient approach to solve the problem by defining unit and counit, after the corresponding functors have been identified. 

{
\def\catAlg{\Category{Alg}}

\BeginLemma{functor-F-for-adjunction}
Define $\funF a := \langle \funT a, \mu_{a}\rangle$ for the object $a\in\objK$, and if $f: a\to b$ is a morphism if $\catK$, define $\funF f := \funT f$. Then $\funF: \catK\to \catAlg$ is a functor.  
\EndLemma

\BeginProof
We have to show that $\funF f: \langle \funT a, \mu_{a}\rangle\to
\langle \funT b, \mu_{b}\rangle$ is an algebra morphism. Since $\mu:
\funT^{2}\to \funT$ is natural, we obtain this commutative diagram
\begin{equation*}
\xymatrix{
\funT a\ar[rr]^{\mu_{a}}\ar[d]_{\funT f} && \funT^{2} a\ar[d]^{\funT^{2}f}\\
\funT b\ar[rr]_{\mu_{b}} && \funT^{2} b
}
\end{equation*}
But this is just the defining definition for an algebra morphism. 
\EndProof

This statement is trivial:
\BeginLemma{functor-G-for-adjunction}
Given an Eilenberg-Moore algebra $\langle x, h\rangle\in\obj{\catAlg}$, define $\funG (x, h) := x$; if $f: \langle x, h\rangle \to \langle x', h'\rangle$ is a morphism in $\catAlg$, put $\funG f := f$. Then $\funG: \catAlg\to \catK$ is a functor. Moreover we have $\funG\circ \funF = \funT$. 
\QED
\EndLemma

We require two natural transformations, which are defined now, and
which are intended to serve as the unit and as the counit,
respectively, for the adjunction. We define for the unit $\eta$ the
originally given $\eta$, so that $\eta: Id_{\catK}\to \funG\circ
\funF$ is a natural transformation. The counit $\varepsilon$ is
defined through $\varepsilon_{\langle x, h\rangle} := h$, so that
$\varepsilon_{\langle x, h\rangle}: (\funF\circ \funG)(x, h)\to
Id_{\catAlg}(x, h)$. This defines a natural transformation
$\varepsilon: \funF\circ \funG\to Id_{\catAlg}$. In fact, let $f:
\langle x, h\rangle\to \langle x', h'\rangle$ be a morphism in
$\catAlg$, then ---~by
expanding definitions~--- the diagram on the left hand side translates to the one on the right hand side, which
commutes:
\begin{equation*}
\xymatrix{
(\funF\circ \funG)(x, h)\ar[d]_{(\funF\circ \funG)f}\ar[rr]^{\varepsilon_{\langle x, h\rangle}}&&\langle x, h\rangle\ar[d]^{f} 
&&& 
\langle \funT x, \mu_{x}\rangle\ar[d]_{\funT f}\ar[rr]^{h} && \langle x, h\rangle\ar[d]^{f}\\
(\funF\circ \funG)(x', h') \ar[rr]_{\varepsilon_{\langle x', h'\rangle}}&&\langle x', h'\rangle 
&&& 
\langle \funT x', \mu_{x'}\rangle\ar[rr]_{h'} && \langle x', h'\rangle
}
\end{equation*}

Now take an object $a\in \objK$, then 
\begin{equation*}
  (\varepsilon\funF\circ \funF\eta)(a) = \varepsilon_{\funF a}(\funF \eta_{a}) = \varepsilon_{\langle \funT a, \mu_{a}\rangle}(\funT \eta_{a}) = \mu_{a}(\funT \eta_{a}) = id_{\funF a}.
\end{equation*}
On the other hand, we have for the algebra $\langle x, h\rangle$
\begin{equation*}
  (\funG\varepsilon\circ \eta\funG)(x, h) = \funG\varepsilon_{\langle x, h\rangle}(\eta_{\funG \langle x, h\rangle})
= \funG \varepsilon_{\langle x, h\rangle}(\eta_{x}) = \varepsilon_{\langle x, h\rangle}(\eta_{x}) = h\eta_{x} \stackrel{(*)}{=} id_{x}= id_{\funG \langle x, h\rangle}
\end{equation*}
where $(*)$ uses that $h: \funT x\to x$ is the structure morphism of an algebra. Taken together, we see that $\eta$ and $\varepsilon$ satisfy the requirements of unit and counit for an adjunction according to Proposition~\ref{conv-char-adjunction-general}.

Hence we have nearly established

\BeginProposition{monad-to-adjunction}
Every monad defines an adjunction. The monad defined by the adjunction is the original one.
\EndProposition

\BeginProof
We have only to prove the last assertion. But this is trivial, because $(\funG\varepsilon\funF) a = (\funG \varepsilon)\langle \funT a, \mu_{a}\rangle = \funG \mu_{a} = \mu_{a}$. 
\EndProof

}

\paragraph{Algebras for discrete probabilities.}
\label{sec:alger-discr-prob}

We identify now the algebras for the functor $\funD$ which
assigns to each set its discrete subprobabilities with finite support,
see Example~\ref{functor-discrete-probs}. Some preliminary and
motivating observations are made first.

Put 
\begin{equation*}
  \Omega := \{\langle \alpha_1, \dots, \alpha_k \rangle \mid  k \in
\Nat, \alpha_i \geq 0, \sum_{i=1}^k \alpha_i \leq 1\}
\end{equation*}
as the set of all positive convex coefficients, and call a subset $V$
of a real vector space\emph{ positive convex} iff $
\sum_{i=1}^k \alpha_i\cdot x_i \in X. $ for $x_1, \dots, x_k \in
X$, $\langle \alpha_1, \dots, \alpha_k\rangle \in \Omega$. Positive convexity appears to
be related to subprobabilities: if $\sum_{i=1}^k \alpha_i\cdot x_i$ is
perceived as an observation in which item $x_{i}$ is assigned
probability $\alpha_{i}$, then clearly $\sum_{i=1}^{k}\alpha_{i}\leq
1$ under the assumption that the observation is incomplete, i.e.,
that not every possible case has been realized.

Suppose a set $X$ over which we formulate subprobabilities is
embedded as a positive convex set into a linear space $V$ over the
reals as a positive convex structure. In this case we could read off a
positive convex combination for an element the probabilities with
which the respective components occurs. 

These observations meet the intuition about positive convexity, but
it has the drawback that we have to look for a linear space $V$
into which $X$ to embed. It has the additional shortcoming that once
we did identify $V$, the positive convex structure on $X$ is fixed
through the vector space, but we will see soon that we need some
flexibility. Consequently, we propose an abstract description of
positive convexity, much in the spirit of Pumplün's
approach~\cite{Pumpluen}. Thus the essential properties (for us,
that is) of positive convexity are described intrinsically for $X$
without having to resort to a vector space. This leads to the
definition of a positive convex structure.

\BeginDefinition{PosConvex} A \emph{positive convex
structure}\index{positive convexity}
 $\wp$ on a set $X$ has for each $ \alpha = \langle \alpha_1, \dots,
\alpha_n\rangle \in \Omega $ a map $
\alpha_\wp: X^n \rightarrow X $ which we write as
\begin{equation*}
\alpha_\wp(x_1, \dots, x_n)
=
\sum_{1\leq i\leq n}^\wp \alpha_i \cdot x_i,
\end{equation*}
such that
\begin{dingautolist}{72}
 \item \label{pc-Eins}
  $
  \sum_{1\leq i \leq n}^\wp \delta_{i, k}\cdot x_i = x_k,
  $
  where $\delta_{i, j}$ is Kronecker's $\delta$ (thus $\delta_{i, j} =
  1$ if $i = j$, and $\delta_{i, j} = 0,$ otherwise),
 \item \label{pc-Zwei} the identity
  $$
  \sum_{1\leq i \leq n}^\wp \alpha_i\cdot\left(\sum_{1\leq k
      \leq m}^\wp\beta_{i, k}\cdot x_k\right)
  =
  \sum_{1 \leq k \leq m}^\wp
  \left(\sum_{1 \leq i \leq n}^\wp \alpha_i\beta_{i,
      k}\right)\cdot x_k
  $$
  holds whenever
  $
  \langle \alpha_1, \dots, \alpha_n\rangle,
  \langle \beta_{i, 1}, \dots, \beta_{i, m}\rangle \in \Omega, 1 \leq
  i \leq n.
  $
\end{dingautolist}
\EndDefinition

Property~\ref{pc-Eins} looks quite trivial, when written down this
way. Rephrasing, it states that the map
$$
\langle \delta_{1, k}, \dots, \delta_{n, k}\rangle_\wp: T^n \rightarrow T,
$$
which is assigned to the $n$-tuple
$
\langle \delta_{1, k}, \dots, \delta_{n, k}\rangle
$
through $\wp$ acts as the projection  to the $k^{th}$ component for
$1 \leq k \leq n$. Similarly, property~\ref{pc-Zwei} may be re-coded in a
formal but less concise way.
Thus we will use freely the notation from vector
spaces, omitting in particular the explicit reference to the structure
whenever possible. Hence simple addition
$
\alpha_1\cdot x_1 + \alpha_2\cdot x_2
$
will be written rather than
$
\sum_{1 \leq i \leq 2}^{\wp} \alpha_i\cdot x_i,
$
with the understanding that it refers to a given positive convex
structure $\wp$ on $X$.

It is an easy exercise to establish that for a positive convex structure
the usual rules for manipulating sums in vector
spaces apply, e.g.,
$
1\cdot x = x,
\sum_{i=1}^n \alpha_i\cdot x_i
=
\sum_{i=1,\alpha_i\not= 0}^n \alpha_i\cdot x_i,
$
or the law of associativity,
$
(\alpha_1\cdot x_1 + \alpha_2\cdot x_2) + \alpha_3\cdot x_3
=
\alpha_1\cdot x_1 + (\alpha_2\cdot x_2 + \alpha_3\cdot x_3).
$
Nevertheless, care should be observed, for of course not all rules
apply: we cannot in general conclude $x = x'$ from
$
\alpha\cdot x = \alpha\cdot x',
$
even if $\alpha \not= 0.$

A morphism\index{positive convexity!morphism}
 $ \theta: \langle X_1,
\wp_1\rangle \rightarrow \langle X_2, \wp_2\rangle
$ between positive convex structures is a 
map $ \theta: X_1 \rightarrow X_2 $ such that
\begin{equation*}
\theta\bigl(\sum_{1\leq i\leq n}^{\wp_1} \alpha_i\cdot x_i\bigr)
=
\sum_{1\leq i \leq n}^{\wp_2} \alpha_i\cdot \theta(x_i)
\end{equation*}
holds for $x_1,\dots, x_n \in X$ and  $\langle \alpha_1, \dots,
\alpha_n\rangle\in\Omega$. In analogy to linear algebra, $\theta$
will be called an \emph{affine} map\index{positive convexity!affine
map}. Positive convex structures with their morphisms form a
category $\Category{StrConv}$.

We need some technical preparations, which are collected in the
following  
\BeginLemma{RechenRegeln}
Let $X$ and $Y$ be sets. 
\begin{enumerate}
 \item Given a map $f: X \rightarrow Y$, let
$
  p = \alpha_1 \cdot \delta_{a_1} + \ldots + \alpha_n \cdot \delta_{a_n}
$
  be the linear combination of Dirac measures for $x_1, \ldots, x_n \in X$
  with positive convex $\langle\alpha_1, \ldots, \alpha_n\rangle \in \Omega.$ Then
$
  \funD(f)(p) = \alpha_1 \cdot \delta_{f(x_1)}
  + \ldots + \alpha_n\cdot \delta_{f(x_n)}.
$
 \item Let $p_1, \ldots, p_n$ be discrete subprobabilities $X$, and let
$
  M = \alpha_1 \cdot \delta_{p_1} + \ldots + \alpha_n \cdot \delta_{p_n}
$
  be the linear combination of the corresponding Dirac measures in
  $
  (\funD\circ \funD)X
  $
  with positive convex coefficients $\langle \alpha_1, \ldots,
  \alpha_n\rangle \in \Omega.$
  Then
$
  \multi_X(M) = \alpha_1 \cdot p_1 + \ldots + \alpha_n \cdot p_n.
$
\end{enumerate}
\EndLemma

\BeginProof
The first part follows directly from the observation
$
\funD(f)(\delta_{x})(B) = \delta_x(\InvBild{f}{B}) = \delta_{f(x)}(B),
$
and the second one is easily inferred from the formula for $\mu$ in Example~\ref{discr-prob-monad}.
\EndProof

The algebras are described now without having to resort to
$\SubProb{X}$ through an intrinsic characterization
using positive convex structures with affine maps. This
characterization is comparable to the one given by Manes for the power
set monad (which also does not resort explicitly to the underlying
monad or its functor); see Example~\ref{manes-monad}.

\BeginLemma{IsPosConvex}
Given an algebra $\langle X, h\rangle$, define for $x_1,\dots, x_n \in X$
and the positive convex coefficients $\langle \alpha_1,\dots, \alpha_n\rangle
\in \Omega$
\begin{equation*}
\langle\alpha_{1}, \dots, \alpha_{n}\rangle_{\wp}
: =
\sum_{i=1}^n \alpha_i\cdot x_i
:=
h(\sum_{i=1}^n \alpha_i\cdot \delta_{x_i}).
\end{equation*}
This defines a positive convex structure $\wp$ on $X.$
\EndLemma

\BeginProof
1.
Because
$$
h\bigl(\sum_{i=1}^n\delta_{i, j}\cdot \delta_{x_i}\bigr)
=
h(\delta_{x_j})
=
x_j,
$$
property~\ref{pc-Eins} in Definition~\ref{PosConvex} is satisfied.

2. Proving property~\ref{pc-Zwei}, we resort to the properties of an
algebra and a monad:
\begin{eqnarray}
  \sum_{i=1}^n\alpha_i\cdot \bigl(\sum_{k=1}^m \beta_{i, k}\cdot x_k\bigr)
  & = &
  h\bigl(\sum_{i=1}^n\alpha_i\cdot
    \delta_{\sum_{k=1}^m\beta_{i, k}\cdot x_k}\bigr)\label{pos-conv-t-1}\\
  & = &
  h\bigl(\sum_{i=1}^n\alpha_i\cdot
    \delta_{h\left(\sum_{k=1}^m\beta_{i, k}\cdot
  \delta_{x_k}\right)}\bigr)\label{pos-conv-t-2}\\
  & = &
  h\bigl(\sum_{i=1}^n\alpha_i\cdot
    \SubProb{h}\bigl(\delta_{\sum_{k=1}^m\beta_{i, k}\cdot
  \delta_{x_k}}\bigr)\bigr)\label{pos-conv-t-3}\\
  & = &
  \bigl(h\circ \SubProb{h}\bigr)\bigl(\sum_{i=1}^n\alpha_i\cdot
    \delta_{\sum_{k=1}^m\beta_{i, k}\cdot
  \delta_{x_k}}\bigr)\label{pos-conv-t-4}\\
  & = &
  \bigl(h\circ \multi_X\bigr)\bigl(\sum_{i=1}^n\alpha_i\cdot
    \delta_{\sum_{k=1}^m\beta_{i, k}\cdot
  \delta_{x_k}}\bigr)\label{pos-conv-t-5}\\
  & = &
  h\bigl(\sum_{i=1}^n\alpha_i\cdot
    \multi_X\bigl(\delta_{\sum_{k=1}^m\beta_{i, k}\cdot
  \delta_{x_k}}\bigr)\bigr)\label{pos-conv-t-6}\\
  & = &
  h\bigl(\sum_{i=1}^n\alpha_i\cdot
    \bigl(\sum_{k=1}^m \beta_{i, k}\cdot \delta_{x_k}\bigr)\bigr)\label{pos-conv-t-7}\\
  & = &
   h\bigl(\sum_{k=1}^m\bigl(\sum_{i=1}^n\alpha_i\cdot
       \beta_{i,k}\bigr) \delta_{x_k}\bigr)\label{pos-conv-t-8}\\
   & = &
   \sum_{k=1}^m\bigl(\sum_{i=1}^n\alpha_i\cdot
       \beta_{i,k}\bigr)x_k. \label{pos-conv-t-9}
\end{eqnarray}

The equations~(\ref{pos-conv-t-1}) and~(\ref{pos-conv-t-2}) reflect
the definition of the structure, equation~(\ref{pos-conv-t-3})
applies $ \delta_{h(\tau)} = \SubProb{h}(\delta_\tau), $
equation~(\ref{pos-conv-t-4}) uses the linearity of $\SubProb{h}$
according to Lemma~\ref{RechenRegeln}, equation~(\ref{pos-conv-t-5})
is due to $h$ being an algebra. Winding down,
equation~(\ref{pos-conv-t-6}) uses Lemma~\ref{RechenRegeln} again,
this time for $\multi_X$, equation~(\ref{pos-conv-t-7}) uses that $
\multi_X \circ \delta_\tau = \tau, $ equation~(\ref{pos-conv-t-8})
is just rearranging terms, and equation~(\ref{pos-conv-t-9}) is the
definition again. 
\EndProof

The converse holds as well. Assume that we have a positive convex
structure $\wp$ on $X$. Put 
\begin{equation*}
  h\bigl(\sum_{i=1}^{n}\alpha_{i}\cdot\delta_{x_{i}}\bigr) := 
\sum_{1\leq i \leq n}^{\wp} \alpha_{i}\cdot x_{i}
\end{equation*}
for $\langle \alpha_{1}, \dots, \alpha_{n}\rangle \in\Omega$ and
$x_{1}, \dots, x_{n}\in X$.
One first checks that $h$ is well-defined: 
This is so since
\begin{equation*}
\sum_{i = 1}^n \alpha_i\cdot \delta_{x_i}
=
\sum_{j = 1}^m \alpha'_j\cdot \delta_{x'_j}
\end{equation*}
implies that
\begin{equation*}
\sum_{i = 1, \alpha_i \not= 0}^n \alpha_i\cdot \delta_{x_i}
=
\sum_{j = 1, \alpha'_j \not= 0}^m \alpha'_j\cdot \delta_{x'_j},
\end{equation*}
hence given $i$ with $\alpha_i \not= 0$ there exists $j$ with
$\alpha'_j \not= 0$ such that $x_i = x'_j$ with $\alpha_{i} =
\alpha'_{j}$ and vice versa. Consequently,
\begin{equation*}
\sum_{1\leq i \leq n}^{\wp} \alpha_i\cdot x_i = \sum_{1\leq i \leq n, \alpha_i \not=0}^\wp
\alpha_i\cdot x_i = \sum_{1\leq j\leq n, \alpha'_j \not= 0}^\wp \alpha'_j\cdot
x'_j = \sum_{1\leq j \leq n}^\wp \alpha'_j\cdot x'_j
\end{equation*}
is inferred from the properties of positive convex structures. Thus
$h: \funD X\to X$. 

An easy induction using property~\ref{pc-Zwei} shows that $h$ is an affine map, i.e., that we have 
\begin{equation}
  \label{eq:easy-induction}
  h\bigl(\sum_{i=1}^{n}\alpha_{i}\cdot\tau_{i}\bigr) = \sum_{1\leq
    i\leq n}^{\wp}\alpha_{i}\cdot h(\tau_{i})
\end{equation}
for $\langle \alpha_{1}, \dots,
\alpha_{n}\rangle \in \Omega$ and $\tau_{1}, \dots, \tau_{n}\in\funD X$. 

Now let $f = \sum_{i=1}^{n}\alpha_{i}\cdot\delta_{\tau_{i}}\in \funD^{2}X$ with $\tau_{1}, \dots, \tau_{n}\in\funD X$. Then we obtain
from Lemma~\ref{RechenRegeln} that $\multi_{X} f =
\sum_{i=1}^{n}\alpha_{i}\cdot \tau_{i}.$ Consequently,
we obtain from~\ref{eq:easy-induction} that 
$
h(\multi_{X} f) = \sum_{1\leq i \leq n}^{\wp} \alpha_{i}\cdot h(\tau_{i}).
$
On the other hand, Lemma~\ref{RechenRegeln} implies together with~\ref{eq:easy-induction} 
\begin{align*}
  (h\circ \funD h) f & = h\bigl(\sum_{1\leq i\leq n}^{\wp}\alpha_{i}\cdot
  (\funD h)(\tau_{i})\bigr)\\
& = \sum_{1\leq i\leq n}^{\wp}\alpha_{i}\cdot
  h\bigl((\funD h)(\tau_{i})\bigr)\\
& = \sum_{1\leq i\leq n}^{\wp}\alpha_{i}\cdot
  h(\delta_{h(\tau_{i})})\\
& =\sum_{1\leq i\leq n}^{\wp}\alpha_{i}\cdot h(\tau_{i}),
\end{align*}
because $h(\delta_{h(\tau_{i})}) = h(\tau_{i})$. We infer
from~\ref{pc-Eins} that $h\circ \multi_{X} = id_{X}$. Hence we have
established

\BeginProposition{pos-convex-algebra}
Each positive convex structure on $X$ induces an algebra for $\funD
X$. \QED
\EndProposition

Thus we obtain a complete characterization of the Eilenberg-Moore
algebras for this monad.

\BeginTheorem{complete-char}
The Eilenberg-Moore algebras for the discrete probability monad are
exactly the positive convex structures.
\QED 
\EndTheorem

This characterization carries over to the probabilistic version of the
monad; we leave the simple formulation to the reader. A similar
characterization is possible for the continuous version of this
functor, at least in Polish spaces. This requires a continuity
condition, however.

\Subsection{Coalgebras}
\label{sec:coalgs}

A coalgebra for a functor $\funF$ is characterized by a carrier object
$c$ and by a morphism $c\to \funF c$. This fairly general structure
can be found in many applications, as we will see. So we will first
define formally what a coalgebra is, and then provide a gallery of
examples, some of them already discussed in another disguise, some of
them new. The common thread is their formulation as a
coalgebra. Bisimilar coalgebras will be discussed, indicating some
interesting facets of the possibilities to describe behavioral
equivalence of some sorts. 

\BeginDefinition{coalgebra}
Given the endofunctor $\funF$ on category $\catK$, an object $a$ on
$\catK$ together with a morphism $f: a\to \funF a$ is called a
\emph{\index{coalgebra}coalgebra} for $\catK$. Morphism $f$ is
sometimes called the \emph{\index{coalgebra!dynamics}dynamics} of the
coalgebra, $a$ its \emph{\index{coalgebra!carrier}carrier}. 
\EndDefinition

Comparing the definitions of an algebra and a coalgebra, we see that
for a coalgebra the functor $\funF$ is a arbitraty endofunctor on
$\catK$, which an algebra requires a monad and compatility with unit
and multiplication. This coalgebras are conceptually simpler by
demanding less resources.

We are going to enter now the gallery of examples and start with
coalgebras for the power set functor. This example will be with us for
quite some time, in particular when we will interpret modal logics. A
refinement of this example will be provided by labelled transition
systems.

\BeginExample{coalg-relations}
We consider the category $\catSET$ of sets with maps as morphisms and
the power set functor $\PowerSenza$. An $\PowerSenza$ coalgebra
consists of a set $A$ and a map $f: A\to \PowerSenza(A)$. Hence we
have $f(a)\subseteq A$ for all $a\in A$, so that a $\catSET$-coalgebra can be
represented as a relation $\{\langle a, b\rangle \mid b\in f(a), a \in
A\}$  over $A$. If, conversely, $R\subseteq A\times A$ is a relation,
then $f(a) := \{b \in A\mid \langle a, b\rangle\in R\}$ is a map $f:
A\to \PowerSenza(A)$. 
\EndExample

A slight extension is to be observed when we introduce actions,
formally, labels for our transitions. Here a transition is dependent
on an action which serves as a label to the corresponding relation.

\BeginExample{ex-transition-syst-act-coalg}
Let us interpret a labeled transition system $\bigl(S,
(\Trans_{a})_{a\in A}\bigr)$ over state space $S$ with set $A$ of
actions, see Example~\ref{ex-transition-syst-act}. Then
$\Trans_{a}\subseteq S\times S$ for all actions $a\in A$. 

Working again in $\catSET$, we define for the set $S$ and for the map $f: S\to T$
\begin{align*}
  \funT S & := \PowerSenza(A\times S),\\
(\funT f)(B) & := \{\langle a, f(x)\rangle  \mid \langle a, x\rangle \in B\}
\end{align*}
(hence $\funT = \PowerSenza(A\times -)$). Define $f(s) := \{\langle a, s'\rangle  \mid s\Trans_{a} s'\}$, thus
$f: S\to \funT S$ is a morphism in $\catSET$. Consequently, a transition system
is interpreted as a coalgebra for the functor $\PowerSenza(A\times
-)$. 
\EndExample

\BeginExample{automata-with-output-coalg}
Let $A$ be the inputs, $B$ the outputs and $X$ the states of an
automaton with output, see Example~\ref{automata-with-output}. Put
$\funF := (-\times B)^{A}$. For $f: X\to Y$ we have this commutative
diagram. 
\begin{equation*}
\xymatrix{
&A\ar[dl]_{t}\ar[dr]^{(\funF f)t}\\
X\times B\ar[rr]_{f\times id_{B}}&& Y\times B
}
\end{equation*}
Let $(S, f)$ be an $\funF$-coalgebra, thus $f: S \to \funF S =
(S\times B)^{A}$. Input $a\in A$ in state $s \in S$ yields $f(s)(a) =
\langle s', b\rangle$, so that $s'$ is the new state, and $b$ is the
output. Hence automata with output are perceived as coalgebras, in
this case for the functor $(-\times B)^{A}$. 
\EndExample

While the automata in Example~\ref{automata-with-output-coalg} are
deterministic (and completely specified), we can also use a similar
approach to modelling nondeterministic automata.

\BeginExample{nond-automata-w-output-coalg}
Let $A, B, X$ be as in Example~\ref{automata-with-output-coalg}, but
take this time $\funF := \PowerSenza(-\times B)^{A}$ as a functor, so
that this diagram commutes for $f: X\to Y$:
\begin{equation*}
\xymatrix{
&A\ar[dl]_{t}\ar[dr]^{(\funF f)t}\\
\PowerSenza(X\times B)\ar[rr]_{\PowerSenza(f\times id_{B})}&& \PowerSenza(Y\times B)
}
\end{equation*} 
Thus $\PowerSenza(f\times B)(D) = \{\langle f(x), b\rangle \in Y\times
B \mid \langle x, b\rangle \in B\}$. Then $(S, g)$ is an $\funF$
coalgebra iff input $a\in A$ in state $s\in S$ gives $g(s)(a) \in
S\times B$ as the set of possible new states and outputs. 

As a variant, we can replace $\PowerSenza(-\times B)$ by
$\PowerSenza_{f}(-\times B)$, so that the automaton presents only a finite
number of alternatives. 
\EndExample

Binary trees may be modelled through coalgebras as well:

\BeginExample{ex-bin-tree-coalg}
Put $\funF X :=  \{\ast\} + X\times X$, where $\ast$ is a new
symbol. If $f: X\to Y$, put 
\begin{equation*}
  \funF(f)(t) := 
  \begin{cases}
    \ast,& \text{ if } t = \ast\\
\langle x_{1}, x_{2}\rangle,& \text{ if } t  = \langle x_{1},
x_{2}\rangle. 
  \end{cases}
\end{equation*}
Then $\funF$ is an endofunctor on $\catSET$. Let $(S, f)$ be an
$\funF$-coalgebra, then $f(s)\in\{\ast\}+S\times S$. This is
interpreted that $s$ is a leaf iff $f(s) = \ast$, and an inner node
with offsprings $\langle s_{1}, s_{2}\rangle$, if $f(s) = \langle
s_{1}, s_{2}\rangle$. Thus such a coalgebra represents a binary tree
(which may be of infinite depth).  
\EndExample

The following example shows that probabilistic transitions may be
modelled as coalgebras as well.
\BeginExample{ex-stoch-rel-coalg}
{
\def\catM{\Category{Meas}}
\def\funM{\SubProbSenza}
Working in the category $\catM$ of measurable spaces with measurable
maps, we have introduced in Example~\ref{prob-comput} the
subprobability functor $\funM$ as an endofunctor on $\catM$. Let $(X,
K)$ be a coalgebra for $\funM$ (we omit here the $\sigma$-algebra from the notation), then $K: X\to \funM X$ is measurable,
so that 
\begin{enumerate}
\item $K(x)$ is a subprobability on (the measurable sets of) $X$,
\item for each measurable set $D \subseteq X$, the map $x\mapsto
  K(x)(D)$ is measurable,
\end{enumerate}
see Example~\ref{ex-cat-stoch-rel} and
Exercise~\ref{ex-meas-generator}. Thus $K$ is a subprobabilistic
transition kernel on $X$.  
}\EndExample

Let us have a look at the upper closed sets introduced in
Example~\ref{upper-closed-functor}. Coalgebras for this functor will be
used for an interpretation of games, see
Example~\ref{ex-spec-rel-6}.

\BeginExample{upper-closed-coalg}
{\def\funV{\FunctorSenza{V}}
\def\funP{\PowerSenza}
Let $\funV S := \{V\subseteq \funP S \mid V\text{ is upper
  closed}\}$. This functor has been studied in
Example~\ref{upper-closed-functor}. A coalgebra $(S, f)$ for $\funV$
is a map $f: S\to \funV S$, so that $f(s)\subseteq\funP(S)$ is upper
closed, hence $A\in f(s)$ and $B\supseteq A$ imply $b\in f(s)$ for
each $s\in S$. We interpret $f(s)$ as the set of states a player may
reach in state $s$, so that if the player can reach $A$ and
$A\subseteq B$, then the player certainly can reach $B$. 

$\funV$ is the basis for neighborhood models in modal logics, see,
e.g., \cite{Chellas, Venema-Handbook} and page~\pageref{def-model-log-nghb-model}.
}\EndExample

It is natural to ask for morphisms of coalgebras, so that coalgebras
can be related to each other. This is a fairly straightforward
definition.

\BeginDefinition{coalg-def-morph}
Let $\funF$ be an endofunctor on category $\catK$, then $t: (a, f)\to
(b, g)$ is a \emph{coalgebra \index{coalgebra!morphism}morphism} for the $\funF$-coalgebras $(a, f)$ and $(b, g)$
iff $t: a \to b$ is a morphism in $\catK$ such that $g\circ t = \funF(t)\circ f$. 
\EndDefinition

Thus  $t: (a, f)\to (b, g)$ is a coalgebra morphism iff $t: a\to b$ is
a morphism so that this diagram commutes:
\begin{equation*}
\xymatrix{
a\ar[d]_{f}\ar[rr]^{t} && b\ar[d]^{g}\\
\funF a\ar[rr]_{\funF t}&&\funF b
}
\end{equation*}

It is clear that $\funF$-coalgebras form a category with coalgebra
morphisms as morphisms. We reconsider some previously discussed
examples and shed some light on the morphisms for these coalgebras.

\BeginExample{ex-bin-tree-coalg-morph}
Continuing Example~\ref{ex-bin-tree-coalg} on binary trees, let $r: (S, f)\to (T, g)$
be a morphism for the $\funF$-coalgebras $(S, f)$ and $(T, g)$. Thus
$g\circ r = \funF(r)\circ f$. This entails
\begin{enumerate}
\item $f(s) = \ast$, then $g(r(s)) = (\funF r)(f(s)) = \ast$ (thus $s$
  is a leaf iff $r(s)$ is one),
\item $f(s) = \langle s_{1}, s_{2}\rangle$, then $g(r(s)) = \langle
  t_{1}, t_{2}\rangle$ with $t_{1} = r(s_{1})$ and $t_{2} =
  r(s_{2})$ (thus $r(s)$ branches out to $\langle r(s_{1}),
  r(s_{2})\rangle$, provided $s$ branches out to $\langle s_{1}, s_{2}\rangle$).
\end{enumerate}
Thus a coalgebra morphism preserves the tree structure.
\EndExample

\BeginExample{det-automata-coalg-morph}
Continuing the discussion of deterministic automata with output from
Example~\ref{automata-with-output-coalg}, let $(S, f)$ and $(T, g)$ be
$\funF$-coalgebras and $r: (S, F)\to (T, g)$ be a morphism. Given
state $s\in S$, let $f(s)(a) = \langle s', b\rangle$ be the new state
and the output, respectively, after input $a\in A$  for automaton $(S, f)$. Then $g(r(s))(a) =
\langle r(s'), b\rangle$, so after input $a\in A$ the automaton $(T,
g)$ will be in state $r(s)$ and give the output $b$, as
expected. Hence coalgebra morphisms preserve the automatas' working. 
\EndExample

\BeginExample{ex-transition-syst-act-coalg-morph}
{\def\funF{\funT}
Continuing the the discussion of transition systems from
Example~\ref{ex-transition-syst-act-coalg}, let $(S, f)$ and $(T, g)$ be
labelled transition systems with $A$ as the set of actions. Thus a
transition from $s$ to $s'$ on action $a$ is given in $(S, f)$ iff
$\langle a, s'\rangle \in f(s)$. Let us just for convenience write
$s\Trans_{a, S} s'$ iff this is the case, similarly, we write
$t\Trans_{a, T} t'$ iff $t, t'\in T$ with $\langle a, t'\rangle \in
g(t)$.  

Now let $r: (S, f)\to (T, g)$ be a coalgebra morphism. We claim that
for given $s\in S$ we have a transition $r(s)\Trans_{a, T} t_{0}$ for
some $t_{0}$ iff we can find $s_{0}$ such that $s\Trans_{a, S} s_{0}$
and $r(s_{0}) = t_{0}$. Because $r: (S, f)\to (T, g)$ is a coalgebra
morphism, we have $g\circ r = (\funF r)\circ f$ with $\funF =
\PowerSenza (A\times -)$. Thus
\begin{equation*}
  g(r(s)) = \PowerSenza(A\times r)(s) = \{\langle a, r(s')\rangle
  \mid \langle a, s'\rangle\in f(s)\}.
\end{equation*}
Consequently,
\begin{align*}
  r(s)\Trans_{a, T} t_{0} 
&\Leftrightarrow 
\langle a, t_{0}\rangle \in g(r(s))\\
&\Leftrightarrow 
\langle a, t_{0}\rangle = \langle a, r(s_{0})\rangle\text{ for some
  $\langle a, s_{0}\rangle\in f(s)$}\\
&\Leftrightarrow 
s\Trans_{a, S} s_{0}\text{ for some $s_{0}$ with $r(s_{0}) = t_{0}$}
\end{align*}
This means that the transitions in $(T, g)$ are essentially controlled
by the morphism $r$ and the transitions in $(S, f)$. Hence a coalgebra
morphism between transition systems is a bounded morphism\index{morphism!bounded} in the sense
of Example~\ref{ex-cat-transition-syst-bound}. 
}\EndExample

\BeginExample{upper-closed-coalg-morph}
{\def\funV{\FunctorSenza{V}}
\def\funP{\PowerSenza}
We continue the discussion of upper closed sets from
Example~\ref{upper-closed-coalg}. Let $(S, f)$ and $(T, g)$ be
$\funV$-coalgebras, so this diagram is commutative for morphism $r:
(S, F)\to (T, g)$:
\begin{equation*}
\xymatrix{
S\ar[rr]^{r}\ar[d]_{f}&&T\ar[d]^{g}\\
\funV S\ar[rr]_{\funV r}&&\funV T
}
\end{equation*}
Consequently, $W\in g(r(s))$ iff $\InvBild{r}{W}\in f(s)$. Taking up
the interpretation of sets of states which may be achieved by a player,
we see that it\footnote{The present author is not really sure about
  the players' gender --- players are female in the overwhelming
  majority of papers in the literature, but on the other hand are addressed as
  \emph{Angel} or \emph{Demon}; this may be politically correct, but does not seem
  to be biblically so with a view toward Matthew 22:30. To be on the safe
  side ---~it is so hopeless to argue with feminists~--- players are neutral in the present treatise.} may achieve $W$
in state $r(s)$ in $(T, g)$ iff it
may achieve in $(S, f)$ the set $\InvBild{r}{W}$ in state $s$.  
}\EndExample

\Subsubsection{Bisimulations}
\label{sec:bism-coalg}

The notion of bisimilarity is fundamental for the application of
coalgebras to system modelling. Bisimilar coalgebras behave in a
similar fashion, witnessed by a mediating system.

\BeginDefinition{bisim-coalg}
Let $\funF$ be an endofunctor on a category $\catK$. The
$\funF$-coalgebras $(S, f)$ and $(T, g)$ are said to be \emph{\index{coalgebra!bisimilar}bisimilar} iff
there exists a coalgebra $(M, m)$ and coalgebra morphisms 
$
\xymatrix{
  (S, f)&(M, m)\ar[l]\ar[r]&(T, g)
}
$
The coalgebra $(M, m)$ is called \emph{\index{coalgebra!mediating}mediating}. 
\EndDefinition
Thus we obtain this characteristic diagram with $\ell$ and $r$ as the
corresponding morphisms.
\begin{equation*}
\xymatrix{
S\ar[d]_{f} && M\ar[d]^{m}\ar[ll]_{\ell}\ar[rr]^{r} && T\ar[d]^{g}\\
\funF F && \funF M\ar[ll]^{\funF \ell}\ar[rr]_{\funF r} && \funF T
}
\end{equation*}
Thus we have
\begin{align*}
  f\circ \ell & = (\funF\ell)\circ m\\
g\circ r & = (\funF r)\circ m
\end{align*}
In this way it is easy to see why $(M, m)$ is called mediating. 

Bisimilarity was originally investigated when concurrent systems
became of interest. The original formulation, however, was not
coalgebraic but rather relational.

\BeginDefinition{bisim-relat}
Let $(S, \Trans_{S})$ and $(T, \Trans_{T})$ be transition
systems. Then $B\subseteq S\times T$ is called a
\emph{\index{transition system!bisimulation}\index{bisimulation}bisimulation} iff for
all $\langle s, t\rangle\in B$ these conditions are satisfied:
\begin{enumerate}
\item if $s\Trans_{S}s'$, then there is a $t'\in T$ such that
  $t\Trans_{T}t'$ and $\langle s', t'\rangle\in B$.
\item if $t\Trans_{T}t'$, then there is a $s'\in S$ such that
  $s\Trans_{S}s'$ and $\langle s', t'\rangle\in B$.
\end{enumerate}
\EndDefinition
Hence a bisimulation simulates transitions in one system through the
other one. On first sight, these notions of bisimilarity are not
related to each other. Recall that transition systems are coalgebras
for the power set functor $\PowerSenza$. This is the connection:

\BeginTheorem{ascel-bisim}
Given the  transition systems $(S, \Trans_{S})$ and $(T, \Trans_{T})$
with the associated $\PowerSenza$-coalge\-bras $(S, f)$ and $(T, g)$, then
these statements are equivalent for $B\subseteq S\times T$: 
\begin{enumerate}
\item\label{ascel-bisim-1} $B$ is a bismulation.
\item\label{ascel-bisim-2} There exists a $\PowerSenza$-coalgebra structure $h$ on
  $B$ such that 
$
\xymatrix{
  (S, f)&(B, h)\ar[l]\ar[r]&(T, g)
}
$
with the projections as morphisms is mediating. 
\end{enumerate}
\EndTheorem

\BeginProof
That 
$
\xymatrix{
  (S, f)&(B, h)\ar[l]_{\pi_{S}}\ar[r]^{\pi_{T}}&(T, g)
}
$
is mediating follows from commutativity of this diagram.
\begin{equation*}
\xymatrix{
S\ar[d]_{f} && B\ar[d]_{h}\ar[ll]_{\pi_{S}}\ar[rr]^{\pi_{T}} && T\ar[d]^{g}\\
\PowerSenza(S) && \PowerSenza(B)\ar[ll]^{\PowerSenza(\pi_{S})}\ar[rr]_{\PowerSenza(\pi_{T})}&&\PowerSenza(T)
}
\end{equation*}

\labelImpl{ascel-bisim-1}{ascel-bisim-2}: We have to construct a map
$h: B\to \PowerSenza(B)$ such that
\begin{align*}
  f(\pi_{S}(s, t)) & = \PowerSenza(\pi_{S})(h(s, t))\\
f(\pi_{T}(s, t)) & = \PowerSenza(\pi_{T})(h(s, t))
\end{align*}
for all $\langle s, t\rangle \in B$. The choice is somewhat obvious:
put for $\langle s, t\rangle \in B$
\begin{equation*}
  h(s, t) := \{\langle s', t'\rangle \mid s\Trans_{S}s', t\Trans_{T}t'\}.
\end{equation*}
Thus $h: B\to \PowerSenza(B)$ is a map, hence $(B, h)$ is a
$\PowerSenza$-coalgebra. 

Now fix $\langle s, t\rangle\in B$, then we claim that $f(s) =
\PowerSenza(\pi_{S})(h(s, t)).$

``$\subseteq$'': Let $s'\in f(s)$, hence $s\Trans_{S}s'$, thus there
exists $t'$ with $\langle s', t'\rangle\in B$ such that
$t\Trans_{T}t'$, hence
\begin{align*}
  s' & \in \{\pi_{S}(s_{0}, t_{0}) \mid \langle s_{0}, t_{0}\rangle \in
  h(s, t)\}\\
& = \{s_{0}\mid \langle s_{0}, t_{0}\rangle \in h(s, t)\text{ for some
  $t_{0}$}\}\\
& = \PowerSenza(\pi_{S})(h(s, t)).
\end{align*}
``$\supseteq$'' If $s'\in\PowerSenza(\pi_{S})(h(s, t))$, then in
particular $s\Trans_{S}s'$, thus $s'\in f(s)$. 

Thus we have shown that $\PowerSenza(\pi_{S})(h(s, t)) = f(s) =
f(\pi_{S}(s, t))$. One shows $\PowerSenza(\pi_{T})(h(s, t)) = g(t) =
f(\pi_{T}(s, t))$ in exactly the same way. We have constructed
$h$ such that $(B, h)$ is a $\PowerSenza$-coalgebra, and such that the
diagrams above commute.

\labelImpl{ascel-bisim-2}{ascel-bisim-1}: Assume that $h$ exists
with the properties described in the assertion, then we have to show
that $B$ is a bisimulation. Now let $\langle s, t\rangle \in B$ and
$s\Trans_{S}s'$, hence $s'\in f(s) = f(\pi_{S}(s, t)) =
\PowerSenza(\pi_{S})(h(s, t))$. Thus there exists $t'$ with $\langle
s', t'\rangle \in h(s, t)\subseteq B$, and hence $\langle s',
t'\rangle \in B$. We claim that $t\Trans_{T}t'$, which is tantamount
to saying $t'\in g(t)$. But $g(t) = \PowerSenza(\pi_{T})(h(s, t))$, and
$\langle s', t'\rangle \in h(s, t)$, hence $t'\in
\PowerSenza(\pi_{T})(h(s, t)) = g(t)$. This establishes
$t\Trans_{T}t'$. A similar argument finds $s'$ with $s\Trans_{S}s'$
with $\langle s', t'\rangle\in B$ in case $t\Trans_{T}t'$.

This completes the proof.
\EndProof

Thus for transition systems we may use bisimulations as relations and
bisimulations as coalgebras interchangeably. The connection to
$\PowerSenza$-coalgebra morphisms and bisimulations is further
strengthened by investigating the graph of a morphism (recall that the
graph $Graph(r)$ of a map $r: S\to T$ is the relation $\{\langle s,
r(s)\rangle \mid s\in S\}$). 

\BeginProposition{morph-as-bisim}
Given coalgebras $(S, f)$ and $(T, g)$ for the power set functor
$\PowerSenza$, $r: (S, f)\to (T, g)$ is a morphism iff $Graph(r)$ is a
bisimulation for $(S, f)$ and $(T, g)$. 
\EndProposition

\BeginProof
1.
Assume that $r: (S, f)\to (T, g)$ is a morphism, so that $g\circ r =
\PowerSenza(r)\circ f$. Now define 
\begin{equation*}
  h(s, t):=\{\langle s', r(s')\rangle \mid s'\in f(s)\}\subseteq Graph(r)
\end{equation*}
for $\langle s, t\rangle \in Graph(r)$. Then
$g(\pi_{T}(s, t)) = g(t) = \PowerSenza(\pi_{T})(h(s, t))$ for $t = r(s)$.

``$\subseteq$'' If $t'\in g(t)$ for $t = r(s)$, then
\begin{align*}
  t'\in g(r(s)) & = \PowerSenza(r)(f(s))\\
& = \{r(s') \mid s'\in f(s)\}\\
& = \PowerSenza(\pi_{T})(\{\langle s', r(s')\rangle \mid s'\in
f(s)\})\\
& = \PowerSenza(\pi_{T})(h(s, t))
\end{align*}
``$\supseteq$'' If $\langle s', t'\rangle \in h(s, t)$, then $s'\in
f(s)$ and $t' = r(s')$, but this implies $t'\in \PowerSenza(r)(f(s)) =
g(r(s)).$ 

Thus $g\circ \pi_{T}= \PowerSenza(\pi_{T})\circ h$. The equation 
$f\circ \pi_{S} = \PowerSenza(\pi_{S})\circ h$ is established similarly.

Hence we have found a coalgebra structure $h$ on $Graph(r)$ such that
\begin{equation*}
\xymatrix{
(S, f)&&(Graph(r), h)\ar[ll]_{\pi_{S}}\ar[rr]^{\pi_{T}}&&(T, g)
}
\end{equation*}
are coalgebra morphisms, so that $(Graph(r), h)$ is a bisimulation. 

2.
If, conversely, $(Graph(r), h)$ is a bisimulation with the projections
as morphisms, then we have $r = \pi_{T}\circ \pi_{S}^{-1}$. Then
$\pi_{T}$ is a morphism, and $\pi_{S}^{-1}$ is a morphism as well
(note that we work on the graph of $r$). So $r$ is a morphism. 
\EndProof

Let us have a look at the situation with the upper closed sets from
Example~\ref{upper-closed-kleisli}. There we find a comparable
situation. 

{
\def\funV{\FunctorSenza{V}}
\BeginDefinition{bisim-upperclosed}
Let 
\begin{equation*}
  \funV S := \{V\subseteq\PowerSenza(S) \mid V\text{ is upper closed}\}
\end{equation*}
be the endofunctor on $\catSET$ which assigns to set $S$ all upper
closed subsets of $\PowerSenza S$. Given $\funV$-coalgebras $(S, f)$
and $(T, g)$, a subset $B\subseteq S\times T$ is called a
\emph{\index{bisimulation}bisimulation} of $(S, f)$ and $(T, g)$ iff for each $\langle s, t\rangle \in B$
\begin{enumerate}
  \item For all $X\in f(s)$ there exists $Y\in g(t)$ such that for
    each $t'\in Y$ there exists $s'\in X$ with $\langle s', t'\rangle \in B$.
  \item For all $Y\in g(t)$ there exists $X\in f(s)$ such that for
    each $s'\in X$ there exists $t'\in Y$ with $\langle s', t'\rangle \in B$.
\end{enumerate}
\EndDefinition

We have then a comparable characterization of bisimilar coalgebras.

\BeginProposition{bisim-through-morph}
Let $(S, f)$ and $(T, g)$ be coalgebras for $\funV$. Then the
following statements are equivalent for $B\subseteq S\times T$ with
$\Bild{\pi_{S}}{B} = S$ and $\Bild{\pi_{T}}{B} = T$
\begin{enumerate}
  \item \label{bisim-is-coalg-1} $B$ is a
    bisimulation of $(S, f)$ and $(T, g)$.
  \item \label{bisim-is-coalg-2} There exists a coalgebra structure
    $h$ on $B$ so that the projections $\pi_S: B\to S, \pi_T: B \to T$
  are morphisms  
  $
  \xymatrix{
  (S, f)&(B, h)\ar[l]_{\pi_S}\ar[r]^{\pi_T}&(T, g).
  }
  $
\end{enumerate}
\EndProposition

\BeginProof
\labelImpl{bisim-is-coalg-1}{bisim-is-coalg-2}: 
Define
\begin{equation*}
h(s, t) := \{D \subseteq B \mid \Bild{\pi_S}{D}\in f(s)\text{ and }\Bild{\pi_T}{D}\in f(t)\},
\end{equation*}
$\langle s, t\rangle\in B$. Hence $h(s, t)\subseteq\PowerSet{S}$, and because both $f(s)$ and  $g(t)$ are upper
closed, so is $h(s, t)$. 

Now fix $\langle s, t\rangle\in B$. We show first that 
$
f(s) = \{\Bild{\pi_{S}}{Z}\mid Z\in h(s, t)\}.
$
From the definition of $h(s, t)$ it follows that $\Bild{\pi_{S}}{Z}\in
f(s)$ for each $Z\in h(s, t)$. So we have to establish the other
inclusion. Let $X\in f(s)$, then 
$
X = \Bild{\pi_S}{\InvBild{\pi_S}{X}},
$
because $\pi_{S}: B\to S$ is onto, so it suffices to show that $\InvBild{\pi_S}{X}\in h(s, t)$, hence that
$
\Bild{\pi_T}{\InvBild{\pi_S}{X}}\in g(t).
$
Given $X$ there exists $Y\in g(t)$ so that
for each $t'\in Y$ there exists $s'\in X$ such that $\langle s', t'\rangle \in B$. 
Thus 
$
Y = \Bild{\pi_T}{(X\times Y)\cap B}.
$
But this implies
$
Y \subseteq \Bild{\pi_T}{\InvBild{\pi_S}{X}},
$
hence 
$
Y\subseteq\Bild{\pi_T}{\InvBild{\pi_S}{X}} \in g(t).
$
One similarly shows that $g(t) = \{\Bild{\pi_{T}}{Z}\mid Z\in h(s,
t)\}$. 

In a second step, we show that 
$
\{\Bild{\pi_{S}}{Z}\mid Z \in h(s, t)\} = \{C \mid
\InvBild{\pi_{S}}{C}\in h(s, t)\}.
$
In fact, if $C = \Bild{\pi_{S}}{Z}$ for some $Z\in h(s, t)$, then $Z
\subseteq \InvBild{\pi_{S}}{C} =
\InvBild{\pi_{S}}{\Bild{\pi_{S}}{Z}}$, hence $\InvBild{\pi_{S}}{C}\in
h(s, t)$. If, conversely, $Z := \InvBild{\pi_{S}}{C}\in h(s, t)$, then
$C = \Bild{\pi_{S}}{Z}$. Thus we obtain for $\langle s, t\rangle \in
B$
\begin{align*}
  f(s) & = \{\Bild{\pi_{S}}{Z}\mid Z\in h(s, t)\}\\
& = \{C \mid \InvBild{\pi_{S}}{C}\in h(s, t)\}\\
& = (\funV \pi_{S})(h(s, t)).
\end{align*}
Summarizing,  this means that 
$
\pi_S: (B, h) \to (S, f)
$
is a morphism. A very similar proof shows that  
$
\pi_T : (B, h) \to (T, g) 
$
is a morphism as well. 

\labelImpl{bisim-is-coalg-2}{bisim-is-coalg-1}: Now assume that
the projections are coalgebra morphisms, and let $\langle s, t\rangle
\in B$. Given $X\in f(s)$, we know that $X =\Bild{\pi_{S}}{Z}$ for
some $Z\in h(s, t)$. Thus we find for any $t'\in Y$ some $s'\in X$
with $\langle s', t'\rangle \in B$. The symmetric property of a bisimulation
is proved exactly in the same way. Hence $B$ is a bisimulation for
$(S, f)$ and $(T, g)$.
\EndProof
}

Encouraged by these observations, we define bisimulations for set
based functors, i.e., for endofunctors on the category $\catSET$ of
sets with maps as morphisms. This is nothing but a specialization of
the general notion of bisimilarity, taking specifically into
account that we  may in $\catSET$ to consider subsets of the
Cartesian product, and that we have projections at our disposal.

\BeginDefinition{bisim-relat-set-based}
Let $\funF$ be an endofunctor on $\catSET$. Then $R\subseteq S\times T$ is called a
\emph{\index{bisimulation}bisimulation} for the $\funF$-coalgebras $(S, f)$ and $(T, g)$ iff there exists a map $h: R\to
\funF(R)$ rendering this diagram commutative:
\begin{equation*}
\xymatrix{
S\ar[d]_{f} && R\ar[d]_{h}\ar[ll]_{\pi_{S}}\ar[rr]^{\pi_{T}} && T\ar[d]^{g}\\
\funF(S)&&\funF(R)\ar[ll]^{\funF \pi_{S}}\ar[rr]_{\funF \pi_{T}}&&\funF(T)
}
\end{equation*}
\EndDefinition

These are immediate consequences:

\BeginLemma{bisim-triv}
$\Delta_{S} := \{\langle s, s\rangle \mid s\in S\}$ is a bisimulation
for every $\funF$-coalgebra $(S, f)$. If $R$ is a bismulation for the
$\funF$-coalgebras  $(S, f)$ and $(T, g)$, then $R^{-1}$ is a
bisimulation for $(T, g)$ and $(S, f)$. 
\QED
\EndLemma

It is instructive to look back and investigate again the graph of a
morphism $r: (S, f)\to (T, g)$, where this time we do not have the
power set functor ---~as
in Proposition~\ref{morph-as-bisim}~--- but a general endofunctor $\funF$
on $\catSET$.

\BeginCorollary{morph-as-bisim-gen}
Given coalgebras $(S, f)$ and $(T, g)$ for the endofunctor $\funF$ on
$\catSET$, $r: (S, f)\to (T, g)$ is a morphism iff $Graph(r)$ is a
bisimulation for $(S, f)$ and $(T, g)$. 
\EndCorollary

\BeginProof
0.
The proof for Proposition~\ref{morph-as-bisim} needs some small adjustments, because we do not know how exactly
functor $\funF$ is operating on maps. 

1.
If $r: (S, f)\to (T, g)$ is a morphism, we know that $g\circ r =
\funF(r)\circ f$. Consider the map $\tau: S\to S\times T$ which is
defined as $s\mapsto \langle s, r(s)\rangle$, thus $\funF(\tau):
\funF(S)\to \funF(S\times T)$. Define 
\begin{equation*}
  h :
  \begin{cases}
   Graph(r) & \to \funF(Graph(r))\\
\langle s, r(s)\rangle & \mapsto \funF(\tau)(f(s))
\end{cases}
\end{equation*}
Then it is not difficult to see that both $g\circ \pi_{T} =
\funF(\pi_{T})\circ h$ and $f\circ \pi_{S} = \funF(\pi_{S})\circ h$
holds. Hence $(Graph(r), h)$ is an $\funF$-coalgebra mediating between
$(S, f)$ and $(T, g)$. 

2.
Assume that $Graph(r)$ is a bisimulation for $(S, f)$ and $(T, g)$,
then both $\pi_{T}$ and $\pi_{S}^{-1}$ are morphisms for the
$\funF$-coalgebras, so the proof proceeds exactly as the corresponding
one for Proposition~\ref{morph-as-bisim}.
\EndProof

We will study some properties of bisimulations now, including a
preservation property of functor $\funF: \catSET\to \catSET$. This
functor will be fixed for the time being. 

We may construct bisimulations from morphisms.

\BeginLemma{phi-psi-property}
Let $(S, f)$, $(T, g)$ and $(U, h)$ be $\funF$-coalgebras with
morphisms $\phi: (S, f)\to (T, g)$ and $\psi: (S, f)\to (U, h)$. Then
the image of $S$ under $\phi\times\psi$, 
\begin{equation*}
  \langle \phi, \psi\rangle[S] := \{\langle\phi(s), \psi(s)\rangle
  \mid s\in S\}
\end{equation*}
is a bisimulation for $(T, g)$ and $(U, h)$. 
\EndLemma

\BeginProof
0.
Look at this diagram
\begin{equation*}
\xymatrix{
&&\langle \phi, \psi\rangle[S]
\ar@/^0.7pc/[d]^{i}\ar[dll]_{\pi_{T}}\ar[drr]^{\pi_{U}}\\
T && S\ar[ll]^{\phi}\ar[rr]_{\psi}\ar@/^0.7pc/[u]^{j} && U
}
\end{equation*}
Here $j(s) := \langle \phi(s),\psi(s)\rangle$, hence $j: S\to
\langle\phi, \psi\rangle[S]$ is surjective. We can find a map $i:
\langle \phi, \psi\rangle[S]\to S$ so that $j\circ i = id_{\langle\phi,
  \psi\rangle[S]}$ using the Axiom of Choice\footnote{For each
  $r\in\langle\phi, \psi\rangle[S]$ there exists at least one $s\in S$ with
  $r = \langle \phi(s), \psi(s)\rangle$. Pick for each $r$ such an $s$
and call it $i(r)$, thus $r = \langle \phi(i(r)),
\psi(i(r))\rangle$.}. So we have a left inverse to $j$, which will
help us in the construction below. 

1.
We want to define a coalgebra structure for $\langle\phi, \psi\rangle[S]$ such that the diagram
below commutes, i.e., forms a bisimulation diagram. Put $k := \funF(j)\circ f\circ i$, then we have
\begin{equation*}
\xymatrix{
T\ar[d]_{g} && \langle\phi,
\psi\rangle[S]\ar[d]^{k}\ar[ll]_{\pi_{T}}\ar[rr]^{\pi_{U}} && U\ar[d]^{h}\\
\funF T&& \funF(\langle\phi, \psi\rangle[S])\ar[ll]^{\funF
  \pi_{T}}\ar[rr]_{\funF \pi_{U}} && \funF U
}
\end{equation*}
Now
\begin{align*}
  \funF (\pi_{T})\circ k & = \funF (\pi_{T})\circ \funF(j)\circ
  f\circ i\\
& = \funF (\pi_{T}\circ  j)\circ f\circ i\\
& = \funF (\phi)\circ f\circ i&&\text{(since $\pi_{T}\circ j = \phi$)}\\
& = g\circ \phi\circ i&&\text{(since $\funF(\phi)\circ f = g\circ \phi$)}\\
& = g\circ \pi_{T}
\end{align*}
Hence the left hand diagram commutes. Similarly
\begin{align*}
  \funF (\pi_{U})\circ k &= \funF (\pi_{U}\circ j)\circ f\circ i\\
& = \funF (\psi)\circ f\circ i\\
& = h\circ \psi\circ i\\
& = h\circ \pi_{U}
\end{align*}
Thus we obtain a commutative diagram on the right hand as well.
\EndProof

This is applied to the composition of relations:

\BeginLemma{compos-relations-bisim}
Let $R\subseteq S\times T$ and $Q\subseteq T\times U$ be relations,
and put $X := \{\langle s, t, u\rangle \mid \langle s, t\rangle \in R,
\langle t, u\rangle \in Q\}$. 
Then
\begin{equation*}
  R\circ Q = \langle \pi_{S}\circ \pi_{R}, \pi_{U}\circ \pi_{Q}\rangle[X].
\end{equation*}
\EndLemma

\BeginProof
Simply trace an element of $R\circ Q$ through this construction:
\begin{align*}
  \langle s, u\rangle \in R 
&\Leftrightarrow
\exists t\in T: \langle s, t\rangle \in R, \langle t, u\rangle \in Q\\
&\Leftrightarrow
\exists t\in T: \langle s, t, u\rangle \in X\\
&\Leftrightarrow
\exists t\in T: s = (\pi_{S}\circ \pi_{R})(s, t, u) \text{ and } u = (\pi_{U}\circ \pi_{Q})(s, t, u).
\end{align*}
\EndProof

Looking at $X$ in its relation to the projections, we see that $X$ is
actually a weak pullback, to be precise:

\BeginLemma{X-is-weak-pullback}
Let $R, Q, X$ be as above, then $X$ is a weak pullback of
$\pi_{T}^{R}: R\to T$ and $\pi_{T}^{Q}: Q\to T$, so that in
particular $\pi_{T}^{Q}\circ \pi_{Q} = \pi_{T}^{R}\circ \pi_{R}$. 
\EndLemma

\BeginProof
1.
It is easy to see that this diagram
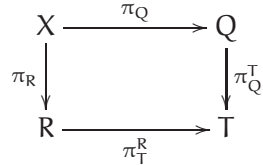
\begin{equation*}
\xymatrix{
X\ar[d]_{\pi_{R}}\ar[rr]^{\pi_{Q}} && Q\ar[d]^{\pi_{Q}^{T}}\\
R\ar[rr]_{\pi_{T}^{R}} && T
}
\end{equation*}
commutes. In fact, given $\langle s, t, u\rangle \in X$, we know that
$\langle s, t\rangle \in R$ and $\langle t, u\rangle \in Q$, hence 
$(\pi_{T}^{Q}\circ \pi_{Q})(s, t, u) = \pi_{T}^{Q}(t, u) = t$ and
$(\pi_{T}^{R}\circ \pi_{R})(s, t, u) = \pi_{T}^{R}(s, t) = t$.

2.
If $f_{1}: Y\to R$ and $f_{2}: Y\to Q$ are maps for some set $Y$ such
that $\pi_{T}^{R}\circ f_{1} = \pi_{T}^{R}\circ f_{2}$, we can write
$f_{1}(y) = \langle f_{1}^{S}(y), f_{2}^{T}(y)\rangle \in R$ and
$f_{2}(y) = \langle f_{2}^{T}(y), f_{2}^{U}(y)\rangle \in Q$. Put
$\sigma(y) := \langle f_{1}^{S}(y), f_{2}^{T}(y), f_{2}^{U}(y)\rangle$,
then $\sigma: Y\to X$ with $f_{1} = \pi_{R}\circ \sigma$ and $f_{2} =
\pi_{q}\circ \sigma$. Thus $X$ is a weak pullback. 
\EndProof

It will turn out that the
functor should preserve the pullback property. Preserving the
uniqueness property of a pullback will be too strong a requirement,
but preserving weak pullbacks will be helpful and not too
restrictive.

\BeginDefinition{preserves-wek-pullback}
Functor $\funF$ \emph{\index{pullback!preserves weak pullbacks}preserves weak pullbacks} iff $\funF$ maps weak
pullbacks to weak pullbacks. 
\EndDefinition

Thus a weak pullback diagram like
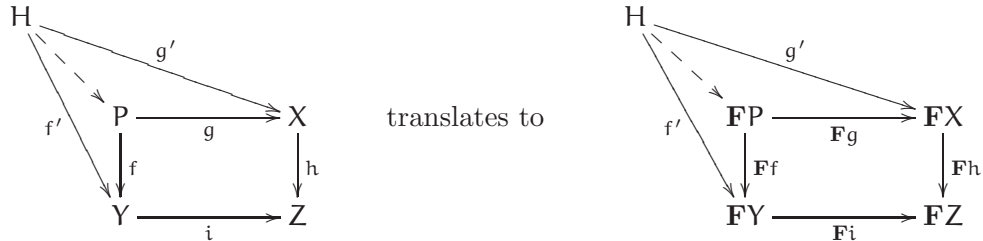
\begin{equation*}
\xymatrix{
H\ar[ddr]_{f'}\ar[drrr]^{g'}\ar@{-->}[dr]&&&&&H\ar[ddr]_{f'}\ar[drrr]^{g'}\ar@{-->}[dr]\\
&P\ar[d]^{f}\ar[rr]_{g}&&X\ar[d]^{h}&\text{translates to}&&\funF
P\ar[d]^{\funF f}\ar[rr]_{\funF g}&&\funF X\ar[d]^{\funF h}\\ 
&Y\ar[rr]_{i}&&Z&&&\funF Y\ar[rr]_{\funF i}&&\funF Z
}
\end{equation*}

We want to show that the composition of bisimulations is a
bisimulation again: this requires that the functor preserves weak
pullbacks. Before we state and prove a corresponding property, we
need an auxiliary statement which is of independent interest, viz.,
that the weak pullback of bisimulations forms a bisimulation again. To
be specific:

\BeginLemma{weak-pullback-is-bisim}
Assume that functor $\funF$ preserves weak pullbacks, and let $r: (S,
f)\to (T, g)$ and $s: (U, h)\to (T, g)$ be morphisms for the
$\funF$-coalgebras $(S, f)$, $(T, g)$ and $(U, h)$. Then there exists
a coalgebra structure $p: P\to \funF P$ for the weak pullback $P$ of
$r$ and $s$ with projections $\pi_{S}$ and $\pi_{T}$ such that $(P,
p)$ is a bismulation for $(S, f)$ and $(U, h)$.
\EndLemma

\BeginProof
We will need these diagrams
\begin{equation}
\label{weak-pullback-is-bisim-1}
\xymatrix{
S\ar[d]_{f}\ar[rr]^{r} && R\ar[d]_{g} && U\ar[d]^{h}\ar[ll]_{s}\\
\funF S\ar[rr]_{\funF r} && \funF T && \funF U\ar[ll]^{\funF s}
}
\end{equation}

\begin{equation}
\label{weak-pullback-is-bisim-2} 
\xymatrix{
P\ar[d]_{\pi_{S}}\ar[rr]^{\pi_{U}} && U\ar[d]^{s}\\
S\ar[rr]_{r} && T
}
\end{equation}
\begin{equation}
\label{weak-pullback-is-bisim-3} 
\xymatrix{
S\ar[d]_{f} && P\ar[ll]_{\pi_{S}}\ar[rr]^{\pi_{U}}\ar@{-->}[d]_{?}^{?} && U\ar[d]^{h}\\
\funF S && \funF P\ar[ll]^{\funF \pi_{S}}\ar[rr]_{\funF \pi_{U}}  && \funF U
}
\end{equation}
While the first  two diagrams are helping with the proof's argument,
the third diagram has a gap in the middle. We want to find an arrow
$P\to \funF P$ so that the diagrams will commute. Actually, the weak
pullback will help us obtaining this information.

Because
\begin{align*}
  \funF(r)\circ f\circ \pi_{S} & = g\circ r\circ
  \pi_{S}&&\text{(diagram~\ref{weak-pullback-is-bisim-1}, left)}\\
& = g\circ s\circ \pi_{U} &&
\text{(diagram~\ref{weak-pullback-is-bisim-2})}\\
& = \funF(s)\circ h\circ \pi_{U}&&\text{(diagram~\ref{weak-pullback-is-bisim-1}, right)}
\end{align*}
we may conclude that $\funF(r)\circ f\circ \pi_{S} = \funF(s)\circ
h\circ \pi_{U}$. Diagram~\ref{weak-pullback-is-bisim-2} is a pullback
diagram. Because $\funF$ preserves weak pullbacks, this diagram can be
complemented by an arrow $P\to \funF P$ rendering the triangles
commutative. 
\begin{equation*}
\xymatrix{
P\ar[drrr]^{h\circ \pi_{U}}\ar[ddr]_{f\circ \pi_{S}}\ar@{-->}[dr]\\
& \funF P\ar[d]^{\funF \pi_{S}}\ar[rr]^{\funF \pi_{U}}&&\funF
U\ar[d]_{\funF s}\\
&\funF S\ar[rr]_{\funF r} && \funF T
}
\end{equation*}
Hence there exists $p:
P\to \funF P$ with $\funF(\pi_{S})\circ p = f\circ \pi_{S}$ and
$\funF(\pi_{U})\circ p = h\circ \pi_{U}$. Thus $p$ makes
diagram~(\ref{weak-pullback-is-bisim-3}) a bismulation diagram.
\EndProof

Now we are in a position to show that the composition of bisimulations
is a bisimulation again, provided the functor $\funF$ behaves
decently.

\BeginProposition{composition-is-bisim}
Let $R$ be a bisimulation of $(S, f)$ and $(T, g)$, $Q$ be a bismulation
of $(T, g)$ and $(U, h)$, and assume that $\funF$ preserves weak
pullbacks. Then $R\circ Q$ is a bisimulation of $(S, f)$ and $(U,
h)$. 
\EndProposition

\BeginProof
We can write 
$
R\circ Q = \langle \pi_{S}\circ \pi_{R}, \pi_{U}\circ \pi_{Q}\rangle
[X]
$
with
$
X:= \{\langle s, t, u\rangle \mid \langle s, t\rangle \in R, \langle t,
u\rangle \in Q\}.
$
Since $X$ is a weak pullback of $\pi_{T}^{R}$ and $\pi_{T}^{Q}$ by Lemma~\ref{X-is-weak-pullback}, we
know that $X$ is a bisimulation of $(R, r)$ and $(Q, q)$, with $r$ and
$q$ as the dynamics of the corresponding
$\funF$-coalgebras. $\pi_{S}\circ \pi_{R}: X\to S$ and $\pi_{U}\circ
\pi_{Q}: X\to U$ are morphisms, thus $\langle \pi_{S}\circ \pi_{R}, \pi_{U}\circ \pi_{Q}\rangle
[X]$ is a bismulation, since $X$ is a weak pullback. Thus the assertion follows from Lemma~\ref{compos-relations-bisim}. 
\EndProof

The proof shows in which way the existence of the morphism $P\to \funF P$ is
used for achieving the desired properties.

Let us have a look at bisimulations on a coalgebra. Here bisimulations
may have an additional structure, viz., they may be equivalence
relations as well. Accordingly, we call these bisimulations \index{bisimulation
  equivalence}\emph{bisimulation equivalences}. Hence given a
coalgebra $(S, f)$, a bisimulation equivalence $\alpha$ for $(S, f)$
is a bisimulation for $(S, f)$ which is also an equivalence
relation. While bisimulations carry properties which are concerned
with the coalgebraic structure, an equivalence relation is purely
related to the set structure. It is, however, fairly natural to ask  in
view of the properties which we did explore so far
(Lemma~\ref{bisim-triv}, Proposition~\ref{composition-is-bisim}) 
whether or not we can take a bisimulation and turn it into an equivalence
relation, or at least do so under favorable conditions on functor
$\funF$. We will deal with this question and some of its cousins now.

Observe first that the factor space of a bisimulation equivalence can
be turned into a coalgebra.

\BeginLemma{factor-will-be-coalgebra}
Let $(S, f)$ be an $\funF$-coalgebra, and $\alpha$ be a bisimulation
equivalence on $(S, f)$. Then there exists a unique dynamics
$\alpha_{R}: \Faktor{S}{\alpha}\to \funF(\Faktor{S}{\alpha})$ with
$\funF(\fMap{\alpha})\circ f = \alpha_{R}\circ \fMap{\alpha}$. 
\EndLemma

\BeginProof
Because $\alpha$ is in particular a bisimulation, we know that there
exists by Theorem~\ref{ascel-bisim} a dynamics $\rho: \alpha\to \funF(\alpha)$ rendering this diagram
commutative.
\begin{equation*}
\xymatrix{
S\ar[d]_{f} && \alpha\ar[d]^{\rho}\ar[ll]_{\pi_{S}^{(1)}}\ar[rr]^{\pi_{S}^{(2)}} && S\ar[d]^{f}\\
\funF S && \funF \alpha\ar[ll]^{\funF \pi_{S}^{(1)}}\ar[rr]_{\funF \pi_{S}^{(2)}} && \funF S
}
\end{equation*}
The obvious choice would be to set $\alpha_{R}(\Klasse{s}{\alpha}) :=
(\funF(\fMap{\alpha})\circ f)(s)$, but this is only possible if we
know that the map is well defined, so we have to check whether
$(\funF(\fMap{\alpha})\circ f)(s_{1})= (\funF(\fMap{\alpha})\circ
f)(s_{2})$ holds, whenever $\isEquiv{s_{1}}{s_{2}}{\alpha}$. 

In fact,
$\isEquiv{s_{1}}{s_{2}}{\alpha}$ means $\langle s_{1}, s_{2}\rangle
\in \alpha$, so that 
$f(s_{1}) = f(\pi_{S}^{(1)}(s_{1}, s_{2})) = (\funF(\pi_{S}^{(1)})\circ
\rho)(s_{1}, s_{2})$, similarly for $f(s_{2})$. 
Because $\alpha$ is an equivalence relation, we have $\eta_{\alpha}\circ \pi_{S}^{(1)} = \eta_{\alpha}\circ \pi_{S}^{(2)}$.
Thus
\begin{align*}
  \funF(\fMap{\alpha})(f(s_{1}))
& = \bigl(\funF(\eta_{\alpha}\circ \pi_{S}^{(1)})\circ \rho\bigr)(s_{1}, s_{2})\\
& = \bigl(\funF(\eta_{\alpha}\circ \pi_{S}^{(2)})\circ \rho\bigr)(s_{1}, s_{2})\\
& = \funF(\fMap{\alpha})(f(s_{2}))
\end{align*}
This means that $\alpha_{R}$ is well defined indeed, and that $\fMap{\alpha}$ is a
morphism. Hence the dynamics $\alpha_{R}$ exists and renders
$\fMap{\alpha}$  a morphism.

Now assume that $\beta_{R}: \Faktor{S}{\alpha}\to
\funF(\Faktor{S}{\alpha})$ satisfies also $\funF(\fMap{\alpha})\circ f
= \beta_{R}\circ \fMap{\alpha}$. But then
$
\beta_{R}\circ \fMap{\alpha} = \funF(\fMap{\alpha})\circ f =
\alpha_{R}\circ \fMap{\alpha},
$
and, since $\fMap{\alpha}$ is onto, it is an epi, so that we may
conclude $\beta_{R} = \alpha_{R}$. Hence $\alpha_{R}$ is uniquely
determined. 
\EndProof

Bisimulations can be transported along morphisms, if the functor
preserves weak pullbacks.

\BeginProposition{transport-bisim}
Assume that $\funF$ preserves weak pullbacks, and let $r: (S, f)\to
(T, g)$ be a morphisms. Then 
\begin{enumerate}
\item \label{transport-bisim:1} If $R$ is a bismulation on $(S, f)$,
  then $\Bild{(r\times r)}{R} = \{\langle r(s), r(s')\rangle \mid
  \langle s, s'\rangle \in R\}$ is a bismulation on $(T, g)$.
\item \label{transport-bisim:2} If $Q$ is a bisimulation on $(T, g)$,
  then $\InvBild{(r\times r)}{Q} = \{\langle s, s'\rangle \mid \langle
  r(r), r(s')\rangle \in Q\} $ is a bismulation on $(S, f)$. 
\end{enumerate}
\EndProposition

\BeginProof
0.
Note that $Graph(r)$ is a bisimulation by
Corollary~\ref{morph-as-bisim-gen}, because $r$ is a morphism. 

1.
We claim that 
\begin{equation*}
\Bild{(r\times r)}{R} = \bigl(Graph(r)\bigr)^{-1}\circ R\circ
Graph(r)
\end{equation*}
holds. Granted that, we can apply
Proposition~\ref{composition-is-bisim} together with
Lemma~\ref{bisim-triv} for establishing the first property. But 
$\langle t, t'\rangle\in \Bild{(r\times r)}{R}$ iff we can find
$\langle s, s'\rangle \in R$ with $\langle t, t'\rangle = \langle
r(s), r(s')\rangle$, hence $\langle r(s), s\rangle \in Graph(r)^{-1}$,
$\langle s, s'\rangle \in R$ and $\langle s', r(s')\rangle \in
Graph(r)$, hence iff $\langle t, t'\rangle \in Graph(r)^{-1}\circ R\circ
Graph(r)$. 

2. Similarly, we show that $\InvBild{(r\times r)}{Q} = Graph(r)\circ R\circ
Graph(r)^{-1}$. This is left to the reader. 
\EndProof

For investigating further structural properties, we need 

\BeginLemma{coalg-on-coproduct}
If $(S, f)$ and $(T, g)$ are $\funF$-coalgebras, then there exists a
unique coalgebraic structure on $S+T$ such that the injections $i_{S}$
and $i_{T}$ are morphisms.
\EndLemma

\BeginProof
We have to find a morphism $S+T\to \funF(S+T)$ such that this diagram
is commutative
\begin{equation*}
\xymatrix{
S\ar[d]_{f}\ar[rr]^{i_{S}} && S+T\ar@{-->}[d]&& T\ar[d]^{g}\ar[ll]_{i_{T}}\\
\funF S\ar[rr]_{\funF i_{S}}  && \funF (S+T)&&\funF T\ar[ll]^{\funF i_{T}}
}
\end{equation*}
Because $\funF(i_{S})\circ f: S\to \funF(S+T)$ and $\funF(i_{T})\circ
g: T\to \funF(S+T)$ are morphisms, there exists a unique morphism $h:
S+T\to \funF(S+T)$ with $h\circ i_{S} = \funF(i_{S})\circ f$ and
$h\circ i_{T} = \funF(i_{T})\circ g$. Thus $(S+T, h)$ is a coalgebra,
and $i_{S}$ as well as $i_{T}$ are morphisms. 
\EndProof

The attempt to establish a comparable property for the product could
not work with the universal property for products, as a look at the
universal property for products will show.

We obtain as a consequence that bisimulations are closed under finite
unions. 

\BeginLemma{bisim-finite-unions}
Let $(S, f)$ and $(T, g)$ be coalgebras with bisimulations $R_{1}$ and
$R_{2}$. Then $R_{1}\cup R_{2}$ is a bismulation. 
\EndLemma

\BeginProof
1.
We can find morphisms $r_{i}: R_{i}\to \funF R_{i}$ for $i = 1, 2$
rendering the corresponding bisimulation diagrams commutative. Then
$R_{1}+ R_{2}$ is an $\funF$-coalgebra with 
\begin{equation*}
\xymatrix{
R_{1}\ar[d]_{r_{1}}\ar[rr]^{j_{1}}&&R_{1}+R_{2}\ar[d]_{r}&&R_{2}\ar[d]^{r_{2}}\ar[ll]_{j_{2}}\\
\funF R_{1}\ar[rr]_{\funF j_{1}} && \funF(R_{1}+R_{2})&&\funF
R_{2}\ar[ll]^{\funF j_{2}}
}
\end{equation*}
as commuting diagram, where $j_{i}: R_{i}\to R_{1}+R_{2}$ is the
respective embedding, $i = 1, 2$.

2.
We claim that the projections $\pi_{S}': R_{1}+R_{2}\to S$ and $\pi_{T}':
R_{1}+R_{2}\to T$ are morphisms. We establish this property only for
$\pi_{S}'$. First note that $\pi_{S}'\circ j_{1} = \pi_{S}^{R_{1}}$,
so that we have 
$
f\circ \pi_{S}'\circ j_{1} = \funF(\pi_{S}')\circ \funF(j_{1})\circ
r_{1} = \funF(\pi_{S}')\circ r\circ j_{1},
$ similarly, 
$
f\circ \pi_{S}'\circ j_{2} = \funF(\pi_{S}')\circ r\circ j_{2}.
$
Thus we may conclude that $f\circ \pi_{S}' = \funF(\pi_{S}') \circ r$,
so that indeed $\pi_{S}': R_{1}+R_{2}\to S$ is a morphism. 

3.
Since $R_{1}+R_{2}$ is a coalgebra, we know from
Lemma~\ref{phi-psi-property} that $\langle \pi_{S}',
\pi_{T}'\rangle[R_{1}+R_{2}]$ is a bisimulation. But this equals
$R_{1}\cup R_{2}$. 
\EndProof

We briefly explore lattice properties for bisimulations on a
coalgebra. For this, we investigate the union of an arbitrary family
of bisimulations. Looking back at the union of two bisimulations, we
used their sum as an intermediate construction. A more general
consideration requires the sum of an arbitrary family. The following
definition describes the coproduct as a specific form of a colimit,
see Definition~\ref{def-colimit} .

\BeginDefinition{arbitrary-coproduct}
Let $(s_{k})_{k\in I}$ be an arbitrary non-empty family of objects on
a category $\catK$. The object $s$ together with morphisms $i_{k}:
s_{k}\to s$ is called the \emph{\index{coproduct}coproduct} of
$(s_{k})_{k\in I}$ iff given morphisms $j_{k}: s_{k}\to t$ for an
object $t$ there exists a unique morphism $j: s\to t$ with $j_{k} =
j\circ i_{k}$ for all $k\in I$. $s$ is denoted as $\sum_{k\in
  I}s_{k}$. 
\EndDefinition

Taking $I = \{1, 2\}$, one sees that the coproduct of two objects is
in fact a special case of the coproduct just defined. The following
diagram gives a general idea.
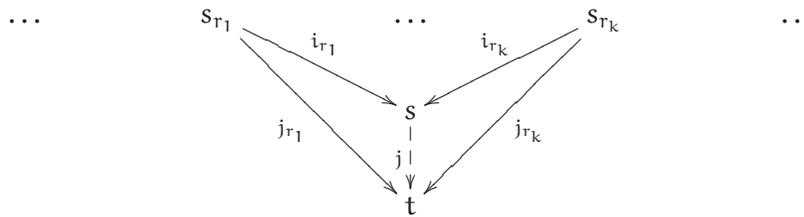
\begin{equation*}
\xymatrix{
\dots &&s_{r_{1}}\ar[drr]^{i_{r_{1}}}\ar[ddrr]_{j_{r_{1}}} && \dots 
&& s_{r_{k}}\ar[dll]_{i_{r_{k}}}\ar[ddll]^{j_{r_{k}}}&& \dots\\
&&&&s\ar@{-->}[d]_{j}\\
&&&&t
}
\end{equation*}

The coproduct is uniquely determined up to isomorphisms.

\BeginExample{coproduct-in-set}
Consider the category $\catSET$ of sets with maps as morphisms, and
let $(S_{k})_{k\in I}$ be a family of sets. Then 
\begin{equation*}
  S := \bigcup_{k\in I}\{\langle s, k\rangle \mid s\in S_{k}\}
\end{equation*}
is a coproduct. In fact, $i_{k}: s \mapsto \langle s, k\rangle$ maps
$S_{k}$ to $S$, and if $j_{k}: S_{k}\to T$, put $j: S\to T$ with
$j(s, k) := j_{k}(s)$, then $j_{k} = j\circ i_{k}$ for all $k$. 
\EndExample
 
We put this new machinery to use right away, returning to our scenario
given by functor $\funF$.

\BeginProposition{big-coprod-bisim}
Assume that $\funF$ preserves weak pullbacks. Let $(R_{k})_{k\in I}$
be a family of bisimulations for coalgebras $(S, f)$ and $(T, g)$. Then
$\bigcup_{k\in I}R_{k}$ is a bisimulation for these coalgebras.
\EndProposition

\BeginProof
1.
Given $k\in I$, let $r_{k}: R_{k}\to \funF R_{k}$ be the morphism on $R_{k}$ such that
$\pi_{S}: (S, f)\to (R_{k}, r_{k})$ and $\pi_{T}: (T, g)\to (R_{k},
r_{k})$ are morphisms for the coalgebras involved. Then there exists a
unique coalgebra structure $r$ on $\sum_{k\in I} R_{k}$ such that
$i_{\ell}: (R_{\ell}, r_{\ell})\to (\sum_{k\in I} R_{k}, r)$ is a
coalgebra structure for all $\ell\in I$. This is shown exactly through the same
argument as in the proof of Lemma~\ref{bisim-finite-unions} (\emph{mutatis mutandis}: replace the coproduct of
two bisimulations by the general coproduct).

2.
The projections $\pi_{S}': \sum_{k\in I} R_{k}\to S$ and $\pi_{T}':
\sum_{k\in I} R_{k}\to T$ are morphisms, and one shows exactly as in
the proof of Lemma~\ref{bisim-finite-unions} that
\begin{equation*}
  \bigcup_{k\in I}R_{k} = \langle\pi_{S}', \pi_{T}'\rangle[\sum_{k\in I} R_{k}].
\end{equation*}
An application of Lemma~\ref{phi-psi-property} now establishes the claim. 
\EndProof

This may be applied to an investigation of the lattice structure on
the set of all bisimulations between coalgebras.

\BeginProposition{bisim-lattice-struct}
Assume that $\funF$ preserves weak pullbacks. Let $(R_{k})_{k\in I}$
be a non-empty family of bisimulations for coalgebras $(S, f)$ and
$(T, g)$. Then
\begin{enumerate}
\item \label{bisim-lattice-struct-1} There exists a smallest
  bisimulation $R^{*}$ with $R_{k}\subseteq R^{*}$ for all $k$.
\item \label{bisim-lattice-struct-2} There exists a largest
  bisimulation $R_{*}$ with $R_{k}\supseteq R_{*}$ for all $k$.
\end{enumerate}
\EndProposition

\BeginProof
1.
We claim that $R^{*} = \bigcup_{k\in I} R_{k}$. It is clear that
$R_{k}\subseteq R^{*}$ for all $k\in I$. If $R'$ is a bisimulation on
$(S, f)$ and $(T, g)$ with $R_{k}\subseteq R'$ for all $k$, then
$\bigcup_{k}R_{k}\subseteq R'$, thus $R^{*}\subseteq R'$. In addition,
$R^{*}$ is a bismulation by Proposition~\ref{big-coprod-bisim}. This
establishes part~\ref{bisim-lattice-struct-1}.

2.
Put
\begin{equation*}
  {\cal R} := \{R \mid R\text{ is a bisimulation for $(S, f)$ and $(T,
    g)$ with $R\subseteq R_{k}$ for all $k$}\}
\end{equation*}
If ${\cal R} = \emptyset$, we put $R_{*} := \emptyset$, so  we
may assume that ${\cal R} \not=\emptyset$. Put $R_{*} := \bigcup{\cal
  R}$. By Proposition~\ref{big-coprod-bisim} this is a bisimulation
for $(S, f)$ and $(T, g)$ with $R_{k}\subseteq R_{*}$ for all
$k$. Assume that $R'$ is a bisimulation for for $(S, f)$ and $(T, g)$
with $R'\subseteq R_{k}$ for all $k$, then $R'\in {\cal R}$, hence
$R'\subseteq R_{*}$, so $R_{*}$ is the largest one. This settles part~\ref{bisim-lattice-struct-2}.
\EndProof

Looking a bit harder at bisimulations for $(S, f)$ alone, we find that
the largest bisimulation is actually an equivalence
relation. But we have to make sure first that a largest bismulation
exists at all.

\BeginProposition{largest-is-equivalence}
If functor $\funF$ preserves weak pullbacks, then there exists a
largest bisimulation $R^{*}$ on coalgebra $(S, f)$. $R^{*}$ is an
equivalence relation.
\EndProposition

\BeginProof
1.
Let 
\begin{equation*}
  {\cal R} := \{R \mid R\text{ is a bisimulation on }(S, f)\}.
\end{equation*}
Then $\Delta_{S}\in {\cal R}$, hence ${\cal R}\not=\emptyset$. We know
from Lemma~\ref{bisim-triv} that $R\in {\cal R}$ entails $R^{-1}\in
{\cal R}$, and from Proposition~\ref{big-coprod-bisim} we infer that
$R^{*}:= \bigcup{\cal R}\in{\cal R}$. Hence $R^{*}$ is a bismulation
on $(S, f)$. 

2.
$R^{*}$ is even an equivalence relation. 
\begin{itemize}
  \item Since $\Delta_{S}\in {\cal R}$, we know that
    $\Delta_{S}\subseteq R^{*}$, thus $R^{*}$ is reflexive.
\item Because $R^{*}\in {\cal R}$ we conclude that
  $(R^{*})^{-1}\in{\cal R}$, thus $(R^{*})^{-1}\subseteq R^{*}$. Hence
  $R^{*}$ is symmetric.
\item Since $R^{*}\in {\cal R}$, we conclude from
  Proposition~\ref{composition-is-bisim} that
  $R^{*}\circ R^{*}\in{\cal R}$, hence $R^{*}\circ R^{*}\subseteq
  R^{*}$. This means that $R^{*}$ is transitive.
\end{itemize}
\EndProof

This has an interesting consequence. Given a bisimulation equivalence
on a coalgebra, we do not only find a larger one which contains it,
but we can also find a morphism between the corresponding factor
spaces. To be specific:

\BeginCorollary{bisim-equiv-morph-factor}
Assume that functor $\funF$ preserves weak pullbacks, and that
$\alpha$ is a bisimulation equivalence on $(S, f)$, then there exists
a unique morphism 
$
\tau_{\alpha}: (\Faktor{S}{\alpha}, f_{\alpha})\to
(\Faktor{S}{R^{*}}, f_{R^{*}}),
$
where $f_{\alpha}: \Faktor{S}{\alpha}\to \funF(\Faktor{S}{\alpha})$ and  $f_{R^{*}}: \Faktor{S}{R^{*}}\to \funF(\Faktor{S}{R^{*}})$ are the induced dynamics.
\EndCorollary

\BeginProof
0.
The dynamics $f_{\alpha}: \Faktor{S}{\alpha}\to
\funF(\Faktor{S}{\alpha})$ and $f_{R^{*}}: \Faktor{S}{R^{*}}\to
\funF(\Faktor{S}{R^{*}})$ exist by the definition of a bisimulation. 

1.
Define 
\begin{equation*}
  \tau(\Klasse{s}{\alpha}) := \Klasse{s}{R^{*}}
\end{equation*}
for $s\in S$. This is well defined. In fact, if
$\isEquiv{s}{s'}{\alpha}$ we conclude by the maximality of $R^{*}$
that $\isEquiv{s}{s'}{R^{*}}$, so
$\Klasse{s}{\alpha}=\Klasse{s'}{\alpha}$ implies
$\Klasse{s}{R^{*}}=\Klasse{s'}{R^{*}}$. 

2. 
We claim that $\tau_{\alpha}$ is a morphism, hence that the right hand side
of this diagram commutes; the left hand side of the diagram is just
for nostalgia.
\begin{equation*}
\xymatrix{
S\ar[d]_{f}\ar[rr]^{\fMap{\alpha}} 
&& \Faktor{S}{\alpha}\ar[d]_{f_{\alpha}} \ar[rr]^{\tau_{\alpha}}
&& \Faktor{S}{R^{*}}\ar[d]^{f_{R^{*}}} \\
\funF S\ar[rr]_{\funF \fMap{\alpha}}  
&& \funF(\Faktor{S}{\alpha})\ar[rr]_{\funF \tau_{\alpha}} 
&& \funF(\Faktor{S}{R^{*}})
}
\end{equation*}
Now $\tau_{\alpha}\circ \fMap{\alpha} = \fMap{R^{*}}$, and the outer
diagram commutes. The left diagram commutes because $\fMap{\alpha}:
(S, f)\to \Faktor{S}{f_{\alpha}}$ is a morphism, moreover,
$\fMap{\alpha}$ is a surjective map. Hence the claim follows from
Lemma~\ref{diagram-chasing}, so that $\tau_{\alpha}$ is a morphism
indeed. 

3.
If $\tau_{\alpha}'$ is another morphism with these properties, then we
have $\tau_{\alpha}'\circ \fMap{\alpha} = \fMap{R^{*}} =
\tau_{\alpha}\circ \fMap{\alpha}$, and since $\fMap{\alpha}$ is
surjective, it is an epi by Proposition~\ref{char-surjective-cancel},
which implies $\tau_{\alpha} = \tau_{\alpha}'$. 
\EndProof

This is all well, but where do we get bisimulation equivalences from?
If we cannot find examples for them, the efforts just spent may run
dry. Fortunately, we are provided with ample bisimulation equivalences
through coalgebra morphisms, specifically through their kernel (for a
definition see page~\pageref{page:def-kernel}). It will turn out that
each of these equivalences can be generated so.

\BeginProposition{coalg-morph-bisim-equiv}
Assume that $\funF$ preserves weak pullbacks, and that $\phi: (S,
f)\to (T, g)$ is a coalgebra morphism. Then $\Kern{\phi}$ is a
bisimulation equivalence on $(S, f)$. Conversely, if $\alpha$ is a
bisimulation equivalence on $(S, f)$, then there exists a coalgebra
$(T, g)$ and a coalgebra morphism $\phi: (S, f)\to (T, g)$ with
$\alpha = \Kern{\phi}$.
\EndProposition

\BeginProof
1.
We know that $\Kern{\phi}$ is an equivalence relation; since
$\Kern{\phi} = Graph(\phi)\circ Graph(\phi)^{-1}$, we conclude from
Corollary~\ref{morph-as-bisim-gen} that $\Kern{\phi}$ is a bismulation. 

2.
Let $\alpha$ be a bisimulation equivalence on $(S, f)$, then the
factor map $\fMap{\alpha}: (S, f)\to (\Faktor{S}{\alpha}, f_{\alpha})$
is a morphism by Lemma~\ref{factor-will-be-coalgebra}, and 
$
  \Kern{\fMap{\alpha}} = \{\langle s, s'\rangle \mid
  \Klasse{s}{\alpha} = \Klasse{s'}{\alpha}\} = \alpha.
$
\EndProof

\Subsubsection{Congruences}
\label{sec:congruence}

Bisimulations compare two systems with each other, while a congruence permits to talk about elements in a coalgebra which behave similar. Let us have a look at Abelian groups. An equivalence relation $\alpha$ on an Abelian group $G$, which is written additively, is a congruence iff $\isEquiv{g}{h}{\alpha}$ and $\isEquiv{g'}{h'}{\alpha}$ together imply $\isEquiv{(g+g')}{(h+h')}{\alpha}$. This means that $\alpha$ is compatible with the group structure; an equivalent formulation says that there exists a group structure on $\Faktor{G}{\alpha}$ such that the factor map $\eta_{\alpha}: G\to \Faktor{G}{\alpha}$ is a group morphism. Thus the factor map is the harbinger of the good news.

\BeginDefinition{congruence-for-coalg}
Let $(S, f)$ be an $\funF$-coalgebra for the endofunctor $\funF$ on the category $\catSET$ of sets. An equivalence relation $\alpha$ on $S$ is called an $\funF$-\index{congruence}\emph{congruence} iff there exists a coalgebra structure $f_{\alpha}$ on $\Faktor{S}{\alpha}$ such that $\eta_{\alpha}: (S, f)\to (\Faktor{S}{\alpha}, f_{\alpha})$ is a coalgebra morphism. 
\EndDefinition

Thus we want that this diagram 
\begin{equation*}
\xymatrix{
S\ar[d]_{f}\ar[rr]^{\eta_{\alpha}} && \Faktor{S}{\alpha}\ar[d]^{f_{\alpha}}\\
\funF S\ar[rr]_{\funF \eta_{a}} && \funF(\Faktor{S}{\alpha})
}
\end{equation*}
is commutative, so that we have 
\begin{equation*}
  f_{\alpha}(\Klasse{s}{\alpha}) = (\funF \eta_{\alpha})(f(s))
\end{equation*}
for each $s\in S$. A brief look at Lemma~\ref{factor-will-be-coalgebra} shows that bisimulation equivalences are congruences, and we see from Proposition~\ref{coalg-morph-bisim-equiv} that the kernels of coalgebra morphisms are congruences, provided the functor $\funF$ preserves weak pullbacks. 

Hence congruences and bisimulations on a coalgebra are actually very closely related. They are, however, not the same, because we have

\BeginProposition{kernels-are-congruences}
Let $\phi: (S, f)\to (T, g)$ be a morphism for the $\funF$-coalgebras  $(S, f)$ and $(T, g)$. Assume that 
$
\Kern{\funF \phi} \subseteq \Kern{\funF \eta_{\Kern{\phi}}}.
$
Then $\Kern{\phi}$ is a congruence for $(S, f)$. 
\EndProposition

{
\def\Kf{\ensuremath{\Kern{\phi}}}
\def\Kl#1{\Klasse{#1}{\Kf}}
\BeginProof
Define 
$
f_{\Kf}(\Kl{s}) := \funF(\eta_{\Kf})(f(s))
$
for $s\in S$. Then $f_{\Kf}: \Faktor{S}{\Kf}\to \funF (\Faktor{S}{\Kf})$ is well defined. In fact, assume that $\Kl{s} = \Kl{s'}$, then $g(\phi(s)) = g(\phi(s'))$, so that $(\funF \phi)(f(s)) = (\funF \phi)(f(s'))$, consequently $\langle f(s), f(s')\rangle \in \Kern{\funF \phi}$. By assumption, $(\funF \eta_{\Kf})(f(s)) = (\funF \eta_{\Kf})(f(s'))$, so that $f_{\Kf}(\Kl{s}) = f_{\Kf}(\Kl{s'})$. It is clear that $\eta_{\alpha}$ is a coalgebra morphism. 

\EndProof

The next example leaves the category of sets and considers the
category of measurable spaces, introduced in
Example~\ref{ex-cat-meas-space}. The subprobability functor
$\SubProbSenza$, introduced in Example~\ref{functor-cont-probs}, is an
endofunctor on $\catMeas$, and we know that the coalgebras for this
functor are just the subprobabilistic transition kernels $K: (S,{\cal
  A})\to (S, {\cal A})$, see Example~\ref{ex-stoch-rel-coalg}. 

Fix measurable spaces $(S, {\cal A})$ and $(T,{\cal B})$. A measurable
map $f: (S, {\cal A})\to (T, {\cal B})$ is called \emph{final} iff
${\cal B}$ is the largest $\sigma$-algebra on $T$ which renders $f$
measurable, so that ${\cal A} = \{\InvBild{f}{B}\mid B\in{\cal
  B}\}$. Thus we conclude from $\InvBild{f}{B}\in {\cal A}$ that
$B\in{\cal B}$. Given an equivalence relation $\alpha$ on $S$, we can
make the factor space $\Faktor{S}{\alpha}$ a measurable space by
endowing it with the final $\sigma$-algebra $\Faktor{{\cal
    A}}{\alpha}$ with respect to $\eta_{\alpha}$, compare Exercise~\ref{ex-em-factorization-in-meas}. 

This is the definition then of a congruence for coalgebras for the Giry functor.
\BeginDefinition{congr-stoch-rel}
Let $(S, {\cal A},
K)$ and $(T, {\cal B}, L)$ be coalgebras for the subprobability
functor, then $\phi: (S, {\cal A}, K\to T, {\cal B}, L)$ is a
\emph{coalgebra morphism} iff $\phi: (S, {\cal A})\to (T,\beta)$ is a
measurable map such that this diagram commutes.
\begin{equation*}
\xymatrix{
S\ar[d]_{K}\ar[rr]^{\phi} && T\ar[d]^{L}\\
\SubProbSenza (S, {\cal A})\ar[rr]_{\SubProbSenza{\phi}} && \SubProbSenza (T, {\cal B})
}
\end{equation*}
\EndDefinition
Thus we have 
\begin{equation*}
  L(\phi(s))(B) = \SubProbSenza(\phi)(K(s))(B) = K(s)(\InvBild{\phi}{B})
\end{equation*}
for each $s\in S$ and for each measurable set $B\in{\cal B}$. We will
investigate the kernel of a morphism now in order to obtain a result
similar to the one reported in
Proposition~\ref{kernels-are-congruences}. The crucial property in
that development has been the comparison of the kernel $\Kern{\funF
  \phi}$ with $\Kern{\funF \eta_{\Kf}}$. We will investigate this
property now.

Call a morphism $\phi$ \emph{strong} iff $\phi$ is surjective and final. Now fix a strong morphism $\phi: K\to L$. A measurable subset $A\in {\cal A}$ is called $\phi$-\emph{\index{invariant subset}invariant} iff $a\in A$ and $\phi(a) = \phi(a')$ together imply $a'\in A$, so that $A\in{\cal A}$ is $\phi$ invariant iff $A$ is the union of $\Kf$-equivalence classes. 

We obtain:

\BeginLemma{invariant-are-inverse}
Let $\Sigma_{\phi} := \{A\in{\cal A}\mid A\text{ is $\phi$-invariant}\}$. Then
\begin{enumerate}
\item $\Sigma_{\phi}$ is a $\sigma$-algebra.
\item $\Sigma_{\phi}$ is isomorphic to $\{\InvBild{\phi}{B}\mid B\in {\cal B}\}$ as a Boolean $\sigma$-algebra.
\end{enumerate}
\EndLemma

\BeginProof
1.
Clearly, both $\emptyset$ and $S$ are $\phi$-invariant, and the complement of an invariant set is invariant again. Invariant sets are closed under countable unions. Hence $\Sigma_{\phi}$ is a $\sigma$-algebra.

2.
Given $B\in{\cal B}$, it is clear that $\InvBild{\phi}{B}$ is $\phi$-invariant; since the latter is also a measurable subset of $S$, we conclude that $\{\InvBild{\phi}{B}\mid B\in{\cal B}\} \subseteq \Sigma_{\phi}$. Now let $A\in\Sigma_{\phi}$, we claim that $A=\InvBild{\phi}{\Bild{\phi}{A}}$. In fact, since $\phi(a)\in\Bild{\phi}{A}$ for $a\in A$, the inclusion $A\subseteq\InvBild{\phi}{\Bild{\phi}{A}}$ is trivial. Let $a\in\InvBild{\phi}{\Bild{\phi}{A}}$, so that there exists $a'\in A$ with $\phi(a) = \phi(a')$. Since $A$ is $\phi$-invariant, we conclude $a\in A$, establishing the other inclusion. Because $\phi$ is final and surjective, we infer from this representation that $\Bild{\phi}{A}\in{\cal B}$, whenever $A\in\Sigma_{\phi}$, and that $\phi^{-1}: {\cal B}\to \Sigma_{\phi}$ is surjective. Since $\phi$ is surjective, $\phi^{-1}$ is injective, hence this yields a bijection. The latter map is compatible with the operations of a Boolean $\sigma$-algebra, so it is an isomorphism. 
\EndProof

This helps in establishing the crucial property for kernels.

\BeginCorollary{kernels-are-contained}
Let $\phi:K\to L$ be a strong morphism, then $\Kern{\SubProbSenza \phi}\subseteq\Kern{\SubProbSenza \eta_{\Kf}}$. 
\EndCorollary

\BeginProof
Let $\langle \mu, \mu'\rangle\in\Kern{\SubProbSenza \phi}$, thus $(\SubProbSenza{\phi})(\mu)(B) = (\SubProbSenza{\phi})(\mu')(B)$ for all $B\in{\cal B}$. Now let $C\in\Faktor{\cal A}{\Kf}$, then $\InvBild{\eta_{\Kf}}{C} \in\Sigma_{\phi}$, so that there exists by Lemma~\ref{invariant-are-inverse} some $B\in{\cal B}$ such that $\InvBild{\eta_{\Kf}}{C} = \InvBild{\phi}{B}$. Hence
\begin{align*}
  (\SubProbSenza\eta_{\Kf})(\mu)(C) & = \mu(\InvBild{\eta_{\Kf}}{C})\\
&  = \mu(\InvBild{\phi}{B})\\
& = (\SubProbSenza{\phi})(\mu)(B)\\
& = (\SubProbSenza{\phi})(\mu')(B) \\
& = (\SubProbSenza\eta_{\Kf})(\mu')(C), 
\end{align*}
so that $\langle \mu, \mu'\rangle\in\Kern{\SubProbSenza \eta_{\Kf}}$. 
\EndProof

Now everything is in place to show that the kernel of a strong morphism is a congruence for the $\SubProbSenza$-coalgebra $(S, {\cal A}, K)$.

\BeginProposition{is-congruence-giry}
Let $\phi: K\to L$ be a strong morphism for the $\SubProbSenza$-coalgebras $(S, {\cal A}, K)$ and $(T, {\cal B}, L)$. Then $\Kern{\phi}$ is a congruence for $(S, {\cal A}, K)$. 
\EndProposition

\BeginProof
1.
We want define the coalgebra $K_{\Kf}$ on $(\Faktor{S}{\Kf}, \Faktor{{\cal A}}{\Kf})$ upon setting
\begin{equation*}
  K_{\Kf}(\Kl{s})(C) := (\SubProbSenza{\eta_{\Kf}})(K(s))(C) \bigl(= K(s)(\InvBild{\eta_{\Kf}}{C})\bigr)
\end{equation*}
for $C\in\Faktor{\cal A}{\Kf}$, but we have to be sure that this is well
defined. In fact, let $\Kl{s} = \Kl{s'}$, which means $\phi(s) =
\phi(s')$, hence $L(\phi(s)) = L(\phi(s'))$, so that
$(\SubProbSenza\phi){K(s)} = (\SubProbSenza\phi){K(s')}$, because
$\phi: K\to L$ is a morphism. But the latter equality implies $\langle
K(s), K(s')\rangle \in\Kern{\SubProbSenza
  \phi}\subseteq\Kern{\SubProbSenza{\eta_{\Kf}}}$, the 
inclusion holding by Corollary~\ref{kernels-are-contained}. Thus we
conclude $(\SubProbSenza{\eta_{\Kf}})(K(s)) =
(\SubProbSenza{\eta_{\Kf}})(K(s'))$, so that $K_{\Kf}$ is well defined
indeed.

2.  It is immediate that $C\mapsto K_{\Kf}(\Kl{s})(C)$ is a
subprobability on $\Faktor{\cal A}{\Kf}$ for fixed $s\in S$, so it
remains to show that $t\mapsto K_{\Kf}(t)(C)$ is a measurable map on
the factor space $(\Faktor{S}{\Kf},\Faktor{\cal A}{\Kf})$. Let
$q\in[0, 1]$, and consider for $C\in\Faktor{\cal A}{\Kf}$ the set $G
:= \{t\in\Faktor{S}{\Kf}\mid K_{\Kf}(t)(C) < q\}$. We have to show
that $G\in\Faktor{\cal A}{\Kf}$. Because $C\in\Faktor{\cal A}{\Kf}$,
we know that $A := \InvBild{\eta_{\Kf}}{C}\in\Sigma_{\phi}$, hence it
is sufficient to show that the set $H := \{s\in S\mid K(s)(A) <
q\}\in\Sigma_{f}$. Since $K$ is the dynamics of a
$\SubProbSenza{}$-coalgebra, we know that $H\in{\cal A}$, so it
remains to show that $H$ is $\phi$-invariant. Because
$A\in\Sigma_{f}$, we infer from Lemma~\ref{invariant-are-inverse} that
$A = \InvBild{\phi}{B}$ for some $B\in{\cal B}$. Now take $s\in H$ and
assume $\phi(s) = \phi(s')$. Thus
\begin{align*}
  K(s')(A) & = K(s')(\InvBild{\phi}{B})\\
& = (\SubProbSenza \phi)(K(s'))(B)\\
& = L(\phi(s'))(B)\\
& = L(\phi(s))(B)\\
& = K(s)(A)\\
& < q,
\end{align*}
so that $H\in\Sigma_{\phi}$ indeed. Because $H =
\InvBild{\eta_{\Kf}}{G}$, it follows that $G\in\Kl{{\cal A}}$, and we
are done.
\EndProof

}

\Subsection{Modal Logics}
\label{sec:modal-logics}

This section will discuss modal logics and have a closer look at the
interface between models for this logics and coalgebras. Thus the
topics of this section may be seen as an application and illustration
of coalgebras. 

We will define the language for the formulas of modal
logics, first for the conventional logics which permits expressing
sentences like ``it is possible that formula $\phi$ holds'' or
``formula $\phi$
holds necessarily'', then for an extended version, allowing for modal
operators that govern more than one formula. The interpretation
through Kripke models is discussed, and it becomes clear at least
elementary elements of the language of categories is helpful in
investigating these logics. For completeness, we also give the
construction for the canonical model, displaying the elegant
construction through the Lindenbaum Lemma.

It shows that coalgebras can be used directly in the interpretation of
modal logics. We demonstrate that a set of predicate liftings define a
modal logics, discuss briefly expressivity for these modal logics, and
display an interpretation of CTL*, one of the basic logics for model
checking, through coalgebras.

We fix a set $\Phi$ of \emph{\index{propositional letters}propositional letters}. 

\BeginDefinition{def-basic-modal-lang}
The \emph{\index{modal language!basic}basic modal language} \index{${\cal L}(\Phi)$}${\cal L}(\Phi)$ over $\Phi$ is given by this
grammar
\begin{equation*}
  \phi ::= \bot~\mid~p~\mid~\phi_{1}\wedge\phi_{2}~\mid~\neg\phi~\mid~\Diamond\phi
\end{equation*}
\EndDefinition
We introduce additional operators
\begin{align*}
  \top & := \neg\bot\\
\phi_{1}\vee\phi_{2} & := \neg(\neg\phi_{1}\wedge\neg\phi_{2})\\
\phi_{1}\to \phi_{2} & := \neg\phi_{1}\vee\phi_{2}\\
\Box\phi & := \neg\Diamond\neg\phi.
\end{align*}
The constant $\bot$ denotes falsehood, consequently, $\top =
\neg\bot$ denotes truth, negation $\neg$ and conjunction $\wedge$ should not come as
a surprise; informally, $\Diamond\phi$ means that it is possible that formula $\phi$
holds, while $\Box\phi$ expresses that $\phi$ holds
necessarily. Syntactically, this looks like propositional logic,
extended by the modal operators $\Diamond$ and $\Box$. 

Before we have a look at the semantics of modal logic, we indicate
that this logic is syntactically sometimes a bit too restricted; after
all, the modal operators operate only on one argument at a time. 

The extension we want should offer modal operators with more
arguments. For this, we introduce the notion of a \emph{modal similarity
type} $\tau = (O, \rho)$, which is a set $O$ of operators, each
operator $\Delta\in O$ has an arity $\rho(\Delta)\in\Nat_{0}$. Note
that $\rho(\Delta) = 0$ is not excluded; these modal constants will
not play a distinguished r\^ole, however, they are sometimes nice to
have. 

Clearly, the set $\{\Diamond\}$ together with $\rho(\Diamond) = 1$ is
an example for such a modal similarity type.

\BeginDefinition{def-extend-modal-lang}
Given a modal similarity type $\tau = (O, \rho)$ and the set $\Phi$ of
propositional letters, the \emph{\index{modal
    language!extended}extended modal language} \index{${\cal L}(\tau,
  \Phi)$}${\cal L}(\tau, \Phi)$ is given by this grammar:
\begin{equation*}
  \phi ::= \bot~\mid~ p ~\mid~ \phi_{1}\wedge\phi_{2}~\mid~ \neg\phi\mid
  \Delta(\phi_{1}, \dots, \phi_{k})
\end{equation*}
with $p\in \Phi$ and $\Delta\in O$ such that $\rho(\Delta) = k$. 
\EndDefinition

We also introduce for the general case operators which negate on the
negation of the arguments of a modal operator; they are called
\emph{\index{modal language!nabla}nablas}, the nabla $\nabla$ of $\Delta$ is defined through
($\Delta\in O, \rho(\Delta) = k$)
\begin{equation*}
  \nabla(\phi_{1}, \dots, \phi_{k}) := \neg\Delta(\neg\phi_{1}, \dots, \neg\phi_{k})
\end{equation*}
Hence $\Box$ is the nabla of $\Diamond$; this is the reason why we did
not mention $\Box$ in the example above --- it is dependent on
$\Diamond$ in a systematic way.

It is time to have a look at some examples.

\BeginExample{ex-future-modal}
Let $O = \{\mathbf{F}, \mathbf{P}\}$ with $\rho(\mathbf{F}) =
\rho(\mathbf{P}) = 1$; the operator $\mathbf{F}$ looks into the future, and $\mathbf{P}$
into the past. This may be useful, e.g., when you are traversing a
tree and are visiting an inner node. The future may  then look at all
nodes in its subtree, the past at all nodes on a path from the root to
this tree. 

Then $\tau_{Fut} := (O, \rho)$ is a modal similarity type. If $\phi$
is a formula in ${\cal L}(\tau_{Fut}, \Phi)$, formula $\mathbf{F}\phi$
is true iff $\phi$ will hold  in the future, and $\mathbf{P}\phi$ is
true iff $\phi$ did hold in the past. The nablas are defined as 
\begin{align*}
  \mathbf{G}\phi & := \neg \mathbf{F} \neg \phi&& \text{ ($\phi$ will
    always be the case)}\\
\mathbf{H}\phi & := \neg \mathbf{P} \neg\phi&& \text{ ($\phi$ has
  always been the case).}
\end{align*}
Look at some formulas:
\begin{description}
\item[$\mathbf{P}\phi\to \mathbf{G}\mathbf{P}\phi$:] If something has
  happened, it will always have happened.
\item[$\mathbf{F}\phi\to \mathbf{FF}\phi$:] If $\phi$ will be true in
  the future, then it will be true in the future that $\phi$ will be
  true.
\item[$\mathbf{GF}\phi\to \mathbf{FG}\phi$:] If $\phi$ will be true in the future, then
  it will at some point be always true. 
\end{description}
\EndExample

The next example deals with a simple model for sequential programs. 

\BeginExample{ex-program-modal}
Take $\Psi$ as a set of atomic programs (think of elements of $\Psi$
as executable program components). The set of programs is defined
through this grammar
\begin{equation*}
  t ::= \psi~\mid~ t_{1}\cup t_{2}~\mid~ t_{1};t_{2}~\mid~ t^{*}~\mid~\phi?
\end{equation*}
with $\psi\in\Psi$ and $\phi$ a formula of the underlying modal
logic. 

Here $t_{1}\cup t_{2}$ denotes the
nondeterministic choice between programs $t_{1}$ and $t_{2}$,
$t_{1};t_{2}$ is the sequential execution of $t_{1}$ and $t_{2}$ in
that order, and $t^{*}$ is iteration of program $t$ a finite number of
times (including zero). The program $\phi?$ tests whether or not formula $\phi$ holds;
$\phi?$ serves as a guard: $(\phi?;t_{1})\cup(\neg\phi?;t_{2})$ tests
whether $\phi$ holds, if it does $t_{1}$ is executed, otherwise,
$t_{2}$ is. So the informal meaning of $\langle t\rangle\phi$ is that
formula $\phi$ holds after program $t$ is executed (we use here and
later an expression like $\langle t\rangle\phi$ rather than the
functional notation or just juxtaposition). 

So, formally we have the modal similarity type $\tau_{PDL} := (O,
\rho)$ with $O := \{\langle t\rangle \mid  t\text{ is a program}\}$. This logic is
known as \index{modal language!PDL}\index{PDL}PDL --- propositional dynamic logic. 
\EndExample

The next example deal with games and a syntax very similar to the one
just explored for PDL.

\BeginExample{ex-game-modal}
We introduce two players, \index{Angel}Angel and \index{Demon}Demon, playing against each
other, taking turns. So Angel starts, then Demon makes the next move,
then Angel replies, etc. 

For modelling  game logic, we assume that we have a set $\Gamma$ of
simple games; the syntax for games looks like this:
\begin{equation*}
  g ::= \gamma~\mid~ g_{1}\cup g_{2}~\mid~g_{1}\cap g_{2}~\mid~g_{1};g_{2}~\mid~g^{d}~\mid~g^{*}~\mid~g^{\times}~\mid~\phi?
\end{equation*}
with $\gamma\in\Gamma$ and $\phi$ a formula of the underlying
logic. The informal interpretation of $g_{1}\cup g_{2}$,
$g_{1};g_{2}$, $g^{*}$ and $\phi?$ are as in PDL
(Example~\ref{ex-program-modal}), but as actions of player Angel. The
actions of player Demon are indicated by 
\begin{description}
\item[$g_{1}\cap g_{2}$:] Demon chooses between games $g_{1}$ and
  $g_{2}$; this is called \emph{demonic \index{choice!demonic}choice} (in contrast to \emph{angelic
  \index{choice!angelic}choice} $g_{1}\cup g_{2}$).
\item[$g^{\times}$:] Demon decides to play game $g$ a finite number of
  times (including not at all). 
\item[$g^{d}$:]  Angel and Demon change places.
\end{description}
Again, we indicate through $\langle g\rangle\phi$ that formula $\phi$
holds after game $g$. We obtain the similarity type $\tau_{GL} := (O,
\rho)$ with $O := \{\langle g\rangle \mid  g\text{ is a game}\}$ and $\rho = 1$. The corresponding
logic is called \index{modal language!game logic}\emph{\index{game!logic}game logic}
\EndExample

Another example is given by arrow logic. Assume that you have arrows
in the plane; you can compose them, i.e., place the beginning of one
arrow at the end of the first one, and you can reverse them. Finally,
you can leave them alone, i.e., do nothing with an arrow.

\BeginExample{ex-arrow-modal}
The set $O$ of operators for arrow logic is given by $\{\circ,
\otimes, \mathtt{skip}\}$ with $\rho(\circ) = 2$, $\rho(\otimes) = 1$ and $\rho(\mathtt{skip})
= 0$. The arrow composed from arrows $a_{1}$ and $a_{2}$ is arrow
$a_{1}\circ a_{2}$, $\otimes a_{1}$ is the reversed arrow $a_{1}$, and
$\mathtt{skip}$ does nothing. 
\EndExample

\Subsubsection{Frames and Models}
\label{sec:frames-models}

For interpreting the basic modal language, we introduce frames. A
frame models transitions, which are at the very heart of modal
logics. Let us have a brief look at a modal formula like $\Box p$ for
some propositional letter $p\in\Phi$. This formula models ``$p$ always
holds'', which implies a transition from the current state to another
one, in which $p$ always holds; without a transition, we would not
have to think whether $p$ always holds --- it would just hold or
not. Hence we need to have transitions at our disposal, thus a
transition system, as in Example~\ref{ex-cat-transition-syst}. In the
current context, we take the disguise of a transition system as a
relation. All this is captured in the notion of a frame.

\BeginDefinition{def-model-log-Kripke-frame}
A \emph{\index{frame!Kripke} Kripke frame} $\fmF := (W, R)$ for the basic modal
language is a set $W\not=\emptyset$ of states together with a relation $R\subseteq
W\times W$. $W$ is sometimes called the \emph{\index{frame!set of worlds}set of worlds}, $R$ the
\emph{\index{frame!accessibility relation}accessibility relation}. 
\EndDefinition

The access-ability relation of a Kripke frame does not yet carry enough
information about the meaning of a modal formula, since the
propositional letters are not captured by the frame. This is the case,
however, in a Kripke model.

\BeginDefinition{def-modal-kripke}
A \emph{\index{model!Kripke}\index{Kripke model}Kripke model} (or simply a \emph{model}) $\fmM = (W, R, V)$
for the basic modal language consists of a Kripke frame $(W, R)$ together
with a map $V: \Phi\to \PowerSet{W}$. 
\EndDefinition

So, roughly speaking, the frame part of a Kripke model caters for the
propositional and the modal part of the logic whereas the map $V$ takes
care of the propositional letters. This now permits us to define the
meaning of the formulas for the basic modal language. We state under
which conditions a formula $\phi$ is true in a world $w\in W$; this is expressed
through $\fmM, w \models \phi$; note that this will depend on the
model $\fmM$, hence we incorporate it usually into the notation. Here
we go.
\begin{align*}
\fmM, w &\models \bot \text{ is always false.} \\
\fmM, w &\models p  \Leftrightarrow w\in V(p), \text{ if } p\in\Phi.\\
\fmM, w &\models \phi_{1}\wedge\phi_{2}\Leftrightarrow \fmM, w \models
\phi_{1}\text{ and }\fmM, w \models \phi_{2}.\\
\fmM, w &\models \neg\phi \Leftrightarrow \fmM, w \models
\phi\text{ is false.}\\
\fmM, w &\models \Diamond\phi \Leftrightarrow \text{ there exists $v$
  with $\langle w, v\rangle\in R$ and } \fmM, v \models \phi.
\end{align*}
The interesting part is of course the last line. We want
$\Diamond\phi$ to hold in state $w$; by our informal understanding
this means that a transition into a state such that $\phi$
holds in this state is possible. But this means that there exists some state $v$
with $\langle w, v\rangle\in R$ such that $\phi$ holds in $v$. This is
just the formulation we did use above. Look at $\Box\phi$; an easy
calculation shows that $\fmM, w \models \Box\phi$ iff $\fmM, w\models
\phi$ for all $v$ with $\langle w, v\rangle\in R$; thus, no matter
what transition from world $w$ to another world $v$ we make, and $\fmM, v
\models \phi$ holds, then $\fmM, w\models \Box\phi$.

We define $\Gilt_{\fmM}$ as the set of all states in which formula
$\phi$ holds. Formally,
\begin{equation*}
\index{$\Gilt_{\fmM}$}\Gilt_{\fmM} := \{w\in W \mid \index{$\fmM, w \models \phi$}\fmM, w \models \phi\}.
\end{equation*}

Let us look at some examples.

\BeginExample{ex-frame-linear-1}
Put $\Phi := \{p, q, r\}$ as the set of propositional letters, $W :=
\{1, 2, 3, 4, 5\}$ as the set of states; relation $R$ is given through 
\begin{equation*}
\xymatrix{
1\ar[r] & 2\ar[r] & 3\ar[r] & 4\ar[r] & 5
}
\end{equation*}
Finally, put 
\begin{equation*}
V(\ell) :=
  \begin{cases}
    \{2, 3\}, & \ell = p\\
\{1, 2, 3, 4, 5\},& \ell = q\\
\emptyset, & \ell = r
  \end{cases}
\end{equation*}
Then we have for the Kripke model $\fmM := (W, R, V)$ for example
\begin{description}
\item[$\fmM, 1 \models \Diamond\Box p$:] This is so since $\fmM, 
  3\models p$ (because $3\in V(p)$), thus $\fmM, 2\models \Box p$, hence $\fmM, 1\models
  \Diamond\Box p$.
\item[$\fmM, 1 \not\models \Diamond\Box p\to p$:] Since $1\not\in
  V(p)$, we have $\fmM, 1 \not\models p$.
\item[$\fmM, 2 \models \Diamond(p\wedge\neg r)$:] The only successor to
  $2$ in $R$ is state $3$, and we see that $3\in V(p)$ and $3\not\in
  V(r)$. 
\item[$\fmM, 1 \models q\wedge
  \Diamond(q\wedge\Diamond(q\wedge\Diamond(q\wedge\Diamond q)))$:]
  Because $1\in V(q)$ and $2$ is the successor to $1$, we investigate whether $\fmM, 2 \models
  q\wedge\Diamond(q\wedge\Diamond(q\wedge\Diamond q))$ holds. Since
  $2\in V(q)$ and $\langle 2, 3\rangle\in R$, we look at $\fmM, 3 \models
  q\wedge\Diamond(q\wedge\Diamond q)$; now $\langle 3, 4\rangle\in R$ 
  and $\fmM, 3\models q$, 
  so we investigate $\fmM, 4 \models q\wedge\Diamond q$. Since $4 \in
  V(q)$ and $\langle 4, 5\rangle\in R$, we find that this is true. Let 
$\phi$ denote the formula $q\wedge
\Diamond(q\wedge\Diamond(q\wedge\Diamond(q\wedge\Diamond q)))$, then
this peeling off layers of parentheses shows that $\fmM, 2 \not\models
\phi$, because $\fmM, 5\models \Diamond p$ does not hold.
\item[$\fmM, 1 \not\models \Diamond\phi\wedge q$:] Since $\fmM,
  2\not\models\phi$, and since state $2$ is the only successor to $1$,
  we see that $\fmM, 1\not\models\phi$. 
\item[$\fmM, w\models \Box q$:] This is true for all worlds $w$,
  because $w'\in V(q)$ for all $w'$ which are successors to some
  $w\in W$.  
\end{description}
\EndExample

\BeginExample{ex-frame-linear-2}
We have two propositional letters $p$ and $q$, as set of states we put
$W := \{1, 2, 3, 4, 6, 8, 12, 24\}$, and we say 
\begin{equation*}
  \isEquiv{x}{y}{R} \Leftrightarrow x \not= y \text{ and $x$ divides $y$}.
\end{equation*}
This is what $R$ looks like without transitive arrows:
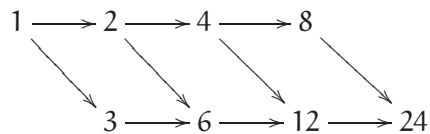
\begin{equation*}
\xymatrix{
1\ar[r]\ar[dr] & 2\ar[r]\ar[dr] & 4\ar[r]\ar[dr] & 8\ar[dr]\\
& 3\ar[r] & 6\ar[r] & 12\ar[r] & 24
}
\end{equation*}
Put $V(p) := \{4, 8, 12, 24\}$ and $V(q) := \{6\}$. 
Define the Kripke model $\fmM := (W, R, V)$, then we obtain for example
\begin{description}
\item[$\fmM, 4 \models \Box p$:] The set of successor to state $4$ is
  just $\{8, 12, 24\}$ which is a subset of $V(p)$.
\item[$\fmM, 6 \models \Box p$:] Here we may reason in the same way.
\item[$\fmM, 2 \not\models \Box p$:] State $6$ is a successor to $2$,
  but $6\not\in V(p)$.
\item[$\fmM, 2 \models \Diamond(q\wedge\Box p)\wedge\Diamond(\neg
  q\wedge \Box p)$:] State $6$ is a successor to state $2$ with $\fmM, 6
  \models q\wedge\Box p$, and state $4$ is a successor to state $2$
  with $\fmM, 4 \models \neg q\wedge\Box p$
\end{description}
\EndExample

Let us introduce some terminology which will be needed later. We say
that a formula $\phi$ is \emph{\index{formula!globally true}globally
  true} in a Kripke model $\fmM$ with state space $W$ iff
$\Gilt_{\fmM} = W$, hence iff $\fmM, w \models \phi$ for all states
$w\in W$; this is indicated by $\fmM\models\phi$. If
$\Gilt_{\fmM}\not=\emptyset$, thus if there exists $w\in W$ with
$\fmM, w\models \phi$, we say that formula $\phi$ is
\emph{\index{formula!satisfiable}satisfiable}; $\phi$ is said to be
\emph{\index{formula!refutable}refutable} or \emph{falsifiable} iff
$\neg\phi$ is satisfiable. A set $\Sigma$ of formulas is said to be
\emph{globally true} iff $\fmM, w\models\Sigma$ for all $w\in W$
(where we put $\fmM, w\models \Sigma$ iff $\fmM, w\models \phi$ for
all $\phi\in\Sigma$). $\Sigma$ is \emph{satisfiable} iff $\fmM,
w\models\Sigma$ for some $w\in W$.

Kripke models are but one approach for interpreting modal
logics. We observe that for a given transition system $(S, \Trans)$
the set $N(s) := \{s'\in S\mid s\Trans s'\}$ may consist of more than
one state; one may consider $N(s)$ as the neighborhood of state
$s$. An external observer may not be able to observe $N(s)$ exactly,
but may determine that $N(s)\subseteq A$ for some subset $A\subseteq
S$. Obviously, $N(s)\subseteq A$ and $A\subseteq B$ implies
$N(s)\subseteq B$, so that the sets defined by containing the
neighborhood $N(s)$ of a state $s$forms an upper closed set. This leads to the
definition of neighborhood frames.

\BeginDefinition{def-model-log-nghb-frame}
Given a set $S$ of states, a \emph{neighborhood \index{frame!neighborhood}\index{neighborhood!frame}frame} $\fmN := (S, N)$ is
defined by a map $N: S\to \funV(S) := \{V\subseteq\PowerSet{S}\mid
V\text{ is upper closed}\}$.  
\EndDefinition

The set $\funV(S)$ of all upper closed families of subsets $S$ was
introduced Example~\ref{upper-closed-functor}.

So if we consider state $s\in S$ in a neighborhood frame, then $N(s)$
is an upper closed set which gives all sets the next state
may be a member of. These frames occur in a natural way in topological
spaces.

\BeginExample{ngbh-frame-top}
Let $(T, \tau)$ be a topological space, then 
\begin{equation*}
  V(t) := \{A\subseteq T\mid U\subseteq A\text{ for some open
    neighborhood $U$ of $t$}\}
\end{equation*}
defines a neighborhood frame $(T, V)$.
\EndExample
 
Another straightforward example is given by ultrafilters.

\BeginExample{ngbh-frame-ultra}
Given a set $S$, define
\begin{equation*}
  U(x) := \{U\subseteq S \mid  x \in U\},
\end{equation*}
the ultrafilter associated with $x$. Then $(S, U)$ is a neighborhood frame.
\EndExample

Each Kripke frame gives rise to a neighborhood frame in this way:

\BeginExample{kripke-to-ngbh}
Let $(W, R)$ be a Kripke frame, and define for the world $w\in W$ the
set
\begin{equation*}
  V_{R}(w) := \{A\in\PowerSet{W}\mid R(w)\subseteq A\},
\end{equation*}
(with $R(w) := \{v\in W \mid \langle w, v\rangle \in R\}$), then plainly $(W, V_{R})$ is a neighborhood frame. 
\EndExample

A neighborhood frame induces a map on the power set of the state
space\label{dress-up-neighbor} into this power set. This map is used
sometimes for an interpretation in lieu of the neighborhood
function. Fix a map $P: S\to \funV{S}$ for illustrating this. Given a
subset $A\subseteq S$, we determine those states $\tau_{P}(A)$ which
can achieve a state in $A$ through $P$; hence $\tau_{P}(A) := \{s\in S
\mid A\in P(s)\}$. This yields a map $\tau_{P}: \PowerSet{S}\to
\PowerSet{S}$, which is monotone since $P(s)$ is upward closed for
each $s$. Conversely, given a monotone map $\theta: \PowerSet{S}\to
\PowerSet{S}$, we define $P_{\theta}: S\to \funV(S)$ through
$R_{\theta}(s) := \{A\subseteq S\mid s\in \theta(A)\}$. It is plain
that $\tau_{R_{\theta}} = \theta$ and $R_{\tau_{P}} = P$.

\BeginDefinition{def-model-log-nghb-model}
Given a set $S$ of states, a neighborhood  frame $(S, N)$, and a map $V:
\Phi\to \PowerSet{S}$, associating each propositional letter with a
set of states. Then
$\calN := (S, N, V)$ is called a \emph{\index{model!neighborhood}\index{neighborhood!model}neighborhood model}
\EndDefinition

We define  validity in a neighborhood model by induction on the structure of a
formula, this time through the validity sets.
\begin{align*}
\Gilt[\top]_{\calN} & := S,\\
\Gilt[p]_{\calN} & := V(p),\text{ if $p\in\Phi$},\\
\Gilt[\phi_{1}\wedge\phi_{2}]_{\calN} & := \Gilt[\phi_{1}]_{{\cal
    N}}\cap \Gilt[\phi_{2}]_{\calN},\\
\Gilt[\neg\phi]_{\mathcal{N}} & := S\setminus\Gilt_{\mathcal{N}}, \\
\Gilt[\Box\phi]_{\calN} & := \{s\in S \mid \Gilt_{\calN} \in N(s)\}.
\end{align*}
In addition, we put $\calN, s \models \phi$ iff $s\in \Gilt_{{\cal
    N}}$. Consider the last line and assume that the neighborhood
frame underlying the model is generated by a Kripke frame (W, R), so
that $A\in N(w)$ iff $R(w)\subseteq A$. Then $ \calN, w'\models
\Box\phi$ translates into $w'\in\{w \in S \mid R(w)\subseteq
\Gilt_{\calN}\}$, so that $\calN, w \models\Box\phi$ iff each
world which is accessible from world $w$ satisfies $\phi$; this is
what we want. Extending the definition above, we put
\begin{equation*}
  \Gilt[\Diamond\phi]_{\calN} := \{s\in S\mid S\setminus\Gilt_{\calN}\not\in N(s)\},
\end{equation*}
so that $\calN, s \models \Diamond\phi$ iff $\calN, s \models \neg\Box\neg\phi$. 

We generalize the notion of a Kripke model for capturing extended
modal languages. The idea for an extension in straightforward --- for
interpreting a modal formula given by a modal operator of arity $n$ we
require a subset of $W^{n+1}$. This leads to the definition of a frame,
adapted to this purpose.

\BeginDefinition{tau-frame}
Given a similarity type $\tau = (O, \rho)$, $\fmF = (W,
(R_{\Delta})_{\Delta\in O})$ is said to be a \emph{$\tau$-\index{frame!$\tau$}frame} iff
$W\not=\emptyset$ is a set of states, and $R_{\Delta}\subseteq
W^{\rho(\Delta)+1}$ for each $\Delta\in O$. A \emph{$\tau$-\index{model!$\tau$}model} $\fmM =
(\fmF, V)$ is a $\tau$-frame $\fmF$ with a map $V: \Phi\to
\PowerSet{W}$. 
\EndDefinition

Given a $\tau$-model $\fmM$, we define the interpretation of formulas
like $\Delta(\phi_{1}, \dots, \phi_{n})$ and its nabla-cousin
$\nabla(\phi_{1}, \dots, \phi_{n})$  in this way:
\begin{itemize}
\item 
  $\fmM, w \models \Delta(\phi_{1}, \dots, \phi_{n})$
 iff there exist  $w_{1}, \dots, w_{n}$ with
 \begin{enumerate}
 \item $\fmM, w_{i}\models \phi_{i}$ for $1\leq i \leq n$,
 \item $\langle w, w_{1},\dots , w_{n}\rangle \in R_{\Delta}$,
 \end{enumerate}
 if $n > 0$,
 \item $\fmM, w \models \Delta$ iff $w\in R_{\Delta}$ for $n = 0$,
 \item $\fmM, w \models \nabla(\phi_{1}, \dots, \phi_{n})$ iff 
$\bigl(\langle w, w_{1}, \dots, w_{n}\rangle \in R_{\Delta}$ implies $\fmM,
w_{i}\models \phi_{i}$ for all $i\in \{1, \dots, n\}\bigr)$
for all $w_{1}, \dots, w_{n}\in W$, if $n > 0$,
\item $\fmM, w \models \nabla$ iff $w\not\in R_{\Delta}$, if $ n =
  0$. 
\end{itemize}
In the last two cases, $\nabla$ is the nabla for modal operator
$\Delta$. 

Just in order to get a grip on these definitions, let us have a look
at some examples.

\BeginExample{ex-spec-rel-1}
The set $O$ of modal operators consists just of the unary operators $\{\langle a \rangle,
\langle b \rangle, \langle c \rangle\}$, the relations on the set $W
:= \{w_{1}, w_{2}, w_{3}, w_{4}\}$ of worlds are given by
\begin{align*}
  R_{a} & := \{\langle w_{1}, w_{2}\rangle, \langle w_{4}, w_{4}\rangle\},\\
R_{b} & := \{\langle w_{2}, w_{3}\rangle\},\\
R_{c} & := \{\langle w_{3}, w_{4}\rangle\}. 
\end{align*}
There is only one propositional letter $p$ and put $V(p) :=
\{w_{2}\}$. This comprises a $\tau$-model $\fmM$. We want to check
whether $\fmM, w_{1}\models \langle a\rangle p \to \langle b\rangle p$
holds. Allora: In order to establish whether or not $\fmM,
w_{1}\models \langle a\rangle p$ holds, we have to find a state $v$
such that $\langle w_{1}, v\rangle\in R_{a}$ and $\fmM, v\models p$;
state $w_{2}$ is the only possible choice. But $\fmM, w_{1}\not\models
p$, because $w_{1}\not\in V(p)$. Hence $\fmM, w_{1}\not\models \langle a\rangle p \to \langle b\rangle p$.
\EndExample

\BeginExample{ex-spec-rel-2}
Let $W = \{u, v, w, s\}$ be the set of worlds, we take $O :=
\{\diamondsuit, \clubsuit\}$ with $\rho(\diamondsuit) = 2$ and
$\rho(\clubsuit) = 3$. Put $R_{\diamondsuit} := \{\langle u, v,
w\rangle\}$ and $R_{\clubsuit} := \{\langle u, v, w, s\rangle\}$. The
set $\Phi$ of propositional letters is $\{p_{0}, p_{1}, p_{2}\}$ with 
$V(p_{0}) :=\{v\}$, $V(p_{1}) := \{w\}$ and $V(p_{2}) := \{s\}$. This
yields a model $\fmM$. 
\begin{enumerate}
\item 
We want to determine $\Gilt[\diamondsuit
(p_{0}, p_{1})]_{\fmM}$. From the definition of $\models$ we see that 
\begin{equation*}
  \fmM, x \models \diamondsuit (p_{0}, p_{1})\text{ iff } \exists
  x_{0}, x_{1}: \fmM, x_{0}\models p_{0}\text{ and } \fmM, x_{1}\models p_{1}\text{ and}
\langle x, x_{0}, x_{1}\rangle \in R_{\diamondsuit}.
\end{equation*}
We obtain by inspection  $\Gilt[\diamondsuit
(p_{0}, p_{1})]_{\fmM} = \{u\}$.
\item We have $\fmM, u \models \clubsuit(p_{0}, p_{1}, p_{2})$. This
  is so since $\fmM, v\models p_{0}$, $\fmM, w\models p_{1}$, and
  $\fmM, s\models p_{2}$ together with $\langle u, v, w, s\rangle\in
  R_{\clubsuit}$. 
\item Consequently, we have $\Gilt[\diamondsuit
(p_{0}, p_{1})\to \clubsuit(p_{0}, p_{1}, p_{2})]_{\fmM} = \{u\}$. 
\end{enumerate}
\EndExample

\BeginExample{ex-spec-rel-3}
Let's look into the future and into the past. We are given the unary
operators $O = \{\mathbf{F}, \mathbf{P}\}$ as in
Example~\ref{ex-future-modal}. The interpretation requires two binary
relations $R_{\mathbf{F}}$ and $R_{\mathbf{P}}$; we have defined the
corresponding nablas $\mathbf{G}$ resp $\mathbf{H}$.  Unless we want to
change the past, we assume that $R_{\mathbf{P}} =
R_{\mathbf{F}}^{-1}$, so just one relation $R := R_{\mathbf{F}}$
suffices for interpreting this logic. Hence
\begin{description}
\item[$\fmM, x \models \mathbf{F}\phi$:] This is the case iff there
  exists $z\in W$ such that $\langle x, z\rangle \in R$ and $\fmM,
  z\models \phi$.
\item[$\fmM, x \models \mathbf{P}\phi$:] This is true iff there exists
  $z\in W$ with $\langle v, x\rangle \in R$ and $\fmM, v\models \phi$.
\item[$\fmM, x \models \mathbf{G}\phi$:] This holds iff we have $\fmM, y\models \phi$ for
  all $y$ with $\langle x, y\rangle\in R$.
\item[$\fmM, x \models \mathbf{H}\phi$:] Similarly, for all $y$ with
  $\langle y, x\rangle\in R$ we have $\fmM, y\models \phi$. 
\end{description}
\EndExample

The next case is a little more complicated since we have to construct
the relations from the information that is available. In the case of
PDL (see Example~\ref{ex-program-modal}), we have only information
about the behavior of atomic programs, and we construct from it the
relations for compound programs. 

\BeginExample{ex-spec-rel-4}
Let $\Psi$ be the set of all atomic programs, and assume that we have
for each $t\in \Psi$ a relation $R_{t}\subseteq W\times W$; so if
atomic program $t$ is executed in state $s$, then $R_{t}(s)$ yields
the set of all possible successor states after execution. Now we
define by induction on the structure of the programs these relations.
\begin{align*}
  R_{\pi_{1}\cup\pi_{2}} & := R_{\pi_{1}}\cup R_{\pi_{2}},\\
R_{\pi_{1};\pi_{2}} & := R_{\pi_{1}}\circ R_{\pi_{2}},\\
R_{\pi^{*}} & := \bigcup_{n\geq 0} R_{\pi^{n}}.
\end{align*}
(here we put $R_{\pi_{0}} :=\{\langle w, w\rangle \mid w
\in W\}$, and $R_{\pi^{n+1}} := R_{\pi}\circ R_{\pi^{n}}$). Then, if
$\langle x, y\rangle\in R_{\pi_{1}\cup\pi_{2}}$, we know that $\langle
x, y\rangle \in R_{\pi_{1}}$ or $\langle x, y\rangle \in R_{\pi_{2}}$, which
reflects the observation that we can enter a new state $y$ upon
choosing between $\pi_{1}$ and $\pi_{2}$. Hence executing
$\pi_{1}\cup\pi_{2}$ in state $x$, we should be able to enter this
state upon executing one of the programs. Similarly, if in state $z$
we execute first $\pi_{1}$ and then $\pi_{2}$, we should enter an
intermediate state $z$ after executing $\pi_{1}$ and then execute
$\pi_{2}$ in state $z$, yielding the resulting state. Executing
$\pi^{*}$ means that we execute $\pi^{n}$ a finite number of times
(probably not at all). This explains the definition for
$R_{\pi^{*}}$. 

Finally, we should define $R_{\phi?}$ for a formula $\phi$. The
intuitive meaning of a program like $\phi?;\pi$ is that we want to
execute $\pi$, provided formula $\phi$ holds. This suggests defining
\begin{equation*}
  R_{\phi?} := \{\langle w, w\rangle \mid \fmM, w\models \phi\}.
\end{equation*}
Note that we rely here on a  model $\fmM$ which is already defined. 

Just to get familiar with these definitions, let us have a look at the
composition operator.
\begin{align*}
  \fmM, x \models \langle \pi_{1};\pi_{2}\rangle \phi
& \Leftrightarrow
\exists v: \fmM, v \models \phi\text{ and } \langle x, v\rangle\in
R_{\pi_{1};\pi_{2}}\\
& \Leftrightarrow
\exists w\in W\exists v \in \Gilt_{\fmM}: \langle x, w\rangle \in R_{\pi_{1}}\text{
  and } \langle w, v\rangle \in R_{\pi_{2}}\\
&\Leftrightarrow 
\exists w\in \Gilt[\langle\pi_{2}\rangle\phi]_{\fmM}: \langle x, w\rangle \in
R_{\pi_{1}}\\
&\Leftrightarrow 
\fmM, x \models \langle\pi_{1}\rangle\langle\pi_{2}\rangle\phi
\end{align*}
This means that $\langle\pi_{1};\pi_{2}\rangle \phi$ and
$\langle\pi_{1}\rangle\langle\pi_{2}\rangle\phi$ are semantically
equivalent, which is intuitively quite clear. 

The test operator is examined in the next formula. We have 
\begin{equation*}
R_{\phi?;\pi} = R_{\phi?}\circ R_{\pi} = \{\langle x, y\rangle \mid \fmM, x\models x\text{
  and }\langle x, y\rangle \in R_{\pi}\} = R_{\phi?}\cap R_{\pi}.
\end{equation*}
hence
$
\fmM, y \models \langle\phi?;\pi\rangle\psi
$
iff 
$
\fmM, y \models \phi
$
and
$
\fmM, y \models \langle\pi\rangle\psi,
$
so that
\begin{equation*}
\fmM, y \models (\langle
\phi?;\pi_{1}\rangle\cup\langle\neg\phi?;\pi_{2})\phi\text{ iff }
  \begin{cases}
    \fmM, y \models \langle \pi_{1}\rangle\phi,& \text{ if } \fmM,
    y\models \phi\\
    \fmM, y \models \langle \pi_{2}\rangle\phi,& \text{ otherwise}\\
  \end{cases}
\end{equation*}
\EndExample

The next example shows that we can interpret PDL in a neighborhood
model as well.

\BeginExample{ex-spec-rel-5}
We associate with each atomic program $t\in\Psi$ of PDL an effectivity
function $E_{t}$ on the state space $W$. Hence if we execute $t$ in
state $w$, then $E_{t}(w)$ is the set of all subsets $A$ of the states
so that the next state is a member of $A$ (we say that the program $t$
can \emph{achieve} a state in $A$). Hence $\bigr(W, (E_{t})_{t\in
  \Psi}\bigl)$ is a neighborhood frame. We have indicated that we can
construct from a neighborhood function a relation (see
page~\pageref{dress-up-neighbor}), so we put
\begin{equation*}
  R_{t}'(A) := \{w \in W \mid A\in R_{t}(w)\},
\end{equation*}
giving a monotone function $R_{t}': \PowerSet{W}\to
\PowerSet{W}$. This function can be extended to programs along the
syntax for programs in the following way, which is very similar to the
one for relations:
\begin{align*}
  R_{\pi_{1}\cup\pi_{2}}' & := R_{\pi_{1}}'\cup R_{\pi_{2}}',\\
R_{\pi_{1};\pi_{2}}' & := R_{\pi_{1}}'\circ R_{\pi_{2}}'\\
R_{\pi^{*}}' & := \bigcup_{n\geq 0}R_{\pi^{n}}'
\end{align*}
with $R_{\pi^{0}}'$ and $R_{\pi^{n}}'$ defined as above. 

Assume that we have again a function $V: \Phi\to \PowerSet{W}$,
yielding a neighborhood model $\calN$. The definition  above
are used now for the interpretation of formulas $\langle\pi\rangle\phi$
through
$
\Gilt[\langle \pi\rangle\phi]_{\calN} := R_{\pi}'(\Gilt_{\calN}).
$
The definition of $R_{\phi?}$ carries over, so that this yields an
interpretation of PDL. 
\EndExample

Turning to game logic (see Example~\ref{ex-game-modal}), we note that
neighborhood models are suited to interpret this logic as well. Assign
for each atomic game $\gamma\in \Gamma$ to Angel the effectivity
function $P_{\gamma}$, then $P_{\gamma}(s)$ indicates what Angel can
achieve when playing $\gamma$ in state $s$. Specifically, $A\in P_{\gamma}(s)$
indicates that Angel has a strategy for achieving that
the next state of the game is a member of $A$ by playing $\gamma$ in
state $s$. We will not formalize
the notion of a strategy here but appeal rather to an informal
understanding. The dual operator permit converting a game into its
dual, where players change r\^oles: the moves of Angel become moves of Demon, and vice
versa. 

Let us just indicate informally by $\langle\gamma\rangle\phi$ that
Angel has a strategy in game $\gamma$ which makes sure that game $\gamma$
results in a state which satisfies formula $\phi$.
We assume the game to be
\emph{\index{game!determined}determined}: if one player does not have
a winning strategy, then the other one has. Thus if Angle does not
have a $\neg\phi$-strategy, then Demon has a $\phi$-strategy, and vice versa.

\BeginExample{ex-spec-rel-6}
As in Example~\ref{ex-game-modal} we assume that games are given
thorough this grammar
\begin{equation*}
   g ::= \gamma~\mid~ g_{1}\cup g_{2}~\mid~g_{1}\cap g_{2}~\mid~g_{1};g_{2}~\mid~g^{d}~\mid~g^{*}~\mid~g^{\times}~\mid~\phi?
\end{equation*}
with $\gamma\in \Gamma$, the set of atomic games. We assume that the
game is determined, hence we may express demonic choice $g_{1}\cap
g_{2}$ through $(g_{1}^{d}\cup g_{2}^{d})^{d}$, and demonic iteration
$g^{\times}$ through angelic iteration
$\bigr((g^{d})^{*}\bigl)^{d})$.

Assign to each
$\gamma\in\Gamma$ an effectivity function $P_{\gamma}$ on the set $W$
of worlds, and put 
\begin{equation*}
  P'_{\gamma}(A) := \{w\in W \mid A\in P_{\gamma}(w)\}.
\end{equation*}
Hence $w\in P'_{\gamma}(A)$ indicates that Angel has a strategy to
achieve $A$ by playing game $\gamma$ in state $w$. We extend $P'$ to
games along the lines of the games' syntax:
\begin{align*}
  P'_{g_{1}\cup g_{2}}(A) & := P'_{g_{1}}(A)\cup P'_{g_{2}}(A),&
P'_{g^{d}}(A) & := W\setminus P'_{g}(W\setminus A),\\
P'_{g_{1};g_{2}}(A) & := P'_{g_{1}}(P'_{g_{2}}(A)),&
P'_{g_{1}\cap g_{2}}(A) & := P'_{(g_{1}^{d}\cup g_{2}^{d})^{d}}(A),\\
P'_{g^{*}}(A) & := \bigcup_{n\geq 0}P'_{g^{n}}(A),&
P'_{g^{\times}}(A) & := P'_{\bigr((g^{d})^{*}\bigl)^{d}}(A),\\
P'_{\phi?}(A) & := \Gilt_{\calN}\cap A.
\end{align*}
The last line refers to a model $\calN$. 
\EndExample

We have finally a look at arrow logic, see Example~\ref{ex-arrow-modal}.

\BeginExample{ex-arrow-logic-mod}
Arrows are interpreted as vectors, hence, e.g., as pairs. Let $W$ be a
set of states, then we take $W\times W$ as the domain of our
interpretation. We have three modal operators.
\begin{itemize}
  \item The nullary operator $\mathtt{skip}$ is interpreted through $R_{\mathtt{skip}} := \{\langle w, w\rangle \mid w\in W\}$.
  \item The unary operator $\otimes$ is interpreted through 
  $
  R_\otimes := \bigl\{\bigl\langle\langle a, b\rangle, \langle b, a\rangle\bigr\rangle\mid a, b\in W\bigr\}.
  $
  \item The binary operator is intended to model composition, thus one end of the first arrow should be the be other end of the second arrow, hence
  $
  R_\circ := \bigl\{\bigl\langle\langle a, b\rangle, \langle b, c\rangle, \langle  a, c\rangle\bigr\rangle \mid a, b, c\in W\bigr\}.
  $
\end{itemize} 
With this, we obtain for example
$
\fmM, \langle w_1, w_2\rangle \models \psi_1\circ\psi_2
$
iff there exists $v$ such that $\fmM, \langle w_1, v\rangle \models \psi_1$ and $\fmM, \langle v, w_2\rangle \models \psi_2$. 
\EndExample

Frames are related through frame morphisms. Take a frame $(W, R)$ for the basic modal language, then $R: W \to \PowerSet{W}$ is perceived as a coalgebra for the power set functor. this helps in defining morphisms.

\BeginDefinition{frame-morphisms}
Let $\fmF = (W, R)$ and $\fmG = (X, S)$ be Kripke frames. A \emph{frame \index{frame!morphism}\index{morphism!frame}morphism} $f: \fmF\to\fmG$ is a map $f: W\to X$ which makes this diagram commutative:
\begin{equation*}
\xymatrix{
W\ar[d]_R\ar[rr]^f && X\ar[d]^S\\
\PowerSet{W}\ar[rr]_{\PowerSenza{f}} && \PowerSet{X}
}
\end{equation*}
\EndDefinition
Hence we have for a frame morphism $f: \fmF\to\fmG$ the condition  
\begin{equation*}
S(f(w)) = (\PowerSenza{f})(R(w)) = \Bild{f}{R(w)} = \{f(w') \mid w'\in R(w)\}.
\end{equation*}
for all $w\in W$.

This is a characterization of frame morphisms.

\BeginLemma{char-frame-morph}
Let $\fmF$ and $\fmG$ be frames, as above. Then $f:\fmF\to \fmG$ is a frame morphism iff these conditions hold
\begin{enumerate}
\item $\isEquiv{w}{w'}{R}$ implies $\isEquiv{f(w)}{f(w')}{S}$.
\item If $\isEquiv{f(w)}{z}{S}$, then there exists $w'\in W$ with $z =
  f(w')$ and $\isEquiv{w}{w'}{R}$.
\end{enumerate}
\EndLemma

\BeginProof
1.
These conditions are necessary. In fact, if $\langle w, w'\rangle\in
R$, then $f(w') \in \Bild{f}{R(w)} = S(f(w))$, so that $\langle f(w),
f(w')\rangle \in S$. Similarly, assume that $\isEquiv{f(w)}{z}{S}$,
thus $z\in S(f(w)) = \PowerSet{f}(R(w)) = \Bild{f}{R(w)}$. Hence there
exists $w'$ with $\langle w, w'\rangle\in R$ and $z = f(w')$. 

2.
The conditions are sufficient. The first condition implies
$\Bild{f}{R(w)}\subseteq S(f(w))$. Now assume $z\in S(f(w))$, hence
$\isEquiv{f(w)}{z}{S}$, thus there exists $w'\in R(w)$ with $f(w')
= z$, consequently, $z = f(w')\in \Bild{f}{R(w)}$. 
\EndProof

We see that the bounded morphisms from
Example~\ref{ex-cat-transition-syst-bound} appear here again in a natural context. 

If we want to compare models for the basic modal language, then we
certainly should be able to compare the underlying frames. But this is
not yet enough, because the interpretation for atomic propositions has
to be taken care of.

\BeginDefinition{mod-morph-simple-modal}
Let $\fmM = (W, R, V)$ and $\fmN = (X, S, Y)$ be models for the basic
modal language and $f: (W, R)\to (X, S)$ be a frame morphism. Then $f:
\fmM\to \fmN$ is said to be a \emph{model \index{model!morphism}\index{morphism!model}morphism} iff $f^{-1}\circ Y = V$.
\EndDefinition

Hence $\InvBild{f}{Y(p)} = V(p)$ for a model morphism $f$ and for each
atomic proposition $p$, thus $\fmM, w \models p$ iff $\fmN,
f(w)\models p$ for each atomic proposition. This extends to all
formulas of the basic modal language.

\BeginProposition{morphs-preserve-valid}
Assume $\fmM$ and $\fmN$ are models as above, and $f: \fmM\to \fmN$ is
a model morphism. Then
\begin{equation*}
  \fmM, w \models \phi \text{ iff } \fmN, f(w)\models \phi
\end{equation*}
for all worlds $w$ of $\fmM$, and for all formulas $\phi$.
\EndProposition

\BeginProof
0.
The assertion is equivalent to 
\begin{equation*}
  \Gilt_{\fmM} = \InvBild{f}{\Gilt_{\fmN}}
\end{equation*}
for all formulas $\phi$. This is established by induction on the
structure of a formula now.

1.
If $p$ is an atomic proposition, then this is just the definition of a
frame morphism to be a model morphism: 
\begin{equation*}
  \Gilt[p]_{\fmM} = V(p) = \InvBild{f}{Y(p)} = \Gilt[p]_{\fmN}.
\end{equation*}
Assume that the assertion holds for $\phi_{1}$ and $\phi_{2}$, then
\begin{multline*}
  \Gilt[\phi_{1}\wedge\phi_{2}]_{\fmM} = 
  \Gilt[\phi_{1}]_{\fmM}\cap\Gilt[\phi_{2}]_{\fmM} = 
\InvBild{f}{\Gilt[\phi_{1}]_{\fmN}}\cap\InvBild{f}{\Gilt[\phi_{2}]_{\fmN}}
=\\
\InvBild{f}{\Gilt[\phi_{1}]_{\fmN}\cap\Gilt[\phi_{2}]_{\fmN}}
=
\InvBild{f}{\Gilt[\phi_{1}\wedge\phi_{2}]_{\fmN}}
\end{multline*}
Similarly, one shows that $\Gilt[\neg\phi]_{\fmM} =
\InvBild{f}{\Gilt[\neg\phi]_{\fmN}}$.

2.
Now consider $\Diamond\phi$, assume that the hypothesis holds for
formula $\phi$, then we have
\begin{align*}
  \Gilt[\Diamond\phi]_{\fmM} 
& = \{w \mid \exists w'\in R(w): w'\in\Gilt_{\fmM}\}\\
& = \{w \mid \exists w'\in R(w): f(w')\in \Gilt_{\fmN}\}&& \text{ (by
  hypothesis)}\\
& = \{w\mid \exists w': f(w')\in S(f(w)), f(w')\in\Gilt_{\fmN}\}
&&\text{ (by Lemma~\ref{char-frame-morph})}\\
& = \InvBild{f}{\{x\mid \exists x'\in S(x): x'\in\Gilt_{\fmN}\}}\\
& = \InvBild{f}{\Gilt[\Diamond\phi]_{\fmN}}
\end{align*}
Thus the assertion holds for all formulas $\phi$. 
\EndProof

This permits comparing worlds in two models. Two worlds are said to be
equivalent iff they cannot be separated by a formula, i.e., iff
they satisfy exactly the same formulas.

\BeginDefinition{modal-equivalence}
Let $\fmM$ and $\fmN$ be models with state spaces $W$
resp. $X$. States $w\in W$ and $x\in X$ are called \emph{\index{equivalence!modal}modally equivalent}
iff we have
\begin{equation*}
  \fmM, w \models \phi \text{ iff } \fmN, x\models \phi
\end{equation*}
for all formulas $\phi$
\EndDefinition

Hence if $f: \fmM\to \fmN$ is a model morphism, then $w$ and $f(w)$
are modally equivalent for each world $w$ of $\fmM$. One might be
tempted to compare models with respect to their transition behavior;
after all, underlying a model is a transition system,
a.k.a. a frame. This leads directly to this notion of bisimilarity for models
---~note that we have to take the atomic propositions into account.

\BeginDefinition{bimulation-models}
Let $\fmM = (W, R, V)$ and $\fmN = (X, S, Y)$ be models for the basic
modal language, then a relation $B\subseteq W\times X$ is called a
\emph{\index{bisimulation}bisimulation} iff
\begin{enumerate}
\item If $\isEquiv{w}{x}{B}$, then $w$ and $x$ satisfy the same
  propositional letters (``\index{atomic harmony}atomic harmony'').
\item If $\isEquiv{w}{x}{B}$ and $\isEquiv{w}{w'}{R}$, then there
  exists $x'$ with $\isEquiv{x}{x'}{S}$ and $\isEquiv{w'}{x'}{B}$
  (forth condition).
\item If $\isEquiv{w}{x}{B}$ and $\isEquiv{x}{x'}{S}$, then there
  exists $w'$ with $\isEquiv{w}{w'}{R}$ and $\isEquiv{w'}{x'}{B}$
  (back condition).
\end{enumerate}
States $w$ and $x$ are called \emph{\index{bisimilar!states}bisimilar} iff there exists a bisimulation
$B$ with $\langle w, x\rangle\in B$. 
\EndDefinition

Hence the forth condition says for a pair of worlds $\langle w,
x\rangle\in B$ that, if $w\Trans_{R}w'$ there exists $x'$ with
$\langle w', x'\rangle\in B$ such that $x\Trans_{S}x'$, similarly for
the back condition. So this rings a bell: we did discuss this in
Definition~\ref{bisim-relat}. Consequently, if models $\fmM$ and
$\fmN$ are bisimilar, then the underlying frames are bisimilar
coalgebras. 

Consider this example for bisimilar states.

\BeginExample{states-bisimilar}
Let relation $B$ be defined through 
\begin{equation*}
  B := \{\langle 1, a\rangle, \langle 2, b\rangle, \langle 2,
  c\rangle, \langle 3, d\rangle, \langle 4, e\rangle, \langle 5, e\rangle\}
\end{equation*}
with $V(p) := \{a, d\}, V(q) := \{b, c, e\}$.
\par\begin{minipage}[t]{.45\linewidth}
The transitions for $\fmM$ are given through
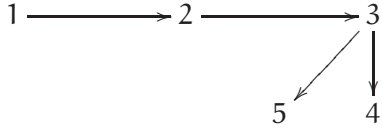
\begin{equation*}
\xymatrix{
1\ar[rr] && 2\ar[rr] && 3\ar[dl]\ar[d]\\
&&&5&4
}
\end{equation*}
\end{minipage}
\hfill
\begin{minipage}[t]{.45\linewidth}
$\fmN$ is given through 
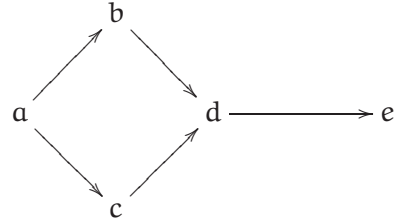
\begin{equation*}
\xymatrix{
&b\ar[dr]\\
a\ar[ur]\ar[dr] && d\ar[rr] && e\\
&c\ar[ur]
}
\end{equation*}
\end{minipage}

Then $B$ is a bisimulation
\EndExample

The first result relating bisimulation and modal equivalence is
intuitively quite clear. Since a bisimulation reflects the structural
similarity of the transition structure of the underlying transition
systems, and since the validity of modal formulas is determined
through this transition structure (and the behavior of the atomic
propositional formulas), it does not come as a surprise that bisimilar
states are modal equivalent.

\BeginProposition{modal-equivalence-bisim}
Let $\fmM$ and $\fmN$ be models with states $w$ and $x$. If $w$ and
$x$ are bisimilar, then they are modally equivalent.
\EndProposition

\BeginProof
0.
Let $B$ be the bisimulation for which we know that $\langle w,
x\rangle \in B$. We have to show that 
\begin{equation*}
  \fmM, w\models \phi \Leftrightarrow \fmN, x\models \phi
\end{equation*}
for all formulas $\phi$.  This is done by induction on the formula.

1.
Because of atomic harmony, the equivalence holds for propositional
formulas. It is also clear that conjunction and negation are preserved
under this equivalence, so that the remaining (and interesting) case
of proving the equivalence for a formula $\Diamond\phi$ under the
assumption that it holds for $\phi$. 

``$\Rightarrow$'' Assume that $ \fmM, w\models \Diamond\phi$
holds. Thus there exists a world $w'$ in $\fmM$ with $\isEquiv{w}{w'}{R}$ and $\fmM,
w'\models \phi$. Hence there exists by the forward condition a world $x'$ in $\fmN$ with
$\isEquiv{x}{x'}{S}$ and $\langle w', x'\rangle \in B$ such that $\fmN,
x'\models \phi$ by the induction hypothesis. Because $x'$ is a
successor to $x$, we conclude $\fmN, x\models \Diamond\phi$. 

``$\Leftarrow$'' This is shown in the same way, using the back
condition for $B$. 
\EndProof

The converse holds only under the restrictive condition that the
models are image finite. Thus each state has only a finite number of
successor states; formally, model $(W, R, V)$ is called \emph{\index{model!image
finite}image
finite} iff for each world $w$ the set $R(w)$ is finite. Then the
famous \index{Theorem!Hennessy-Milner}Hennessy-Milner Theorem says

\BeginTheorem{hennessy-milner}
If the models $\fmM$ and $\fmN$ are image finite, then modal
equivalent states are bisimilar.
\EndTheorem

\BeginProof
1.
Given two modal equivalent states $w^{*}$ and $x^{*}$, we have to find a
bisimulation $B$ with $\langle w^{*}, x^{*}\rangle\in B$. The only thing we
know about the states is that they are modally equivalent, hence that
they satisfy exactly the same formulas. This suggests to define
\begin{equation*}
  B := \{\langle w', x'\rangle \mid \text{ $w'$ and $x'$ are modally equivalent}\}
\end{equation*}
and to establish $B$ as a bisimulation. Since by assumption $\langle w^{*},
x^{*}\rangle\in B$, this will then prove the claim.

2.  If $\langle w, x\rangle \in B$, then both satisfy the same atomic
propositions by the definition of modal equivalence. Now let $\langle
w, x\rangle \in B$ and $\isEquiv{w}{w'}{R}$. Assume that we cannot
find $x'$ with $\isEquiv{x}{x'}{S}$ and $\langle w', x'\rangle \in
B$. We know that $\fmM, w\models \Diamond\top$, because this says
that there exists a successor to $w$, viz., $w'$. Since $w$ and $x$
satisfy the same formulas, $\fmN, x\models \Diamond\top$ follows,
hence $S(x)\not=\emptyset$. Let $S(x) = \{x_{1}, \dots,
x_{k}\}$. Then, since $w$ and $x_{i}$ are not modally equivalent, we
can find for each $x_{i}\in S(x)$ a formula $\psi_{i}$ such that
$\fmM, w'\models \psi_{i}$, but $\fmN, x_{i}\not\models
\psi_{i}$,. Hence $\fmM, w\models
\Diamond(\psi_{1}\wedge\dots\wedge\psi_{k})$, but $\fmN, w\not\models
\Diamond(\psi_{1}\wedge\dots\wedge\psi_{k})$. This is a contradiction,
so the assumption is false, and we can find $x'$ with
$\isEquiv{x}{x'}{S}$ and $\langle w', x'\rangle\in B$.

The other conditions for a bisimulation are shown in exactly the same way.
\EndProof


Neighborhood models can be compared through morphisms as well. Recall that the functor $\funV$ underlies a neighborhood frame,see Example~\ref{upper-closed-functor}.

\BeginDefinition{def-nghb-morph}
Let $\calN = (W, N, V)$ and $\calM = (X, M, Y)$ be neighborhood models for the basic modal language. A map $f: W\to X$ is called a \emph{\index{neighborhood morphism}\index{neighborhood!morphism}neighborhood morphism} $f: \calN\to \calM$ iff 
\begin{itemize}
\item $N\circ f = (\funV{f})\circ M$,
\item $V = f^{-1}\circ Y$.
\end{itemize}
\EndDefinition

A neighborhood morphism is a morphism for the neighborhood frame
(the definition of which is straightforward), respecting the validity of
atomic propositions. In this way, the definition follows the pattern
laid out for morphisms of Kripke models. 

Expanding the definition, $f: \calN\to \calM$ is a neighborhood
morphism iff $B\in N(f(w))$ iff $\InvBild{f}{B}\in M(w)$ for all
$B\subseteq X$ and all worlds $w\in W$, and iff $V(p) =
\InvBild{f}{Y(p)}$ for all atomic sentences $p\in\Phi$. Morphisms for
neighborhood models preserve validity in the same way as morphisms for
Kripke models do:

\BeginProposition{morphs-preserve-valid-nghb}
Let $f: \calN\to \calM$ be a neighborhood morphism for the neighborhood models $\calN = (W, N, V)$ and $\calM = (X, M, Y)$. Then
\begin{equation*}
  \calN, w\models\phi \Leftrightarrow \calM, f(w)\models\phi
\end{equation*}
for all formulas $\phi$ and for all states $w\in W$.
\EndProposition

\BeginProof
The proof proceeds by induction on the structure of formula
$\phi$. The induction starts with $\phi$ an atomic proposition.The
assertion is true in this case because of atomic harmony, see the
proof of Proposition~\ref{morphs-preserve-valid}. We pick only the
interesting modal case for the induction step. Hence assume the
assertion is established for formula $\phi$, then
\begin{align*}
  \calM, f(w)\models\Box\phi 
& \Leftrightarrow
\Gilt_{\calM}\in M(f(w))&& \text{ (by definition)}\\
& \Leftrightarrow 
\InvBild{f}{\Gilt_{\calM}}\in N(w)&&\text{ ($f$ is a morphism)}\\
& \Leftrightarrow
\Gilt_{\calN}\in N(w)&&\text{ (by induction hypothesis)}\\
& \Leftrightarrow
\calN, w \models \Box\phi
\end{align*}
\EndProof

We will not pursue this observation further at this point but rather
turn to the construction of a canonic model. When we will discuss
coalgebraic logics, however, this striking structural similarity of
models and their morphisms will be shown to be the instance of more
general phenomenon.

Before proceeding, we introduce the notion of a
\emph{\index{substitution}substitution}, which is a map $\sigma:
\Phi\to {\cal L}(\tau, \Phi)$. We extend a substitution in a
natural way to formulas. Define by induction on the structure of a
formula
\begin{align*}
  p^{\sigma} & := \sigma(p),\text{ if } p\in\Phi,\\
(\neg\phi)^{\sigma} & := \neg(\phi^{\sigma}),\\
(\phi_{1}\wedge\phi_{2})^{\sigma} & := \phi_{1}^{\sigma}\wedge\phi_{2}^{\sigma},\\
\bigl(\Delta(\phi_{1}, \dots, \phi_{k})\bigr)^{\sigma} & := \Delta(\phi_{1}^{\sigma}, \dots, \phi_{k}^{\sigma}), \text{ if }\Delta \in{O}\text{ with } \rho(\Delta) = k.
\end{align*}

\Subsubsection{The Lindenbaum Construction}
\label{sec:lind-constr}

We will show now how we obtain from a set of formulas a model which
satisfies exactly these formulas. The scenario is the basic modal
language, and it is clear that not every set of formulas is in a
position to generate such a model. 

Let $\Lambda$ be a set of formulas,
then we say that
\begin{itemize}
\item $\Lambda$ is \emph{closed under \index{logic!closed!modus
      ponens}modus ponens} iff $\phi\in \Lambda$ and $\phi\to \psi$
  together imply $\psi\in \Lambda$; 
\item $\Lambda$ is \emph{closed under
    \index{logic!closed!uniform substitution}uniform substitution} iff
  given $\phi\in\Lambda$ we may conclude that
  $\phi^{\sigma}\in\Lambda$ for all substitutions
  $\sigma$.
\end{itemize}

These two closure properties turn out to be crucial for the
generation of a model from a set of formulas. Those sets which
satisfy them will be called modal logics, to be precise:

\BeginDefinition{modal-logic-lang}
Let $\Lambda$ be a set of formulas of the basic modal
language. $\Lambda$ is called a \emph{\index{logic!modal}modal logic} iff these conditions are
satisfied:
\begin{enumerate}
\item $\Lambda$ contains all propositional tautologies.
\item $\Lambda$ is closed under modus ponens and under uniform
  substitution. 
\end{enumerate}
If formula $\phi\in \Lambda$, then $\phi$ is called a theorem of
$\Lambda$; we write this as $\vdash_{\Lambda}\phi$. 
\EndDefinition

\BeginExample{modal-logics}
These are some instances of elementary properties for modal logics.
\begin{enumerate}
\item If $\Lambda_{i}$ is a modal logic for each $i\in
  I\not=\emptyset$, then $\bigcap_{i\in I}\Lambda_{i}$ is a modal
  logic. This is fairly easy to check.
\item We say for a formula $\phi$ and a
  frame $\fmF$ over $W$ as a set of states that $\phi$ \emph{holds in this
  frame} (in symbols $\fmF\models\phi$) iff
  $\fmM, w\models \phi$ for each $w\in W$ and each model $\fmM$ which
  is based on $\fmF$. Let $\mathbb{S}$ be a class of frames, then
  \begin{equation*}
    \Lambda_{\mathbb{S}} := \bigcap_{\fmF\in \mathbb{S}}\{\phi\mid \fmF\models\phi\}
  \end{equation*}
is a modal logic. We abbreviate $\phi\in\Lambda_{\mathbb{S}}$ by
$\mathbb{S}\models\phi$. 
\item Define similarly $\fmM\models\phi$ for a model $\fmM$ iff $\fmM,
  w\models\phi$ for each world $w$ of $\fmM$. Then put for a class $\mathbb{M}$
  of models 
  \begin{equation*}
    \Lambda_{\mathbb{M}} := \bigcap_{\fmM\in \mathbb{M}}\{\phi\mid \fmM\models\phi\}.
  \end{equation*}
Then there are sets $\mathbb{M}$ for which $\Lambda_{\mathbb{M}}$ is
not a modal language. In fact, take a model $\fmM$ with world $W$ and two
propositional letters $p, q$ with $V(p) = W$ and $V(q)\not= W$, then
$\fmM, w\models p$ for all $w$, hence $\fmM\models p$, but
$\fmM\not\models q$. On the other hand, $q = p^{\sigma}$ under the
substitution $\sigma: p\mapsto q$. Hence $\Lambda_{\{\fmM\}}$ is not
closed under uniform substitution. 
\end{enumerate}
\EndExample

This formalizes the notion of deduction:

\BeginDefinition{deducible}
Let $\Lambda$ be a logic, and $\Gamma\cup\{\phi\}$ a set of modal
formulas. 
\begin{itemize}
\item $\phi$ is \emph{deducible} in $\Lambda$ from $\Gamma$ iff either
  $\vdash_{\Lambda}$, or if there exist formulas $\psi_{1}, \dots,
  \psi_{k}\in\Gamma$ such that $\vdash_{\Lambda}
  (\psi_{1}\wedge\dots\wedge\psi_{k})\to \phi$. We write this down as
  $\Gamma\vdash_{\Lambda} \phi$. 
\item $\Gamma$ is
  \emph{$\Lambda$-\index{consistent!$\Lambda$}consistent} iff
  $\Gamma\not\vdash_{\Lambda}\bot$, otherwise $\Gamma$ is called\emph{
    $\Lambda$-inconsistent}.
\item  $\phi$ is called
  $\Lambda$-consistent iff $\{\phi\}$ is $\Lambda$-consistent.
\end{itemize}
\EndDefinition

This is a simple and intuitive criterion for inconsistency. We fix for
the discussions below a modal logic $\Lambda$. 

\BeginLemma{crit-inconsist}
Let $\Gamma$ be a set of formulas. Then
these statements are equivalent
\begin{enumerate}
\item\label{crit-inconsist-1} $\Gamma$ is $\Lambda$-inconsistent.
\item\label{crit-inconsist-2}
  $\Gamma\vdash_{\Lambda}\phi\wedge\neg\phi$ for some formula $\phi$.
\item\label{crit-inconsist-3} $\Gamma\vdash_{\Lambda}\psi$ for all
  formulas $\psi$.
\end{enumerate}
\EndLemma

\BeginProof
\labelImpl{crit-inconsist-1}{crit-inconsist-2}: Because
$\Gamma\vdash_{\Lambda}\bot$, we know that
$\psi_{1}\wedge\dots\wedge\psi_{k}\to \bot$ is in $\Lambda$ for some
formulas $\psi_{1}, \dots, \psi_{k}\in\Gamma$. But $\bot\to
\phi\wedge\neg\phi$ is a tautology, hence
$\Gamma\vdash_{\Lambda}\phi\wedge\neg\phi$. 

\labelImpl{crit-inconsist-2}{crit-inconsist-3}: By assumption there
exists $\psi_{1}, \dots, \psi_{k}\in\Gamma$ such that
$\vdash_{\Lambda}\psi_{1}\wedge\dots\wedge\psi_{k}\to\phi\wedge\neg\phi$,
and $\phi\wedge\neg\phi \to \psi$ is a tautology for an arbitrary
formula $\psi$, hence $\vdash_{\Lambda}\phi\wedge\neg\phi \to
\psi$. Thus $\Gamma\vdash_{\Lambda}\psi$. 

\labelImpl{crit-inconsist-3}{crit-inconsist-1}: We have in particular
$\Gamma\vdash_{\Lambda}\bot$. 
\EndProof

$\Lambda$-consistent sets have an interesting compactness property.

\BeginLemma{compact-lambda}
A set $\Gamma$ of formulas is $\Lambda$-consistent iff each finite
subset is $\Gamma$ is $\Lambda$-consistent. 
\EndLemma

\BeginProof
If $\Gamma$ is $\Lambda$-consistent, then certainly each finite subset
if. If, on the other hand, each finite subset is $\Lambda$-consistent,
then the whole set must be consistent, since consistency is tested
with finite witness sets. 
\EndProof

Proceeding on our path to finding a model for a modal logic, we define
normal logics. They are closed under some properties which appear as
fairly, well, normal, so it is not surprising that these normal logics
will play an important r\^ole.

\BeginDefinition{normal-modal-logics}
Modal logic $\Lambda$ is called \emph{\index{modal logic!normal}\index{logic!modal!normal}normal} iff it satisfies these
conditions for all propositional letters $p, q\in\Phi$ and all formulas $\phi$:
\begin{description}
\item[(K)] $\vdash_{\Lambda}\Box(p\to q)\to (\Box p\to \Box q)$,
\item[(D)] $\vdash_{\Lambda}\Diamond p \leftrightarrow \neg\Box\neg
  p$,
\item[(G)] If $\vdash_{\Lambda}\phi$, then $\vdash_{\Lambda}\Box\phi$.
\end{description}
\EndDefinition

Property \textbf{(K)} states that if it is necessary that $p$ implies $q$, then
the fact that $p$ is necessary will imply that $q$ is necessary. Note
that the formulas in $\Lambda$ do not have a semantics yet, they are for
the time being just syntactic entities. Property \textbf{(D)} connects the
constructors $\Diamond$ and $\Box$ in the desired manner. Finally, \textbf{(G)}
states that, loosely speaking, if something is the case, then is is
necessarily the case. We should finally note that \textbf{(K)} and \textbf{(D)} are both
formulated for propositional letters only. This, however, is
sufficient for modal logics, since they are closed under uniform
substitution. 

In a normal logic, the equivalence of formulas is preserved by the
diamond.

\BeginLemma{diamond-preserves-equiv}
Let $\Lambda$ be a normal modal logic, then
$\vdash_{\Lambda}\phi\leftrightarrow\psi$ implies
$\vdash_{\Lambda}\Diamond\phi\leftrightarrow\Diamond\psi$. 
\EndLemma

\BeginProof
We show that $\vdash_{\Lambda}\phi\to \psi$ implies
$\vdash_{\Lambda}\Diamond\phi\to \Diamond\psi$, the rest will follow
in the same way. 
\begin{align*}
  \vdash_{\Lambda}\phi\to \psi 
& \Rightarrow 
\vdash_{\Lambda}\neg\psi\to \neg\phi&&\text{ (contraposition)}\\
& \Rightarrow
\vdash_{\Lambda}\Box(\neg\psi\to \neg\phi) &&\text{ (by \textbf{(G)})}\\
& \Rightarrow
\vdash_{\Lambda}(\Box(\neg\psi\to \neg\phi))\to (\Box\neg\psi\to
\Box\neg\phi)&&\text{ (uniform substitution, \textbf{(K)})}\\
& \Rightarrow
\vdash_{\Lambda}\Box\neg\psi\to\Box\neg\phi&&\text{ (modus ponens)}\\
& \Rightarrow
\vdash_{\Lambda}\neg\Box\neg\phi\to\neg\Box\neg\psi&&\text{ (contraposition)}\\
& \Rightarrow
\vdash_{\Lambda}\Diamond\phi\to \Diamond\psi&&\text{ (by \textbf{(D)})}
\end{align*}
\EndProof

Let us define a semantic counterpart to $\Gamma\vdash_{\Lambda}$. Let
$\fmF$ be a frame and $\Gamma$ be a set of formulas, then we say that
$\Gamma$ holds on $\fmF$ (written as $\fmF\models\Gamma$) iff each
formula in $\Gamma$ holds in each model which is based on frame $\fmF$
(see Example~\ref{modal-logics}). We say that $\Gamma$ entails formula
$\phi$ ($\Gamma\models_{\fmF}\phi$) iff $\fmF\models\Gamma$ implies
$\fmF\models\phi$. This carries over to classes of frames in an
obvious way. Let $\mathbb{S}$ be a class of frames, then
$\Gamma\models_{\mathbb{S}}\phi$ iff we have $\Gamma\models_{\fmF}\phi$
for all frames $\fmF\in \mathbb{S}$.

\BeginDefinition{class-of-frames-soundness}
Let $\mathbb{S}$ be a class of frames, then the normal logic $\Lambda$ is
called \emph{$\mathbb{S}$-sound} iff $\Lambda\subseteq\Lambda_{\mathbb{S}}$. If $\Lambda$ is
$\mathbb{S}$-sound, then $\mathbb{S}$ is called a \emph{class of
  \index{frames!class of}frames for $\Lambda$}. 
\EndDefinition
 
Note that $\mathbb{S}$-soundness indicates that $\vdash_{\Lambda}\phi$
implies $\fmF\models\phi$ for all frames $\fmF\in\mathbb{S}$ and for all
formulas $\phi$. 

This example dwells on traditional names.

\BeginExample{k4-vs-4}
Let $\Lambda_{4}$ be the smallest modal logic which contains
$\Diamond\Diamond p\to \Diamond p$ (if it is possible that $p$ is
possible, then $p$ is possible), and let $K4$ be
the class of transitive frames. Then $\Lambda_{4}$ is $K4$-sound. In
fact, it is easy to see that $\fmM, w\models \Diamond\Diamond p\to
\Diamond p$ for all worlds $w$, whenever $\fmM$ is a model the frame
of which carries a transitive relation. 
\EndExample

Thus $\mathbb{S}$-soundness permits us to conclude that a formula which is
deducible from $\Gamma$ holds also in all frames from $\mathbb{S}$.
Completeness goes the other way: roughly, if we know that a formula holds in a
class of frames, then it is deducible. To be more precise:

\BeginDefinition{S-complete}
Let $\mathbb{S}$ be a class of frames and $\Lambda$ a normal modal logic. 
\begin{enumerate}
\item $\Lambda$ is \emph{strongly $\mathbb{S}$-\index{complete!strongly $\mathbb{S}$}complete} iff for any set
  $\Gamma\cup\{\phi\}$ of formulas $\Gamma\models_{\mathbb{S}}\phi$
  implies $\Gamma\vdash_{\Lambda}\phi$.
\item $\Lambda$ is \emph{weakly $\mathbb{S}$-\index{complete!weakly $\mathbb{S}$}complete} iff 
  $\mathbb{S}\models\phi$ implies $\vdash_{\Lambda}\phi$ for any
  formula $\phi$. 
\end{enumerate}
\EndDefinition

This is a characterization of completeness.

\BeginProposition{complete-vs-weakly-complete}
Let $\Lambda$ and $\mathbb{S}$ be as above. 
\begin{enumerate}
\item $\Lambda$ is strongly $\mathbb{S}$-complete iff every
  $\Lambda$-consistent set of formulas is satisfiable for some
  $\fmF\in\mathbb{S}$.
\item $\Lambda$ is weakly $\mathbb{S}$-complete iff every
  $\Lambda$-consistent formula is satisfiable for some
  $\fmF\in\mathbb{S}$. 
\end{enumerate}
\EndProposition

\BeginProof
1.
If $\Lambda$ is not strongly $\mathbb{S}$-complete, then we can find a
set $\Gamma$ of formulas and a formula $\phi$ with
$\Gamma\models_{\mathbb{S}}\phi$, but
$\Gamma\not\vdash_{\Lambda}\phi$. Then $\Gamma\cup\{\neg\phi\}$ is
$\Lambda$-consistent, but this set cannot be satisfied on
$\mathbb{S}$. So the condition for strong completeness is
sufficient. It is also necessary. In fact, we may assume by
compactness that $\Gamma$ is finite. Thus by consistency
$\Gamma\not\vdash_{\Lambda}\bot$, hence
$\Gamma\not\models_{\mathbb{S}}\bot$ by completeness, thus there exists
a frame $\fmF\in\mathbb{S}$ with $\fmF\models\Gamma$ but
$\fmF\not\models\bot$. 

2.
This is but a special case of cardinality 1. 
\EndProof

Consistent sets are not yet sufficient for the construction of a
model, as we will see soon. We need consistent sets which cannot be
extended further without jeopardizing their consistency. To be
specific:

\BeginDefinition{max-consistent}
The set $\Gamma$ of formulas is \emph{maximal $\Lambda$-\index{consistent!maximal $\Lambda$}consistent} iff
$\Gamma$ is $\Lambda$-consistent, and it is is not properly contained
in a $\Lambda$ consistent set. 
\EndDefinition

Thus if we have a maximal $\Lambda$-consistent set $\Gamma$, and if we
know that $\Gamma\subset\Gamma_{0}$ with $\Gamma\not=\Gamma_{0}$, then
we know that $\Gamma_{0}$ is not $\Lambda$-consistent. This criterion
is sometimes a bit unpractical, but we have

\BeginLemma{max-cons-crit}
Let $\Lambda$ be a normal logic and $\Gamma$ be a maximally
$\Lambda$-consistent set of formulas. Then
\begin{enumerate}
\item\label{max-cons-crit-1} $\Gamma$ is closed under modus ponens.
\item\label{max-cons-crit-2} $\Lambda\subseteq\Gamma$.
\item\label{max-cons-crit-3} $\phi\in\Gamma$ or $\neg\phi\in\Gamma$ for all formulas $\phi$.
\item\label{max-cons-crit-4} $\phi\vee\psi\in\Gamma$ iff $\phi\in\Gamma$ or $\psi\in\Gamma$
  for all formulas $\phi, \psi$. 
\item\label{max-cons-crit-5} $\phi_{1}\wedge\phi_{2}\in\Gamma$ if $\phi_{1}, \phi_{2}\in\Gamma$.
\end{enumerate}
\EndLemma

\BeginProof
1.
Assume that $\phi\in\Gamma$ and $\phi\to \psi\in\Gamma$, but
$\psi\not\in\Gamma$. Then $\Gamma\cup\{\psi\}$ is inconsistent, hence
$\Gamma\cup\{\psi\}\vdash_{\Lambda}\bot$ by
Lemma~\ref{crit-inconsist}. Thus we can find formulas $\psi_{1},
\dots, \psi_{k}\in\Gamma$ such that
$\vdash_{\Lambda}\psi\wedge\psi_{1}\wedge\dots\wedge\psi_{k}\to
\bot$. Because
$\vdash_{\Lambda}\phi\wedge\psi_{1}\wedge\dots\wedge\psi_{k}\to
\psi\wedge\psi_{1}\wedge\dots\wedge\psi_{k}$, we conclude
$\Gamma\vdash_{\Lambda}\bot$. This contradicts $\Lambda$-consistency
by Lemma~\ref{crit-inconsist}.

2.
In order to show that $\Lambda\subseteq\Gamma$, we assume that there
exists $\psi\in\Lambda$ such that $\psi\not\in\Gamma$, then
$\Gamma\cup\{\psi\}$ is inconsistent, hence
$\vdash_{\Lambda}\psi_{1}\wedge\dots\wedge\psi_{k}\to \neg\psi$ for
some $\psi_{1}, \dots, \psi_{k}\in\Lambda$ (here we use
$\Gamma\cup\{\psi\}\vdash_{\Lambda}\psi$ and
Lemma~\ref{crit-inconsist}). By propositional logic,
$\vdash_{\Lambda}\psi\to \neg(\psi_{1}\wedge\dots\wedge\psi_{k})$,
hence $\psi\in\Lambda$ implies
$\Gamma\vdash_{\Lambda}\neg(\psi_{1}\wedge\dots\wedge\psi_{k})$. But
$\Gamma\vdash_{\Lambda}\psi_{1}\wedge\dots\wedge\psi_{k}$,
consequently, $\Gamma$ is $\Lambda$-inconsistent. 

3.
If both $\phi\not\in\Gamma$ and $\neg\phi\not\in\Gamma$, $\Gamma$ is
$\Lambda$-inconsistent. 

4.
Assume first that $\phi\vee\psi\in\Gamma$, but $\phi\not\in\Gamma$ and
$\psi\not\in\Gamma$, hence both $\Gamma\cup\{\phi\}$ and
$\Gamma\cup\{\psi\}$ are inconsistent. Thus we can find $\psi_{1},
\dots, \psi_{k}, \phi_{1}, \dots, \phi_{n}\in \Gamma$ with
$\vdash_{\Lambda}\psi_{1}\wedge\dots\wedge\psi_{k}\to \neg\psi$ and
$\vdash_{\Lambda}\phi_{1}\wedge\dots\wedge\phi_{n}\to \neg\phi$. This
implies 
$\vdash_{\Lambda}\psi_{1}\wedge\dots\wedge\psi_{k}\wedge
\phi_{1}\wedge\dots\wedge\phi_{n}\to \neg\psi\wedge\neg\phi$, and by
arguing propositionally, 
$\vdash_{\Lambda}(\psi\vee\phi)\wedge\psi_{1}\wedge\dots\wedge\psi_{k}\wedge
\phi_{1}\wedge\dots\wedge\phi_{n}\to \bot$, which contradicts
$\Lambda$-consistency of $\Gamma$. For the converse, assume that $\phi\in\Gamma$. Since $\phi\to \phi\vee\psi$ is a tautology, we obtain $\phi\vee\psi$ from modus ponens. 

5. Assume $\phi_{1}\wedge\phi_{2}\not\in\Gamma$, then $\neg\phi_{1}\vee\neg\phi_{2}\in\Gamma$ by part~\ref{max-cons-crit-3} Thus $\neg\phi_{1}\in\Gamma$ or $\neg\phi_{2}\in\Gamma$ by part~\ref{max-cons-crit-4}, hence $\phi_{1}\not\in\Gamma$ or $\phi_{2}\not\in\Gamma$.  
\EndProof

Hence consistent sets have somewhat convenient properties, but how do
we construct them? The famous Lindenbaum Lemma states that we may
obtain them by enlarging consistent sets. 

From now on we fix a normal modal logic $\Lambda$.

\BeginLemma{Lindenbaum}
If $\Gamma$ is a $\Lambda$-consistent set, then there exists a maximal
$\Lambda$-consistent set $\Gamma^{+}$ with
$\Gamma\subseteq\Gamma^{+}$. 
\EndLemma

We will give two proofs for the Lindenbaum Lemma, depending on the
cardinality of the set of all formulas. If the set $\Phi$ of
propositional letters is countable, the set of all formulas is
countable as well, so the first proof may be applied. If, however, we
have more than a countable number of formulas, then this proof will
fail to exhaust all formulas, and we have to apply another method, in
this case transfinite induction (in the disguise of Tuckey's Lemma).

\BeginProof (First --- countable case)
Assume that the set of all formulas is countable, and let
$\{\phi_{n}\mid n\in \Nat\}$ be an enumeration of them. Define by
induction 
\begin{align*}
  \Gamma_{0} & := \Gamma,\\
\Gamma_{n+1} & := \Gamma_{n}\cup \{\psi_{n}\},
\end{align*}
where
\begin{equation*}
\psi_{n} :=
  \begin{cases}
    \phi_{n}, & \text{ if $\Gamma_{n}\cup\{\phi_{n}\}$ is
      consistent,}\\
\neg\phi_{n}, & \text{ otherwise.}
  \end{cases}
\end{equation*}
Put
\begin{equation*}
  \Gamma^{+} := \bigcup_{n\in\Nat}\Gamma_{n}.
\end{equation*}
Then these properties are easily checked:
\begin{itemize}
\item $\Gamma_{n}$ is consistent for all $n\in\Nat_{0}$.
\item Either $\phi\in\Gamma^{+}$ or $\neg\phi\in\Gamma^{+}$ for all
  formulas $\phi$.
\item If $\Gamma^{+}\vdash_{\Lambda}\phi$, then $\phi\in\Gamma^{+}$.
\item $\Gamma^{+}$ is maximal.
\end{itemize}
\EndProof

\BeginProof (Second --- general case)
Let
\begin{equation*}
  \mathbb{C} := \{\Gamma' \mid \Gamma'\text{ is $\Lambda$-consistent
    and }\Gamma\subseteq\Gamma'\}.
\end{equation*}
Then $\mathbb{C}$ contains $\Gamma$, hence $\mathbb{C}\not=\emptyset$,
and $\mathbb{C}$ is ordered by inclusion. By Tuckey's Lemma, it
contains a maximal chain $\mathbb{C}_{0}$. Let $\Gamma^{+}:=
\bigcup\mathbb{C}_{0}$. Then $\Gamma^{+}$ is a $\Lambda$-consistent set which
contains $\Gamma$ as a subset. While the latter is evident, we have
to take care of the former. Assume that $\Gamma^{+}$ is not
$\Lambda$-consistent, hence
$\Gamma^{+}\vdash_{\Lambda}\phi\wedge\neg\phi$ for some formula
$\phi$. Thus we can find $\psi_{1}, \dots, \psi_{k}\in\Gamma^{+}$ with
$\vdash_{\Lambda}\psi_{1}\wedge\dots\wedge\psi_{k}\to
\phi\wedge\neg\phi$. Given $\psi_{i}\in\Gamma^{+}$, we can find
$\Gamma_{i}\in\mathbb{C}_{0}$ with $\psi_{i}\in\Gamma_{i}$. Since
$\mathbb{C}_{0}$ is linearly ordered, we find some $\Gamma'$ among
them such that $\Gamma_{i}\subseteq\Gamma'$ for all $i$. Hence
$\psi_{1}, \dots, \psi_{k}\in\Gamma'$, so that $\Gamma'$ is not
$\Lambda$-consistent. This is a contradiction. Now assume that
$\Gamma^{+}$ is not maximal, then there exists $\phi$ such that
$\phi\not\in\Gamma^{+}$ and $\neg\phi\not\in\Gamma^{+}$. If
$\Gamma^{+}\cup\{\phi\}$ is not consistent,
$\Gamma^{+}\cup\{\neg\phi\}$ is, and vice versa, so either one of
$\Gamma^{+}\cup\{\phi\}$ and $\Gamma^{+}\cup\{\neg\phi\}$ is
consistent. But this means that $\mathbb{C}_{0}$ is not maximal. 
\EndProof

We are in a position to construct a model now, specifically, we will define a set of
states, a transition relation and the validity sets for the
propositional letters. Put
\begin{align*}
  W^{\sharp} & := \{\Sigma \mid \Sigma \text{ is $\Lambda$-consistent
    and maximal}\},\\
R^{\sharp} & := \{\langle w, v\rangle\in W^{\sharp}\times W^{\sharp}
\mid \text{for all formulas }\psi, \psi\in v\text{
  implies }\Diamond\psi\in w\},\\
V^{\sharp}(p) & := \{w\in W^{\sharp}\mid p\in w\} \text{ for }p\in\Phi.
\end{align*}

Then $\fmM^{\sharp} := (W^{\sharp}, R^{\sharp}, V^{\sharp})$ is called
the \emph{\index{model!canonical}canonical model} for $\Lambda$. 

This is another view of relation $R^{\sharp}$:
\BeginLemma{another-r-sharp}
Let $v, w\in W^{\sharp}$, then $wR^{\sharp}v$ iff $\Box\psi\in w$
implies $\psi\in v$ for all formulas $\psi$.
\EndLemma

\BeginProof
1.
Assume that $\langle w, v\rangle\in R^{\sharp}$, but that $\psi\not\in
v$ for some formula $\psi$. Since $v$ is maximal, we conclude from
Lemma~\ref{max-cons-crit} that $\neg\psi\in v$, hence the definition
of $R^{\sharp}$ tells us that $\Diamond\neg\psi\in w$, which in turn
implies by the maximality of $w$ that $\neg\Diamond\neg\psi\not\in w$,
hence $\Box\psi\not\in w$. 

2.
If $\Diamond\psi\not\in w$, then by maximality $\neg\Diamond\psi\in
w$, so $\Box\neg\psi\in w$, which means by assumption that
$\neg\psi\in v$. Hence $\psi\not\in v$. 
\EndProof

The next lemma gives a more detailed look at the transitions which are
modelled by $R^{\sharp}$.

\BeginLemma{look-at-transitions}
Let $w\in W^{\sharp}$ with $\Diamond\phi\in w$. Then there exists a
state $v\in W^{\sharp}$ such that $\phi\in v$ and $\isEquiv{w}{v}{R^{\sharp}}$. 
\EndLemma

\BeginProof
Because we can extend $\Lambda$-consistent sets to maximal consistent
ones by the \index{Lindenbaum Lemma}Lindenbaum Lemma~\ref{Lindenbaum}, it is enough to show
that 
$
v_{0} := \{\phi\}\cup\{\psi \mid \Box\psi\in w\}
$
is $\Lambda$-consistent. Assume it is not. Then we have 
$
\vdash_{\Lambda}(\psi_{1}\wedge\dots\wedge\psi_{k})\to \neg\phi
$
for some $\psi_{1}, \dots, \psi_{k}\in v_{0}$, from which we obtain with
\textbf{(G)} and \textbf{(K)} that 
$
\vdash_{\Lambda}\Box(\psi_{1}\wedge\dots\wedge\psi_{k})\to
\Box\neg\phi.
$
Because 
$
\Box\psi_{1}\wedge\dots\wedge\Box\psi_{k}\to
\Box(\psi_{1}\wedge\dots\wedge\psi_{k}),
$
this implies
$
\vdash_{\Lambda}\Box\psi_{1}\wedge\dots\wedge\Box\psi_{k}\to \Box\neg\phi.
$
Since $\Box\psi_{1}, \dots, \Box\psi_{k}\in w$, we conclude from Lemma~\ref{max-cons-crit} that $\Box\psi_{1}\wedge \dots\wedge\Box\psi_{k}\in w$, thus we have
$\Box\neg\phi\in w$ by modus ponens, hence $\neg\Diamond\phi\in w$. Since $w$ is
maximal, this implies $\Diamond\phi\not\in w$. This is a contradiction. So $v_{0}$ is consistent, thus there exists by the Lindenbaum Lemma a maximal consistent set $v$ with $v_{0}\subseteq v$. We have in particular $\phi\in v$, and we know that $\Box\psi\in w$ implies $\psi\in v$, hence $\langle w, v\rangle\in R^{\sharp}.$
\EndProof

This helps in characterizing the model, in particular the validity
relation $\models$ by the well-known Truth Lemma.

\BeginLemma{truth-lemma}
$\fmM^{\sharp}, w\models \phi$ iff $\phi\in w$
\EndLemma

\BeginProof
The proof proceeds by induction on formula $\phi$. The statement is
trivially true if $\phi=p\in \Phi$ is a propositional letter. The set
of formulas for which the assertion holds is certainly closed under
Boolean operations, so the only interesting case is the case that the
formula in question has the shape $\Diamond\phi$, and that the
assertion is true for $\phi$.

``$\Rightarrow$'': If $\fmM^{\sharp}, w\models \Diamond\phi$, then
we can find some $v$ with $\isEquiv{w}{v}{R^{\sharp}}$ and
$\fmM^{\sharp}, v\models \phi$. Thus there exists $v$ with $\langle w,
v\rangle\in R^{\sharp}$ such that $\phi\in v$ by hypothesis, which in
turn means $\Diamond\phi\in w$. 

``$\Leftarrow$'': Assume $\Diamond\phi\in w$, hence there exists
$v\in W^{\sharp}$ with $\isEquiv{w}{v}{R^{\sharp}}$ and $\phi\in v$,
thus $\fmM^{\sharp}, v\models \phi$. But this means $\fmM^{\sharp},
w\models \Diamond\phi$. 
\EndProof

Finally, we obtain

\BeginTheorem{completeness-normal}
Any normal logic is complete with respect to its canonical model. 
\EndTheorem

\BeginProof
Let $\Sigma$ be a $\Lambda$-consistent set for the normal logic
$\Lambda$. Then there exists by Lindenbaum's Lemma~\ref{Lindenbaum} a
maximal $\Lambda$-consistent set $\Sigma^{+}$ with
$\Sigma\subseteq\Sigma^{+}$. By the Truth Lemma we have now
$\fmM^{\sharp}, \Sigma^{+}\models \Sigma$. 
\EndProof

\Subsubsection{Coalgebraic Logics}
\label{sec:coalg-logic}

We have seen several points where coalgebras and modal logics touch
each other, for example, morphisms for Kripke models are based on
morphisms for the underlying $\PowerSenza$-coalgebra, as a comparison
of Example~\ref{ex-cat-transition-syst-bound} and
Lemma~\ref{char-frame-morph} demonstrates. Let $\fmM = (W, R, V)$ be a
Kripke model, then the accessibility relation $R \subseteq W\times W$
can be seen as a map, again denoted by $R$, with the signature $W\to
\PowerSet{W}$. Map $V: \Phi\to \PowerSet{W}$, which indicates the
validity of atomic propositions, can be decoded through a map $V_{1}:
W\to \PowerSet{\Phi}$ upon setting $ V_{1}(w) := \{p\in\Phi\mid w\in
V(p)\}.  $ Both $V$ and $V_{1}$ describe the same relation $ \{\langle
p, w\rangle \in \Phi\times W\mid \fmM, w \models p\}, $ albeit from
different angles. One can be obtained from the other one. This new
representation has the advantage of describing the model from
vantage point $w$.

Define $\funF X := \PowerSet{X}\times\PowerSet{\Phi}$ for the set $X$,
and put, given map $f:X\to Y$, $ (\funF f)(A, Q) := \langle
\Bild{f}{A}, Q\rangle = \langle(\PowerSenza{f}) A, Q\rangle $ for
$A\subseteq X, Q\subseteq\Phi$, then $\funF $ is an endofunctor on
$\catSET$. Hence we obtain from the Kripke model $\fmM$ the
$\funF$-coalgebra $(W, \gamma)$ with $\gamma(w) := R(w)\times
V_{1}(w)$. This construction can easily be reversed: given a
$\funF$-coalgebra $(W, \gamma)$, we put $R(w) := \pi_{1}(\gamma(w))$
and $V_{1}(w) := \pi_{2}(\gamma(w))$ and construct $V$ from $V_{1}$,
then $(W, R, V)$ is a Kripke model (here $\pi_{1}, \pi_{2}$ are the
projections). Thus Kripke models and $\funF$-coalgebras are in an
one-to-one correspondence with each other. This correspondence goes a
bit deeper, as can be seen when considering morphisms.

\BeginProposition{coalg-mod-morph}
Let $\fmM = (W, R, V)$ and $\fmN = (X, S, Y)$ be Kripke models with associated $\fmF$-coalgebras $(W, \gamma)$ resp. $(X, \delta)$. Then these statements are equivalent for a map $f: W\to X$
\begin{enumerate}
\item\label{coalg-mod-morph-2} $f: (W, \gamma)\to (X, \delta)$ is a morphism of coalgebras.
\item\label{coalg-mod-morph-1} $f:\fmM \to \fmN$ is a morphism of Kripke models.
\end{enumerate}
\EndProposition

\BeginProof
\labelImpl{coalg-mod-morph-2}{coalg-mod-morph-1}:
We obtain for each $w\in W$ from the defining equation $(\funF f)\circ \gamma = \delta\circ f$ these equalities
\begin{align*}
  \Bild{f}{R(w)} & = S(f(w)),\\
V_{1}(w) & = Y_{1}(f(w)).
\end{align*}
Since $\Bild{f}{R(w)} = (\PowerSenza{f})(R(w))$, we conclude that $(\PowerSenza{f})\circ R = S\circ f$, so $f$ is a morphism of the $\PowerSenza$-coalgebras. We have moreover for each atomic sentence $p\in \Phi$
\begin{equation*}
  w \in V(p) \Leftrightarrow p \in V_{1}(w) \Leftrightarrow p \in Y_{1}(f(w)) \Leftrightarrow f(w)\in Y(p).
\end{equation*}
This means $V = f^{-1}\circ Y$, so that $f: \fmM\to \fmN$ is a morphism. 

\labelImpl{coalg-mod-morph-1}{coalg-mod-morph-2}:
Because we know that $S\circ f = (\PowerSenza{f})\circ R$, and because one shows as above that $V_{1} = Y_{1}\circ f$, we obtain for $w\in W$
\begin{align*}
  (\delta\circ f)(w) & = \langle S(f(w)), Y_{1}(f(w))\rangle\\
& = \langle (\PowerSenza{f})(R(w)), V_{1}(w)\rangle\\
& = \bigl((\funF f)\circ \gamma\bigr)(w).
\end{align*}
Hence $f: (W,\gamma)\to (X, \delta)$ is a morphism for the $\funF$-coalgebras. 
\EndProof
 
Given a world $w$, the value of $\gamma(w)$ represents the worlds
which are accessible from $w$, making sure that the validity of
the atomic propositions is maintained; recall that they are not affected by a
transition. This information is to be extracted in a variety of
ways. We need predicate liftings for this.

Before we define liftings, however, we observe that the same mechanism works for neighborhood models.

\BeginExample{nbhmod-as-coalgs}
Let $\calN = (W, N, V)$ be a neighborhood model. Define functor $\funG$ by putting $\funG(X) := \funV(X)\times \PowerSet{\Phi}$ for sets, and if $f: X\to Y$ is a map, put $(\funG f)(U, Q) := \langle (\funV f) U, Q\rangle$. Then $\funG$ is an endofunctor on $\catSET$. The $\funG$-coalgebra $(W, \nu)$ associated with $\calN$ is defined through $\nu(w) := \langle N(w), V_{1}(w)\rangle$ (with $V_{1}$ defined through $V$ as above). 

Let $\calM = (X, M, Y)$ be another neighborhood model with associated coalgebra $(X, \mu)$. Exactly the same proof as the one for Proposition~\ref{coalg-mod-morph} shows that $f: \calN\to \calM$ is a neighborhood morphism iff $f: (W, \nu)\to (X, \mu)$ is a coalgebra morphism. 
\EndExample

{ \def\PSo{\PowerSenza^{op}} 
  
  Proceeding to define predicate liftings, let $\PowerSenza^{op}: \catSET\to \catSET$ be the
  contravariant power set functor, i.e., given the set $X$,
  $\PowerSenza^{op}(X)$ is the power set $\PowerSet{X}$ of $X$, and if
  $f: X\to Y$ is a map, then $(\PowerSenza^{op} f):
  \PowerSenza^{op}(Y)\to  \PowerSenza^{op}(Y)$ works as $B \mapsto
  \InvBild{f}{B}$.

\BeginDefinition{predicate-lifting}
Given a (covariant) endofunctor $\funT$ on $\catSET$, a
\emph{\index{predicate lifting}\index{logic!coalgebraic!predicate
    lifting}predicate lifting $\lambda$ for $\funT$} is a monotone
natural transformation $\lambda: \PSo\to \PSo\circ T$.
\EndDefinition

Interpret $A\in\PSo(X)$ as a predicate on $X$, then
$\lambda_{X}(A)\in\PSo(\funT X)$ is a predicate on $\funT X$, hence
$\lambda_{X}$ lifts the predicate into the realm of functor $\funT$;
the requirement of naturalness is intended to reflect compatibility
with morphisms, as we will see below. Thus a predicate lifting helps
in specifying a requirement on the level of sets, which it then
transports onto the level of those sets that are controlled by functor
$\funT$. Technically, this requirement means that this diagram
commutes, whenever $f: X\to Y$ is a map:
\begin{equation*}
\xymatrix{
\PowerSenza X\ar[rr]^{\lambda_{X}}&&\PowerSenza(\funT X)\\
\PowerSenza Y\ar[u]^{f^{-1}}\ar[rr]_{\lambda_{Y}} && \PowerSenza(\funT Y)\ar[u]_{(\funT f)^{-1}}
}
\end{equation*}
Hence we have $\lambda_{X}(\InvBild{f}{G}) = \InvBild{(\funT f)}{\lambda_{Y}(G)}$ for any $G\subseteq Y$. 

Finally, monotonicity says that
$\lambda_{X}(D)\subseteq\lambda_{X}(E)$, whenever $D\subseteq
E\subseteq X$; this condition models the requirement that informations
about states should only depend on their precursors. Informally it is
reflected in the rule $\vdash (\phi\to \psi)\to (\Box\phi\to
\Box\psi)$ 

This example illuminates the idea.

\BeginExample{pred-lift-for-modLog}
Let $\funF = \PowerSet{-}\times\PowerSenza{\Phi}$ be defined as above, put for the set $X$ and for $D\subseteq X$
\begin{equation*}
  \lambda_{X}(D) := \{\langle D', Q\rangle \in\PowerSet{X}\times\PowerSet{\Phi}\mid D'\subseteq D\}.
\end{equation*}
This defines a predicate lifting $\lambda: \PSo\to \PSo\circ
\funF$. In fact, let $f: X\to Y$ be a map and $G\subseteq Y$, then
\begin{align*}
  \lambda_{X}(\InvBild{f}{G}) 
& =
\{\langle D', Q\rangle \mid  D'\subseteq \InvBild{f}{G}\}\\
& =
\{\langle D', Q\rangle \mid \Bild{f}{D'}\subseteq G\}\\
& =
\InvBild{(\funF f)}{\{\langle G', Q\rangle\in\PowerSet{Y}\times\PowerSet{\Phi}\mid G'\subseteq G\}}\\
& =
\InvBild{(\funF f)}{\lambda_{Y}(G)}
\end{align*}
(remember that $\funF f$ leaves the second component of a pair alone). It is clear that $\lambda_{X}$ is monotone for each set $X$. 

Let $\gamma: W\to \funF W$ be the coalgebra associated with Kripke model $\fmM := (W, R, V)$, and look at this ($\phi$ is a formula)
\begin{align*}
  w\in \InvBild{\gamma}{\lambda_{W}(\Gilt_{\fmM})}
& \Leftrightarrow
\gamma(w)\in \lambda_{W}(\Gilt_{\fmM})\\
& \Leftrightarrow
\langle R(w), V_{1}(w)\rangle \in\lambda_{W}(\Gilt_{\fmM})\\
& \Leftrightarrow
R(w)\subseteq \Gilt_{\fmM}\\
& \Leftrightarrow
w \in\Gilt[\Box\phi]_{\fmM}
\end{align*}
This means that we can describe the semantics of the $\Box$-operator through a predicate lifting, which cooperates with the coalgebra's dynamics. 

Note that it would be equally possible to do this for the $\Diamond$-operator: define the lifting through
$
D \mapsto\{\langle D', Q\rangle \mid  D'\cap D\not=\emptyset\}.
$
But we'll stick to the $\Box$-operator, keeping up with tradition.
\EndExample

\BeginExample{pred-lift-for-nghb-mod}
The same technique works for neighborhood models. In fact, let $(W, \nu)$ be the $\funG$-coalgebra associated with neighborhood model $\calN = (W, N, V)$ as in Example~\ref{nbhmod-as-coalgs}, and define
\begin{equation*}
  \lambda_{X}(D) := \{\langle V, Q\rangle\in\funV(X)\times\PowerSet{\Phi} \mid D\in V\}.
\end{equation*}
Then $\lambda_{X}: \PowerSet{X}\to
\PowerSet{\funV(X)\times\PowerSet{\Phi}}$ is monotone, because the
elements of $\funV{X}$ are upward closed. If $f: (W, \nu)\to (X, \mu)$ is a $\funG$-coalgebra morphism, we obtain for $D\subseteq X$ 
\begin{align*}
  \lambda_{W}(\InvBild{f}{D})
& = \{\langle V, Q\rangle \in \funV(W)\times\PowerSet{\Phi}\mid \InvBild{f}{D}\subseteq V\}\\
& = \{\langle V, Q\rangle \in \funV(W)\times\PowerSet{\Phi}\mid D\in (\funV f)(V)\}\\
& = \InvBild{(\funG f)}{\{\langle V', Q\rangle\in\funV(X)\times\PowerSet{\Phi}\mid  D\in V'\}}\\
& = \InvBild{(\funG f)}{\lambda_{X}(D)}
\end{align*}
Consequently, $\lambda$ is a predicate lifting for $\funG$. We see also for formula $\phi$
\begin{align*}
  w\in\lambda_{W}(\Gilt_{\calN}) 
& \Leftrightarrow
\langle \Gilt_{\calN}, V_{1}(w)\rangle \in \lambda_{X}(\Gilt_{\calN})\\
&\Leftrightarrow
\Gilt_{\calN}\in N(w)&&\text{ (by definition of $\nu$)}\\
&\Leftrightarrow
w\in \Gilt[\Box\phi]_{\calN}
\end{align*}
Hence we can define the semantics of the $\Box$-operator also in this case through a predicate lifting. 
\EndExample

There is a general mechanism permitting us to define predicate liftings, which is outlined in the next lemma.

\BeginLemma{nat-trans-lift}
Let $\eta: \funT\to \PowerSenza$ be a natural transformation, and define 
\begin{equation*}
  \lambda_{X}(D) := \{c\in\funT X\mid \eta_{X}(c)\subseteq D\}
\end{equation*}
for $D\subseteq X$.  Then $\lambda$ defines a predicate lifting for $\funT$.
\EndLemma

\BeginProof
It is clear from the construction that $D\mapsto \lambda_{X}(D)$ defines a monotone map, so we have to show that the diagram below is commutative for $f: X\to Y$.
\begin{equation*}
\xymatrix{
\PowerSenza X\ar[rr]^{\lambda_{X}} && \PowerSenza\funT X\\
\PowerSenza Y\ar[rr]_{\lambda_{Y}}\ar[u]^{f^{-1}} && \PowerSenza\funT Y\ar[u]_{(\funT f)^{-1}}
}
\end{equation*}
We note that 
\begin{equation*}
  \eta_{X}(c)\subseteq \InvBild{f}{E} \Leftrightarrow\Bild{f}{\eta_{X}(c)}\subseteq E \Leftrightarrow (\PowerSenza f)(\eta_{X}(c))\subseteq E
\end{equation*}
and 
\begin{equation*}
  (\PowerSenza f) \circ \eta_{X} = \eta_{Y}\circ (\funT f),
\end{equation*}
because $\eta$ is natural. Hence we obtain for $E\subseteq Y$:
\begin{align*}
  \eta_{X}(\InvBild{f}{E}) 
& = \{c \in \funT X \mid \eta_{X}(c)\subseteq \InvBild{f}{E}\}\\
& = \{c \in \funT X \mid \bigl((\PowerSenza f)\circ \eta_{X}\bigr)(c)\subseteq E\}\\
& = \{c \in \funT X \mid (\eta_{Y}\circ \funT f)(c)\subseteq E\}\\
& = \InvBild{(\funT f)}{\{d \in \funT Y \mid \eta_{Y}(d)\subseteq E\}}\\
& = \bigl((\funT f)^{-1}\circ \eta_{Y}\bigr)(E).
\end{align*}
\EndProof

Let us return to the endofunctor $\funF = \PowerSet{-}\times\PowerSet{\Phi}$ and fix for the moment an atomic proposition $p\in \Phi$. Define the constant function
\begin{equation*}
  \lambda_{p, X}(D) := \{\langle D', Q\rangle \in \funF X \mid p\in Q\}.
\end{equation*}
Then an easy calculation shows that $\lambda_{p}: \PSo\to \PSo\circ \funF$ is a natural transformation, hence a predicate lifting for $\funF$. Let $\gamma: W\to \funF W$ be a coalgebra with carrier $W$ which corresponds to the Kripke model $\fmM = (W, R, V)$, then
\begin{equation*}
w \in (\gamma^{-1}\circ \lambda_{p, W})(D) 
\Leftrightarrow 
\gamma(w) \in \lambda_{p, W}(D) 
\Leftrightarrow p \in \pi_{2}(\gamma(w))
\Leftrightarrow w\in V(p),
\end{equation*}
which means that we can use $\lambda_{p}$ for expressing the meaning
of formula $p\in \Phi$. A very similar construction can be made for
functor $\funG$, leading to the same conclusion.

Let us cast this into a more general framework. Let
$\ell_{X}: X\to \{0\}$ be the unique map from set $X$ to the singleton set $\{0\}$. Given
$A\subseteq \funT(\{0\})$, define $\lambda_{A, X}(D) := \{c\in \funT X\mid
(\funT\ell_{X})(c)\in A\} = \InvBild{(\funT\ell_{X})}{A}$. This defines a predicate lifting for $\funT$. In fact, let $f: X\to Y$ be a map, then $\ell_{X} = \ell_{Y}\circ f$, so $(\funT f)^{-1}\circ (\funT \ell_{Y})^{-1} = \bigl((\funT\ell_{Y})\circ (\funT f)\bigl)^{-1} = \bigl(\funT (\ell_{Y}\circ f)\bigr)^{-1} = (\funT\ell_{X})^{-1}$, hence $\lambda_{A, X}(\InvBild{f}{B}) = \InvBild{(\funT f)}{\lambda_{A, Y}(B)}$. As we have seen, this construction is helpful for capturing the semantics of atomic propositions. 

Negation can be treated as well in this framework. Given a predicate lifting $\lambda$ for $\funT$, we define for the set $X$ and $A\subseteq X$ the set
\begin{equation*}
  \lambda^{\neg}_{X}(A) := (\funT X)\setminus\lambda_{X}(X\setminus A),
\end{equation*}
then this defines a predicate lifting for $\funT$. This is easily checked: monotonicity of $\lambda^{\neg}$ follows from $\lambda$ being monotone, and since $f^{-1}$ is compatible with the Boolean operations, naturality follows. 

Summarizing, those operations which are dear to us when interpreting
modal logics through a Kripke model or through a neighborhood model can be represented using predicate liftings. 

We now take a family $\lift$ of predicate liftings and define a logic for it.

\BeginDefinition{coalg-log-for-L}
Let $\funT$ be an endofunctor on the category $\catSET$ of sets with maps, and let $\lift$ be a set of predicate listings for $\funT$. The formulas for the language \index{${\cal L}(\lift)$}${\cal L}(\lift)$ are defined through 
\begin{equation*}
  \phi ::= \bot \mid  \phi_{1}\wedge \phi_{2} \mid \neg\phi \mid [\lambda]\phi
\end{equation*}
with $\lambda\in\lift$. 
\EndDefinition

The semantics of a formula in ${\cal L}(\lift)$ in a $\funT$-coalgebra
$(W, \gamma)$ is defined recursively through fixing the sets of worlds
$\Gilt_{\gamma}$ in which formula $\phi$ holds (with
\index{$w\models_{\gamma}\phi$}$w\models_{\gamma}\phi$ iff $w\in\Gilt_{\gamma}$):
\begin{align*}
  \Gilt[\bot]_{\gamma} & :=\emptyset\\
\Gilt[\phi_{1}\wedge\phi_{2}]_{\gamma} & := \Gilt[\phi_{1}]_{\gamma}\cap\Gilt[\phi_{2}]_{\gamma}\\
\Gilt[\neg\phi]_{\gamma} & := W\setminus\Gilt_{\gamma}\\
\Gilt[[\lambda]\phi]_{\gamma} & := (\gamma^{-1}\circ \lambda_{C})(\Gilt_{\gamma}).
\end{align*}
The most interesting definition is of course the last one. It is
defined through a modality for the predicate lifting $\lambda$, and it
says that formula $[\lambda]\phi$ holds in world $w$ iff the
transition $\gamma(w)$ achieves a state which is lifted by $\lambda$
from one in which $\phi$ holds. Hence each successor to $w$ satisfies
the predicate for $\phi$ lifted by $\lambda$.

\BeginExample{pred-lift-for-modLog-1}
Continuing Example~\ref{pred-lift-for-modLog}, we see that the simple modal logic can be defined as the modal logic for $\lift = \{\lambda\}\cup\{\lambda_{p}\mid p\in\Phi\}$, where $\lambda$ is defined in Example~\ref{pred-lift-for-modLog}, and $\lambda_{p}$ are the constant liftings associated with $\Phi$. 
\EndExample

We obtain also in this case the invariance of validity under morphisms.

\BeginProposition{validity-morphism}
Let $f: (W, \gamma)\to (X, \delta)$ be a $\funT$-coalgebra morphism. Then
\begin{equation*}
  w\models_{\gamma}\phi\Leftrightarrow f(w)\models_{\delta}\phi
\end{equation*}
holds for all formulas $\phi\in{\cal L}(\lift)$ and all worlds $w\in W$. 
\EndProposition

\BeginProof
The interesting case occurs for a modal formula $[\lambda]\phi$ with $\lambda\in\lift$; so assume that the hypothesis is true for $\phi$, then we have 
\begin{align*}
  \InvBild{f}{\Gilt[[\lambda]\phi]_{\delta}}
& =
\bigl((\delta\circ f)^{-1}\circ \lambda_{D}\bigr)(\Gilt_{\delta})\\
& = 
\bigl((\funT(f)\circ \gamma)^{-1}\circ \lambda_{D}\bigr)(\Gilt_{\delta}) && f\text{ is a morphism}\\
& =
\bigl(\gamma^{-1}\circ (\funT f)^{-1}\circ \lambda_{D}\bigr) (\Gilt_{\delta})\\
& =
\bigl(\gamma^{-1}\circ \lambda_{C}\circ f^{-1}\bigr) (\Gilt_{\delta})&& \lambda\text{ is natural}\\
& =
\bigl(\gamma^{-1}\circ \lambda_{C}\bigr)(\Gilt_{\gamma})&& \text{by hypothesis}\\
& = 
\Gilt[[\lambda]\phi]_{\gamma}
\end{align*}
\EndProof

Let $(C, \gamma)$ be a $\funT$-coalgebra, then we define the \emph{\index{world!theory}\index{state!theory}\index{theory}theory of $c$}
\begin{equation*}
  \index{$\theTheory{\gamma}{c}$}\theTheory{\gamma}{c} := \{\phi\in{\cal L}(\lift) \mid c\models_{\gamma}\phi\}
\end{equation*}
for $c\in C$. Two worlds which have the same theory cannot be distinguished through formulas of the logic ${\cal L}(\lift)$.

\BeginDefinition{log-equiv}
Let $(C, \gamma)$ and $(D, \delta)$ be $\funT$-coalgebras, $c\in C$ and $d\in D$. 
\begin{itemize}
\item We call $c$ and $d$ are \emph{\index{world!logical
      equivalence}\index{state!logical equivalence}\index{logic!coalgebraic!logical
    equivalence}logically
    equivalent} iff $\theTheory{\gamma}{c} = \theTheory{\delta}{d}$.
\item The states $c$ and $d$ are called\emph{
    \index{world!behavioral equivalence}behaviorally
    \index{state!behavioral equivalence}equivalent\index{logic!coalgebraic!behavioral equivalence}} iff there exists a
  $\funT$-coalgebra $(E, \epsilon)$ and morphisms $(C,
  \gamma)\stackrel{f}{\to}(E, \epsilon)\stackrel{g}{\leftarrow}(D,
  \delta)$ such that $f(c) = g(d)$.
\end{itemize}
\EndDefinition
Thus, logical equivalence looks locally at all the formulas which are
true in a state, and then compares two states with each
other. Behavioral equivalence looks for an external instance, viz., a
mediating coalgebra, and at morphisms; whenever we find states the
image of which coincide, we know that the states are behaviorally
equivalent.

This implication is fairly easy to obtain.

\BeginProposition{beh-equiv-impl-log}
Behaviorally equivalent states are logically equivalent.
\EndProposition

\BeginProof
Let $c\in C$ and $d\in D$ be behaviorally equivalent for the
$\funT$-coalgebras $(C, \gamma)$ and $(D, \delta)$, and assume that we
have a mediating $\funT$-coalgebra $(E, \epsilon)$ with morphisms 
\begin{equation*}
(C,
\gamma)\stackrel{f}{\longrightarrow}(E, \epsilon)\stackrel{g}{\longleftarrow}(D,
\delta).
\end{equation*}
and $f(c) = g(d)$. Then we obtain
\begin{equation*}
  \phi\in \theTheory{\gamma}{c}
\Leftrightarrow
c\models_{\gamma}\phi
\Leftrightarrow
f(c)\models_{\epsilon}\phi
\Leftrightarrow
g(d)\models_{\epsilon}\phi
\Leftrightarrow
d\models_{\delta}\phi
\Leftrightarrow
\phi\in\theTheory{\delta}{d}
\end{equation*}
from Proposition~\ref{validity-morphism}.
\EndProof

We have seen that coalgebras are useful when it comes to generalize
modal logics to coalgebraic logics. Morphisms arise in a fairly
natural way in this context, giving rise to defining behaviorally
equivalent coalgebras. It is quite clear that bisimilarity can be
treated on this level as well, by introducing a mediating coalgebra
and morphisms from it; bisimilar states are logically equivalent, the
argument to show this is exactly as in the case above through
Proposition~\ref{validity-morphism}. In each case the question arises
whether the implications can be reversed --- are logically equivalent
states behaviorally equivalent? Bisimilar? Answering this question
requires a fairly elaborate machinery and depends strongly on the
underlying functor. We will not discuss this question here but rather
point to the literature, e.g., to~\cite{Pattinson}. For the
subprobability functor some answers and some techniques can be found
in~\cite{EED-CS-Survey}.

The following example discusses the basic modal language with no atomic propositions.

\BeginExample{hennessy-milner-lookalike}
{
\def\sR{\sim_R}
\def\sL{\sim_L}
We interpret ${\cal L}(\{\Diamond\}$ with $\Phi = \emptyset$ through $\PowerSenza$-coalgebras, i.e., through transition systems. Given a transition system $(S, R)$, denote by $\sim$ the equivalence provided by logical equivalence, so that $\isEquiv{s}{s'}{\sim}$ iff states $s$ and $s'$ cannot be separated through a formula in the logic, i.e., iff $\theTheory{R}{s} = \theTheory{R}{s'}$. Then $\fMap{\sim}: (S, R)\to (\Faktor{S}{\sim}, \Faktor{R}{\sim})$ is a coalgebra morphism. Here
\begin{equation*}
  \Faktor{R}{\sim} := \{\langle\Klasse{s_{1}}{}, \Klasse{s_{2}}{}\rangle\mid \langle s_{1}, s_{2}\rangle \in R\}.
\end{equation*}
In fact, look at this diagram
\begin{equation*}
\xymatrix{
S\ar[d]_{R}\ar[rr]^{\fMap{\sim}}
&&\Faktor{S}{\sim}\ar[d]^{\Faktor{R}{\sim}}
\\
\PowerSet{S}\ar[rr]_{\PowerSet{\fMap{\sim}}}&&\PowerSet{\Faktor{S}{\sim}}
}
\end{equation*}
Then 
\begin{equation*}
  \Klasse{s_{2}}{}\in\Faktor{R}{\sim}(\Klasse{s_{1}}{})\Leftrightarrow 
\Klasse{s_{2}}{}\in\{\Klasse{s}{}\mid s\in\Klasse{s_{1}}{}\} = \PowerSet{\fMap{\sim}}\bigl(\Faktor{R}{\sim}(\Klasse{s_{1}}{})\bigr),
\end{equation*}
which means that the diagram commutes. We denote the factor model $(\Faktor{S}{\sim}, \Faktor{R}{\sim})$ by $(S', R')$, and denote the class of an element without an indication of the equivalence relation. It will be clear from the context from which set of worlds a state will be taken. 

{
\def\Faktor#1#2{{#1}'}
\def\Klasse#1#2{[#1]}
Call the transition systems $(S, R)$ and $(T, L)$ \emph{\index{transition system!logical equivalence}logically equivalent} iff for each state in one system there exists a logically equivalent state in the other one. We carry over behavioral equivalence and bisimilarity from individual states to systems, taking the discussion for coalgebras in Section~\ref{sec:bism-coalg} into account. Call the transition systems $(S, R)$ and $(T, L)$ \emph{\index{transition system!behavioral equivalence}behaviorally equivalent} iff there exists a transition system $(U, M)$ with surjective morphisms
\begin{equation*}
\xymatrix{
(S, R)\ar[r]^{f} & (U, M)& (T, L) \ar[l]_{g}.
}
\end{equation*}
Finally, they are called \emph{\index{transition system!bisimilar}bisimilar} iff there exists a transition system $(U, M)$ with surjective morphisms
\begin{equation*}
\xymatrix{
(S, R) & (U, M)\ar[l]_{f}\ar[r]^{g} & (T, L).
}
\end{equation*}

We claim that logical equivalent transition systems have isomorphic factor spaces under the equivalence induced by the logic, provided both are image finite. Consider this diagram
\begin{equation*}
\xymatrix{
\Faktor{S}{\sR}
\ar[d]_{\Faktor{R}{\sR}}
\ar[rr]^{\zeta}&&
\Faktor{T}{\sL}
\ar[d]^{\Faktor{L}{\sL}}
\\
\PowerSet{\Faktor{S}{\sR}}
\ar[rr]_{\PowerSet{\zeta}}
&&
\PowerSet{\Faktor{T}{\sL}}
}
\end{equation*}
with $\zeta(\Klasse{s}{\sR}) := \Klasse{t}{\sL}$ iff $\theTheory{R}{s} = \theTheory{L}{t}$, thus $\zeta$ preserves classes of logically equivalent states. It is immediate that $\zeta: \Faktor{S}{\sR}\to \Faktor{T}{\sL}$ is a bijection, so commutativity has to be established.

Before working on the diagram, we show first that for any $\langle t,
t'\rangle\in L$ and for any $s\in S$ with
$\theTheory{R}{s}=\theTheory{L}{t}$ there exists $s'\in S$ with
$\langle s, s'\rangle\in R$ and $\theTheory{R}{s'}=\theTheory{L}{t'}$
(written more graphically in terms of arrows, we claim that
$t\to_{L}t'$ and $\theTheory{R}{s}=\theTheory{L}{t}$ together imply
the existence of $s'$ with $s\to_{R}s'$ and
$\theTheory{s'}{R}=\theTheory{L}{t'}$). This is established by
adapting the idea from the proof of the Hennessy-Milner
Theorem~\ref{hennessy-milner} to the situation at hand. Well, then:
Assume that such a state $s'$ cannot be found. Since $L, t'\models
\top$, we know that $L, t\models \Diamond\top$, thus $\theTheory{L}{t}
= \theTheory{R}{s}\not=\emptyset$. Let $R(s) = \{s_{1}, \dots,
s_{k}\}$ for some $k\geq 1$, then we can find for each $s_{i}$ a
formula $\psi_{i}$ with $L, t'\models \psi_{i}$ and $R,
s_{i}\not\models \psi_{i}$. Thus $L, t\models
\Diamond(\psi_{1}\wedge\dots\wedge\psi_{k})$, but $R, s\not\models
\Diamond(\psi_{1}\wedge\dots\wedge\psi_{k})$, which contradicts the
assumption that $\theTheory{R}{s}=\theTheory{L}{t}$. This uses only
image finiteness of $(S, R)$, by the way.

Now let $s\in S$ with $\Klasse{t_{1}}{\sL}\in\Faktor{L}{\sL}\bigl(\zeta(\Klasse{s}{\sR})\bigr) = \Faktor{L}{\sL}\bigl(\Klasse{t}{\sL}\bigr)$ for some $t\in T$. Thus $\langle t, t_{1}\rangle \in L$, so we find $s_{1}\in S$ with $\theTheory{R}{s_{1}} = \theTheory{L}{t_{1}}$ and $\langle s, s_{1}\rangle \in R$. Consequently, $\Klasse{t_{1}}{\sL} = \zeta(\Klasse{s_{1}}{\sR})\in \PowerSenza(\zeta)\bigl(\Faktor{R}{\sR}(\Klasse{s}{\sR})\bigr)$. Hence $\Faktor{L}{\sL}\bigl(\zeta(\Klasse{s}{\sR})\bigr)\subseteq \PowerSet{\zeta}\bigl(\Faktor{R}{\sR}(\Klasse{s}{\sR})\bigr)$.

Working on the other inclusion, we take
$\Klasse{t_{1}}{\sL}\in\PowerSet{\zeta}\bigl(\Faktor{R}{\sR}(\Klasse{s}{\sR}\bigr)$,
and we want to show that
$\Klasse{t_{1}}{\sL}\in\Faktor{L}{\sL}\bigl(\zeta(\Klasse{s}{\sR})\bigr)$. Now
$\Klasse{t_{1}}{\sL} = \zeta(\Klasse{s_{1}}{\sR})$ for some $s_{1}\in
S$ with $\langle s, s_{1}\rangle \in R$, hence $\theTheory{R}{s_{1}} =
\theTheory{L}{t_{1}}$. Put $\Klasse{t}{\sL} = \zeta(\Klasse{s}{\sR})$,
thus $\theTheory{R}{s} = \theTheory{L}{t}$. Because $(T, L)$ is image
finite as well, we may conclude from the Hennessy-Milner argument
above ---~by interchanging the r\^oles of the transition systems~--- that we
can find $t_{2}\in T$ with $\langle t, t_{2}\rangle \in L$ so that
$\theTheory{L}{t_{2}} = \theTheory{R}{s_{1}} = \theTheory{L}{t_{1}}$.
This impies $\Klasse{t_{2}}{\sL}=\Klasse{t_{1}}{\sL}$ and
$\Klasse{t_{1}}{\sL}\in\Faktor{L}{\sL}(\Klasse{t}{\sL}) =
\Faktor{L}{\sL}\bigl(\zeta(\Klasse{s}{\sR})\bigr)$. Hence $\Faktor{L}{\sL}\bigl(\zeta(\Klasse{s}{\sR})\bigr)\supseteq
\PowerSet{\zeta}\bigl(\Faktor{R}{\sR}(\Klasse{s}{\sR})\bigr)$. 

Thus the diagram above commutes, and we have shown that the factor
models are isomorphic. Consequently, two image finite transition
systems which are logically equivalent are behaviorally equivalent
with one of the factors acting as a mediating system. Clearly,
behaviorally equivalent systems are bisimilar, so that we obtain these relationships
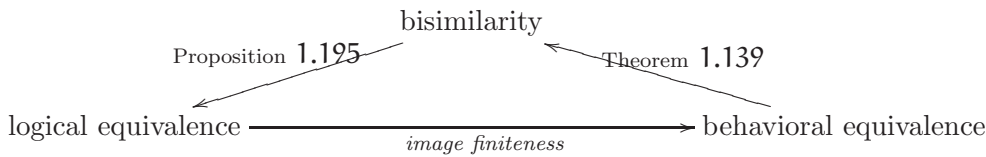
\begin{equation*}
\xymatrix{
&&\mathrm{bisimilarity}\ar[dll]_{\mathrm{Proposition~\ref{morphs-preserve-valid}}}\\
\mathrm{logical~equivalence}\ar[rrrr]_{\emph{image~finiteness}}&&&&\mathrm{behavioral~equivalence}\ar[ull]_{\mathrm{Theorem~\ref{ascel-bisim}}}
} 
\end{equation*}

}}\EndExample


We finally give an idea of modelling CTL* as a popular logic for model
checking coalgebraically. This shows how this modelling technique is
applied, and it shows also that some additional steps become
necessary, since things are not always straightforward.

\BeginExample{ctl-coalg}
{
\def\opp#1{\ensuremath{\mathbf{#1}}}
\def\fA{\opp{A}}
\def\fE{\opp{E}}
\def\fF{\opp{F}}
\def\fG{\opp{G}}
\def\fU{\opp{U}}
\def\fX{\opp{X}}
\def\catCoAlg{\Category{CoAlg}}
The logic \index{CTL*}CTL* is used for model checking~\cite{Clarke-Grumberg-Peled}. The abbreviation \emph{CTL} stands for \emph{computational tree logic}. CTL* is actually one of the simpler members of this family of tree logics used for this purpose, some of which involve continuous time~\cite{Baier+Haverkort+Hermanns+Katoen, EED-HennessyMilner}. The logic has state formulas and path formulas, the former ones are used to describe a particular state in the system, the latter ones express dynamic properties. Hence CTL* operates on two levels. 

These operators are used
\begin{description}
\item[State operators] They include the operators $\fA$  and $\fE$, indicating that a property holds in a state iff it holds on all paths resp. on at least one path emanating from it,
\item[Path operators] They include the operators
  \begin{itemize}
  \item $\fX$ for \emph{next time} --- a property holds in the next, i.e., second state of a path,
  \item $\fF$ for \emph{in the future} --- the specified property holds for some state on the path,
  \item $\fG$ for \emph{globally} --- the property holds always on a path,
  \item $\fU$ for \emph{until} --- this requires two properties as arguments; it holds on a path if there exists a state on the path for which the second property holds, and the first one holds on each preceding state.
  \end{itemize}
\end{description}

State formulas are given through this syntax:
\begin{equation*}
  \phi ::= \bot \mid p\mid \neg \phi\mid \phi_{1}\wedge\phi_{2}\mid \fE\psi\mid \fA\psi
\end{equation*}
with $p\in \Phi$ an atomic proposition and $\psi$ a path formula. Path formulas are given through
\begin{equation*}
  \psi ::= \phi\mid \neg\psi\mid \psi_{1}\wedge\psi_{2}\mid \fX\psi\mid \fF\psi\mid \fG\psi\mid \psi_{1}\fU\psi_{2}
\end{equation*}
with $\phi$ a state formula. So both state and path formulas are closed under the usual Boolean operations, each atomic proposition is a state formula, and state formulas are also path formulas. Path formulas are closed under the operators $\fX, \fF, \fG, \fU$, and the operators $\fA$ and $\fE$ convert a path formula to a state formula. 

Let $W$ be the set of all states, and assume that $V: \Phi\to
\PowerSet{W}$ assigns to each atomic formula the states for which it
is valid. We assume also that we are given a transition relation
$R\subseteq W\times W$; it is sometimes assumed that $R$ is left total, but this is mostly for computational reasons, so we will not make this assumption here.  Put
\begin{equation*}
S := \{\langle w_{1}, w_{2}, \dots\rangle \in W^{\infty}\mid \isEquiv{w_{i}}{w_{i+1}}{R} \text{ for all $i\in \Nat$}\}
\end{equation*}
as the set of all infinite $R$-paths over $W$. The interpretation of formulas is then defined as follows:
\begin{description}
\item[State formulas] Let $w\in W$, $\phi, \phi_{1}, \phi_{2}$ be state formulas and $\psi$ be a path formula, then
  \begin{align*}
w&\models\top   \Leftrightarrow \text{ always}\\
    s&\models p   \Leftrightarrow w \in V(p)\\
w&\models \neg\phi    \Leftrightarrow w\models\phi\text{ is false}\\
w&\models\phi_{1}\wedge\phi_{2}  \Leftrightarrow w\models\phi_{1}\text{ and }w\models\phi_{2}\\
w&\models\fE\psi   \Leftrightarrow \sigma\models\psi \text{ for some path $\sigma$ starting from $w$}\\
w&\models\fA\psi  \Leftrightarrow \sigma\models\psi \text{ for all paths $\sigma$ starting from $w$}\\
  \end{align*}
\item[Path formulas] Let $\sigma\in S$ be an infinite path with first node $\sigma_{1}$, $\sigma^{k}$ is the path with the first $k$ nodes deleted; $\psi$ is a path formula, and  $\phi$ a state formula, then
  \begin{align*}
   \sigma &\models\phi    \Leftrightarrow  \sigma_{1}\models\phi\\
\sigma&\models \neg\psi   \Leftrightarrow \sigma\models\psi\text{ is false}\\
\sigma&\models \psi_{1}\wedge\psi_{2}  \Leftrightarrow \sigma\models\psi_{1}\text{ and }\sigma\models\psi_{2}\\
\sigma&\models\fX\psi  \Leftrightarrow  \sigma^{1}\models\psi\\
\sigma&\models\fF\psi  \Leftrightarrow \sigma^{k}\models \psi\text{ for some $k\geq0$}\\
\sigma&\models\fG\psi  \Leftrightarrow \sigma^{k}\models\psi\text{ for all $k\geq0$}\\
\sigma&\models\psi_{1}\fU\psi_{2}  \Leftrightarrow \exists k\geq0: \sigma^{k}\models\psi_{2}\text{ and }\forall 0\leq j < k: \sigma^{j}\models\psi_{1}.
  \end{align*}
\end{description}
Thus a state formula holds on a path iff it holds on the first node,
$\fX\psi$ holds on path $\sigma$ iff $\psi$ holds on $\sigma$ with its first
node deleted, and $\psi_{1}\fU\psi_{2}$ holds on path $\sigma$ iff
$\psi_{2}$ holds on $\sigma^{k}$ for some $k$, and iff $\psi_{1}$ holds on
$\sigma^{i}$ for all $i$ preceding $k$.

We would have to provide interpretations only for conjunction, negation, for $\fA$, $\fX$, and $\fU$. This is so since $\fE$ is the nabla of $\fA$, $\fG$ is the nabla of $\fF$, and $\fF \psi$ is equivalent to $(\neg\bot)\fU\psi$. Conjunction and negation are easily interpreted, so we have to take care only of the temporal operators $\fA$, $\fX$ and $\fU$. 

A coalgebraic interpretation reads as follows. The $\PowerSenza$-coalgebras together with their morphisms form a category $\catCoAlg$. Let $(X, R)$ be a $\PowerSenza$-coalgebra, then 
\begin{equation*}
\funR(X, R) := \{\Folge{x}\in X^{\infty} \mid \isEquiv{x_{n}}{x_{n+1}}{R}\text{ for  all } n\in\Nat\}
\end{equation*} 
is the object part of a functor, $(\funR f)(\Folge{x}) :=
(f(x_{n})_{n\in \Nat})$ sends each coalgebra morphism $f: (X, R)\to
(Y, S)$ to a map $(\funR f): \funR(X, R)\to \funR(Y, S)$, which maps
$\Folge{x}$ to $f(x_{n})_{n\in\Nat}$; recall that $\isEquiv{x}{x'}{R}$
implies $\isEquiv{f(x)}{f(x')}{S}$. Thus $\funR: \catCoAlg\to \catSET$ is a functor. Note that the transition structure of the underlying Kripke model is already encoded through functor $\funR$. This is reflected in the definition of the dynamics $\gamma: X\to \funR(X, R)\times\PowerSet{\Phi}$ upon setting 
\begin{equation*}
  \gamma(x) := \bigl\langle\{w\in\funR(X, R) \mid w_{1}=x\}, V_{1}(x)\rangle,
\end{equation*}
where $V_{1}: X\to \PowerSet{\Phi}$ is defined according to $V: \Phi\to \PowerSet{X}$ as above. 
Define for the model $\fmM := (W, R, V)$ the map $\lambda_{\funR(W, R)}: C\mapsto \{\langle C', A\rangle \in\PowerSet{\funR(W, R)}\times\PowerSet{\Phi}\mid C'\subseteq C\}$, then $\lambda$ defines a natural transformation $\PSo\circ \funR\to \PSo\circ \funF\circ \funR$ (the functor $\funF$ has been defined in Example~\ref{pred-lift-for-modLog}); note that we have to check naturality in terms of model morphisms, which are in particular morphisms for the underlying $\PowerSenza$-coalgebra.  Thus we can define for $w\in W$
\begin{equation*}
  w\models_{\fmM}\fA\psi \Leftrightarrow w\in \gamma^{-1}\circ \lambda_{\funR(W, R)}(\Gilt[\psi]_{\fmM})
\end{equation*}
In a similar way we define $w\models_{\fmM}p$ for atomic propositions $p\in\Phi$; this is left to the reader. 

The interpretation of path formulas requires a slightly different approach. We define 
\begin{align*} 
\mu_{\funR(X, R)}(A) & := \{\sigma\in \funR(X, R) \mid \sigma^{1}\in A\},\\
\vartheta_{\funR(X, R)}(A, B) & := \bigcup_{k\in\Nat}\{\sigma \in \funR(X, R) \mid \sigma^{k}\in B, \sigma^{i}\in A\text{ for }0 \leq i < k\},
\end{align*}
whenever $A, B\in \funR(X, R)$. Then $\mu: \PSo\circ \funR\to \PSo\circ \funR$ and $\vartheta: (\PSo\circ \funR)\times (\PSo\circ \funR)\to \PSo\circ \funR$ are natural transformations, and we put 
\begin{align*}
  \Gilt[\fX \psi]_{\fmM} & := \mu_{\funR(M, R)}(\Gilt[\psi]_{\fmM}),\\
\Gilt[\psi_{1}\fU\psi_{2}]_{\fmM} & := \vartheta_{\funR(X, R)}(\Gilt[\psi_{1}]_{\fmM}, \Gilt[\psi_{2}]_{\fmM}).
\end{align*}

The example shows that a two level logics can be interpreted as well
through a coalgebraic approach, provided the predicate liftings which
characterize this approach are complemented by additional natural
transformations (which are called
\emph{\index{logic!coalgebraic!bridge operators}bridge operators}
in~\cite{EED-CoalgLogic-Book}) }
\EndExample

}

\Subsection{Bibliographic Notes}
\label{sec:bibliographic-notes-cat}

The monograph by Mac~Lane~\cite{MacLane} discusses all the definitions
and basic constructions; the text~\cite{Barr+Wells} takes much of its
motivation for categorical constructions from applications in computer
science. Monads are introdced following essentially Moggi's
seminal paper~\cite{Moggi-Inf+Control}. The text book~\cite{Pumpluen-Kategorien} is an exposition
fine tuned towards students interested in categories; the proof of
Lemma~\ref{yoneda-lemma} and the discussion on Yoneda's construction
follows its exposition rather closely. The discrete probability functor has been studied
extensively in~\cite{Sokolova}, it continuous step twin
in~\cite{Giry,EED-Book}. The use of upper closed subsets for the
interpretation of game logic is due to Parikh~\cite{Parikh-Games1985},
\cite{Pauly-Parikh} defines bisimilarity in this context.
The coalgebraic interpretation is investigated
in~\cite{EED-Game-Coalg}. Coalgebras are carefully discussed at length
in~\cite{Rutten}, by which the present discussion has been inspired.

The programming language \texttt{Haskell} is discussed in a growing
number of accessible books, a personal selection
includes~\cite{Real-World-Haskell,Learn-You}; the present short discussion is
taken from~\cite{EED-Haskell}. The representation of modal logics
draws substantially from~\cite{Blackburn-Rijke-Venema}, and the
discussion on coalgebraic logic is strongly influenced
by~\cite{Pattinson} and by the survey paper~\cite{EED-CS-Survey} as well as
the monograph~\cite{EED-CoalgLogic-Book}.

\renewcommand\Solution[2]{}
\Subsection{Exercises}
\label{sec:cat-exercises}
\Exercise{ex-graph-cat}{ The category $\Category{uGraph}$ has as
  objects \index{graph!undirected}undirected graphs. A morphism $f:
  (G, E)\to (H, F)$ is a map $f: G\to H$ such that $\{f(x), f(y)\}\in
  F$ whenever $\{x, y\}\in E$ (hence a morphism respects edges). Show
  that the laws of a category are satisfied.  }
\Solution{ex-graph-cat}{ The identity $id_{(G, E)}: (G, E)\to (G, E)$
  is a morphism. Let $f: (G, E)\to (H, F)$ and $g: (H, F)\to (I, K)$
  be morphisms, and take an edge $\{x, y\}\in E$, then $\{f(x),
  f(y)\}\in F$ (because $f$ preserves edges) and $\{(g\circ f)(x),
  (g\circ f)(y)\} = \{g(f(x)), g(f(y))\}\in K$ (because $g$ preserves
  edges), so morphisms are closed under composition. Because morphisms
  are in particular maps, their composition is associative.  
}
\Exercise{ex-split-morphisms}{ 
A morphism $f: a \to b$ in a category
  $\catK$ is a \emph{split \index{category!monomorphism!split}monomorphism} iff it has a left inverse,
  i.e. there exists $g: b \to a$ such that $g \circ f =
  id_{a}$. Similarly, $f$ is a \emph{split \index{category!epimorphism!split}epimorphism} iff it has a
  right inverse, i.e. there exists $g: b \to  a$ such that $f
  \circ g = id_{b}$.\label{split}
\begin{enumerate}
\item Show that every split monomorphism is monic and every split epimorphism is epic.
\item Show that a split epimorphism that is monic must be an isomorphism.
\item Show that for a morphism $f: a \to  b$, it holds that:
\begin{enumerate}
\item[(i)] $f$ is a split monomorphism $\Leftrightarrow$ $\homK{f, x}$ is surjective for every object $x$,
\item[(ii)] $f$ is a split epimorphism $\Leftrightarrow$ $\homK{x, f}$ is surjective for every object $x$,
\end{enumerate}
\item Characterize the split monomorphisms in $\catSET$. What can you say about split epimorphisms in $\catSET$?
\end{enumerate}
}
\Exercise{ex-par-maps}{
\def\catP{\Category{Par}}
The category $\catP$ of sets and partial functions is defined as follows:
\begin{itemize}
\item Objects are sets.
\item A morphism in $\hom{\catP} (A,B)$ is a partial function $f: A \rightharpoonup B$, i.e. it is a set-theoretic function $f: \mathsf{car}(f) \rightarrow B$ from a subset $\mathsf{car}(f) \subseteq A$ into $B$. $\mathsf{car} (f)$ is called the \emph{carrier} of $f$.
\item The identity $\mathsf{id}_A: A \rightharpoonup A$ is the usual identity function with $\mathsf{car} (\mathsf{id}_A) = A$.
\item For $f: A \rightharpoonup B$ and $g: B \rightharpoonup C$ the composition $g \circ f$ is defined as the usual composition $g(f(x))$ on the carrier:
\begin{equation*}
\mathsf{car} (g \circ f) := \{x \in \mathsf{car}(f) |\: f(x) \in
\mathsf{car}(g) \}.
\end{equation*}
\end{itemize}
\begin{enumerate}
\item Show that $\catP$ is a category and characterize its monomorphisms
  and epimorphisms.
\item Show that the usual set-theoretic Cartesian product you know is not the categorical product in $\catP$. Characterise binary products in $\catP$.
\end{enumerate}
}
\Exercise{ex-order-again-snd}{
{\def\catPos{\Category{Pos}}
Define the category $\catPos$ of ordered sets and monotone maps. The
objects are ordered sets $(P,\leq)$, morphisms are monotone maps $f:
(P, \leq) \rightarrow (Q, \sqsubseteq)$, i.e. maps $f: P\to Q$ such
that $x \leq y$ implies $f(x) \sqsubseteq f(y)$. Composition and identities are inherited from $\catSET$.
\begin{enumerate}
\item Show that under this definition $\catPos$  is a category.
\item Characterize monomorphisms and epimorphisms in $\catPos$.
\item Give an example of an ordered set $(P,\leq)$ which is isomorphic (in $\catPos$) to $(P,\leq)^{op}$ but $(P,\leq) \neq (P,\leq)^{op}$.
\end{enumerate}

An ordered set $(P,\leq)$ is called \textit{totally ordered} if for all $x,y \in P$ it holds that $x \leq y$ or $y \leq x$.

Show that if $(P,\leq)$ is isomorphic (in $\mathbf{Pos}$) to a totally
ordered set $(Q,\sqsubseteq)$, then $(P,\leq)$ is also totally
ordered. Use this result to give an example of a monotone map $f:
(P,\leq) \rightarrow (Q,\sqsubseteq)$ that is monic and epic but not
an isomorphism.  
} 
} 
\Exercise{ex-kleene-closure}{ Given a set $X$,
the set of (finite) strings of elements of $X$ is again denoted by
$X^{*}$.
\begin{enumerate}
\item Show that $X^*$ forms a monoid under concatenation, the \emph{free monoid} over $X$.
\item Given a map $f: X \rightarrow Y$, extend it uniquely to a
  monoid morphism $f^*: X^* \rightarrow Y^*$. In particular for all
  $x \in X$, it should hold that $f^*(\langle x \rangle) = \langle
  f(x)\rangle$, where $\langle x \rangle$ denotes the string consisting only of the character $x$.
\item Under what conditions on $X$ is $X^*$ a commutative monoid, i.e. has a commutative operation?
\end{enumerate}
}
\Exercise{ex-kleene-closure-monoid}{
Let $(M, \ast)$ be a monoid. We define a category $\Category{M}$ as
follows: it has only one object $\star$, $hom_{\Category{M}}
(\star,\star) = M$ with $id_{\star}$ as the unit of the monoid, and
composition is defined through $m_2 \circ m_1 := m_2 \ast m_1$.

\begin{enumerate}
\item Show that  $\Category{M}$ indeed forms a category.
\item Characterize the dual category $\mathbf{M}^{op}$. When are $\mathbf{M}$ and $\mathbf{M}^{op}$ equal?
\item Characterize monomorphisms, epimorphisms and isomorphisms for finite $M$. (What happens in the infinite case?)
\end{enumerate}
}
\Exercise{ex-meas-generator}{
Let $(S, {\cal A})$ and $(T, {\cal B})$ be measurable spaces, and
assume that the $\sigma$-algebra ${\cal B}$ is generated by ${\cal B}_{0}$. Show
that a map $f: S\to T$ is ${\cal A}$-${\cal B}$-measurable iff
$\InvBild{f}{B_{0}}\in{\cal A}$ for all $B_{0}\in{\cal B}_{0}$. 
}
\Solution{ex-meas-generator}{
Put
\begin{equation*}
  {\cal G} := \{B\in{\cal B} \mid \InvBild{f}{B}\in{\cal A}\}.
\end{equation*}
Then ${\cal G}$ is a $\sigma$-algebra. In fact, $T\in{\cal G}$, since
$\InvBild{f}{T} = S\in{\cal A}$; since $\InvBild{f}{T\setminus B} =
S\setminus\InvBild{f}{B}$, ${\cal G}$ is closed under complementation,
and because $\InvBild{f}{\bigcup_{n\in\Nat}B_{n}} =
\bigcup_{n\in\Nat}\InvBild{f}{B_{n}}$, ${\cal G}$ is closed under
countable unions. Moreover, we know that ${\cal B}_{0}\subseteq{\cal
  G}$, hence
\begin{equation*}
  {\cal B}=\sigma({\cal B}_{0})\subseteq\sigma({\cal G}) = {\cal G}\subseteq{\cal B},
\end{equation*}
so that ${\cal G} = {\cal B}$.
}
\Exercise{ex-meas-maps-indiced}{
Let $(S, {\cal A})$ and $(T, {\cal B})$ be measurable spaces and
$f: S\to T$ be ${\cal A}$-${\cal B}$-measurable. Define 
$
f_{*}(\mu)(B) := \mu(\InvBild{f}{B})
$
for 
$\mu\in\Prob{S, {\cal A}}, B\in{\cal B}$,
then $f_{*}: \Prob{S, {\cal A}}\to \Prob{T, {\cal B}}$. Show that
$f_{*}$ is $w({\cal A})$-$w({\cal B})$-measurable. Hint: Use Exercise~\ref{ex-meas-generator}. 
}
\Solution{ex-meas-maps-indiced}{
Because 
$
w({\cal B}) = \sigma(\{\beta_{T}(B, r) \mid 0 \leq r \leq 1, B\in{\cal
  B}\})
$
according to Example~\ref{ex-cat-stoch-rel}, it is sufficient by Exercise~\ref{ex-meas-generator} to show that 
$
\InvBild{f_{*}}{\beta_{T}(B, r)} \in w({\cal A}).
$
Now let $B\in{\cal B}$ and $0 \leq r \leq 1$, then 
\begin{equation*}
  \mu\in\InvBild{f_{*}}{\beta_{T}(B, r)}
\Leftrightarrow
f_{*}(\mu)(B) \geq r
\Leftrightarrow
\mu(\InvBild{f}{B}) \geq r
\Leftrightarrow
\mu\in\beta_{S}(\InvBild{f}{B}, r),
\end{equation*}
and $\beta_{S}(\InvBild{f}{B}, r)\in w({\cal A})$ on account of $\InvBild{f}{B}\in{\cal A}$.
}
\Exercise{ex-discr-probs}{
Let $S$ be a countable sets with  $p: S\to[0, 1]$ as a
\emph{discrete probability distribution}, thus $\sum_{s\in S} p(s) = 1$;
denote the corresponding probability measure on $\PowerSet{S}$ by
$\mu_{p}$, hence $\mu_{p}(A) = \sum_{s\in A}p(s)$. Let $T$ be an at most countable set with a
discrete probability distribution $q$. Show that a map $f: S\to T$ is a
morphism for the probability spaces $(S, \PowerSet{S}, \mu_{p})$ and
$(T, \PowerSet{T}, \mu_{q})$ iff $q(t) = \sum_{f(s)=t} p(s)$ holds for
all $t\in T$.
}
\Solution{ex-discr-probs}{
This follows immediately from $\mu_{p}(\InvBild{f}{\{t\}}) = \mu_{p}(\{s\in S \mid
f(s) = t\}) = \sum_{f(s) = t}p(s)$.
}
\Exercise{ex-aus-measurability}{
Show that 
$\{x\in [0, 1] \mid \langle x, x\rangle \in E\}\in \Borel{[0, 1]}$,
whenever $E\in \Borel{[0, 1]}\otimes\Borel{[0, 1]}$. }
\Solution{ex-aus-measurability}{
Define for $E\subseteq[0, 1]\times[0, 1]$ just as an abbreviation the set
$
H(E) :=  \{x\in [0, 1] \mid \langle x, x\rangle \in E\},
$
thus the claim is that 
\begin{equation*}
  {\cal G} := \{E\in\Borel{[0, 1]}\otimes\Borel{[0, 1]} \mid
  H(E)\in\Borel{[0, 1]}\} = \Borel{[0, 1]}\otimes\Borel{[0, 1]}.
\end{equation*}
Apparently, ${\cal G}$ is a $\sigma$-algebra, and $H(A\times B) =
A\cap B$, so ${\cal G}$ contains all measurable rectangles.
Thus ${\cal G} = \Borel{[0, 1]}\otimes\Borel{[0, 1]}$.
}
\Exercise{ex-diagram-chasing}{
Let's chase some diagrams. Consider the following diagram in a category $\catK$:

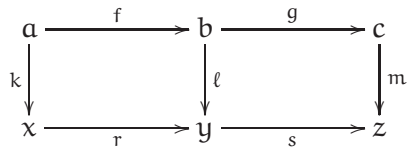
\begin{equation*}
\xymatrix{
a\ar[rr]^{f}\ar[d]_{k} && b\ar[rr]^{g}\ar[d]^{\ell}&& c\ar[d]^{m}\\
x\ar[rr]_{r} && y\ar[rr]_{s}&& z
}
\end{equation*}

\begin{enumerate}
\item Show that if the left inner and right inner diagrams commute, then the outer diagram commutes as well.
\item Show that if the outer and right inner diagrams commute and $s$ is a monomorphism, then the left inner diagram commutes as well.
\item Give examples in $\catSET$ such that:
\begin{enumerate}
\item the outer and left inner diagrams commute, but not the right inner diagram,
\item the outer and right inner diagrams commute, but not the left inner diagram.
\end{enumerate}
\end{enumerate}
}
\Exercise{ex-proj-not-epic}{
Give an example of a product in a category $\catK$ such that one
of the projections is not epic.
}
\Exercise{es-prod-sum-in-monoids}{
What can you say about products and sums in the category $\Category{M}$
given by a finite monoid $(M, \ast)$, as defined in Exercise~\ref{ex-kleene-closure}? (Consider the case that $(M,
\ast)$ is commutative first.)
}
\Exercise{ex-cont-univ-prop-prod}{
Show that the product topology has this universal property:   $f: (D,
{\cal D})\to (S\times T, {\cal G}\times{\cal H})$ is continuous iff
$\pi_{S}\circ f: (D, {\cal D})\to (S, {\cal G})$ and $\pi_{T}\circ f:
(D, {\cal D})\to (T, {\cal H})$ are continuous. Formulate and prove
the corresponding property for morphisms in $\Category{Meas}$. 
}

\Exercise{ex-jointly-monic-epic}{ 
A collection of morphisms $\{f_i : a \to  b_i\}_{i
    \in I}$ with the same domain in category $\catK$  is called \emph{\index{monic!jointly}jointly monic}
  whenever the following holds: If $g_1: x \to  a$ and $g_2: x
  \to  a$ are morphisms such that $f_i \circ g_1 = f_i \circ g_2$ for
  all $i \in I$, then $g_1 = g_2$. Dually one defines a collection of morphisms to be \emph{\index{epic!jointly}jointly epic}.

Show that the projections from a categorical product are jointly monic and the injections into a categorical sum are jointly epic.
}
\Exercise{ex-chase-pullback-again}{
Assume the following diagram in a category $\catK$ commutes:
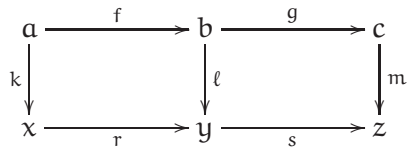
\begin{equation*}
\xymatrix{
a\ar[rr]^{f}\ar[d]_{k} && b\ar[rr]^{g}\ar[d]^{\ell}&& c\ar[d]^{m}\\
x\ar[rr]_{r} && y\ar[rr]_{s}&& z
}
\end{equation*}
Prove or disprove: if the outer diagram is a pullback, one of the
inner diagrams is a pullback as well. Which inner diagram has to be a
pullback for the outer one to be also a pullback?  
}
\Exercise{ex-limits-simple}{ Suppose $f,g: a \to b$ are morphisms in a
  category $\mathbf{C}$. An \emph{\index{equalizer}equalizer} of $f$
  and $g$ is a morphism $e: x \to a$ such that $f \circ e = g \circ
  e$, and whenever $h: y \to a$ is a morphism with $f \circ h = g
  \circ h$, then there exists a unique $j: y \to x$ such that $h = e
  \circ j$.

This is the diagram:
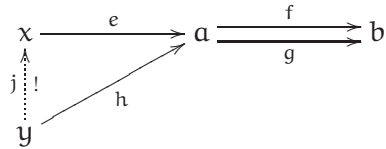
\begin{equation*}
\xymatrix{
x\ar[rr]^{e} && a\ar@<0.6ex>[rr]^{f}\ar@<-0.6ex>[rr]_{g} && b\\
y\ar@{..>}[u]^{j}_{!}\ar[urr]_{h}}
\end{equation*}

\begin{enumerate}
\item Show that equalizers are uniquely determined up to isomorphism.
\item Show that the morphism $e: x \to  a$ is a monomorphism. 
\item Show that a category has pullbacks if it has  products and equalizers.
\end{enumerate}
}
\Exercise{ex-terminal-objects}{
A \emph{\index{terminal object}terminal object} in  category $\catK$ is an object
$\mathbf{1}$ such that for every object $a$ there exists a unique
morphism $!: a \to  \mathbf{1}$.
\begin{enumerate}
\item Show that terminal objects are uniquely determined up to isomorphism.
\item Show that a category has (binary) products and equalizers if it has pullbacks and a terminal object.
\end{enumerate}

}
\Solution{ex-cont-univ-prop-prod}{
Given $f$ with the properties above, we know that there exists a
unique continuous map $h: (D, {\cal D})\to (S\times T, {\cal
  G}\times{\cal H})$ with $\pi_{S}\circ h = \pi_{S}\circ f$ and
$\pi_{T}\circ h = \pi_{T}\circ f$. By uniqueness in $\catSET$, the
maps $f$ and $h$ must be identical. So $f:(D,
{\cal D})\to (S\times T, {\cal G}\times{\cal H})$ is continuous.

For category $\Category{Meas}$, replace ``continuous'' by
``measurable'' in the statement and in the proof.
}
\Exercise{ex-cont-univ-prop-coprod}{
Show that the coproduct $\sigma$-algebra has this universal property:  
$f: (S+T, {\cal A}+{\cal B})\to (R, {\cal X})$ is ${\cal A}+{\cal B}$-${\cal X}$-measurable iff
$f\circ i_{S}$ and $f\circ i_{T}$  are ${\cal A}$-${\cal X}$- resp.${\cal B}$-${\cal X}$-measurable. Formulate and prove
the corresponding property for morphisms in $\Category{Top}$. 
}
\Solution{ex-cont-univ-prop-coprod}{
Dualize the solution to Exercise~\ref{ex-cont-univ-prop-prod}.
}
\Exercise{ex-prod-is-associative}{
Assume that in category $\catK$ any two elements have a product. Show
that $a\times(b\times c)$ and $(a\times b)\times c$ are isomorphic.
}
\Exercise{ex-list-products}{
Prove Lemma~\ref{double-product-constr}.
}
\Exercise{ex-list-coproducts}{
Assume that the coproducts $a+ a'$ and $b+ b'$ exist in
category $\catK$. Given morphisms $f: a\to b$ and $f': a'\to b'$, show that there exists a unique morphism $q: a+ a'\to
b+ b'$ such that this diagram commutes

\begin{equation*}
\xymatrix{
a\ar[d]_{f} && a + a'\ar[ll]_{i_{a}}\ar[rr]^{i_{a'}}\ar[d]^{q} && a'\ar[d]^{f'}\\
b && b + b'\ar[ll]^{i_{b}}\ar[rr]_{i_{b'}} && b'
}
\end{equation*}

}
\Exercise{ex-prob-no-coproducts}{
Show that the category $\Category{Prob}$ has no coproducts (Hint:
Considering $(S, {\cal C}) + (T, {\cal D})$, show
that, e.g., $\InvBild{i_{S}}{\Bild{i_{S}}{A}}$ equals $A$ for
$A\subseteq S$).
}
\Solution{ex-prob-no-coproducts}{
Let $A\subseteq S$, then $A\subseteq\InvBild{i_{S}}{\Bild{i_{S}}{A}}$
is trivial. On the other hand, given $x$ with $i_{S}(x)\in i_{S}(A)$,
we conclude $x\in A$. Hence
$A\supseteq\InvBild{i_{S}}{\Bild{i_{S}}{A}}$. Similar for $i_{T}$. But
this means that $i_{S}(S) \in {\cal C}+{\cal D}$, similarly,
$i_{T}(T)\in{\cal C}+{\cal D}$. If $\kappa$ is the probability in the
coproduct of $(S, {\cal C}, \mu)$ and $(T, {\cal D}, \nu)$, then
$\kappa = i_{S, *}(\mu)$ and $\kappa = i_{T, *}(\nu)$, thus
$1 = \kappa(S+T) = \kappa(i_{S}(S)) + \kappa(i_{T}(T)) = \mu(S) + \nu(T) =
2$. 
}

\Exercise{ex-prod-cat-rel}{
Identify the product of two objects in the category $\Category{Rel}$ of relations. 
}

\Exercise{ex-em-factorization-in-meas}{
We investigate the epi-mono factorization in the category
$\Category{Meas}$ of measurable spaces. Fix two measurable spaces
$(S, {\cal A})$ and $(T, {\cal B})$ and a morphism $f: (S, {\cal A})\to
(T, {\cal B})$. 

\begin{enumerate}
  \item
Let $\Faktor{{\cal A}}{\Kern{f}}$ be the largest
$\sigma$-algebra ${\cal X}$ on $\Faktor{S}{\Kern{f}}$ rendering the
factor map $\fMap{\Kern{f}}: S\to \Faktor{S}{\Kern{f}}$ ${\cal
  A}$-${\cal X}$-measurable. Show that $\Faktor{{\cal A}}{\Kern{f}} =
\{C\subseteq\Faktor{S}{\Kern{f}} \mid
\InvBild{\fMap{\Kern{f}}}{C}\in{\cal A}\}$, and show that
$\Faktor{{\cal A}}{\Kern{f}}$ has this universal property: given a
measurable space $(Z, {\cal C})$, a map $g:
\Faktor{S}{\Kern{f}}\to Z$ is $\Faktor{{\cal A}}{\Kern{f}}$-${\cal
  C}$ measurable iff $g\circ \fMap{\Kern{f}}: S\to Z$ is ${\cal
  A}$-${\cal C}$-measurable. 

\item
Show that $\fMap{\Kern{f}}$ is an epimorphism in $\Category{Meas}$,
and that $f_{\bullet}: \Klasse{x}{\Kern{f}} \mapsto f(x)$ is a
monomorphism in $\Category{Meas}$. 

\item
Let $f = m\circ e$ with an epimorphism $e: (S, {\cal A})\to (Z, {\cal
  C})$ and a monomorphism $m: (Z, {\cal C})\to (T, {\cal B})$, and
define $b: \Faktor{S}{\Kern{f}}\to Z$ through $\Klasse{s}{\Kern{f}}
\mapsto e(s)$, see Corollary~\ref{cat-epi-mono-decomp-cor}. Show that
$b$ is $\Faktor{{\cal A}}{\Kern{f}}$-${\cal C}$-measurable, and prove or disprove measurability of $b^{-1}$.
\end{enumerate}}

\Solution{ex-em-factorization-in-meas}{
  \begin{enumerate}
  \item If $\fMap{\Kern{f}}$ is ${\cal A}$-${\cal X}$-measurable, then
    $\InvBild{\fMap{\Kern{f}}}{C}\in{\cal A}$ for all $C\in {\cal X}$,
    thus ${\cal X}\subseteq\{C\subseteq\Faktor{S}{\Kern{f}} \mid
\InvBild{\fMap{\Kern{f}}}{C}\in{\cal A}$. On the other hand, the latter
set is a $\sigma$-algebra, hence the claim follows. Now assume that  $g\circ \fMap{\Kern{f}}: S\to Z$ is ${\cal
  A}$-${\cal C}$-measurable, and let $D\in{\cal D}$. Since 
$
\InvBild{\fMap{\Kern{f}}}{\InvBild{g}{D}} = \InvBild{(g\circ
  \fMap{\Kern{f}})}{D}\in{\cal A},
$
it follows that $\InvBild{g}{D}\in\Faktor{{\cal A}}{\Kern{f}}.$ Thus
$g$ is $\Faktor{{\cal A}}{\Kern{f}}$-${\cal D}$-measurable.
\item $\fMap{\Kern{f}}$ is ${\cal A}$-$\Faktor{{\cal
      A}}{\Kern{f}}$-measurable by construction, since $f = f_{\bullet}\circ \fMap{\Kern{f}}T$ is
  ${\cal A}$-${\cal B}$-measurable, we conclude from part 1. that
  $f_{\bullet}$  is $\Faktor{{\cal A}}{\Kern{f}}$-${\cal
    B}$-measurable. Because morphisms in $\Category{Meas}$ are
  in particular maps, we conclude that $\fMap{\Kern{f}}$ is an
  epimorphism and $f_{\bullet}$ is a monomorphism in
  $\Category{Meas}$. 
\item Because $b\circ \fMap{\Kern{f}} = e$, and $e$
  is ${\cal A}$-${\cal C}$-measurable, we conclude from the first part that $b$
  is $\Faktor{{\cal A}}{\Kern{f}}$-${\cal C}$-measurable. Now let
  $C\in\Faktor{{\cal A}}{\Kern{f}}$, then there exists $B\in{\cal B}$
  such that $C = \InvBild{f_{\bullet}}{B}$. This is so since
  $f_{\bullet}^{-1}: \PowerSet{T}\to \PowerSet{\Faktor{S}{\Kern{f}}}$ is onto due to $f_{\bullet}$
  being injective. Hence $\Bild{b}{C} =
  \Bild{b}{\InvBild{f_{\bullet}}{B}} = \InvBild{m}{B}\in{\cal C}$,
  since $b\circ f_{\bullet}^{-1} = (f_{\bullet}\circ b^{-1})^{-1}
  = m^{-1}.$ Consequently, $b^{-1}$ is ${\cal C}$-$\Faktor{{\cal
      A}}{\Kern{f}}$-measurable as well. Thus category
  $\Category{Meas}$ inherits an epi-mono-factorization from category $\catSET$.
  \end{enumerate}
}
\Exercise{ex-abel-group}{
Let \index{$\Category{AbGroup}$}$\Category{AbGroup}$ be the category of Abelian groups. Its
objects are commutative groups, a morphism $\phi: (G, +) \to (H, *)$
is a map $\phi: G\to H$ with $\phi(a+b) = \phi(a)*\phi(b)$ and
$\phi(-a) = -\phi(a)$. Each subgroup $V$ of an Abelian group $(G, *)$
defines an equivalence relation $\rho_{V}$ through
$\isEquiv{a}{b}{\rho_{V}}$ iff $a - b\in V$. Characterize the pushout
of $\fMap{\rho_{V}}$ and $\fMap{\rho_{W}}$ for subgroups $V$ and $W$
in $\Category{AbGroup}$. 
}

\Exercise{ex-binary-systems}{
Given a set $X$, define $\funF(X) := X\times X$, for a map $f: X\to
Y$, $\funF(f)(x_{1}, x_{2}) := \langle f(x_{1}), f(x_{2})\rangle$ is
defined. Show that $\funF$ is an endofunctor on $\catSET$.
}
\Solution{ex-binary-systems}{
$\funF(id_{X}) = id_{\funF X}$ and with $f: X\to Y, g: Y\to Z$ we have 
$
\funF(g\circ f)(x_{1}, x_{2}) = \langle g(f(x_{1})),
g(f(x_{2}))\rangle = \funF(g)\bigl(\funF(f)(x_{1}, x_{2})\bigr),
$ 
thus 
$
\funF(g\circ f) = \funF(g)\circ \funF(f).
$
}
\Exercise{ex-term-or-lab-outp}{
Fix a set $A$ of labels; define $\funF(X) := \{*\}\cup A\times X$ for
the set $X$, if $f: X\to Y$ is a map, put 
$
\funF(f)(*) := *
$
and
$
\funF(f)(a, x) := \langle a, f(x)\rangle.
$
Show that $\funF: \catSET\to \catSET$ defines an endofunctor. 

This endofunctor models termination or labeled output. 
}
\Exercise{ex-finite-branching}{
Fix a set $A$ of labels, and put for the set $X$
\begin{equation*}
  \funF(X) := \PowerSenza_{f}(A\times X),
\end{equation*}
where $\PowerSenza_{f}$ denotes all finite
subsets of its argument. Thus $G\subseteq\funF(X)$ is a finite subset
of $A\times X$, which models finite branching, with $\langle a,
x\rangle\in G$ as one of the possible branches, which is  in this case
labeled by $a\in A$. Define 
\begin{equation*}
  \funF(f)(B) :=\{\langle a, f(x)\rangle \mid \langle a, x\rangle\in B\}
\end{equation*}
for the map $f: X\to Y$ and $B\subseteq A\times X$. Show that $\funF:
\catSET\to \catSET$ is an endofunctor.
}
\Exercise{unique-limit}{
Show that the limit cone for a functor $\funF:\catK\to \catL$ is
unique up to isomorphisms, provided it exists.
}
\Exercise{product-as-limit}{
Let $I\not=\emptyset$ be an arbitrary index set, and let $\catK$ be
the discrete category over $I$. Given a family $(X_{i})_{i\in I}$,
define $\funF: I\to \catSET$ by $\funF i := X_{i}$. Show that 
\begin{equation*}
  X := \prod_{i\in I} X_{i} := \{x: I\to \bigcup_{i\in I} X_{i}\mid x(i)\in
  X_{i}\text{ for all }i\in I\}
\end{equation*}
with $\pi_{i}: x \mapsto x(i)$ is a limit $(X, (\pi_{i})_{i\in I})$ of $\funF$. 
}
\Exercise{equalizer-as-limit}{
Formulate the equalizer of two morphisms
(cp. Exercise~\ref{ex-limits-simple}) as a limit.
}
\Exercise{free-semi}{
Define for the set $X$ the free monoid $X^*$ generated by $X$ through
\begin{equation*}
X^* := \{\langle x_1, \dots, x_k\rangle \mid x_i\in X, k\geq 0\}
\end{equation*}
with juxtaposition as multiplication, i.e., $\langle x_1, \dots, x_k\rangle*\langle
x_1', \dots, x_r'\rangle := \langle x_1, \dots, x_k, x_1', \dots,
x_r'\rangle$; the neutral element $\epsilon$ is $\langle x_1, \dots,
x_k\rangle$ with $k = 0$. 
Define 
\begin{align*}
  f^{*}(x_{1}* \dots* x_{k}) & := f(x_{1})* \dots * f(x_{k})\\
\eta_{X}(x) & := \langle x \rangle
\end{align*}
for the map $f: X\to Y^{*}$ and $x\in X$. Put $\funF X := X^{*}$. Show
that $(\funF, \eta, -^{*})$ is a Kleisli tripel, and compare it with
the \texttt{list} monad, see page~\pageref{the-list-monad}. Compute
$\mu_{X}$ for this monad. 
}
\Exercise{ex-2-syst}{
Given are the systems $S$ and $T$. 

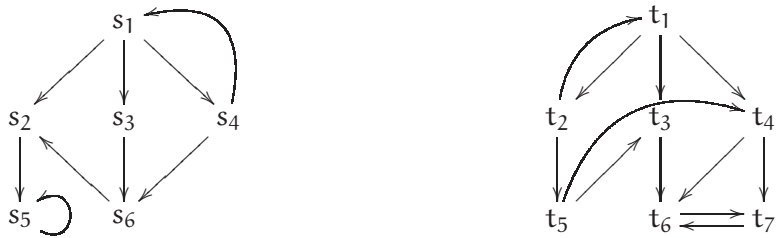
\begin{minipage}{.45\linewidth}
\begin{equation*}
\xymatrix{
&s_{1}\ar[dr]\ar[d]\ar[dl]\\
s_{2}\ar[d] & s_{3}\ar[d]&s_{4}\ar[dl]\ar@/_{2pc}/[ul]\\
s_{5}\ar@(dr, ur) & s_{6}\ar[ul]
}
\end{equation*}
\end{minipage}
\begin{minipage}{.45\linewidth}
\begin{equation*}
\xymatrix{
&t_{1}\ar[dl]\ar[d]\ar[dr]\\
t_{2}\ar[d]\ar@/_{-1pc}/[ur]& t_{3}\ar[d]& t_{4}\ar[d]\ar[dl]\\
t_{5}\ar[ur]\ar@/_{-2pc}/[urr]& t_{6}\ar@<.5ex>[r] & t_{7}\ar@<.5ex>[l]
}
\end{equation*}
\end{minipage}
\begin{itemize}
\item Consider the transition systems $S$ and $T$ as coalgebras
  for a suitable functor $\funF: \catSET \rightarrow \catSET$,
  $X \mapsto \mathcal{P}(X)$. Determine the dynamics of the respective
  coalgebras.
\item Show that there is no coalgebra morphism $S \rightarrow T$.
\item Construct a coalgebra morphism $T \rightarrow S$.
\item Construct a bisimulation between $S$ and $T$ as a coalgebra on the carrier
  \begin{equation*}
\{\langle s_2,t_3 \rangle, \: \langle s_2,t_4 \rangle, \: \langle s_4,t_2 \rangle, \: \langle s_5,t_6 \rangle, \: \langle s_5,t_7 \rangle, \: \langle s_6,t_5 \rangle  \}.
\end{equation*}
\end{itemize}
}
\Exercise{ex-3-syst}{
Characterize this nondeterministic transition system $S$ as a coalgebra for a suitable functor $\funF: \catSET\to \catSET$.
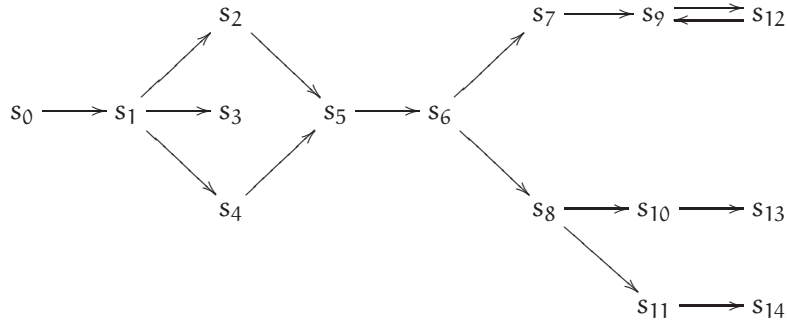
\begin{equation*}
\xymatrix{&&s_{2}\ar[dr]&&&s_{7}\ar[r]&s_{9}\ar@<.5ex>[r]&s_{12}\ar@<.5ex>[l]\\
s_{0}\ar[r]&s_{1}\ar[r]\ar[ur]\ar[dr]&s_{3}&s_{5}\ar[r]&s_{6}\ar[ur]\ar[dr]\\
&&s_{4}\ar[ur]&&&s_{8}\ar[r]\ar[dr]&s_{10}\ar[r]&s_{13}\\
&&&&&&s_{11}\ar[r]&s_{14}
}
\end{equation*}
Show that
\begin{equation*}
\alpha := \{\langle s_i, s_i \rangle |\: 0 \leq i \leq 12 \} \cup
\{\langle s_{2}, s_{4} \rangle, \langle s_{4}, s_{2} \rangle, \langle
s_{9}, s_{12} \rangle, \langle s_{12}, s_{9} \rangle, \langle s_{13},
s_{14} \rangle, \langle s_{14}, s_{13} \rangle \}
\end{equation*}
is a bisimulation equivalence on $S$. Simplify $S$ by giving a coalgebraic characterisation of the factor system $\Faktor{S}{\alpha}$. Furthermore, determine whether $\alpha$ is the largest bisimulation equivalence on $S$.
}
\Exercise{ex-automata-bisim}{
The deterministic finite automata $A_1, A_2$ with input and output
alphabet $\{0,1\}$ and the following transition tables are given:
{\footnotesize
$$
\begin{array}{c||c|c||c|c|c||c||c||c|c||c|c|c}
  \mathbf{A_1} & \text{state} & \text{input} & \text{output} & \text{next state} & & \mathbf{A_2} & \text{state} & \text{input} & \text{output} & \text{next state} \\
\hline
 & s_0 & 0 & 0 & s_1 & & & s'_0 & 0 & 0 & s'_0\\
 & s_0 & 1 & 1 & s_0 & & & s'_0 & 1 & 1 & s'_1\\
\hline
 & s_1 & 0 & 0 & s_2 & & & s'_1 & 0 & 0 & s'_0\\
 & s_1 & 1 & 1 & s_3 & & & s'_1 & 1 & 1 & s'_2\\
\hline
 & s_2 & 0 & 1 & s_4 & & & s'_2 & 0 & 1 & s'_3\\
 & s_2 & 1 & 0 & s_2 & & & s'_2 & 1 & 0 & s'_2\\
\hline
 & s_3 & 0 & 0 & s_1 & & & s'_3 & 0 & 1 & s'_4\\
 & s_3 & 1 & 1 & s_3 & & & s'_3 & 1 & 0 & s'_2\\
\hline
 & s_4 & 0 & 1 & s_3 & & & s'_4 & 0 & 0 & s'_5\\
 & s_4 & 1 & 0 & s_2 & & & s'_4 & 1 & 1 & s'_4\\
\hline
 & & & & & & & s'_5 & 0 & 0 & s'_2\\
 & & & & & & & s'_5 & 1 & 1 & s'_4\\
\hline
\end{array}$$}

\begin{enumerate}
\item Formalize the automata as coalgebras for a suitable functor
  $\funF: \catSET \rightarrow \catSET$, $\funF(X) = (X \times
  O)^I$. (You have to choose $I$ and $O$ first.)
\item Construct a coalgebra morphism from $A_1$ to $A_2$ and use this
  to find a bisimulation $R$ between $A_1$ and $A_2$. Describe the
  dynamics of $R$ coalgebraically.
\end{enumerate}
}
\Exercise{ex-dual-eff-fnct}{
Let $P$ be an effectivity function on $X$, and define $\partial P(A)
:= X\setminus P(X\setminus A)$. Show that $\partial P$ defines an
effectivity function on $X$. Given an effectivity function $Q$ on $Y$
and a morphism $f: P\to Q$, show that $f: \partial P\to \partial Q$ is
a morphism as well.
}
\Exercise{ex-no-waek-pullback}{
Show that the power set functor $\mathcal{P} : \catSET\to \catSET$ does not preserve pullbacks.
(Hint: You can use the fact, that in $\catSET$ the pullback
of the left diagram is explicitly given as $P := \{\langle x,y
\rangle\mid  f(x) = g(y) \}$ with $\pi_X$ and $\pi_Y$ being the usual
projections.)
}
\Exercise{ex-product-wpullb}{
Suppose $\funF, \funG: \catSET\to \catSET$ are functors.
\begin{enumerate}
\item Show that if $\funF$ and $\funG$ both preserve weak pullbacks, then also the product functor $\funF \times \funG:\catSET \rightarrow \catSET$, defined as $(\funF \times \funG) (X) = \funF(X) \times \funG(X)$ and $(\funF \times \funG) (f) = \funF(f) \times \funG(f)$, preserves weak pullbacks.
\item Generalize to arbitrary products, i.e show the following: If $I$ is a set and for every $i \in I$, $\funF_i : \catSET \rightarrow \catSET$ is a functor preserving weak pullbacks, then also the product functor $\prod_{i \in I} \funF_i:\catSET \rightarrow \catSET$ preserves pullbacks.

Use this to show that the exponential functor $(-)^A: \catSET
\rightarrow \catSET$, given by $X \mapsto X^A = \prod_{a \in A} X$ and
$f \mapsto f^A = \prod_{a \in A} f$ preserves weak pullbacks.
\item Show that if $\funF$ preserves weak pullbacks and there exist natural transformations $\eta: \funF \rightarrow \funG$ and $\nu: \funG \rightarrow \funF$, then also $\funG$ preserves weak pullbacks.

\item Show that if both $\funF$ and $\funG$ preserve weak
  pullbacks, then also $\funF + \funG:\catSET \rightarrow
  \catSET$, defined as $X \mapsto \funF(X) + \funG(X)$ and $f
  \mapsto \funF(f) + \funG(f)$, preserves weak pullbacks. (Hint: Show
  first that for every morphism $f: X \rightarrow A + B$, one has a
  decomposition $X \cong X_A + X_B$ and $f_A: X_A \rightarrow A$,
  $f_B: X_B \rightarrow B$ such that $f \cong (f_A \circ i_A) + (f_B
  \circ i_B)$.)
\end{enumerate}
} 
\Exercise{ex-modal-logics-1}{ Consider the modal similarity type
  $\tau =(O,\rho)$, with $O := \{ \langle a \rangle, \langle b \rangle
  \}$ and $\rho( \langle a \rangle) = \rho( \langle b \rangle) = 1$,
  over the propositional letters $\{p,q \}$. Let furthermore $[a],
  [b]$ denote the nablas of $\langle a \rangle$ and $\langle b
  \rangle$.

Show that the following formula is a tautology, i.e. it holds in every possible $\tau$-model:
\begin{equation*}
( \langle a \rangle p \vee  \langle a \rangle q \vee [b] (\neg p \vee q)) \rightarrow ( \langle a \rangle (p \vee q) \vee \neg [b] p \vee [b] q)
\end{equation*}

A \emph{frame morphism} between frames $(X, (R_{\langle a \rangle},
R_{\langle b \rangle}))$ and $(Y, (S_{\langle a \rangle}, S_{\langle b
  \rangle}) )$ is given for this modal similarity type by a map $f: X
\rightarrow Y$ which satisfies the following properties:
\begin{itemize}
\item If $\langle x,x_1 \rangle \in R_{\langle a \rangle}$, then
  $\langle f(x),f(x_1) \rangle \in S_{\langle a \rangle}$. Moreover,
  if $\langle f(x),y_1 \rangle \in S_{\langle a \rangle}$, then there
  exists $x_1 \in X$ with $\langle x,x_1 \rangle \in R_{\langle a
    \rangle}$ and $y_1 = f(x_1)$.

\item If $\langle x,x_1 \rangle \in R_{\langle b \rangle}$, then
  $\langle f(x),f(x_1) \rangle \in S_{\langle b \rangle}$. Moreover,
  if $\langle f(x),y_1 \rangle \in S_{\langle b \rangle}$, then there
  exists $x_1 \in X$ with $\langle x,x_1 \rangle \in R_{\langle b
    \rangle}$ and $y_1 = f(x_1)$.
\end{itemize}

Give a coalgebraic definition of frame morphisms for this modal
similarity type, i.e. find a functor $\funF: \catSET\to \catSET$ such that frame morphisms correspond to $\funF$-coalgebra morphisms.
}
\Exercise{ex-pdl-fragment}{
Consider the fragment of PDL defined mutually recursive by:
\begin{description}
\item[Formulas] $\varphi ::= p\mid  \varphi_1 \wedge \varphi_2\mid
  \neg \varphi_1 \mid \langle \pi \rangle \varphi$ (where $p \in \Phi$ for a set of basic propositions $\Phi$, and $\pi$ is a program).

\item[Programs] $\pi::= t\mid  \pi_1 ; \pi_2\mid  \varphi ?$ (where $t \in \mathsf{Bas}$ for a set of basic programs $\mathsf{Bas}$ and $\varphi$ is a formula).
\end{description}

Suppose you are given the set of basic programs $\mathsf{Bas} := \{\mathsf{init}, \mathsf{run}, \mathsf{print}\}$ and basic propositions $\Phi:= \{is \text{\_} init, did \text{\_} print\}$.

We define a model $\mathfrak{M}$ for this language as follows:
\begin{itemize}
\item The basic set of $\mathfrak{M}$ is $X := \{-1,0,1\}$.

\item The modal formulas for basic programs are interpreted by the
  relations 
  \begin{align*}
    R_{\mathsf{init}} & := \{\langle -1,0 \rangle, \langle 0,0 \rangle,
    \langle 1,1 \rangle \},\\
   R_{\mathsf{run}}  & := \{\langle -1,-1
    \rangle, \langle 0,0 \rangle, \langle 1,1 \rangle \},\\
    R_{\mathsf{print}} & := \{\langle -1,-1 \rangle, \langle 0,1
    \rangle, \langle 1,1 \rangle \}.
  \end{align*}

\item The modal formulas for composite programs are defined by $R_{\pi_1 ; \pi_2} := R_{\pi_1} \circ R_{\pi_2}$ and $R_{\varphi?} := \{\langle x,x \rangle |\: \mathfrak{M},x \models \varphi \}$, as usual.

\item The valuation function is given by $V(is \text{\_} init) := \{0,1\}$ and $V(did \text{\_} print) := \{1\}$.
\end{itemize}

Show the following:
\begin{enumerate}
\item $\mathfrak{M}, -1 \nvDash \langle \mathsf{run}; \mathsf{print} \rangle did \text{\_} print$,
\item $\mathfrak{M}, x \models \langle \mathsf{init}; \mathsf{run}; \mathsf{print} \rangle did \text{\_} print$ (for all $x \in X$),
\item $\mathfrak{M}, x \nvDash \langle (\neg is \text{\_} init) ?; \mathsf{print} \rangle did \text{\_} print$ (for all $x \in X$).
\end{enumerate}

Informally speaking, the model above allows one to determine whether a program composed of initialization ($\mathsf{init}$), doing some kind of work ($\mathsf{run}$), and printing ($\mathsf{print}$) is initialized or has printed something.

Suppose we want to modify the logic by counting how often we have
printed, i.e. we extend the set of basic propositional letters by $\{
did \textbf{\_} print_n |\: n \in \mathbb{N} \}$. Give an appropriate model for the new logic.
}


\newpage
\addcontentsline{toc}{section}{References}

\newpage
\addcontentsline{toc}{section}{Index}
\begin{theindex}

  \item $F({\cal G})$, 5
  \item $\Category{Meas}$, 7
  \item $\Category{Prob}$, 7
  \item $\Category{P}$, 5
  \item $\Category{Rel}$, 10
  \item $\Category{Top}$, 7
  \item $\Gilt_{\fmM}$, 89
  \item $\catK^{op}$, 10
  \item $\catSET$, 4
  \item $\ensuremath  {\mathbf  {AbGroup}}$, 121
  \item $\fmM, w \models \phi$, 89
  \item $\sigma $-algebra
    \subitem product, 17
    \subitem weak-*, 8
  \item $\theTheory{\gamma}{c}$, 112
  \item $w\models_{\gamma}\phi$, 111
  \item ${\cal L}(\Phi)$, 86
  \item ${\cal L}(\lift)$, 111
  \item ${\cal L}(\tau,   \Phi)$, 87

  \indexspace

  \item adjoint
    \subitem left, 53
    \subitem right, 53
  \item adjunction, 53
    \subitem counit, 58
    \subitem unit, 58
  \item algebra
    \subitem Eilenberg-Moore, 60
    \subitem free, 61
    \subitem morphism, 60
    \subitem structure morphism, 60
  \item Angel, 88
  \item atomic harmony, 98
  \item automaton, 9
    \subitem with output, 30

  \indexspace

  \item bisimilar
    \subitem states, 98
  \item bisimulation, 71, 73, 74, 98
  \item bisimulation   equivalence, 78

  \indexspace

  \item category
    \subitem comma, 34
    \subitem composition, 3
    \subitem coproduct, 19
      \subsubitem injection, 19
    \subitem discrete, 5
    \subitem dual, 10
    \subitem epimorphism, 12
      \subsubitem split, 118
    \subitem free, 5
    \subitem monomorphism, 11
      \subsubitem split, 118
    \subitem morphism, 3
    \subitem natural transformation, 34
      \subsubitem component, 34
    \subitem object, 3
    \subitem product, 16
      \subsubitem projection, 16
    \subitem pullback, 22
    \subitem pushout, 26
    \subitem slice, 10
    \subitem small, 5
    \subitem sum, 19
    \subitem weak pullback, 22
  \item choice
    \subitem angelic, 88
    \subitem demonic, 88
  \item coalgebra, 67
    \subitem bisimilar, 70
    \subitem carrier, 67
    \subitem dynamics, 67
    \subitem mediating, 70
    \subitem morphism, 69
  \item cocone, 42
    \subitem colimit, 42
  \item colimit, 42
  \item complete
    \subitem strongly $\mathbb  {S}$, 104
    \subitem weakly $\mathbb  {S}$, 104
  \item cone, 40
    \subitem limit, 41
  \item congruence, 83
  \item consistent
    \subitem $\Lambda $, 102
    \subitem maximal $\Lambda $, 104
  \item coproduct, 80
  \item CTL*, 115
  \item currying, 54

  \indexspace

  \item Demon, 88
  \item diagram
    \subitem chasing, 15
    \subitem commutative, 9
  \item Dirac measure, 48, 49

  \indexspace

  \item Eilenberg-Moore algebra, 60
    \subitem morphism, 60
  \item epic, 12
    \subitem jointly, 120
  \item equalizer, 120
  \item equivalence
    \subitem modal, 98
  \item exception, 43

  \indexspace

  \item formula
    \subitem globally true, 91
    \subitem refutable, 91
    \subitem satisfiable, 91
  \item frame
    \subitem $\tau $, 92
    \subitem accessibility relation, 89
    \subitem Kripke, 89
    \subitem morphism, 96
    \subitem neighborhood, 91
    \subitem set of worlds, 89
  \item frames
    \subitem class of, 103
  \item functor, 28
    \subitem constant, 28
    \subitem contravariant, 29
    \subitem covariant, 29
    \subitem endofunctor, 28
    \subitem forgetful, 29
    \subitem identity, 28
    \subitem power set, 28

  \indexspace

  \item Galois connection, 55
  \item game
    \subitem determined, 95
    \subitem logic, 88
  \item Giry monad, 50
  \item Godement product, 37
  \item graph
    \subitem free category, 5
    \subitem path, 5
    \subitem undirected, 117

  \indexspace

  \item invariant subset, 84

  \indexspace

  \item Kleisli
    \subitem category, 44
    \subitem tripel, 43
  \item Kripke model, 89

  \indexspace

  \item left adjoint, 53
  \item limit, 41
  \item Lindenbaum Lemma, 106
  \item logic
    \subitem closed
      \subsubitem modus ponens, 101
      \subsubitem uniform substitution, 101
    \subitem coalgebraic
      \subsubitem behavioral equivalence, 112
      \subsubitem bridge operators, 117
      \subsubitem logical equivalence, 112
      \subsubitem predicate lifting, 108
    \subitem modal, 101
      \subsubitem normal, 102

  \indexspace

  \item map
    \subitem continuous, 7
    \subitem kernel, 12
    \subitem measurable, 7
  \item modal language
    \subitem basic, 86
    \subitem extended, 87
    \subitem game logic, 88
    \subitem nabla, 87
    \subitem PDL, 88
  \item modal logic
    \subitem normal, 102
  \item model
    \subitem $\tau $, 92
    \subitem canonical, 106
    \subitem image finite, 99
    \subitem Kripke, 89
    \subitem morphism, 97
    \subitem neighborhood, 92
  \item monad, 44
    \subitem \texttt{Haskell}, 51
    \subitem Manes, 61
    \subitem multiplication, 44
    \subitem unit, 44
  \item monic, 11
    \subitem jointly, 120
  \item monoid, 13
  \item morphism
    \subitem automaton, 9
    \subitem bounded, 6, 70
    \subitem codomain, 4
    \subitem domain, 4
    \subitem epi/mono factorization, 13
    \subitem epimorphism, 12
    \subitem frame, 96
    \subitem isomorphism, 14
    \subitem model, 97
    \subitem monomorphism, 11

  \indexspace

  \item natural transformation, 34
    \subitem component, 34
    \subitem Godement product, 37
    \subitem horizontal composition, 37
    \subitem vertical composition, 37
  \item neighborhood
    \subitem frame, 91
    \subitem model, 92
    \subitem morphism, 100
  \item neighborhood morphism, 100

  \indexspace

  \item PDL, 88
  \item positive convexity, 63
    \subitem affine map, 64
    \subitem morphism, 64
  \item predicate lifting, 108
  \item probability
    \subitem discrete, 31
  \item probability functor
    \subitem continuous, 48
    \subitem continuous space, 33
    \subitem discrete, 32, 47
  \item propositional letters, 86
  \item pullback, 22
    \subitem preserves weak pullbacks, 76
    \subitem weak, 22

  \indexspace

  \item right adjoint, 53

  \indexspace

  \item semi lattice, 61
  \item space
    \subitem measurable, 7
    \subitem probability, 7
    \subitem topological, 7
  \item state
    \subitem behavioral equivalence, 112
    \subitem logical equivalence, 112
    \subitem theory, 112
  \item substitution, 101

  \indexspace

  \item terminal object, 120
  \item Theorem
    \subitem Hennessy-Milner, 99
    \subitem Manes, 45
  \item theory, 112
  \item transition system, 6
    \subitem behavioral equivalence, 113
    \subitem bisimilar, 114
    \subitem bisimulation, 71
    \subitem labeled, 31
    \subitem logical equivalence, 113

  \indexspace

  \item upper closed, 33

  \indexspace

  \item world
    \subitem behavioral equivalence, 112
    \subitem logical equivalence, 112
    \subitem theory, 112

  \indexspace

  \item Yoneda isomorphism, 38

\end{theindex}

\end{document}